\documentclass[11pt,oneside,reqno]{amsart}
\usepackage[T1]{fontenc}
\usepackage[latin9]{inputenc}
\usepackage[a4paper]{geometry}
\usepackage{color}
\usepackage{float}
\usepackage{amsbsy}
\usepackage{amstext}
\usepackage{amsthm}
\usepackage{amssymb}
\usepackage{graphicx}
\usepackage{setspace}
\usepackage[unicode=true,pdfusetitle,
 bookmarks=true,bookmarksnumbered=false,bookmarksopen=false,
 breaklinks=true,pdfborder={0 0 0},pdfborderstyle={},backref=page,colorlinks=true]
 {hyperref}

\makeatletter
\usepackage[natbibapa]{apacite}
\theoremstyle{plain}
\newtheorem{assumption}{\protect\assumptionname}
\theoremstyle{remark}
\newtheorem{rem}{\protect\remarkname}
\theoremstyle{definition}
\newtheorem{defn}{\protect\definitionname}
\theoremstyle{plain}
\newtheorem{thm}{\protect\theoremname}
\theoremstyle{plain}
\newtheorem{lem}{\protect\lemmaname}
\theoremstyle{plain}
\newtheorem{cor}{\protect\corollaryname}

\usepackage[dvipsnames]{xcolor}
\definecolor{darkblue}{rgb}{0.0, 0.0, 0.55}
\definecolor{teal}{rgb}{0.0, 0.5, 0.5}
\hypersetup{linkcolor=teal, urlcolor=blue, citecolor=darkblue}

\renewcommand*{\backrefalt}[4]{%
     \ifcase #1 %
     \or
         [Cited on page #2]
     \else
         [Cited on pages #2]
     \fi}

\usepackage{threeparttable}

\usepackage{array}
\newcolumntype{H}{>{\iffalse}c<{\fi}@{}}

\makeatother

\providecommand{\assumptionname}{Assumption}
\providecommand{\corollaryname}{Corollary}
\providecommand{\definitionname}{Definition}
\providecommand{\lemmaname}{Lemma}
\providecommand{\remarkname}{Remark}
\providecommand{\theoremname}{Theorem}

\begin{document}
\title{Nonparametric Regression under Cluster Sampling}
\author{Yuya Shimizu}
\address{Department of Economics, University of Wisconsin, Madison. 1180 Observatory Drive, Madison, WI 53706-1393, USA.}
\email{\href{mailto:}{yuya.shimizu@wisc.edu}}
\thanks{\textit{This version: \today}}
\thanks{I am grateful to Bruce Hansen for his invaluable advice and encouragement. I thank Naoki Aizawa, Harold Chiang, Junho Choi, Jack Collison, Kenta Fukuda, Woosik Gong, Michael Jansson, J.C. Lazzaro, Taisuke Otsu, Jack Porter, Xiaoxia Shi, Gonzalo Vazquez-Bare, Kohei Yata, and seminar participants at Wisconsin, Kobe, Keio, EEA-ESEM, CESG, and SEA for helpful comments and suggestions. This paper won the Kanematsu Prize 2023 from the Research Institute for Economics and Business Administration at Kobe University, Japan. I thank the co-editor Xiaohong Chen, the associate editor, and an anonymous referee for comments that helped improve this paper. I thank the anonymous reviewers of the Kanematsu Prize for their valuable comments. This work is supported by the summer research fellowship from the University of Wisconsin-Madison. }
\begin{abstract}
This paper develops a general asymptotic theory for nonparametric kernel regression in the presence of cluster dependence. We examine nonparametric density estimation, Nadaraya-Watson kernel regression, and local linear estimation. Our theory accommodates growing and heterogeneous cluster sizes. We derive asymptotic conditional bias and variance, establish uniform consistency, and prove asymptotic normality. Our findings reveal that under heterogeneous cluster sizes, the asymptotic variance includes a new term reflecting within-cluster dependence, which is overlooked when cluster sizes are presumed to be bounded. We propose valid approaches for bandwidth selection and inference, introduce estimators of the asymptotic variance, and demonstrate their consistency. In simulations, we verify the effectiveness of the cluster-robust bandwidth selection and show that the derived cluster-robust confidence interval improves the coverage ratio. We illustrate the application of these methods using a policy-targeting dataset in development economics.
\end{abstract}

\maketitle

\section{\textbf{Introduction}}

Nonparametric regression is widely used in economics for its flexibility. Typically, data are assumed to be independently and identically distributed; however, in reality, observations may exhibit dependence within a group structure called a cluster. Examples of clusters are classrooms, schools, families, hospitals, firms, industries, villages, regions, and so on. The cluster sampling framework assumes independence between observations from different clusters but allows dependence within each cluster. 

The previous literature on nonparametric regression under cluster sampling assumes a bounded and homogeneous number of observations per cluster. This assumption may not hold in real data due to heterogeneous cluster sizes. To fill this gap, this paper studies nonparametric kernel regressions that accommodate heterogeneous cluster sizes, including those that grow to infinity asymptotically. Our approach is general, allowing for both bounded and growing clusters simultaneously, and includes cluster-level regressors.

We develop a comprehensive asymptotic theory for nonparametric density estimation, Nadaraya-Watson kernel regression, and local linear estimation. Our results on asymptotic conditional bias and variance, uniform consistency, and asymptotic normality enable us to propose valid methods for bandwidth selection and inference. 

For clusters of growing sizes, the asymptotic variance contains a novel term for within-cluster dependence, which does not appear under the assumption of bounded cluster sizes. This term becomes significant due to the potential for a cluster to contain a growing number of observations within a local neighborhood, making cluster dependence non-negligible asymptotically. We propose consistent estimators of the asymptotic variance that account for cluster dependence and validate its importance through simulation. Our cluster-robust confidence interval achieves improved coverage ratios, while conventional confidence intervals could suffer from under-coverage in our simulated datasets.

Nonparametric regression, while significant on its own, also serves as an intermediate tool for other estimators, such as regression discontinuity design, nonparametric auction estimation, and semiparametric models under cluster sampling. Our results could extend to these areas as well.

\subsection*{Related literature}

There is a substantial body of literature on cluster sampling in econometrics. C. \citet{hansen2007asymptotic} provides an asymptotic theory for parametric regression with homogeneous cluster sizes. \citet{djogbenou2019asymptotic} and B. \citet{hansen2019asymptotic} extend this theory to heterogeneous cluster sizes. \citet{bugni2022inference} considers heterogeneous and random cluster sizes for cluster-level randomized experiments. For further literature on parametric models under cluster sampling, the reader can refer to \citet{cameron2015practitioner} and \citet{mackinnon2022cluster}.

Conversely, the theory on nonparametric regression under cluster dependence, even with homogeneous cluster sizes, is limited. \citet{lin2000nonparametric} and \citet{wang2003marginal} examine local polynomial and local linear regressions, assuming fixed and homogeneous cluster sizes and focusing primarily on asymptotic efficiency. \citet{bhattacharya2005asymptotic} offers an asymptotic theory for local constant estimators under multi-stage samples, analogous to cluster sampling. When the number of first-stage strata is set to one, his setup becomes a standard cluster sampling with fixed and homogeneous cluster sizes. He puts a similar structure on error terms as this paper, but the fixed cluster sizes render the term reflecting within-cluster dependence asymptotically negligible. For the regression discontinuity literature, \citet{bartalotti2017regression} has derived asymptotic theories for local polynomial regression under bounded and homogeneous cluster sizes.

\citet{menzel2024transfer} proposes a method for estimating nonparametric regressions in the presence of cluster dependence, aiming to extrapolate treatment effects across clusters. He considers \textit{independent }but not identical observations between clusters, with a fixed number of clusters exhibiting\textit{ uniformly growing} size. Our approach differs by incorporating general dependence within a cluster and allowing for both bounded and growing cluster sizes simultaneously, leading to distinct asymptotic results and theories.

To the best of our knowledge, there is no literature on nonparametric models with \textit{growing and heterogeneous size} clusters except for \citet{hu2024some}, which became available online after the working paper version of our paper was posted. While both our paper and theirs accommodate flexible cluster sizes, their work concentrates on series regression and does not include any theory on inference, model selection, and kernel regression. Their Assumption 2(i) excludes cluster-level covariates. Our paper adopts the same cluster size framework as \citet{djogbenou2019asymptotic} and \citet{hansen2019asymptotic}. The presence of clusters with growing sizes complicates the proofs for asymptotic theories, as cluster dependence becomes non-negligible. Consequently, this paper introduces new technical results for nonparametric regressions under cluster sampling, notably developing Bernstein's inequality for cluster sampling to demonstrate uniform consistency. These novel contributions are believed to offer valuable theoretical tools for future research.

This research also sheds new light on the literature regarding nonparametric regressions with dependence. Following the foundational work on i.i.d. datasets (e.g., \citealp{stone1982optimal}, \citealp{fan1992design}, \citealp{ruppert1994multivariate}), the results have been extended to time series (\citealp{robinson1983nonparametric}, \citealp{hansen2008uniform}, \citealp{kristensen2009uniform}, \citealp{vogt2012nonparametric}, \citealp{vogt2020multiscale}) and spatial datasets (\citealp{robinson2011asymptotic}, \citealp{lee2016series}), as well as to the cluster dependence framework discussed above. 

\medskip{}

The remainder of this paper is organized as follows: Section $\text{\ref{sec:Cluster-sampling}}$ introduces the cluster sampling framework under consideration. Sections $\text{\ref{sec:Nonparametric-density-estimation}}$-$\text{\ref{sec:Local-linear-estimator}}$ discuss asymptotic theories for nonparametric density estimators, Nadaraya-Watson estimators, and local linear estimators, respectively. Section $\text{\ref{sec:Uniform-convergence}}$ demonstrates uniform convergence of these estimators. Section $\text{\ref{sec:Bandwidth-selection}}$ provides guidelines for selecting bandwidth in nonparametric regressions. Section $\text{\ref{sec:Cluster-robust-variance-estimati}}$ addresses cluster-robust inference. Section $\text{\ref{sec:Monte-Carlo-simulation}}$ presents Monte Carlo simulations for bandwidth selections and inference. Section $\text{\ref{sec:Empirical}}$ illustrates our methods with an application in development economics using a dataset by \citet{alatas2012targeting}. The paper concludes with Section $\text{\ref{sec:Conclusion}}$. All proofs, technical lemmas, technical discussions, and additional simulation results are included in the Appendix. 

\section{\textbf{Cluster sampling\label{sec:Cluster-sampling}}}

The researcher observes $\left(Y_{i},X_{i}\right)\in\mathbb{R}\times\mathbb{R}^{d}$ for $i=1,\ldots,n$, with cluster sizes given by $n_{g}\in\{1,2,\cdots\}$ for $g=1,\ldots,G$. Here, $Y_{i}$ represents a dependent variable, and regressors $X_{i}$ are continuous random variables with the Lebesgue density $f(x)$. Assume that each observation can be grouped into one cluster.\footnote{Formally, we assume that for any $i$, we know a function $g(i)\in\{1,\cdots,G\}$.} Thus, the total number of observations is $n=\sum_{g=1}^{G}n_{g}$. To explicitly represent the cluster structure, we also use the notation $\left(Y_{gj},X_{gj}\right)$ for $g=1,\ldots,G$ and $j=1,\ldots,n_{g}$. We treat cluster size $n_{g}$ as nonrandom and possibly heterogeneous across clusters. We assume that observations belonging to different clusters are mutually independent but permit general dependence within the same cluster. We decompose $X_{gj}$ into $X_{gj}=\left(X_{gj}^{(\mathrm{ind})\top},X_{g}^{(\mathrm{cls})\top}\right)^{\top}\in\mathbb{R}^{d}$ where $X_{gj}^{(\mathrm{ind})}\in\mathbb{R}^{d_{\mathrm{ind}}}$ represents individual-level regressors and $X_{g}^{(\mathrm{cls})}\in\mathbb{R}^{d_{\mathrm{cls}}}$ represents cluster-level regressors. We assume that the regressors contain at least one individual-level regressors, $d_{\mathrm{ind}}\geq1$. By construction, $d=d_{\mathrm{ind}}+d_{\mathrm{cls}}$ holds.

We denote $\mathbf{X}_{g}=\left(X_{g1},\dots,X_{gn_{g}}\right)$ and aim to estimate the nonparametric regression model:
\begin{align}
Y_{gj} & =m\left(X_{gj}\right)+e_{gj},\label{eq:estimand}\\
\mathbb{E}\left[e_{gj}\mid\mathbf{X}_{g}\right] & =\mathbb{E}\left[e_{gj}\mid X_{gj}\right]=0.\label{eq:cond_error_model}
\end{align}
We also assume 
\begin{align}
\mathbb{E}\left[e_{gj}^{2}\mid\mathbf{X}_{g}\right] & =\mathbb{E}\left[e_{gj}^{2}\mid X_{gj}\right]=\sigma^{2}\left(X_{gj}\right),\label{eq:cond_var_model}\\
\mathbb{E}\left[e_{gj}e_{g\ell}\mid\mathbf{X}_{g}\right] & =\mathbb{E}\left[e_{gj}e_{g\ell}\mid X_{gj}^{(\mathrm{ind})},X_{g\ell}^{(\mathrm{ind})};X_{g}^{(\mathrm{cls})}\right]\nonumber \\
 & =\sigma\left(X_{gj}^{(\mathrm{ind})},X_{g\ell}^{(\mathrm{ind})};X_{g}^{(\mathrm{cls})}\right)\text{ for }j\neq\ell.\label{eq:cond_cov_model}
\end{align}

The model specified through $\text{\eqref{eq:estimand}}$-$\text{\eqref{eq:cond_cov_model}}$ exhibits greater flexibility than initially apparent. The constraint imposed by $\text{\eqref{eq:cond_var_model}}$ is that the conditional variance of the error term for an individual is dependent only on the individual's own regressors, both at the individual and cluster levels. Additionally, $\text{\eqref{eq:cond_cov_model}}$ states that the conditional covariance of the error terms between any two individuals within the same cluster is a function only of their individual-level regressors and shared cluster-level regressors. This framework accommodates the inclusion of cluster random effects in $e_{gj}$ and allows for the dependence of regressors within clusters. In economic applications, cluster dependencies often arise from strategic interactions within the cluster or from cluster-level unobserved shocks, including measurement errors.
\begin{assumption}
\label{assu:dgp} We assume the following data-generating process:
\begin{enumerate}
\item The pairs $\left(Y_{gj},X_{gj}\right)$ and $\left(Y_{g^{\prime}\ell},X_{g^{\prime}\ell}\right)$ are mutually independent for any $g\neq g^{\prime}$, $j=1,\cdots,n_{g}$, and $\ell=1,\cdots,n_{g^{\prime}}$.
\item The data is generated according to the model described through $\text{\eqref{eq:estimand}}$-$\text{\eqref{eq:cond_cov_model}}$. 
\item The variables $X_{gj}$ are identically distributed across all $g$ and $j$, possessing a common marginal density $f(x)$. For any $\underline{n}_{g}\in\{2,3,4\}$, and for any cluster \textup{$g$ with }$n_{g}\geq\underline{n}_{g}$, the random vector $\left(X_{gj_{1}}^{(\mathrm{ind})},\cdots,X_{gj_{\underline{n}_{g}}}^{(\mathrm{ind})};X_{g}^{(\mathrm{cls})}\right)$ is identically distributed across all $g$ and $j_{1},\dots,j_{\underline{n}_{g}}$, with a common joint density represented by:
\[
f_{\underline{n}_{g}}\left(x_{1}^{(\mathrm{ind})},\cdots,x_{\underline{n}_{g}}^{(\mathrm{ind})};x^{(\mathrm{cls})}\right).
\]
\end{enumerate}
\end{assumption}
\begin{rem}
\label{rem:joint_density}The conditions in Assumption $\text{\ref{assu:dgp}}$ (iii) for $f(x)$ and $f_{2}\left(x_{1}^{(\mathrm{ind})},x_{2}^{(\mathrm{ind})};x^{(\mathrm{cls})}\right)$ are sufficient for their consistent estimation. On the other hand, since we are not interested in estimating $f_{3}\left(x_{1}^{(\mathrm{ind})},x_{2}^{(\mathrm{ind})},x_{3}^{(\mathrm{ind})};x^{(\mathrm{cls})}\right)$ and $f_{4}\left(x_{1}^{(\mathrm{ind})},x_{2}^{(\mathrm{ind})},x_{3}^{(\mathrm{ind})},x_{4}^{(\mathrm{ind})};x^{(\mathrm{cls})}\right)$, the associated conditions in Assumption $\text{\ref{assu:dgp}}$ (iii) could be weakened. For a detailed discussion, refer to Appendix $\text{\ref{sec:Technical-discussion}}$.
\end{rem}
\begin{rem}
\label{rem:mixture}In nonparametric regressions, unobserved cluster heterogeneity is equivalent to a mixture structure. Consider a scenario where the true data-generating process is defined as follows: 
\begin{align*}
Y_{gj} & =m\left(X_{gj},U_{g}\right)+e_{gj},\\
\mathbb{E}\left[e_{gj}\mid\mathbf{X}_{g},U_{g}\right] & =0,
\end{align*}
where $U_{g}$ is an unobserved cluster-level variable. The critical condition here is that $U_{g}$ and $e_{gj}$ are separable, and $U_{g}$ is exogenous. Under these conditions, the estimand, derived through the law of iterated expectations, is expressed as:
\begin{align*}
m\left(X_{gj}\right) & =\mathbb{E}\left[Y_{gj}\mid\mathbf{X}_{g}\right]=\mathbb{E}\left[\mathbb{E}\left[Y_{gj}\mid\mathbf{X}_{g},U_{g}\right]\mid\mathbf{X}_{g}\right]\\
 & =\mathbb{E}\left[m\left(X_{gj},U_{g}\right)\mid\mathbf{X}_{g}\right]=\int m\left(X_{gj},U_{g}\right)f_{U_{g}\mid\mathbf{X}_{g}}(U_{g}\mid\mathbf{X}_{g})\mathrm{d}U_{g}.
\end{align*}
This formulation implies that $m\left(X_{gj}\right)$ is essentially a mixture of $m\left(X_{gj},U_{g}\right)$, integrated over the unknown conditional density $f_{U_{g}\mid\mathbf{X}_{g}}(U_{g}\mid\mathbf{X}_{g})$. Additionally, the condition $\mathbb{E}\left[e_{gj}\mid\mathbf{X}_{g},U_{g}\right]=0$ ensures $\mathbb{E}\left[e_{gj}\mid\mathbf{X}_{g}\right]=0$, allowing us to treat $m\left(X_{gj}\right)$ as homogeneous across clusters without loss of generality. \\
Similarly, consider a scenario where the true density of $X_{gj}$ exhibits cluster heterogeneity, represented by the marginal density $f_{X,V_{g}}(X_{gj},V_{g})$, with $V_{g}$ being an unobserved cluster-level variable. In this context, our estimand becomes a mixture of $f_{X,V_{g}}(X_{gj},V_{g})$, which can be formally expressed as:
\[
f(X_{gj})=\int f_{X,V_{g}}\left(X_{gj},V_{g}\right)f(V_{g})\mathrm{d}V_{g}.
\]
This integral representation implies that the regressors possess identical marginal distributions across clusters. Analogously to the treatment of marginal densities, cluster heterogeneities within joint densities can be conceptualized as mixture structures.
\end{rem}
\begin{rem}
Although the majority of research on cluster sampling treats cluster sizes as deterministic, as does this paper, \citet{bugni2022inference} treat cluster sizes as a random variable in a cluster-level randomized experiment setup. Their investigation primarily focuses on estimating treatment effects across clusters of varying sizes and developing inference methods that account for the randomness of cluster sizes.\footnote{An important limitation of assuming deterministic cluster sizes is that it is difficult to incorporate cluster size effects into the nonparametric regression function. One practical approach is to approximate these effects by deterministically binning cluster sizes (e.g., categorizing clusters into ``large'' size cluster group with $n_{g}\geq20$, ``intermediate'' size cluster group with $10\leq n_{g}<20$, and ``small'' size cluster group with $n_{g}<10$). We can estimate the nonparametric regression separately for each group if each group contains a sufficient number of clusters.} This methodological divergence stems from differing concepts of the data-generating process. \citet{bugni2022inference} address scenarios where researchers sample \textit{clusters} in an experiment, viewing cluster sizes as one of the attributes. \citet{abadie2023should} propose an alternate sampling framework wherein clusters are sampled from a larger population of cluster, followed by the sampling of individuals from these selected clusters' subpopulations.\footnote{\citet{bugni2022inference} and \citet{abadie2023should} adopt a design-based approach with a finite population, whereas our paper takes a model-based approach with a hyperpopulation. These two approaches consider different sources of randomness. Please refer to the next footnote for further details.} Deterministic cluster sizes, as assumed in our paper, are justified in two ways. First, if the researcher specifies target cluster sizes ex-ante and conducts a survey to draw $\left(Y_{gj},X_{gj}\right)$ from a hyperpopulation, then cluster sizes can be considered nonrandom. Some survey samplings use this type of procedure (e.g., sampling exactly 20 households from each state). See \citeauthor{cochran1977sampling} (\citeyear{cochran1977sampling}, Chapter 9) for details. Second, even if the first justification does not apply, we can treat cluster sizes as nonrandom by conditioning each theoretical result on the realized cluster sizes $\{n_{g}\}_{g=1}^{G}$. In this way, the researcher can use our model-based approach when cluster sizes are random. This perspective is often used in a model-based approach, where the focus is on the data generated given the observed cluster sizes.
\end{rem}
\begin{rem}
This research assumes that the researcher knows the appropriate level of clustering. However, in practice, one has to choose the level of cluster. For example, in terms of the regional cluster level, one may need to choose among zip code, city, county, or state levels. One approach is to assume the structure of the error term, such as cluster random effects, at the most plausible level. See \citeauthor{mackinnon2022cluster} (\citeyear{mackinnon2022cluster}, Section 3.3) for further discussion.\footnote{Our paper adopts a model-based approach, where the source of randomness is the data-generating process. Hence, the arguments on cluster level by \citet{abadie2023should} are not directly applicable to this paper because they consider a design-based approach, where the sources of randomness are treatment assignment uncertainty and sampling uncertainty. Specifically, the design-based approach usually treats the population error term as fixed. The model-based approach is also useful in (quasi-)experiment setting. For example, \citet{eckles2020noise} provide justifications of identification in regression discontinuity designs using a model-based approach. They assume that the running variable is a measure of some latent variable with exogenous measurement error.} 
\end{rem}

\section{\textbf{Nonparametric density estimation\label{sec:Nonparametric-density-estimation}}}

In this section, we show the consistency of nonparametric density estimators. In this paper, we will use kernel functions satisfying the following definitions. 
\begin{defn}
\label{def:kernel} A \textit{univariate kernel function} $k:\mathbb{R}\rightarrow\mathbb{R}$ is defined to satisfy the following criteria:
\begin{enumerate}
\item $0\leq k(u)\leq\overline{k}<\infty$.
\item $k(u)=k(-u)$.
\item $\int_{-\infty}^{\infty}k(u)\mathrm{d}u=1$.
\item $\kappa_{2}\equiv\int_{-\infty}^{\infty}u^{2}k(u)\mathrm{d}u<\infty$ and $\int_{-\infty}^{\infty}u^{4}k(u)\mathrm{d}u<\infty$.
\end{enumerate}
\end{defn}
\begin{defn}
\label{def:kernel_d} A \textit{multivariate kernel function} $K:\mathbb{R}^{d}\rightarrow\mathbb{R}$ is constructed as the product of univariate kernel functions across dimensions, 
\[
K\left(X\right)=\prod_{q=1}^{d}k\left(X^{(q)}\right),
\]
where $k(\cdot)$ is a univariate kernel function and $X^{(q)}$ is the $q$-th component of $X$. The upper bound of the multivariate kernel is $K\left(X\right)\leq\overline{k}^{d}\equiv\overline{K}$.\footnote{Without loss of generality, we assume $\overline{k}\geq1$.}
\end{defn}
The kernel density estimator for $f(x)$ is: 
\begin{equation}
\ensuremath{\widehat{f}(x)=\frac{1}{nh^{d}}\sum_{g=1}^{G}\sum_{j=1}^{n_{g}}K\left(\frac{X_{gj}-x}{h}\right)},\label{eq:f_hat}
\end{equation}
where $h>0$ is a bandwidth.

\begin{rem}
The kernel density estimator given in $\text{\eqref{eq:f_hat}}$ can be rewritten as $\widehat{f}(x)=\frac{1}{nh^{d}}\sum_{i=1}^{n}K\left(\frac{X_{i}-x}{h}\right)$ as in the i.i.d. case. Thus, at least for the estimation, we can use a standard software package. This also applies to nonparametric regression.
\end{rem}
For the sake of simplicity, our discussion will focus on scenarios where a single bandwidth is used for all components of $X$. However, our theory can be generalized to accommodate multivariate bandwidths by substituting $h$ with a bandwidth matrix, as discussed by \citet{ruppert1994multivariate}.
\begin{assumption}
\label{assu:f}
\end{assumption}
\begin{enumerate}
\item $nh^{d}\rightarrow\infty$.
\item $h\rightarrow0$ and $\left(\max_{g\leq G}n_{g}\right)h^{d_{\mathrm{ind}}}=O(1)$.
\item There exists some neighborhood $\mathcal{N}$ of $x=\left(x^{(\mathrm{ind})\top},x^{\mathrm{(cls)}\top}\right)^{\top}$ such that $f(x)$ is twice continuously differentiable and $f_{2}\left(x^{\mathrm{(ind)}},x^{\mathrm{(ind)}};x^{\mathrm{(cls)}}\right)$ is continuously differentiable.
\end{enumerate}
\begin{rem}
Assumption $\text{\ref{assu:f}}$ (ii) notably extends the i.i.d. case to cluster-dependent settings, introducing a novel condition for bandwidth in the presence of cluster heterogeneity. This condition necessitates a more cautious selection of bandwidth under cluster sampling, balancing the need for $nh^{d}\rightarrow\infty$ against the constraint of $\left(\max_{g\leq G}n_{g}\right)h^{d_{\mathrm{ind}}}=O(1)$. The condition $\left(\max_{g\leq G}n_{g}\right)h^{d_{\mathrm{ind}}}=O(1)$ requires that the maximum cluster size is not growing faster than the shrinking speed of the $h$ neighborhood for the individual-level regressors.\\
Furthermore, Assumption $\text{\ref{assu:f}}$ (iii) underscores the importance of smoothness in both marginal and joint densities within clusters, emphasizing the need for careful examination of density shapes affecting within-cluster observation relationships.\\
To be precise, Assumption $\text{\ref{assu:f}}$ (iii) means that $f(\widetilde{x})$ is twice continuously differentiable at any $\widetilde{x}\in\mathcal{N}$ and $f_{2}\left(\widetilde{x}_{1}^{\mathrm{(ind)}},\widetilde{x}_{2}^{\mathrm{(ind)}};\widetilde{x}^{\mathrm{(cls)}}\right)$ is continuously differentiable at any $\left(\widetilde{x}_{1}^{(\mathrm{ind})\top},\widetilde{x}^{\mathrm{(cls)}\top}\right)^{\top}$, $\left(\widetilde{x}_{2}^{(\mathrm{ind})\top},\widetilde{x}^{\mathrm{(cls)}\top}\right)^{\top}\in\mathcal{N}$. Assumption $\text{\ref{assu:f}}$ (iii) limits our analysis to interior points. Although we focus on interior points $x$, the results could be extended to boundary points.
\end{rem}
\begin{rem}
$nh^{d}\rightarrow\infty$ and $\left(\max_{g\leq G}n_{g}\right)h^{d_{\mathrm{ind}}}=O(1)$ together imply that $\left(\max_{g\leq G}n_{g}\right)/n\rightarrow0$, which is a key assumption of \citet{hansen2019asymptotic} for parametric models under cluster sampling. Moreover, $\left(\max_{g\leq G}n_{g}\right)/n\rightarrow0$ implies $G\rightarrow\infty$. Thus, our theory requires $G\rightarrow\infty$ implicitly. If we only have the bounded size of clusters $\max_{g\leq G}n_{g}=O(1)$, then, $n$ has the same asymptotic order as $G$. 
\end{rem}
\begin{thm}
\label{thm:density_consistency}\textbf{(Pointwise consistency)} Suppose that Assumptions \ref{assu:dgp} and \ref{assu:f} hold. Then, $\widehat{f}(x)\overset{p}{\rightarrow}f(x)$.
\end{thm}
\begin{rem}
Beyond pointwise consistency, it is possible to derive expressions for the asymptotic conditional bias and variance, as well as establish the asymptotic normality of $\widehat{f}(x)$. These derivations, while omitted for brevity, follow directly from analogous proofs for the Nadaraya-Watson estimator discussed subsequently.
\end{rem}

\section{\textbf{Nadaraya-Watson estimator\label{sec:Nadaraya-Watson-estimator}}}

In this section, we derive an asymptotic theory for the Nadaraya-Watson estimator (a.k.a. local constant estimator) for estimating the conditional expectation $\mathbb{E}\left[Y_{gj}\mid X_{gj}=x\right]$. The estimator is:
\begin{equation}
\hat{m}_{\text{nw}}\left(x\right)=\frac{\sum_{g=1}^{G}\sum_{j=1}^{n_{g}}K\left(\frac{X_{gj}-x}{h}\right)Y_{gj}}{\sum_{g=1}^{G}\sum_{j=1}^{n_{g}}K\left(\frac{X_{gj}-x}{h}\right)}.\label{eq:nw}
\end{equation}

\begin{assumption}
\label{assu:nw} 
\end{assumption}
\begin{enumerate}
\item The density function is strictly positive at $x$, $f(x)>0$.
\item There exists some neighborhood $\mathcal{N}$ of $x=\left(x^{(\mathrm{ind})\top},x^{\mathrm{(cls)}\top}\right)^{\top}$ such that $m(x)$ and $f(x)$ are twice continuously differentiable, $f_{2}\left(x^{\mathrm{(ind)}},x^{\mathrm{(ind)}};x^{\mathrm{(cls)}}\right)$ is continuously differentiable, and $f_{3}\left(x^{\mathrm{(ind)}},x^{\mathrm{(ind)}},x^{\mathrm{(ind)}};x^{\mathrm{(cls)}}\right)$, $f_{4}\left(x^{\mathrm{(ind)}},x^{\mathrm{(ind)}},x^{\mathrm{(ind)}},x^{\mathrm{(ind)}};x^{\mathrm{(cls)}}\right)$, $\sigma^{2}(x)$, and $\sigma\left(x^{\mathrm{(ind)}},x^{\mathrm{(ind)}};x^{\mathrm{(cls)}}\right)$ are continuous.
\end{enumerate}
\begin{rem}
Assumption $\text{\ref{assu:nw}}$ (i) is standard for the Nadaraya-Watson estimator. Assumption $\text{\ref{assu:nw}}$ (ii) generalizes the assumption for the i.i.d. case. It requires smoothness for joint densities of observations within the same cluster and the conditional covariance as well as the marginal density and the conditional variance.\\
\end{rem}
\begin{thm}
\label{thm:nw_bias}\textbf{(Asymptotic bias)} Suppose that Assumptions \ref{assu:dgp}-\ref{assu:nw} hold. Then, 

\[
\mathbb{E}\left[\hat{m}_{\mathrm{nw}}(x)\mid\mathbf{X}_{1},\cdots,\mathbf{X}_{G}\right]=m(x)+h^{2}B_{\mathrm{nw}}(x)+o_{p}\left(h^{2}\right)+O_{p}\left(\sqrt{\frac{1}{nh^{d-2}}}\right),
\]
where
\[
B_{\mathrm{nw}}(x)=\kappa_{2}\sum_{q=1}^{d}\left(\frac{1}{2}\partial_{qq}m(x)+f(x)^{-1}\partial_{q}f(x)\partial_{q}m(x)\right),
\]
 $\partial_{q}f(x)=\partial f(x)/\partial x^{(q)}$, $\partial_{q}m(x)=\partial m(x)/\partial x^{(q)}$, and $\partial_{qq}m(x)=\partial^{2}m(x)/\partial\left(x^{(q)}\right)^{2}$.
\end{thm}
We use the following assumption to derive the asymptotic variance. 
\begin{assumption}
\label{assu:lambda} $\left(\frac{1}{n}\sum_{g=1}^{G}n_{g}^{2}\right)h^{d_{\mathrm{ind}}}\rightarrow\lambda\in[0,\infty)$.
\end{assumption}
\begin{rem}
$\left(\frac{1}{n}\sum_{g=1}^{G}n_{g}^{2}\right)h^{d_{\mathrm{ind}}}=O(1)$ is implied by $\left(\max_{g\leq G}n_{g}\right)h^{d_{\mathrm{ind}}}=O(1)$ since $\sum_{g=1}^{G}n_{g}=n$. Assumption \ref{assu:lambda} guarantees its convergence.
\end{rem}
\begin{thm}
\label{thm:nw_var}\textbf{(Asymptotic variance) }Suppose that Assumptions \ref{assu:dgp}-\ref{assu:lambda} hold. Then,
\begin{eqnarray}
 &  & \operatorname{Var}\left[\hat{m}_{\mathrm{nw}}(x)\mid\mathbf{X}_{1},\cdots,\mathbf{X}_{G}\right]\nonumber \\
 &  & \qquad=\frac{R_{k}^{d}\sigma^{2}(x)}{f(x)nh^{d}}+\frac{\lambda R_{k}^{d_{\mathrm{cls}}}f_{2}\left(x^{\mathrm{(ind)}},x^{\mathrm{(ind)}};x^{\mathrm{(cls)}}\right)\sigma\left(x^{\mathrm{(ind)}},x^{\mathrm{(ind)}};x^{\mathrm{(cls)}}\right)}{f(x)^{2}nh^{d}}+o_{p}\left(\frac{1}{nh^{d}}\right),\label{eq:asy_cond_var}
\end{eqnarray}
where $R_{k}=\int_{-\infty}^{\infty}k\left(u\right)^{2}\mathrm{d}u$. In particular, if $\lambda=0$,
\[
\operatorname{Var}\left[\hat{m}_{\mathrm{nw}}(x)\mid\mathbf{X}_{1},\cdots,\mathbf{X}_{G}\right]=\frac{R_{k}^{d}\sigma^{2}(x)}{f(x)nh^{d}}+o_{p}\left(\frac{1}{nh^{d}}\right).
\]
\end{thm}
In the special case of $\lambda=0$, the asymptotic conditional variance  is equivalent to the i.i.d. case. A sufficient condition for $\lambda=0$ is $\left(\max_{g\leq G}n_{g}\right)h^{d_{\mathrm{ind}}}=o(1)$. In a finite sample, it is more precise to consider $\lambda>0$. The sign of the second term of $\text{\eqref{eq:asy_cond_var}}$ depends on the sign of $\sigma\left(x^{\mathrm{(ind)}},x^{\mathrm{(ind)}};x^{\mathrm{(cls)}}\right)$. In economic applications, it usually takes a positive value, indicating positive conditional covariance of error terms within clusters. Neglecting this term will lead to under-coverage in empirical applications.

\begin{rem}
The pivotal condition for this theorem is Assumption \ref{assu:lambda}. Note that we can calculate $\left(\frac{1}{n}\sum_{g=1}^{G}n_{g}^{2}\right)h^{d_{\mathrm{ind}}}$ directly. The part $\left(\frac{1}{n}\sum_{g=1}^{G}n_{g}^{2}\right)$ can be interpreted as follows. Although we are considering \textit{deterministic} cluster sizes $n_{g}$, the value $\frac{1}{n}\sum_{g=1}^{G}n_{g}^{2}=\left(\frac{1}{G}\sum_{g=1}^{G}n_{g}^{2}\right)/\left(\frac{n}{G}\right)$ can be interpreted as the second moment of the cluster sizes over the first moment of the cluster sizes ``$\mathbb{E}\left[n_{g}^{2}\right]/\mathbb{E}\left[n_{g}\right]$'', where expectations are taken over $\left\{ n_{g}\right\} _{g=1}^{G}$.
\end{rem}
\begin{rem}
\label{rem:simple_cov}In the following two special cases, the second term of $\text{\eqref{eq:asy_cond_var}}$ has a simpler form. Firstly, if we assume the conditional independence $f_{2}\left(x^{\mathrm{(ind)}},x^{\mathrm{(ind)}}\mid x^{\mathrm{(cls)}}\right)=f\left(x^{\mathrm{(ind)}}\mid x^{\mathrm{(cls)}}\right)^{2}$, $\text{\eqref{eq:asy_cond_var}}$ simplifies to $\lambda R_{k}^{d_{\mathrm{cls}}}\sigma\left(x^{\mathrm{(ind)}},x^{\mathrm{(ind)}};x^{\mathrm{(cls)}}\right)/\left(f\left(x^{\mathrm{(cls)}}\right)nh^{d}\right)$. Secondly, if we assume the independence between individual and cluster-level regressors (or assume that there are no cluster-level regressors, $d_{\mathrm{cls}}=0$), $\text{\eqref{eq:asy_cond_var}}$ simplifies to 
\[
\frac{\lambda R_{k}^{d_{\mathrm{cls}}}\sigma\left(x^{\mathrm{(ind)}},x^{\mathrm{(ind)}};x^{\mathrm{(cls)}}\right)f\left(x^{\mathrm{(ind)}}\mid x^{\mathrm{(ind)}}\right)}{f\left(x\right)nh^{d}}
\]
 (or $\lambda\sigma\left(x^{\mathrm{(ind)}},x^{\mathrm{(ind)}}\right)f\left(x^{\mathrm{(ind)}}\mid x^{\mathrm{(ind)}}\right)/\left(f\left(x^{\mathrm{(ind)}}\right)nh^{d}\right),$ respectively). 
\end{rem}
\begin{thm}
\textbf{\label{thm:nw_cons}(Pointwise consistency)} Suppose that Assumptions \ref{assu:dgp}-\ref{assu:nw} hold. Then, 
\begin{equation}
\widehat{m}_{\mathrm{nw}}\left(x\right)\overset{p}{\rightarrow}m\left(x\right).\label{eq:nw_cons}
\end{equation}
\end{thm}
\begin{assumption}
\label{assu:asy_dist}$ $
\begin{enumerate}
\item There exists some $r\geq2$ such that
\begin{enumerate}
\item for any $\widetilde{x}=\left(\widetilde{x}^{(\mathrm{ind})\top},\widetilde{x}^{\mathrm{(cls)}\top}\right)^{\top}\in\mathcal{N}$, 
\begin{equation}
\mathbb{E}\left[|e|^{2r}\mid X=\widetilde{x}\right]\leq\overline{v}^{2}<\infty,\label{eq:e^r}
\end{equation}
\item for some constant $C>0$,
\begin{equation}
\frac{\left(\sum_{g=1}^{G}n_{g}^{r}\right)^{1/r}}{n^{1/4}}\leq C<\infty,\label{eq:n_g/n_bound}
\end{equation}
\item and
\begin{equation}
\frac{1}{n^{r/2}h^{dr-d}}=O(1).\label{eq:r_bound}
\end{equation}
\end{enumerate}
\item We also assume 
\begin{equation}
nh^{d+4}=O(1),\label{eq:nh^5}
\end{equation}
\[
R_{k}^{d}f(x)\sigma^{2}(x)+\lambda R_{k}^{d_{\mathrm{cls}}}f_{2}\left(x^{\mathrm{(ind)}},x^{\mathrm{(ind)}};x^{\mathrm{(cls)}}\right)\sigma\left(x^{\mathrm{(ind)}},x^{\mathrm{(ind)}};x^{\mathrm{(cls)}}\right)>0,
\]
and
\begin{equation}
\max_{g\leq G}\frac{n_{g}^{4}}{n}\rightarrow0\label{eq:n_g^4/(nhs)}
\end{equation}
as $n\rightarrow\infty$.
\end{enumerate}
\end{assumption}
\begin{thm}
\textbf{\label{thm:nw_asy_dist}(Asymptotic Normality)} Suppose that Assumptions \ref{assu:dgp}-\ref{assu:asy_dist} hold. Then,
\begin{eqnarray}
 &  & \sqrt{nh^{d}}\left(\widehat{m}_{\mathrm{nw}}(x)-m(x)-h^{2}B_{\mathrm{nw}}(x)\right)\nonumber \\
 &  & \qquad\overset{d}{\longrightarrow}\mathrm{N}\left(0,\frac{R_{k}^{d}\sigma^{2}(x)}{f(x)}+\frac{\lambda R_{k}^{d_{\mathrm{cls}}}f_{2}\left(x^{\mathrm{(ind)}},x^{\mathrm{(ind)}};x^{\mathrm{(cls)}}\right)\sigma\left(x^{\mathrm{(ind)}},x^{\mathrm{(ind)}};x^{\mathrm{(cls)}}\right)}{f(x)^{2}}\right).\label{eq:nw_asydist}
\end{eqnarray}
\end{thm}
The asymptotic distribution has the same bias and the same convergence rate as in the i.i.d. case. The asymptotic variance is a scaled value of the primal terms of asymptotic conditional variance that include the conditional covariance term due to the cluster dependence. Our simulation in Section $\text{\ref{sec:Monte-Carlo-simulation}}$ shows the importance of considering this term in inference. 

The asymptotic variance in the previous literature with bounded cluster sizes (e.g., \citealp{bhattacharya2005asymptotic}) has only the first term of $\text{\eqref{eq:nw_asydist}}$. Under bounded cluster sizes, cluster dependence is asymptotically negligible since an observation in the $g$-th cluster has a negligible number of observations belonging to the same cluster around the local neighborhood. On the other hand, under growing cluster sizes $n_{g}\rightarrow\infty$, the observation could have a non-negligible number of neighboring observations belonging to the same cluster. Thus, the conditional covariance of error terms matters in our general setup.

Theorem $\text{\ref{thm:nw_asy_dist}}$ is an asymptotic result that holds pointwise-in-the-underlying distribution. Consequently, the asymptotic bias pertains to a specific data-generating process, which influences how it should be used for bandwidth selection (see Remark $\text{\ref{rem:uniformity}}$ for derails). Also, Theorem $\text{\ref{thm:nw_asy_dist}}$ does not provide guidance on how the asymptotic bias should be addressed in inference. Remark $\text{\ref{rem:bias_handling}}$ discusses a practical approach for handling this issue in inference.

\begin{rem}
Conditions $\text{\eqref{eq:e^r}}$ and $\text{\eqref{eq:nh^5}}$ are standard in the kernel regressions. Replacing $\text{\eqref{eq:nh^5}}$ by $nh^{d+4}=o(1)$ eliminates the asymptotic bias (undersmoothing). Conditions $\text{\eqref{eq:n_g/n_bound}}$ and $\text{\eqref{eq:n_g^4/(nhs)}}$ require smaller cluster sizes than conditions in \citet{hansen2019asymptotic}. Indeed, they require $\left(\sum_{g=1}^{G}n_{g}^{r}\right)^{1/r}/n^{1/2}\leq C<\infty$ and $\max_{g\leq G}n_{g}^{2}/n\rightarrow0$, which are implied by $\text{\eqref{eq:n_g/n_bound}}$ and $\text{\eqref{eq:n_g^4/(nhs)}}$. \\
Condition $\text{\eqref{eq:r_bound}}$ is not strict if regressors have small dimension $d$. For example, the AIMSE-optimal bandwidth in Section $\text{\ref{sec:Bandwidth-selection}}$ satisfies $nh^{d+4}$ is bounded away from zero. In this case, $\text{\eqref{eq:r_bound}}$ is always satisfied if $d\leq4$ and equivalent to $r\leq2d/(d-4)$ if $d>4$.
\end{rem}

\section{\textbf{Local linear estimator\label{sec:Local-linear-estimator}}}

In this section, we consider the local linear estimator
\begin{equation}
\hat{m}_{\text{LL}}\left(x\right)=\sum_{g=1}^{G}\sum_{j=1}^{n_{g}}K_{\text{LL}}\left(X_{gj},x\right)Y_{gj},\label{eq:LL}
\end{equation}
where
\begin{align*}
K_{\text{LL}}\left(u,x\right) & =\mathbf{e}_{1}^{\top}\left(\mathbf{X}_{x}^{\top}\mathbf{W}_{x}\mathbf{X}_{x}\right)^{-1}\left[\begin{array}{c}
1\\
u-x
\end{array}\right]K_{h}\left(u-x\right),\\
\underbrace{\mathbf{e}_{1}}_{(d+1)\times1}=\left[\begin{array}{c}
1\\
0\\
\vdots\\
0
\end{array}\right],\quad & \underbrace{\mathbf{X}_{x}}_{n\times(d+1)}=\left[\begin{array}{cc}
1 & \left(X_{1}-x\right)^{\top}\\
\vdots & \vdots\\
1 & \left(X_{n}-x\right)^{\top}
\end{array}\right],\quad\underbrace{\mathbf{W}_{x}}_{n\times n}=\left[\begin{array}{ccc}
K_{h}\left(X_{1}-x\right) &  & O\\
 & \ddots\\
O &  & K_{h}\left(X_{n}-x\right)
\end{array}\right],
\end{align*}
and $K_{h}\left(\cdot\right)=\frac{1}{h^{d}}K\left(\frac{\cdot}{h}\right)$. We will assume an additional condition for the simplicity of proofs.
\begin{assumption}
\label{assu:LL} $K$ has a compact support.
\end{assumption}
\begin{rem}
Assumption $\text{\ref{assu:LL}}$ is a standard technical assumption for local linear estimators. It can be replaced by a tail decay assumption for $K$ (see e.g., \citealp{fan1992variable}).
\end{rem}
We can establish similar asymptotic theories for local linear estimators as we derived for Nadaraya-Watson estimators. As in the i.i.d. case, the asymptotic bias of a local linear estimator does not include the term of first-order derivatives.
\begin{thm}
\label{thm:LL_bias}\textbf{(Asymptotic bias)} Suppose that Assumptions \ref{assu:dgp}-\ref{assu:nw} and \ref{assu:LL} hold. Then,

\[
\mathbb{E}\left[\hat{m}_{\mathrm{LL}}(x)\mid\mathbf{X}_{1},\cdots,\mathbf{X}_{G}\right]=m(x)+h^{2}B_{\mathrm{LL}}(x)+o_{p}\left(h^{2}\right),
\]
where
\[
B_{\mathrm{LL}}(x)=\frac{\kappa_{2}}{2}\sum_{q=1}^{d}\partial_{qq}m(x).
\]
\end{thm}
\begin{thm}
\label{thm:LL_var}\textbf{(Asymptotic variance)} Suppose that Assumptions \ref{assu:dgp}-\ref{assu:lambda} and \ref{assu:LL} hold. Then,
\begin{eqnarray*}
 &  & \operatorname{Var}\left[\hat{m}_{\mathrm{LL}}(x)\mid\mathbf{X}_{1},\cdots,\mathbf{X}_{G}\right]\\
 &  & \qquad=\frac{R_{k}^{d}\sigma^{2}(x)}{f(x)nh^{d}}+\frac{\lambda R_{k}^{d_{\mathrm{cls}}}f_{2}\left(x^{\mathrm{(ind)}},x^{\mathrm{(ind)}};x^{\mathrm{(cls)}}\right)\sigma\left(x^{\mathrm{(ind)}},x^{\mathrm{(ind)}};x^{\mathrm{(cls)}}\right)}{f(x)^{2}nh^{d}}+o_{p}\left(\frac{1}{nh^{d}}\right).
\end{eqnarray*}
In particular, if $\lambda=0$,
\[
\operatorname{Var}\left[\hat{m}_{\mathrm{LL}}(x)\mid\mathbf{X}_{1},\cdots,\mathbf{X}_{G}\right]=\frac{R_{k}^{d}\sigma^{2}(x)}{f(x)nh^{d}}+o_{p}\left(\frac{1}{nh^{d}}\right).
\]
\end{thm}
\begin{thm}
\textbf{\label{thm:LL_cons}(Pointwise} \textbf{consistency)} Suppose that Assumptions \ref{assu:dgp}-\ref{assu:nw} and \ref{assu:LL} hold. Then,
\begin{equation}
\widehat{m}_{\mathrm{LL}}\left(x\right)\overset{p}{\rightarrow}m\left(x\right).\label{eq:LL_cons}
\end{equation}
\end{thm}
\begin{thm}
\textbf{\label{thm:LL_asy_dist}(Asymptotic normality)} Suppose that Assumptions \ref{assu:dgp}-\ref{assu:LL} hold. Then,
\begin{eqnarray}
 &  & \sqrt{nh^{d}}\left(\widehat{m}_{\mathrm{LL}}(x)-m(x)-h^{2}B_{\mathrm{LL}}(x)\right)\nonumber \\
 &  & \qquad\overset{d}{\longrightarrow}\mathrm{N}\left(0,\frac{R_{k}^{d}\sigma^{2}(x)}{f(x)}+\frac{\lambda R_{k}^{d_{\mathrm{cls}}}f_{2}\left(x^{\mathrm{(ind)}},x^{\mathrm{(ind)}};x^{\mathrm{(cls)}}\right)\sigma\left(x^{\mathrm{(ind)}},x^{\mathrm{(ind)}};x^{\mathrm{(cls)}}\right)}{f(x)^{2}}\right).\label{eq:LL_asydist}
\end{eqnarray}
\end{thm}

\section{\textbf{Uniform convergence\label{sec:Uniform-convergence}}}

If we impose further assumptions, our pointwise consistency result can be strengthened to uniform consistency. Before proving uniform consistency for nonparametric estimators, we will show uniform consistency for the generic function 
\begin{equation}
\widehat{\psi}\left(x\right)=\frac{1}{nh^{d}}\sum_{g=1}^{G}\sum_{j=1}^{n_{g}}K\left(\frac{X_{gj}-x}{h}\right)W_{gj}\label{eq:psi}
\end{equation}
to its expectation, where $X_{gj}\in\mathbb{R}^{d}$ and $W_{gj}\in\mathbb{R}$.

We assume the cluster samples $\left\{ W_{gj},X_{gj}\right\} $ satisfy the following assumptions.
\begin{assumption}
\label{assu:psi_var}There exists a constant $\overline{V}$ such that
\[
\sup_{x}\operatorname{Var}\left(\widehat{\psi}\left(x\right)\right)\leq\frac{\overline{V}}{nh^{d}}
\]
for sufficiently large $n$.
\end{assumption}
\begin{assumption}
\label{assu:Wgj}For every $i=1,\dots,n$ and for some $s>2$, we have
\begin{equation}
\mathbb{E}\left[\left|W_{i}\right|^{s}\right]<B_{1}<\infty\label{eq:Ws_bound}
\end{equation}
and

\begin{equation}
\sup_{x}\mathbb{E}\left[\left|W_{i}\right|^{s}\mid X_{i}=x\right]f\left(x\right)<B_{2}<\infty.\label{eq:Ws_cond}
\end{equation}
We also assume that
\begin{equation}
\frac{\left(\max_{g\leq G}n_{g}\right)^{2}\log n}{n^{1-(2/s)}h^{d}}=O(1).\label{eq:theta_logn}
\end{equation}
\end{assumption}
\begin{rem}
The conditions are standard to establish uniform convergence except for $\text{\eqref{eq:theta_logn}}$. Equation $\text{\eqref{eq:theta_logn}}$ has an additional component $\left(\max_{g\leq G}n_{g}\right)^{2}$ in cluster sampling. If we focus on bounded size clusters, $\text{\eqref{eq:theta_logn}}$ can be reduced to the standard assumption for the i.i.d. case.\\
For some applications, $W_{i}$ has a bounded support. In this case, Assumption $\text{\ref{assu:Wgj}}$ is satisfied with $s=\infty$ after rescaling $W_{i}\in[-1,1]$.
\end{rem}
We also require a further assumption on the kernel function.
\begin{assumption}
For some $0<L<\infty$, $K$ has a compact support, that is, $K(u)=0$ for $\left\Vert u\right\Vert >L$. Furthermore, $K$ is Lipschitz, i.e., for some constant $\Lambda<\infty$ and for all $u,u^{\prime}\in\mathbb{R}$, $\left|K(u)-K\left(u^{\prime}\right)\right|\leq\Lambda\left\Vert u-u^{\prime}\right\Vert $.

\label{assu:kernel}
\end{assumption}
\begin{thm}
\label{thm:psi_unifconv}\textbf{(Uniform consistency for the general estimator)} Suppose that $\left\{ W_{gj},X_{gj}\right\} $ satisfies Assumption $\text{\ref{assu:dgp}}$ and Assumptions $\text{\ref{assu:psi_var}}$, $\text{\ref{assu:Wgj}}$, and $\text{\ref{assu:kernel}}$ hold.

Then, for any 
\begin{equation}
c_{n}=O\left(\left(\max_{g\leq G}n_{g}\right)^{2/d}\left(\log n\right)^{1/d}\right)\label{eq:cn}
\end{equation}
 and 
\begin{equation}
a_{n}=\left(\frac{\log n}{nh^{d}}\right)^{1/2},\label{eq:an}
\end{equation}
$\widehat{\psi}\left(x\right)$ converges in probability to $\mathbb{E}\left[\widehat{\psi}\left(x\right)\right]$ uniformly on $\left\Vert x\right\Vert \leq c_{n}$, i.e.,
\begin{equation}
\sup_{\left\Vert x\right\Vert \leq c_{n}}\left|\widehat{\psi}\left(x\right)-\mathbb{E}\left[\widehat{\psi}\left(x\right)\right]\right|=O_{p}\left(a_{n}\right),\label{eq:psi_unifconv}
\end{equation}
as $nh^{d}\rightarrow\infty$, $h\rightarrow0$, and $\left(\max_{g\leq G}n_{g}\right)h^{d_{\mathrm{ind}}}=O(1)$.
\end{thm}
The proof for Theorem $\text{\ref{thm:psi_unifconv}}$ relies on the following cluster sampling version of Bernstein's inequality, which could be of independent interest.
\begin{lem}
\textbf{(Bernstein's inequality for cluster sampling)}\label{lem:Bernstein}

For random variables under cluster sampling $\left\{ \left\{ Y_{gj}\right\} _{j=1}^{n_{g}}\right\} _{g=1}^{G}$ with bounded ranges $[-B,B]$ and zero means,

\[
\mathbb{P}\left[\left|\widetilde{\mathbf{Y}}_{1}+\cdots+\widetilde{\mathbf{Y}}_{G}\right|>\varepsilon\right]\leq2\exp\left\{ -\frac{1}{2}\frac{\varepsilon^{2}}{v+\left(\max_{g\leq G}n_{g}\right)B\varepsilon/3}\right\} 
\]
 for every $\varepsilon>0$ and $v\geq\operatorname{Var}\left(\widetilde{\mathbf{Y}}_{1}+\cdots+\widetilde{\mathbf{Y}}_{G}\right)$, where $\widetilde{\mathbf{Y}}_{g}=\sum_{j=1}^{n_{g}}Y_{gj}$.
\end{lem}
Based on Theorem $\text{\ref{thm:psi_unifconv}},$ we will show the uniform consistency of the nonparametric density estimator and nonparametric regressions. It requires the following conditions, including uniform smoothness.
\begin{assumption}
\label{assu:nonpara_unif}
\end{assumption}
\begin{enumerate}
\item $nh^{d}\rightarrow\infty$.
\item $h\rightarrow0$ and $\left(\max_{g\leq G}n_{g}\right)h^{d_{\mathrm{ind}}}=O(1)$.
\item $m(x)$ and $f(x)$ have uniformly continuous second-order derivatives and they are uniformly bounded up to second-order derivatives, $f_{2}\left(x^{\mathrm{(ind)}},x^{\mathrm{(ind)}};x^{\mathrm{(cls)}}\right)$ has uniformly continuous first-order derivative and is uniformly bounded up to first-order derivative, and $f_{3}\left(x^{\mathrm{(ind)}},x^{\mathrm{(ind)}},x^{\mathrm{(ind)}};x^{\mathrm{(cls)}}\right)$, $f_{4}\left(x^{\mathrm{(ind)}},x^{\mathrm{(ind)}},x^{\mathrm{(ind)}},x^{\mathrm{(ind)}};x^{\mathrm{(cls)}}\right)$, $\sigma^{2}(x)$, and $\sigma\left(x^{\mathrm{(ind)}},x^{\mathrm{(ind)}};x^{\mathrm{(cls)}}\right)$ are uniformly continuous and uniformly bounded.
\end{enumerate}
\begin{thm}
\label{thm:density_unifconv}\textbf{(Uniform consistency for the nonparametric density estimator)} Suppose that Assumptions $\text{\ref{assu:dgp}}$, $\text{\ref{assu:kernel}}$, and \ref{assu:nonpara_unif} hold. We also assume that
\[
\frac{\left(\max_{g\leq G}n_{g}\right)^{2}\log n}{nh^{d}}=O(1).
\]

Then, for any sequence $c_{n}$ satisfying the condition $\eqref{eq:cn}$, 
\begin{equation}
\sup_{\left\Vert x\right\Vert \leq c_{n}}\left|\widehat{f}\left(x\right)-f\left(x\right)\right|=O_{p}\left(a_{n}+h^{2}\right).\label{eq:denisty_unifconv}
\end{equation}
\end{thm}
\begin{thm}
\label{thm:nw_unifconv}\textbf{(Uniform consistency for the Nadaraya-Watson estimator)} Suppose that the assumptions for Theorem $\text{\ref{thm:density_unifconv}}$ hold. We also also assume that Assumption $\text{\ref{assu:Wgj}}$ holds for the cluster observations $\left\{ Y_{gj},X_{gj}\right\} $. If $c_{n}$ is a sequence satisfying the condition $\eqref{eq:cn}$, 
\begin{equation}
\delta_{n}=\inf_{\left\Vert x\right\Vert \leq c_{n}}f(x)>0,\label{eq:delta_n}
\end{equation}
and 
\[
\delta_{n}^{-1}\left(a_{n}+h^{2}\right)=o(1),
\]

then, 
\begin{equation}
\sup_{\left\Vert x\right\Vert \leq c_{n}}\left|\widehat{m}_{*}\left(x\right)-m\left(x\right)\right|=O_{p}\left(\delta_{n}^{-1}\left(a_{n}+h^{2}\right)\right)\label{eq:nw_unifconv}
\end{equation}
for $\widehat{m}_{*}(x)=\widehat{m}_{\mathrm{nw}}(x)$ or $\widehat{m}_{\mathrm{LL}}(x)$.
\end{thm}
The range $\left\{ x:\left\Vert x\right\Vert \leq c_{n}\right\} $ expands slowly to $\mathbb{R}^{d}$ since our condition $\text{\eqref{eq:cn}}$ can cover a sequence $\left\{ c_{n}\right\} $ such that $c_{n}\rightarrow\infty$ slowly as $n\rightarrow\infty$. This expansion is useful to establish asymptotic theories for semiparametric estimation with a nonparametric kernel estimator in the first-stage. 

Suppose that $c_{n}=c$ (constant) and $\delta_{n}$ is far away zero. Then, the uniform convergence rate for kernel regressions is $a_{n}+h^{2}=\left(\log n/(nh^{d})\right)^{1/2}+h^{2}$. By choosing $h=\left(\log n/n\right)^{1/(d+4)}$, the optimal rate $\ensuremath{\left(\log n/n\right)^{2/(d+4)}}$ is attained. This convergence rate is equivalent to \citet{stone1982optimal}'s optimal rate in the i.i.d. case.

\section{\textbf{Bandwidth selection\label{sec:Bandwidth-selection}}}

In this section, we provide guidelines for selecting bandwidth in nonparametric regressions. We suggest three types of methods: the asymptotic integrated mean squared error (AIMSE) optimal bandwidth selection, the cluster-robust rule-of-thumb, and the cluster-robust cross-validation.

\subsection{AIMSE-optimal bandwidth}

Let $B_{*}(x)=B_{\mathrm{nw}}(x)$ or $B_{\mathrm{LL}}(x)$. The asymptotic integrated mean squared error of the estimator $\widehat{m}_{*}\left(x\right)$ is
\begin{eqnarray}
 &  & \int_{\mathbb{R}^{d}}h^{4}B_{*}(x)^{2}f(x)w(x)\mathrm{d}x\nonumber \\
 &  & +\int_{\mathbb{R}^{d}}\left\{ \frac{R_{k}^{d}\sigma^{2}(x)}{f(x)nh^{d}}+\frac{\lambda R_{k}^{d_{\mathrm{cls}}}f_{2}\left(x^{\mathrm{(ind)}},x^{\mathrm{(ind)}};x^{\mathrm{(cls)}}\right)\sigma\left(x^{\mathrm{(ind)}},x^{\mathrm{(ind)}};x^{\mathrm{(cls)}}\right)}{f(x)^{2}nh^{d}}\right\} f(x)w(x)\mathrm{d}x\nonumber \\
 & = & h^{4}\overline{B}+\frac{R_{k}^{d}\overline{\sigma}^{2}}{nh^{d}}+\frac{\left(\frac{1}{n}\sum_{g=1}^{G}n_{g}^{2}\right)}{n}R_{k}^{d_{\mathrm{cls}}}\overline{\sigma}_{\mathrm{cls}}+o_{p}\left(\frac{1}{nh^{d}}\right),\label{eq:aimse_derivation}
\end{eqnarray}
where $w(x)$ is some integrable weight function which ensures that $\overline{B}\equiv\int_{\mathbb{R}^{d}}B_{*}(x)^{2}f(x)w(x)\mathrm{d}x$, $\overline{\sigma}^{2}\equiv\int_{\mathbb{R}^{d}}\sigma^{2}(x)w(x)\mathrm{d}x$, and 
\[
\overline{\sigma}_{\mathrm{cls}}\equiv\int_{\mathbb{R}^{d}}\frac{f_{2}\left(x^{\mathrm{(ind)}},x^{\mathrm{(ind)}};x^{\mathrm{(cls)}}\right)\sigma\left(x^{\mathrm{(ind)}},x^{\mathrm{(ind)}};x^{\mathrm{(cls)}}\right)}{f(x)}w(x)\mathrm{d}x
\]
 are finite. We define
\begin{equation}
\operatorname{AIMSE}\equiv h^{4}\overline{B}+\frac{R_{k}^{d}\overline{\sigma}^{2}}{nh^{d}}\label{eq:aimse}
\end{equation}
as an objective function for bandwidth selection since the third term in $\text{\eqref{eq:aimse_derivation}}$ does not depend on $h$ and the fourth term in $\text{\eqref{eq:aimse_derivation}}$ is asymptotically negligible.

\begin{thm}
\label{thm:bw}The AIMSE-optimal bandwidth that minimizes the AIMSE $\text{\eqref{eq:aimse}}$ is 
\begin{equation}
h_{0}=\left(\frac{dR_{k}^{d}\overline{\sigma}^{2}}{4\overline{B}}\right)^{1/(d+4)}n^{-1/(d+4)}.\label{eq:h0}
\end{equation}
\end{thm}
Our asymptotic theorems rely on the assumption $\left(\max_{g\leq G}n_{g}\right)h^{d_{\mathrm{ind}}}=O(1)$.\\
When $\left(\max_{g\leq G}n_{g}\right)n^{-d_{\mathrm{ind}}/(d+4)}\rightarrow\infty$, the AIMSE-optimal $h_{0}$ does not satisfy this order. In this case, the AIMSE-optimal bandwidth does not make sense since the AIMSE criterion itself relies on the assumption $\left(\max_{g\leq G}n_{g}\right)h^{d_{\mathrm{ind}}}=O(1)$. Thus, when the largest cluster size is large compared to the sample size $n$, we recommend using the cross-validation criterion (see Section $\text{\ref{subsec:Cross-validation}}$).
\begin{rem}
\label{rem:uniformity}This paper treats the data-generating process as fixed and considers asymptotic bias as pointwise-in-the-underlying distribution. An caveat of this pointwise-in-the-underlying distribution is that it permits the selection of a kernel function with no asymptotic bias while simultaneously choosing an arbitrarily large bandwidth to reduce variance, as discussed in \citeauthor{Tsybakov2009} (\citeyear{Tsybakov2009}, Chapter 1.2.4). Consequently, the optimality of the bandwidth proposed in this paper is guaranteed solely for the underlying distribution. \citet{armstrong2018optimal} argue that the pointwise optimal bandwidth performs poorly over a class of data-generating processes, both analytically and numerically. 
\end{rem}

\subsection{Rule-of-thumb}

In practice, it is not easy to compute the AIMSE-optimal bandwidth since $\text{\eqref{eq:h0}}$ contains unknown parameters. As suggested by \citeauthor{fan1996local} (\citeyear{fan1996local}, Section 4.2) for the i.i.d. case, we provide a cluster-robust Rule-of-Thumb (CR-ROT) bandwidth choice for a one-dimensional individual-level regressor $x\in\mathbb{R}$. This bandwidth could be a crude estimator of the AIMSE-optimal bandwidth, but the primary purpose of it is to give a guess of the bandwidth requiring little computational effort. Let 
\begin{equation}
\check{m}_{-g}(x)=\check{\alpha}_{0,-g}+\cdots+\check{\alpha}_{4,-g}x^{4}\label{eq:global-g}
\end{equation}
be a fitted 4th-order \textit{global} polynomial regression leaving out the $g$-th cluster. Given this parametric model and a user-specified integrable weight function $w(x)$, the CR-ROT bandwidth is calculated by
\begin{equation}
h_{\text{CR-ROT}}=\left(\frac{dR_{k}^{d}\check{\sigma}^{2}}{4\check{B}}\right)^{1/(d+4)}n^{-1/(d+4)},\label{eq:CR-ROT}
\end{equation}
where 
\begin{eqnarray*}
\check{B} & = & \frac{1}{n}\sum_{g=1}^{G}\sum_{j=1}^{n_{g}}\left\{ \frac{1}{2}\check{m}_{-g}^{\prime\prime}\left(X_{gj}\right)\right\} ^{2}w\left(X_{gj}\right)\\
 & = & \frac{1}{n}\sum_{g=1}^{G}\sum_{j=1}^{n_{g}}\left\{ \check{\alpha}_{2,-g}+3\check{\alpha}_{3,-g}X_{gj}+6\check{\alpha}_{4,-g}X_{gj}^{2}\right\} ^{2}w\left(X_{gj}\right),\\
\check{\sigma}^{2} & = & \left(\frac{1}{n}\sum_{g=1}^{G}\sum_{j=1}^{n_{g}}\check{e}_{gj}^{2}\right)\int_{\mathbb{R}^{d}}w(x)\mathrm{d}x,
\end{eqnarray*}
and $\check{e}_{gj}=Y_{gj}-\check{m}_{-g}\left(X_{gj}\right)$. In words, $\check{B}$ and $\check{\sigma}^{2}$ are computed by the parametric model $\text{\eqref{eq:global-g}}$ and the homoskedastic standard error assumption for local linear estimators. For Nadaraya-Watson estimators, we also assume that $X$ has a uniform distribution for simplicity. Then, we have $f^{\prime}(x)=0$ and can compute $\bar{B}$ as for local linear estimators by $B_{\text{nw}}(x)=B_{\text{LL}}(x)$. A common choice of $w(x)$ is an indicator function of some interval.

Equation $\text{\eqref{eq:CR-ROT}}$ is different from the standard Rule-of-Thumb (ROT) bandwidth choice by \citet{fan1996local} since it uses $\check{m}_{-g}(x)$ instead of $\check{m}(x)$, which is estimated by the full sample. We use $\check{m}_{-g}(x)$ to eliminate dependence between the estimator $\check{m}_{-g}(\cdot)$ and $\left(Y_{gj},X_{gj}\right)$. This modification should provide a better estimation of out-of-sample prediction error.

\subsection{Cross-validation\label{subsec:Cross-validation}}

A heuristic cross-validation function for clustered sampling is

\begin{equation}
\mathrm{CV}(h)\equiv\frac{1}{n}\sum_{g=1}^{G}\sum_{j=1}^{n_{g}}\tilde{e}_{gj}\left(h\right)^{2}w\left(X_{gj}\right),\label{eq:cv}
\end{equation}
where $\widetilde{e}_{gj}=Y_{gj}-\widetilde{m}_{-g}\left(X_{gj},h\right)$, and $\widetilde{m}_{-g}\left(X_{gj},h\right)$ is the leave-one-cluster-out nonparametric estimator computed with bandwidth $h$ and without cluster $g$. For example, \citeauthor{hansen2022econometrics} (\citeyear{hansen2022econometrics}, Section 19.20) suggests this form of cross-validation, but he does not provide any theoretical guarantees. For Nadaraya-Watson estimators, the leave-one-cluster-out nonparametric estimator is defined by
\begin{equation}
\widetilde{m}_{\mathrm{nw},-g}\left(x,h\right)=\frac{\sum_{g^{\prime}\neq g}\sum_{j=1}^{n_{g^{\prime}}}K\left(\frac{X_{g^{\prime}j}-x}{h}\right)Y_{g^{\prime}j}}{\sum_{g^{\prime}\neq g}\sum_{j=1}^{n_{g^{\prime}}}K\left(\frac{X_{g^{\prime}j}-x}{h}\right)}.\label{eq:nw_m-g}
\end{equation}
Similarly, for local linear estimators, the leave-one-cluster-out nonparametric estimator is defined by 
\begin{equation}
\widetilde{m}_{\mathrm{LL},-g}\left(x,h\right)=\sum_{g^{\prime}\neq g}\sum_{j=1}^{n_{g^{\prime}}}K_{\text{LL},-g}\left(X_{g^{\prime}j},x\right)Y_{g^{\prime}j},\label{eq:LL_m-g}
\end{equation}
where 
\begin{align*}
K_{\text{LL},-g}\left(u,x\right) & =\mathbf{e}_{1}^{\top}\left(\mathbf{X}_{x,-g}^{\top}\mathbf{W}_{x,-g}\mathbf{X}_{x,-g}\right)^{-1}\left[\begin{array}{c}
1\\
u-x
\end{array}\right]K_{h}\left(u-x\right),
\end{align*}
$\mathbf{X}_{x,-g}$ and $\mathbf{W}_{x,-g}$ are defined by the same way as $\mathbf{X}_{x}$ and $\mathbf{W}_{x}$, but without using the variables in the $g$-th cluster. We will show that this cross-validation criterion works appropriately.

\begin{thm}
\label{thm:cr-cv} Let $\overline{\sigma}_{w}^{2}=\mathbb{E}\left[e_{gj}^{2}w\left(X_{gj}\right)\right]=\mathbb{E}\left[\sigma^{2}\left(X_{gj}\right)w\left(X_{gj}\right)\right]$ and $w(x)$ be some integrable weight function. Under Assumption $\text{\ref{assu:dgp}}$, we can decompose the expectation of the cross-validation function over $\left\{ \mathbf{Y}_{g},\mathbf{X}_{g}\right\} _{g=1}^{G}$ as

\begin{equation}
\mathbb{E}\left[\mathrm{CV}(h)\right]=\overline{\sigma}_{w}^{2}+\operatorname{IMSE}_{G-1}(h)\label{eq:e_cv}
\end{equation}
 where 
\begin{equation}
\operatorname{IMSE}_{G-1}(h)\equiv\sum_{g=1}^{G}\frac{n_{g}}{n}\mathbb{E}_{-g}\left[\int_{\mathbb{R}^{d}}\left\{ m\left(x\right)-\widetilde{m}_{-g}\left(x,h\right)\right\} ^{2}f\left(x\right)w\left(x\right)\mathrm{d}x\right],\label{eq:imse}
\end{equation}
and the last expectation is taken over the sample except for the $g$-th cluster $\left(\mathbf{Y}_{-g},\mathbf{X}_{-g}\right)=\left\{ \mathbf{Y}_{g^{\prime}},\mathbf{X}_{g^{\prime}}\right\} _{g^{\prime}\neq g}$.
\end{thm}
Since $\overline{\sigma}_{w}^{2}$ does not depend on $h$, minimizing $\mathbb{E}\left[\mathrm{CV}(h)\right]$ on $h$ is equivalent to minimizing $\operatorname{IMSE}_{G-1}(h)$, which is a sum of the expected mean squared errors weighted by cluster sizes. Thus, this theorem justifies the use of the leave-one-cluster-out cross-validation. We can choose the bandwidth by minimizing a cluster-robust cross-validation function $\mathrm{CV}(h)$ over some finite grid points $H=[h_{1},\cdots,h_{J}]$,
\begin{equation}
h_{\text{CR-CV}}=\underset{h\in H}{\operatorname{argmin}}\ \mathrm{CV}(h).\label{eq:h_cv}
\end{equation}
Note that the decomposition theorem holds for finite samples and does not rely on assumptions such as $\left(\max_{g\leq G}n_{g}\right)h^{d_{\mathrm{ind}}}=O(1)$.

\section{\textbf{A new cluster-robust variance estimation\label{sec:Cluster-robust-variance-estimati}}}

Since the asymptotic variance of $\text{\eqref{eq:nw_asydist}}$ contains the joint density $f\left(x^{\mathrm{(ind)}},x^{\mathrm{(ind)}};x^{\mathrm{(cls)}}\right)$, the conditional variance $\sigma^{2}(x)$, and the conditional covariance $\sigma\left(x^{\mathrm{(ind)}},x^{\mathrm{(ind)}};x^{\mathrm{(cls)}}\right)$, we need to estimate each of them for inference. Alternatively, \citet{calonico2019nprobust} and \citeauthor{hansen2022econometrics} (\citeyear{hansen2022econometrics}, Section 19.20) propose to use a finite sample conditional variance of $\widehat{m}\left(X_{gj}\right)$ with estimated error terms as an estimator of the asymptotic variance. To the best of our knowledge, there is no theoretical guarantee of their methods, and this paper is the first research providing asymptotic theories of inference for nonparametric regressions under general cluster sizes.

For the joint density estimation, we propose to use 
\begin{eqnarray}
 &  & \widehat{f}_{2}\left(x^{\mathrm{(ind)}},x^{\mathrm{(ind)}};x^{\mathrm{(cls)}}\right)\nonumber \\
 & = & \frac{1}{Nb^{2d_{\mathrm{ind}}+d_{\mathrm{cls}}}}\nonumber \\
 &  & \times\sum_{g:n_{g}\geq2}\sum_{1\leq j<\ell\leq n_{g}}K\left(\frac{\left(X_{gj}^{(\mathrm{ind})\top},X_{g\ell}^{(\mathrm{ind})\top},X_{g}^{\mathrm{(cls)}\top}\right)^{\top}-\left(x^{(\mathrm{ind})\top},x^{(\mathrm{ind})\top},x^{\mathrm{(cls)}\top}\right)^{\top}}{b}\right),\label{eq:joint_density_estimator}
\end{eqnarray}
where $b$ is a bandwidth and $N=\sum_{g:n_{g}\geq2}n_{g}(n_{g}-1)/2$. 

The expression $\text{\eqref{eq:joint_density_estimator}}$ can be interpreted as a standard nonparametric density estimator. We estimate the density using $\left(2d_{\mathrm{ind}}+d_{\mathrm{cls}}\right)$-dimensional regressors $\left(X_{gj}^{(\mathrm{ind})\top},X_{g\ell}^{(\mathrm{ind})\top},X_{g}^{\mathrm{(cls)}\top}\right)^{\top}$, thus we have $b^{2d_{\mathrm{ind}}+d_{\mathrm{cls}}}$ in the denominator in $\text{\eqref{eq:joint_density_estimator}}$. For clusters larger than $2$ (i.e., $n_{g}\geq2$), there are $\sum_{1\leq j<\ell\leq n_{g}}1=n_{g}(n_{g}-1)/2$ possible combinations of $X_{gj}^{(\mathrm{ind})}$ and $X_{g\ell}^{(\mathrm{ind})}$. Each cluster has a $n_{g}(n_{g}-1)/2$ effective size observations, and we have the $N=\sum_{g:n_{g}\geq2}n_{g}(n_{g}-1)/2$ effective size sample in total. In these senses, $\text{\eqref{eq:joint_density_estimator}}$ is a standard nonparametric density estimator for $\left(2d_{\mathrm{ind}}+d_{\mathrm{cls}}\right)$-dimensional regressors and $n_{g}(n_{g}-1)/2$ size clusters. 
\begin{rem}
Note that we use the kernel $K\left(\frac{\left(X_{gj}^{(\mathrm{ind})\top},X_{g\ell}^{(\mathrm{ind})\top},X_{g}^{\mathrm{(cls)}\top}\right)^{\top}-\left(x^{(\mathrm{ind})\top},x^{(\mathrm{ind})\top},x^{\mathrm{(cls)}\top}\right)^{\top}}{b}\right)$ instead of $K\left(\frac{X_{gj}-x}{b}\right)K\left(\frac{X_{g\ell}-x}{b}\right)$. The latter is the kernel to estimate $\left.f_{2^{\prime}}\left(x_{gj},x_{g\ell}\right)\right|_{\left(x_{gj},x_{g\ell}\right)=\left(x,x\right)}$, which is not continuous around $\left(x_{gj},x_{g\ell}\right)=\left(x,x\right)$. Indeed, this joint density is ``degenerate'' in coordinates of cluster-level regressors.
\end{rem}
We make the following assumptions to estimate the joint density consistently.
\begin{assumption}
\label{assu:condvar} 

Define $\varsigma^{2}\left(X_{gj}\right)\equiv\mathbb{E}\left[e_{gj}^{4}\mid\mathbf{X}_{g}\right]=\mathbb{E}\left[e_{gj}^{4}\mid X_{gj}\right]$ and 
\begin{align*}
\varsigma\left(X_{gj}^{(\mathrm{ind})},X_{gj}^{(\mathrm{ind})},X_{g\ell}^{(\mathrm{ind})},X_{g\ell}^{(\mathrm{ind})};X_{g}^{(\mathrm{cls})}\right) & \equiv\mathbb{E}\left[e_{gj}e_{g\ell}e_{gt}e_{gs}\mid\mathbf{X}_{g}\right]\\
 & =\mathbb{E}\left[e_{gj}e_{g\ell}e_{gt}e_{gs}\mid X_{gj}^{(\mathrm{ind})},X_{g\ell}^{(\mathrm{ind})},X_{gt}^{(\mathrm{ind})},X_{gs}^{(\mathrm{ind})};X_{g}^{(\mathrm{cls})}\right].
\end{align*}

\begin{enumerate}
\item $Nb^{2d_{\mathrm{ind}}+d_{\mathrm{cls}}}\rightarrow\infty$.
\item $b\rightarrow0$ and $\left(\max_{g\leq G}n_{g}^{2}\right)b^{2d_{\mathrm{ind}}}=O(1)$\textup{.}
\item $f_{2}\left(x^{\mathrm{(ind)}},x^{\mathrm{(ind)}};x^{\mathrm{(cls)}}\right)>0$.
\item There exists some neighborhood $\mathcal{N}$ of $x=\left(x^{(\mathrm{ind})\top},x^{(\mathrm{ind})\top}\right)^{\top}$ such that $\sigma^{2}\left(x\right)$, $\sigma\left(x^{\mathrm{(ind)}},x^{\mathrm{(ind)}};x^{\mathrm{(cls)}}\right)$, and $f_{2}\left(x^{\mathrm{(ind)}},x^{\mathrm{(ind)}};x^{\mathrm{(cls)}}\right)$ are twice continuously differentiable, $f_{3}\left(x^{\mathrm{(ind)}},x^{\mathrm{(ind)}},x^{\mathrm{(ind)}};x^{\mathrm{(cls)}}\right)$ and $f_{4}\left(x^{\mathrm{(ind)}},x^{\mathrm{(ind)}},x^{\mathrm{(ind)}},x^{\mathrm{(ind)}};x^{\mathrm{(cls)}}\right)$ are continuously differentiable, and $\varsigma^{2}\left(x\right)$ and $\varsigma\left(x^{\mathrm{(ind)}},x^{\mathrm{(ind)}},x^{\mathrm{(ind)}},x^{\mathrm{(ind)}};x^{\mathrm{(cls)}}\right)$ are continuous. Also, joint densities of up to 8 individual-level regressors and cluster-level regressors within the same cluster follow common distributions, and these joint densities 
\[
f_{5}\left(x^{\mathrm{(ind)}},x^{\mathrm{(ind)}},x^{\mathrm{(ind)}},x^{\mathrm{(ind)}},x^{\mathrm{(ind)}};x^{\mathrm{(cls)}}\right),\cdots,
\]
\[
f_{8}\left(x^{\mathrm{(ind)}},x^{\mathrm{(ind)}},x^{\mathrm{(ind)}},x^{\mathrm{(ind)}},x^{\mathrm{(ind)}},x^{\mathrm{(ind)}},x^{\mathrm{(ind)}},x^{\mathrm{(ind)}};x^{\mathrm{(cls)}}\right)
\]
 are continuous on the neighborhood $\mathcal{N}$.
\item There exits some sequence $\left\{ c_{n}\right\} $ satisfying the condition $\eqref{eq:cn}$ such that for any $g=1,\cdots,G$ and for any $j=1,\cdots,n_{g}$, we have $\left\Vert X_{gj}\right\Vert \leq c_{n}$ with probability approaching one.
\end{enumerate}
\end{assumption}
\begin{rem}
Assumption $\text{\ref{assu:condvar}}$ (i) and (ii) correspond to $nh^{d}\rightarrow\infty$, $h\rightarrow0$, and $\left(\max_{g\leq G}n_{g}\right)h^{d_{\mathrm{ind}}}=O(1)$ in the marginal density estimation. Assumption $\text{\ref{assu:condvar}}$ (iii) and (iv) are stronger than Assumption $\text{\ref{assu:nw}}$ so that we can cover regressors $\left(X_{gj}^{(\mathrm{ind})\top},X_{g\ell}^{(\mathrm{ind})\top},X_{g}^{\mathrm{(cls)}\top}\right)^{\top}$ constructed by two observations $X_{gj}$ and $X_{g\ell}$. A sufficient condition for Assumption $\text{\ref{assu:condvar}}$ (v) is $\mathbb{E}\left\Vert X_{gj}\right\Vert <\infty$ since Markov's inequality implies
\[
\Pr\left(\left\Vert X_{gj}\right\Vert \geq c_{n}\right)\leq\mathbb{E}\left\Vert X_{gj}\right\Vert /c_{n}\rightarrow0
\]
if we choose $c_{n}\rightarrow\infty$.
\end{rem}
\begin{thm}
\textbf{\label{thm:joint_density_consistency}(Consistency of the joint density estimator)} Suppose that Assumption $\ref{assu:condvar}$ holds. Then,
\[
\widehat{f}_{2}\left(x^{\mathrm{(ind)}},x^{\mathrm{(ind)}};x^{\mathrm{(cls)}}\right)\overset{p}{\rightarrow}f_{2}\left(x^{\mathrm{(ind)}},x^{\mathrm{(ind)}};x^{\mathrm{(cls)}}\right).
\]
\end{thm}
Next, we will consider conditional variance and covariance estimation. We only provide Nadaraya-Watson type estimators, but they can be easily extended to local linear type ones. Since the goal here is to estimate $\sigma^{2}(x)$, we can estimate it as we did for $m(x)$. The infeasible Nadaraya-Watson estimator is
\[
\widehat{\sigma}_{\mathrm{nw}}^{2*}\left(x\right)=\frac{\sum_{g=1}^{G}\sum_{j=1}^{n_{g}}K\left(\frac{X_{gj}-x}{h}\right)e_{gj}^{2}}{\sum_{g=1}^{G}\sum_{j=1}^{n_{g}}K\left(\frac{X_{gj}-x}{h}\right)},
\]
This estimator is infeasible because $e_{gj}$ is unknown. We can replace it by $\widehat{e}_{gj}=Y_{gj}-\widehat{m}_{*}\left(X_{gj}\right)$ with $\widehat{m}_{*}(x)=\widehat{m}_{\mathrm{nw}}(x)$ or $\widehat{m}_{\mathrm{LL}}(x)$. The feasible variance estimator of the conditional variance is 
\begin{equation}
\widehat{\sigma}_{\mathrm{nw}}^{2}\left(x\right)=\frac{\sum_{g=1}^{G}\sum_{j=1}^{n_{g}}K\left(\frac{X_{gj}-x}{h}\right)\widehat{e}_{gj}^{2}}{\sum_{g=1}^{G}\sum_{j=1}^{n_{g}}K\left(\frac{X_{gj}-x}{h}\right)}.\label{eq:nw_condvar}
\end{equation}
 The following theorem shows $\widehat{\sigma}_{\mathrm{nw}}^{2}\left(x\right)$ is a consistent estimator. 
\begin{thm}
\textbf{\label{thm:nw_condvar_cons}(Consistency of the variance estimator)} Let $\widehat{e}_{gj}=Y_{gj}-\widehat{m}_{*}\left(X_{gj}\right)$ and $\widehat{m}_{*}(x)=\widehat{m}_{\mathrm{nw}}(x)$ or $\widehat{m}_{\mathrm{LL}}(x)$. Suppose that the assumptions for Theorem $\text{\ref{thm:nw_unifconv}}$ and Assumption $\ref{assu:condvar}$ hold. Then,
\begin{equation}
\widehat{\sigma}_{\mathrm{nw}}^{2}\left(x\right)\overset{p}{\rightarrow}\sigma^{2}\left(x\right).\label{eq:nw_condvar_cons}
\end{equation}
\end{thm}
Similar to the joint density estimator, we can construct a Nadaraya-Watson type estimator for the conditional covariance using $\left(2d_{\mathrm{ind}}+d_{\mathrm{cls}}\right)$-dimensional regressors $\left(X_{gj}^{(\mathrm{ind})\top},X_{g\ell}^{(\mathrm{ind})\top},X_{g}^{\mathrm{(cls)}\top}\right)^{\top}$:
\begin{eqnarray*}
 &  & \widehat{\sigma}_{\mathrm{nw}}^{*}\left(x^{\mathrm{(ind)}},x^{\mathrm{(ind)}};x^{\mathrm{(cls)}}\right)\\
 & = & \frac{\sum_{g:n_{g}\geq2}\sum_{1\leq j<\ell\leq n_{g}}K\left(\frac{\left(X_{gj}^{(\mathrm{ind})\top},X_{g\ell}^{(\mathrm{ind})\top},X_{g}^{\mathrm{(cls)}\top}\right)^{\top}-\left(x^{(\mathrm{ind})\top},x^{(\mathrm{ind})\top},x^{\mathrm{(cls)}\top}\right)^{\top}}{b}\right)e_{gj}e_{g\ell}}{\sum_{g:n_{g}\geq2}\sum_{1\leq j<\ell\leq n_{g}}K\left(\frac{\left(X_{gj}^{(\mathrm{ind})\top},X_{g\ell}^{(\mathrm{ind})\top},X_{g}^{\mathrm{(cls)}\top}\right)^{\top}-\left(x^{(\mathrm{ind})\top},x^{(\mathrm{ind})\top},x^{\mathrm{(cls)}\top}\right)^{\top}}{b}\right)}.
\end{eqnarray*}
Because $e_{gj}$ is unknown, it is infeasible as $\widehat{\sigma}_{\mathrm{nw}}^{2*}\left(x\right)$. The feasible version of $\widehat{\sigma}_{\mathrm{nw}}^{2*}$ is estimated by replacing $e_{gj}$ with $\widehat{e}_{gj}=Y_{gj}-\widehat{m}_{*}\left(X_{gj}\right)$,
\begin{eqnarray}
 &  & \widehat{\sigma}_{\mathrm{nw}}\left(x^{\mathrm{(ind)}},x^{\mathrm{(ind)}};x^{\mathrm{(cls)}}\right)\nonumber \\
 & = & \frac{\sum_{g:n_{g}\geq2}\sum_{1\leq j<\ell\leq n_{g}}K\left(\frac{\left(X_{gj}^{(\mathrm{ind})\top},X_{g\ell}^{(\mathrm{ind})\top},X_{g}^{\mathrm{(cls)}\top}\right)^{\top}-\left(x^{(\mathrm{ind})\top},x^{(\mathrm{ind})\top},x^{\mathrm{(cls)}\top}\right)^{\top}}{b}\right)\widehat{e}_{gj}\widehat{e}_{g\ell}}{\sum_{g:n_{g}\geq2}\sum_{1\leq j<\ell\leq n_{g}}K\left(\frac{\left(X_{gj}^{(\mathrm{ind})\top},X_{g\ell}^{(\mathrm{ind})\top},X_{g}^{\mathrm{(cls)}\top}\right)^{\top}-\left(x^{(\mathrm{ind})\top},x^{(\mathrm{ind})\top},x^{\mathrm{(cls)}\top}\right)^{\top}}{b}\right)}.\label{eq:nw_condcov}
\end{eqnarray}

\begin{thm}
\textbf{\label{thm:nw_condcov_cons}(Consistency of the covariance estimator)} Let $\widehat{e}_{gj}=Y_{gj}-\widehat{m}_{*}\left(X_{gj}\right)$ and $\widehat{m}_{*}(x)=\widehat{m}_{\mathrm{nw}}(x)$ or $\widehat{m}_{\mathrm{LL}}(x)$. Suppose that the assumptions for Theorem $\text{\ref{thm:nw_unifconv}}$ and Assumption $\text{\ref{assu:condvar}}$ hold. Then,
\begin{equation}
\widehat{\sigma}_{\mathrm{nw}}\left(x^{\mathrm{(ind)}},x^{\mathrm{(ind)}};x^{\mathrm{(cls)}}\right)\overset{p}{\rightarrow}\sigma\left(x^{\mathrm{(ind)}},x^{\mathrm{(ind)}};x^{\mathrm{(cls)}}\right).\label{eq:nw_condcov_cons}
\end{equation}
\end{thm}
\begin{cor}
\textbf{\label{cor:nw_asydist_condvar} }Let $\widehat{\lambda}=\left(\frac{1}{n}\sum_{g=1}^{G}n_{g}^{2}\right)h^{d_{\mathrm{ind}}}$. Let $\widehat{m}_{*}(x)=\widehat{m}_{\mathrm{nw}}(x)$ and $B_{*}(x)=B_{\mathrm{nw}}(x)$ (or $\widehat{m}_{*}(x)=\widehat{m}_{\mathrm{LL}}(x)$ and $B_{*}(x)=B_{\mathrm{LL}}(x)$). Suppose that the assumptions for Theorem $\text{\ref{thm:nw_asy_dist}}$ (or Theorem $\text{\ref{thm:LL_asy_dist}}$, respectively), Theorem $\text{\ref{thm:nw_condvar_cons}}$, and Theorem $\text{\ref{thm:nw_condcov_cons}}$ hold. Then,
\begin{eqnarray}
 &  & \left(\frac{R_{k}^{d}\widehat{\sigma}_{\mathrm{nw}}^{2}\left(x\right)}{\widehat{f}(x)}+\frac{\widehat{\lambda}R_{k}^{d_{\mathrm{cls}}}\widehat{f}_{2}\left(x^{\mathrm{(ind)}},x^{\mathrm{(ind)}};x^{\mathrm{(cls)}}\right)\widehat{\sigma}_{\mathrm{nw}}\left(x^{\mathrm{(ind)}},x^{\mathrm{(ind)}};x^{\mathrm{(cls)}}\right)}{\left(\widehat{f}(x)\right)^{2}}\right)^{-1/2}\nonumber \\
 &  & \times\sqrt{nh^{d}}\left(\widehat{m}_{*}(x)-m(x)-h^{2}B_{*}(x)\right)\nonumber \\
 &  & \qquad\overset{d}{\longrightarrow}\mathrm{N}\left(0,1\right).\label{eq:asydist_condvar}
\end{eqnarray}
\end{cor}
\label{rem:cov_issue}Corollary $\text{\ref{cor:nw_asydist_condvar}}$ suggests to use 
\begin{equation}
\sqrt{\frac{1}{nh^{d}}}\sqrt{\frac{R_{k}^{d}\widehat{\sigma}_{\mathrm{nw}}^{2}\left(x\right)}{\widehat{f}(x)}+\frac{\widehat{\lambda}R_{k}^{d_{\mathrm{cls}}}\widehat{f}_{2}\left(x^{\mathrm{(ind)}},x^{\mathrm{(ind)}};x^{\mathrm{(cls)}}\right)\widehat{\sigma}_{\mathrm{nw}}\left(x^{\mathrm{(ind)}},x^{\mathrm{(ind)}};x^{\mathrm{(cls)}}\right)}{\left(\widehat{f}(x)\right)^{2}}}\label{eq:se_est}
\end{equation}
 as a standard error. The estimator $\widehat{\lambda}R_{k}^{d_{\mathrm{cls}}}\widehat{f}_{2}\left(x^{\mathrm{(ind)}},x^{\mathrm{(ind)}};x^{\mathrm{(cls)}}\right)\widehat{\sigma}_{\mathrm{nw}}\left(x^{\mathrm{(ind)}},x^{\mathrm{(ind)}};x^{\mathrm{(cls)}}\right)/\left(\widehat{f}(x)\right)^{2}$ could be too difficult to estimate in practice for the following two main reasons. First, it contains $\widehat{f}_{2}\left(x^{\mathrm{(ind)}},x^{\mathrm{(ind)}};x^{\mathrm{(cls)}}\right)$ and $\widehat{\sigma}_{\mathrm{nw}}\left(x^{\mathrm{(ind)}},x^{\mathrm{(ind)}};x^{\mathrm{(cls)}}\right)$, which put most kernel weights for observations that $X_{gj}^{(\mathrm{ind})}$ and $X_{g\ell}^{(\mathrm{ind})}$ are both in the neighborhood of $x^{\mathrm{(ind)}}$. In a finite sample, such observations could be rarely observed, and these estimators could be imprecise. Second, it contains a density ratio $\widehat{f}_{2}\left(x^{\mathrm{(ind)}},x^{\mathrm{(ind)}};x^{\mathrm{(cls)}}\right)/\widehat{f}(x)$, which is difficult to estimate, especially nonparametrically.

To overcome these difficulties, we provide a parametric compromise under additional assumptions. We assume that $f_{2}\left(x^{\mathrm{(ind)}},x^{\mathrm{(ind)}};x^{\mathrm{(cls)}}\right)$ follows a multivariate normal distribution, $x^{\mathrm{(ind)}}$ and $x^{\mathrm{(cls)}}$ are independent or there are no cluster-level regressors (see also Remark $\text{\ref{rem:simple_cov}}$), and the conditional covariance is homoskedastic. Then, we can simplify
\begin{eqnarray}
 &  & \frac{\widehat{\lambda}R_{k}^{d_{\mathrm{cls}}}\widehat{f}_{2}\left(x^{\mathrm{(ind)}},x^{\mathrm{(ind)}};x^{\mathrm{(cls)}}\right)\widehat{\sigma}_{\mathrm{nw}}\left(x^{\mathrm{(ind)}},x^{\mathrm{(ind)}};x^{\mathrm{(cls)}}\right)}{\left(\hat{f}(x)\right)^{2}}\nonumber \\
 & = & \widehat{\lambda}R_{k}^{d_{\mathrm{cls}}}\left(\frac{1}{N}\sum_{g:n_{g}\geq2}\sum_{1\leq j<\ell\leq n_{g}}\check{e}_{gj}\check{e}_{g\ell}\right)\frac{p\left(x^{\mathrm{(ind)}}\mid x^{\mathrm{(ind)}},\widehat{\mu},\widehat{\Sigma}\right)}{\widehat{f}(x)},\label{eq:cov_simplified}
\end{eqnarray}
where $\check{e}_{gj}=Y_{gj}-\check{m}_{-g}\left(X_{gj}\right)$, $\check{m}_{-g}\left(x\right)$ is estimated by the \textit{global} polynomial regression as $\text{\eqref{eq:global-g}}$, and $p\left(x_{1}\mid x_{2},\mu,\Sigma\right)$ is a conditional density function of $x_{1}$ given $x_{2}$ with the joint distribution $\left(x_{1}^{\top},x_{2}^{\top}\right)^{\top}\sim\mathrm{N}\left(\mu,\Sigma\right)$. We can estimate $\widehat{\mu}=\left(\widehat{\mu}_{1}^{\top},\widehat{\mu}_{1}^{\top}\right)^{\top}$ and $\widehat{\Sigma}=\left(\begin{array}{cc}
\widehat{\Sigma}_{11} & \widehat{\Sigma}_{12}\\
\widehat{\Sigma}_{12} & \widehat{\Sigma}_{11}
\end{array}\right)^{\top}$ easily by using sample moments. Note that the expectation $\widehat{\mu}_{1}$ and the variance matrix $\widehat{\Sigma}_{11}$ are the same for $x_{1}$ and $x_{2}$ since we initially assumed identical marginal densities in Assumption $\text{\ref{assu:dgp}}$.

In practice, we can estimate $\sigma^{2}\left(x\right)$ and $\sigma\left(x^{\mathrm{(ind)}},x^{\mathrm{(ind)}};x^{\mathrm{(cls)}}\right)$ by using clustered-level jackknife estimators. \citet{hansen2022jackknife} shows that for parametric linear regressions, clustered-level jackknife variance estimators are better than conventional variance estimators with respect to the worst-case bias. Clustered-level jackknife variance estimators $\widetilde{\sigma}_{\mathrm{nw}}^{2}\left(x\right)$ and $\widetilde{\sigma}_{\mathrm{nw}}\left(x^{\mathrm{(ind)}},x^{\mathrm{(ind)}};x^{\mathrm{(cls)}}\right)$ are estimated by replacing $e_{gj}$ with $\widetilde{e}_{gj}=Y_{gj}-\widetilde{m}_{-g}\left(X_{gj}\right)$, where $\widetilde{m}_{-g}\left(\cdot\right)$ is a nonparametric estimator estimated leaving out the $g$-th cluster observations. In the simulation section, we will compare coverage ratios of confidence intervals constructed by the conventional standard error $\widehat{\sigma}_{\mathrm{nw}}^{2}\left(x\right)$ and the cluster-robust standard error $\widetilde{\sigma}_{\mathrm{nw}}^{2}\left(x\right)$.

\begin{rem}
The theorems use the same bandwidth $h$ for $\widehat{m}_{*}(x)$, $\widehat{f}(x)$, and $\widehat{\sigma}_{\mathrm{nw}}^{2}\left(x\right)$, and the same bandwidth $b$ for $\widehat{f}_{2}\left(x^{\mathrm{(ind)}},x^{\mathrm{(ind)}};x^{\mathrm{(cls)}}\right)$ and $\widehat{\sigma}_{\mathrm{nw}}\left(x^{\mathrm{(ind)}},x^{\mathrm{(ind)}};x^{\mathrm{(cls)}}\right)$ for notational simplicity. However, we can easily extend these results to the case where different bandwidths are used (denoted by $h_{m},h_{f},h_{\sigma^{2}},b_{f},b_{\sigma}$) as long as $h_{m},h_{f},h_{\sigma^{2}}$ and $b_{f_{2}},b_{\sigma}$ have the same asymptotic orders as $h$ and $b$, respectively.
\end{rem}
\begin{rem}
\label{rem:bias_handling}Because the estimator must be centered by the unknown bias $h^{2}B_{*}(x)$ as well as the true value $m(x)$ in $\text{\eqref{eq:asydist_condvar}}$, the bias term should be considered in inference. There are three main ways to handle it. The first way is to ignore it. This ignorance could be justified by an undersmoothing assumption $nh^{d+4}=o(1)$. It is the simplest way but not ideal since the bias exists in a finite sample. Second, in the context of regression discontinuity designs, \citet{calonico2014robust} suggest estimating $B_{*}(x)$ nonparametrically and using a new standard error to take the randomness due to the bias estimation into account. Third, \citet{armstrong2018optimal} characterize finite sample optimal confidence intervals with the worst-case bias correction for i.i.d. observations. Comparing these procedures in the cluster dependence case is important, though it is outside of the scope of this paper. For simplicity, this paper uses the undersmoothing bandwidth in simulation and empirical illustration to disregard the asymptotic bias. The practical choice we recommend is $h_{\text{CR-CV}}\times n^{1/5}\times n^{-2/7}$, where $h_{\text{CR-CV}}$ is computed using the cluster-robust cross-validation. The factor $n^{1/5}\times n^{-2/7}$ is included for undersmoothing, as utilized, for example, in \citet{chernozhukov2013intersection}.
\end{rem}

\section{\textbf{Monte Carlo simulation\label{sec:Monte-Carlo-simulation}}}

In this section, we will check the validity of bandwidth selections and confidence intervals in simulated datasets under cluster sampling. For both simulation studies, we consider the following setup. We fix the number of clusters $G=100$ and cluster sizes $n_{g}=20$ for $g=1,\dots G-1$. To evaluate the effect of the largest cluster size, we try two cluster sizes $n_{G}\in\left\{ 20,100\right\} $ for cluster $g=G$. Thus, we try two scenarios with $\left(\max_{g\leq G}n_{g}\right)/n\approx\left\{ 0.02,0.09\right\} $, also corresponding to homogeneous or heterogeneous size clusters. We generated 2000 datasets for replication. For the data-generating process, the following two models are considered.

\textbf{Setup 1 (homoskedastic errors):}

\[
Y_{gj}=\sin\left(2X_{gj}\right)+2\exp\left(-16X_{gj}^{2}\right)+0.5e_{gj},
\]
where $X_{gj}=\sqrt{\rho_{X}}\left(X_{1}\right)_{g}+\sqrt{1-\rho_{X}}\left(X_{2}\right)_{gj}$, $e_{gj}=\sqrt{\rho_{e}}c_{g}+\sqrt{1-\rho_{e}}u_{gj}$, and we generate $\left(X_{1}\right)_{g}\sim\mathcal{N}\left(0,1\right)$, $\left(X_{2}\right)_{gj}\sim\mathcal{N}\left(0,1\right)$, $c_{g}\sim\mathcal{N}\left(0,1\right)$, $u_{gj}\sim\mathcal{N}\left(0,1\right)$ independently. We set $\rho_{X},\rho_{e}\in\{0.2,0.5\}$. Note that larger $\rho_{X}$ and $\rho_{e}$ imply stronger cluster dependence on the regressor and the error term, respectively.

\textbf{Setup 2 (heteroskedastic errors):}
\begin{align*}
Y_{gj} & =X_{gj}\sin\left(2\pi X_{gj}\right)+\sigma\left(X_{gj}\right)e_{gj},\\
\sigma\left(X_{gj}\right) & =\frac{2+\cos\left(2\pi X_{gj}\right)}{5},
\end{align*}
and $X_{gj}$ and $e_{gj}$ are generated in the same way as Setup 1.

A key feature is that Setup 1 has homoskedastic errors, and Setup 2 has heteroskedastic errors. We adopted the functional form $m(\cdot)$ for Setup 1 from \citet{fan1992variable} and Setup 2 from \citet{kai2010local}. The data-generating process for $X_{gj}$ and $e_{gj}$ are standard in the cluster dependence literature (\citealp{cameron2008bootstrap}; \citealp{bartalotti2017regression}). We set the weight function $w(x)$ for cross-validation and IAMSE equals to $w(x)=\mathbb{I}\left\{ \xi_{\mathrm{L}}\leq x\leq\xi_{\mathrm{U}}\right\} $, where we set $\xi_{\mathrm{L}}=-1.5$ and $\xi_{\mathrm{U}}=1.5$ for Setup 1, and $\xi_{\mathrm{L}}=0$ and $\xi_{\mathrm{U}}=1$ for Setup 2, respectively. For nonparametric regression, we use the Epachenikov kernel and local linear estimators. Results when using Nadaraya-Watson estimators are presented in Appendix $\text{\ref{sec:Add_sim}}$ because their values are similar to the ones by local linear estimators. 

\subsection{Bandwidth selection\label{subsec:sim_bw}}

We will compare four methods of bandwidth choice: (i) rule-of-thumb (ROT), (ii) cluster-robust rule-of-thumb (CR-ROT), (iii) cross-validation (CV), and cluster-robust cross-validation (CR-CV). $h_{\text{CR-ROT}}$ (Equation $\text{\ref{eq:CR-ROT}}$) and $h_{\text{CR-CV}}$ (Equation $\text{\ref{eq:h_cv}}$) are what we suggested. The ROT bandwidth choice $h_{\text{ROT}}$ is proposed by \citet{fan1996local} for i.i.d. observations. Instead of leave-one-cluster-out global fit as $\text{\eqref{eq:global-g}}$ for $h_{\text{CR-ROT}}$, it uses the global fit using the entire sample. $h_{\text{CV}}$ minimizes the cross-validation function. The difference between $h_{\text{CV}}$ and $h_{\text{CR-CV}}$ is that $h_{\text{CV}}$ minimizes a criterion based on leave-one-out prediction errors, while $h_{\text{CR-CV}}$ minimizes a criterion based on leave-one-\textit{cluster}-out prediction errors.

In simulation, we first compute $h_{\text{ROT}}$ and $h_{\text{CR-ROT}}$. Then, $h_{\text{CV}}$ and $h_{\text{CR-CV}}$ are found by the grid search for $50$ points over $[h_{\text{CR-ROT}}/3,3h_{\text{CR-ROT}}]$. The performance of the methods of bandwidth selection is evaluated by the average squared error (ASE):
\[
\operatorname{ASE}(h)=\frac{1}{n_{\text{grid }}}\sum_{k=1}^{n_{\text{grid }}}\left\{ \widehat{m}_{\mathrm{LL}}\left(u_{k},h\right)-m\left(u_{k}\right)\right\} ^{2},
\]
where $\widehat{m}_{\mathrm{LL}}\left(u_{k},h\right)$ is the local linear estimator with the bandwidth $h$, and $\text{\ensuremath{\left\{  u_{1},\dots,u_{n_{\text{grid }}}\right\} } }$ are the grid points to evaluate the performance. We set the number of the grid $n_{\text{grid }}=50$ and $\text{\ensuremath{\left\{  u_{1},\dots,u_{n_{\text{grid }}}\right\} } }$ are evenly distributed over $\left[\xi_{\mathrm{L}},\xi_{\mathrm{U}}\right]$. 

Tables $\text{\ref{tab:bw_LL1}}$ and $\text{\ref{tab:bw_LL2}}$ show means of the ASE for the local linear estimator and means of selected bandwidths (in curly brackets) across each simulation draw for Setup 1 and 2, respectively. Each table contains four methods of bandwidth choice in several scenarios. We consider combinations of homogeneous or heterogeneous size clusters, high or low cluster dependence on regressors, and high or low cluster dependence on error terms. In Setup 1 (Table $\text{\ref{tab:bw_LL1}}$, homoskedastic errors), $h_{\text{ROT}}$ and $h_{\text{CR-ROT}}$ have similar values of the ASE and the selected bandwidth, and $h_{\text{CV}}$ and $h_{\text{CR-CV}}$ have the similar values of them, but $h_{\text{CV}}$ and $h_{\text{CR-CV}}$ work better than $h_{\text{ROT}}$ and $h_{\text{CR-ROT}}$ in terms of the ASE. Within the same method of bandwidth choice, heterogeneous size clusters $n_{G}=100$ and high cluster dependence on regressors $\rho_{X}=0.5$ give a slightly larger ASE. Compared to them, high cluster dependence on error terms $\rho_{e}=0.5$ gives a much larger ASE. 

In Setup 2 (Table $\text{\ref{tab:bw_LL2}}$, heteroskedastic errors), $h_{\text{ROT}}$ and $h_{\text{CR-ROT}}$ work poorly because they assume homoskedasticity. Different from Setup 1, $h_{\text{ROT}}$ has a larger ASE than $h_{\text{CR-ROT}}$. As Setup 1, $h_{\text{CV}}$ and $h_{\text{CR-CV}}$ work well and have similar values of the ASE and the selected bandwidth. The good performance of $h_{\text{CV}}$ can not be explained by our theoretical results. We probably need an asymptotic analysis of $h_{\text{CV}}$ under the cluster dependence, which is outside of the scope of this paper.

\begin{table}
\caption{Mean of ASE and mean of selected bandwidth ($m_{\mathrm{LL}}$, Setup 1)\label{tab:bw_LL1}}

\centering
\begin{center} 
\begin{threeparttable}
{\small

\begin{tabular}{l c c c c c c c c}
\hline
 & \multicolumn{4}{c}{$\max n_g=20$} & \multicolumn{4}{c}{$\max n_g=100$} \\
\cline{2-5} \cline{6-9}
 & $h_{\text{ROT}}$ & $h_{\text{CR-ROT}}$ & $h_{\text{CV}}$ & $h_{\text{CR-CV}}$ & $h_{\text{ROT}}$ & $h_{\text{CR-ROT}}$ & $h_{\text{CV}}$ & $h_{\text{CR-CV}}$ \\
\hline
$(\rho_{X},\rho_{e})$=(0.2,0.2) & $0.0054$   & $0.0053$   & $0.0041$   & $0.0041$   & $0.0053$   & $0.0053$   & $0.0041$   & $0.0041$   \\
                                & $\{0.0297\}$ & $\{0.0302\}$ & $\{0.0482\}$ & $\{0.0483\}$ & $\{0.0292\}$ & $\{0.0297\}$ & $\{0.0477\}$ & $\{0.0479\}$ \\
$(\rho_{X},\rho_{e})$=(0.2,0.5) & $0.0062$   & $0.0061$   & $0.0049$   & $0.0049$   & $0.0063$   & $0.0062$   & $0.0050$   & $0.0050$   \\
                                & $\{0.0297\}$ & $\{0.0302\}$ & $\{0.0482\}$ & $\{0.0484\}$ & $\{0.0292\}$ & $\{0.0297\}$ & $\{0.0479\}$ & $\{0.0479\}$ \\
$(\rho_{X},\rho_{e})$=(0.5,0.2) & $0.0055$   & $0.0054$   & $0.0042$   & $0.0042$   & $0.0056$   & $0.0055$   & $0.0042$   & $0.0042$   \\
                                & $\{0.0292\}$ & $\{0.0300\}$ & $\{0.0484\}$ & $\{0.0486\}$ & $\{0.0288\}$ & $\{0.0295\}$ & $\{0.0482\}$ & $\{0.0484\}$ \\
$(\rho_{X},\rho_{e})$=(0.5,0.5) & $0.0066$   & $0.0065$   & $0.0052$   & $0.0052$   & $0.0068$   & $0.0067$   & $0.0054$   & $0.0054$   \\
                                & $\{0.0292\}$ & $\{0.0300\}$ & $\{0.0486\}$ & $\{0.0486\}$ & $\{0.0288\}$ & $\{0.0295\}$ & $\{0.0482\}$ & $\{0.0483\}$ \\
\hline
\end{tabular}
}
\begin{tablenotes}
\vspace{-0.5cm}
\footnotesize
\item \hspace{-0.5cm} \textit{Note: Means of selected bandwidths are shown in curly brackets.}
\end{tablenotes}\end{threeparttable}\end{center}
\end{table}
\begin{table}
\caption{Mean of ASE and mean of selected bandwidth ($m_{\mathrm{LL}}$, Setup 2)\label{tab:bw_LL2}}

\centering
\begin{center} 
\begin{threeparttable}
{\small

\begin{tabular}{l c c c c c c c c}
\hline
 & \multicolumn{4}{c}{$\max n_g=20$} & \multicolumn{4}{c}{$\max n_g=100$} \\
\cline{2-5} \cline{6-9}
 & $h_{\text{ROT}}$ & $h_{\text{CR-ROT}}$ & $h_{\text{CV}}$ & $h_{\text{CR-CV}}$ & $h_{\text{ROT}}$ & $h_{\text{CR-ROT}}$ & $h_{\text{CV}}$ & $h_{\text{CR-CV}}$ \\
\hline
$(\rho_{X},\rho_{e})$=(0.2,0.2) & $0.0096$   & $0.0080$   & $0.0028$   & $0.0028$   & $0.0090$   & $0.0076$   & $0.0027$   & $0.0028$   \\
                                & $\{0.0890\}$ & $\{0.0865\}$ & $\{0.0461\}$ & $\{0.0462\}$ & $\{0.0876\}$ & $\{0.0853\}$ & $\{0.0457\}$ & $\{0.0458\}$ \\
$(\rho_{X},\rho_{e})$=(0.2,0.5) & $0.0104$   & $0.0087$   & $0.0033$   & $0.0033$   & $0.0098$   & $0.0083$   & $0.0034$   & $0.0034$   \\
                                & $\{0.0893\}$ & $\{0.0868\}$ & $\{0.0461\}$ & $\{0.0462\}$ & $\{0.0878\}$ & $\{0.0855\}$ & $\{0.0457\}$ & $\{0.0459\}$ \\
$(\rho_{X},\rho_{e})$=(0.5,0.2) & $0.0098$   & $0.0084$   & $0.0029$   & $0.0029$   & $0.0096$   & $0.0082$   & $0.0029$   & $0.0029$   \\
                                & $\{0.0896\}$ & $\{0.0877\}$ & $\{0.0465\}$ & $\{0.0467\}$ & $\{0.0889\}$ & $\{0.0869\}$ & $\{0.0463\}$ & $\{0.0465\}$ \\
$(\rho_{X},\rho_{e})$=(0.5,0.5) & $0.0103$   & $0.0091$   & $0.0036$   & $0.0036$   & $0.0104$   & $0.0090$   & $0.0037$   & $0.0037$   \\
                                & $\{0.0892\}$ & $\{0.0874\}$ & $\{0.0464\}$ & $\{0.0466\}$ & $\{0.0886\}$ & $\{0.0866\}$ & $\{0.0461\}$ & $\{0.0463\}$ \\
\hline
\end{tabular}
}
\begin{tablenotes}
\vspace{-0.5cm}
\footnotesize
\item \hspace{-0.5cm} \textit{Note: Means of selected bandwidths are shown in curly brackets.}
\end{tablenotes}\end{threeparttable}\end{center}
\end{table}

\begin{figure}
\includegraphics[scale=0.6]{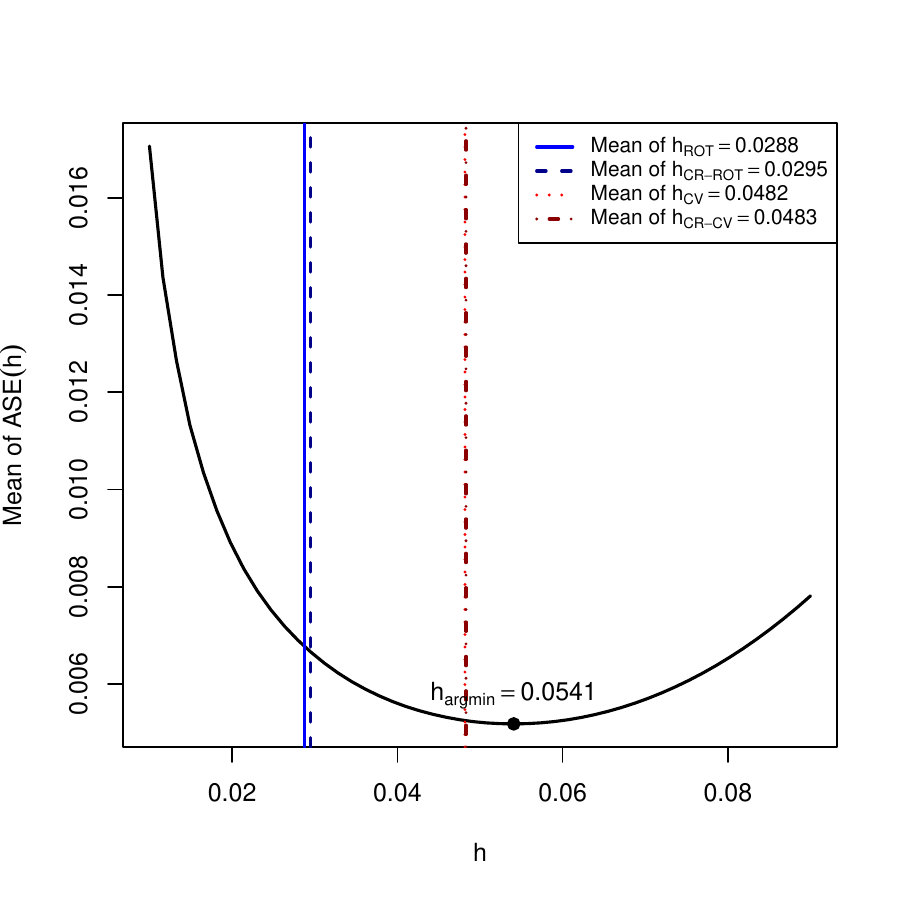}\caption{Mean of $\mathrm{ASE}(h)$ for $m_{\mathrm{LL}}$ in Setup 1 with $\max_{g\protect\leq G}n_{g}=100$ and $\rho_{X}=\rho_{e}=0.5$\label{fig:bw_LL1}}
\end{figure}

\begin{figure}
\includegraphics[scale=0.6]{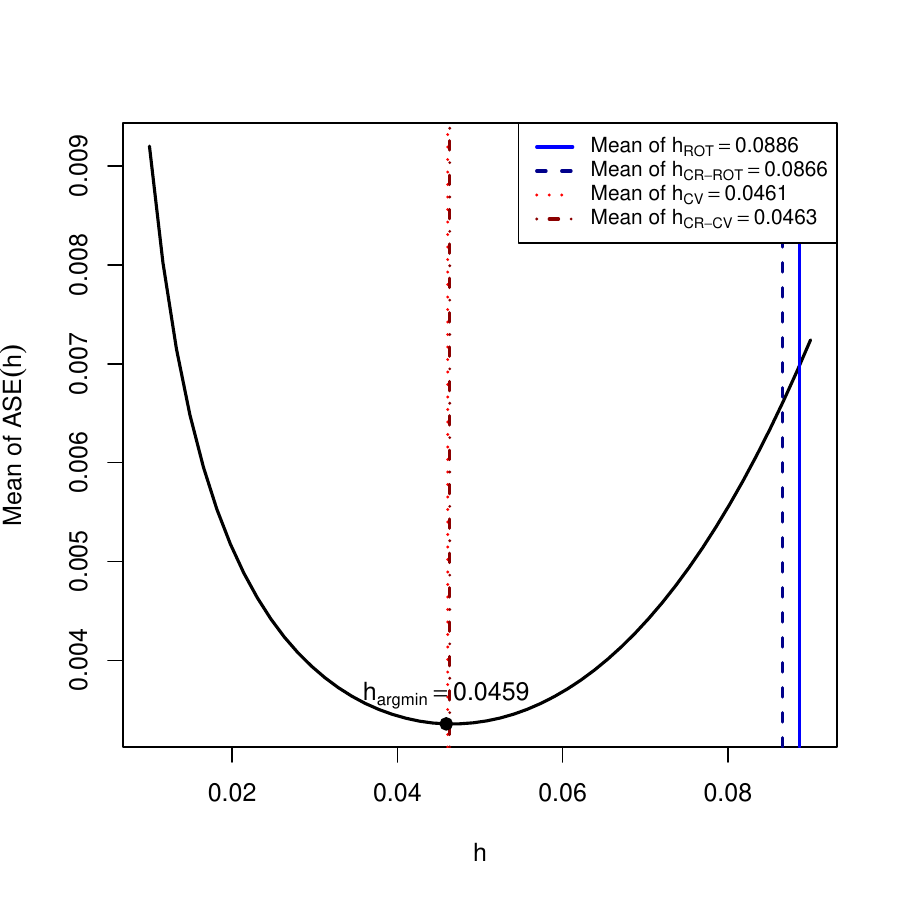}\caption{Mean of $\mathrm{ASE}(h)$ for $m_{\mathrm{LL}}$ in Setup 2 with $\max_{g\protect\leq G}n_{g}=100$ and $\rho_{X}=\rho_{e}=0.5$\label{fig:bw_LL2}}
\end{figure}

To investigate how close the selected bandwidths are to the bandwidth that minimizes the ASE, we plot two figures for a scenario with $n_{G}=100$ and $\rho_{X}=\rho_{e}=0.5$. Figures $\text{\ref{fig:bw_LL1}}$ and $\text{\ref{fig:bw_LL2}}$ have values of bandwidth $h$ in the $x$-axis and means of the function $\mathrm{ASE}(h)$ in the $y$-axis, which are calculated from simulation draws for Setup 1 and 2, respectively. These figures also contain means of selected bandwidths by four selection methods and $h_{\mathrm{argmin}}$ minimizing $\mathrm{ASE}(h)$. We find that $h_{\text{CV}}$ and $h_{\text{CR-CV}}$ are very close to $h_{\mathrm{argmin}}$ in both setups.


We recommend $h_{\text{CR-CV}}$ because it has a theoretical guarantee (Theorem $\text{\ref{thm:cr-cv}}$) and because it performs the best in our simulation, although the difference of ASEs between $h_{\text{CV}}$ and $h_{\text{CR-CV}}$ is subtle. In terms of the computational cost, $h_{\text{CR-CV}}$ is also better than $h_{\text{CV}}$ since leave-one-cluster-out estimators use smaller sample sizes than leave-one-out estimators do. $h_{\text{CR-ROT}}$ is useful for a rough estimation and for choosing the range of the grid search in cross-validation. We recommend $h_{\text{CR-ROT}}$ over $h_{\text{ROT}}$ for these purposes because it has a smaller ASE.

\subsection{Inference\label{subsec:sim_CI}}

We will compare three methods to calculate 95\% confidence intervals: (i) using the conventional standard error as for i.i.d. datasets ($CI$), (ii) using the cluster-robust standard error without the term related to the conditional covariance ($CI_{\mathrm{CR}}$), and (iii) using the cluster-robust standard error with the term related to the conditional covariance ($CI_{\lambda}$). More precisely, we calculate $CI$ with the standard error $\sqrt{R_{k}^{d}\widehat{\sigma}_{\mathrm{nw}}^{2}\left(x\right)/\left(nh^{d}\widehat{f}(x)\right)}$, $CI_{\mathrm{CR}}$ with the standard error $\sqrt{R_{k}^{d}\widetilde{\sigma}_{\mathrm{nw}}^{2}\left(x\right)/\left(nh^{d}\widehat{f}(x)\right)}$, and $CI_{\mathrm{\lambda}}$ with the standard error 
\[
\sqrt{\frac{1}{nh^{d}}}\sqrt{\frac{R_{k}^{d}\widetilde{\sigma}_{\mathrm{nw}}^{2}\left(x\right)}{\widehat{f}(x)}+\frac{\widehat{\lambda}\widehat{f}_{2}\left(x,x\right)\widehat{\sigma}_{\mathrm{nw}}\left(x,x\right)}{\left(\widehat{f}(x)\right)^{2}}},
\]
where $\widehat{\sigma}_{\mathrm{nw}}^{2}\left(x\right)$ and $\widetilde{\sigma}_{\mathrm{nw}}^{2}\left(x\right)$ are nonparametrically estimated with $\widehat{e}_{gj}=Y_{gj}-\widehat{m}_{\mathrm{LL}}\left(X_{gj}\right)$ and $\widetilde{e}_{gj}=Y_{gj}-\widetilde{m}_{\mathrm{LL},-g}\left(X_{gj}\right)$, and $\widehat{\lambda}\widehat{f}\left(x,x\right)\widehat{\sigma}_{\mathrm{nw}}\left(x,x\right)$ is calculated parametrically as $\text{\eqref{eq:cov_simplified}}$. Note that in our data-generating processes, we have no cluster-level regressor $x^{\mathrm{(cls)}}$. In nonparametric regressions, bandwidths are selected as follows. The bandwidth $h_{f}$ for $\widehat{f}\left(x\right)$ is calculated by the reference bandwidth of the Epanechnikov kernel $h_{f}\approx1.049\cdot S_{X}\cdot n^{-1/5}$ where $S_{X}$ is a standard deviation of $X$ (e.g., see \citealp{li2007nonparametric}, Section 1.2). The bandwidth $h_{\sigma^{2}}$ for $\widehat{\sigma}_{\mathrm{nw}}^{2}\left(x\right)$ and $\widetilde{\sigma}_{\mathrm{nw}}^{2}\left(x\right)$ is set to $h_{f}$. Choosing $h_{\sigma^{2}}=h_{f}$ is a conventional choice, for example, used by \citet{imbens2012optimal}. In this simulation, $\ensuremath{\widehat{\lambda}=20\cdot h_{m}}$ for $n_{G}=20$ and $\widehat{\lambda}\approx23.846\cdot h_{m}$ for $n_{G}=100$. To ignore the asymptotic bias, we set the bandwidth for $\hat{m}_{\mathrm{LL}}(x)$ using an undersmoothing approach defined as $h_{m}=h_{\text{CR-CV}}\times n^{1/5}\times n^{-2/7}$, where $h_{\text{CR-CV}}$ is computed using the CR-CV method as outlined in Section $\text{\ref{subsec:sim_bw}}$.

Appendix $\text{\ref{sec:Add_sim}}$ contains results without undersmoothing, where we set $h_{m}=h_{\text{CR-CV}}$. In the appendix, we explore two additional strategies on the bias: an infeasible analytical bias correction using knowledge of the true data-generating process, and simply ignoring the bias term.

The CIs are constructed at $x=0.75$ for Setup 1 and at $x=0.8$ and $0.4$ for Setup 2. Performances of confidence intervals are measured by the coverage ratio across each simulation draw. 

Tables $\text{\ref{tab:CI_LL1}}$-$\text{\ref{tab:CI_LL2_0.4}}$ show the coverage ratio for local linear estimators and means of the length of confidence intervals (in curly brackets) across each simulation draw for Setup 1, Setup 2 with $x=0.8$, and Setup 2 with $x=0.4$, respectively. Each table contains results for three types of confidence intervals in several scenarios. As in Section $\text{\ref{subsec:sim_bw}}$, we consider 8 different scenarios with all possible combinations of $n_{G}\in\left\{ 20,100\right\} $, $\rho_{X}\in\{0.2,0.5\}$ and $\rho_{e}\in\{0.2,0.5\}$.

\begin{table}
\caption{Coverage and mean of length of 95\% CI for each standard error ($m_{\mathrm{LL}}$, Setup 1, undersmoothing)\label{tab:CI_LL1}}

\centering
\begin{center} 
\begin{threeparttable}
{\small

\begin{tabular}{l c c c c c c}
\hline
 & \multicolumn{3}{c}{$\max n_g=20$} & \multicolumn{3}{c}{$\max n_g=100$} \\
\cline{2-4} \cline{5-7}
 & $CI$ & $CI_{\text{CR}}$ & $CI_{\lambda}$ & $CI$ & $CI_{\text{CR}}$ & $CI_{\lambda}$ \\
\hline
$(\rho_{X},\rho_{e})$=(0.2,0.2) & $0.930$   & $0.936$   & $0.948$   & $0.923$   & $0.933$   & $0.951$   \\
                                & $\{0.261\}$ & $\{0.268\}$ & $\{0.284\}$ & $\{0.257\}$ & $\{0.264\}$ & $\{0.283\}$ \\
$(\rho_{X},\rho_{e})$=(0.2,0.5) & $0.908$   & $0.918$   & $0.955$   & $0.903$   & $0.913$   & $0.945$   \\
                                & $\{0.260\}$ & $\{0.268\}$ & $\{0.307\}$ & $\{0.256\}$ & $\{0.264\}$ & $\{0.309\}$ \\
$(\rho_{X},\rho_{e})$=(0.5,0.2) & $0.920$   & $0.930$   & $0.954$   & $0.923$   & $0.931$   & $0.954$   \\
                                & $\{0.260\}$ & $\{0.267\}$ & $\{0.292\}$ & $\{0.256\}$ & $\{0.263\}$ & $\{0.291\}$ \\
$(\rho_{X},\rho_{e})$=(0.5,0.5) & $0.901$   & $0.908$   & $0.959$   & $0.887$   & $0.897$   & $0.958$   \\
                                & $\{0.260\}$ & $\{0.268\}$ & $\{0.320\}$ & $\{0.256\}$ & $\{0.264\}$ & $\{0.323\}$ \\
\hline
\end{tabular}
}
\begin{tablenotes}
\vspace{-0.5cm}
\footnotesize
\item \hspace{-0.5cm} \textit{Note: Lengths of confidence intervals are shown in curly brackets.}
\end{tablenotes}\end{threeparttable}\end{center}
\end{table}

\begin{table}
\caption{Coverage and mean of length of 95\% CI for each standard error ($m_{\mathrm{LL}}$, Setup 2, $x=0.8$, undersmoothing)\label{tab:CI_LL2_0.8}}

\centering
\begin{center} 
\begin{threeparttable}
{\small

\begin{tabular}{l c c c c c c}
\hline
 & \multicolumn{3}{c}{$\max n_g=20$} & \multicolumn{3}{c}{$\max n_g=100$} \\
\cline{2-4} \cline{5-7}
& $CI$ & $CI_{\text{CR}}$ & $CI_{\lambda}$ & $CI$ & $CI_{\text{CR}}$ & $CI_{\lambda}$ \\
\hline
$(\rho_{X},\rho_{e})$=(0.2,0.2) & $0.893$   & $0.904$   & $0.920$   & $0.897$   & $0.904$   & $0.921$   \\
                                & $\{0.230\}$ & $\{0.238\}$ & $\{0.250\}$ & $\{0.228\}$ & $\{0.235\}$ & $\{0.249\}$ \\
$(\rho_{X},\rho_{e})$=(0.2,0.5) & $0.873$   & $0.885$   & $0.922$   & $0.867$   & $0.881$   & $0.919$   \\
                                & $\{0.230\}$ & $\{0.238\}$ & $\{0.267\}$ & $\{0.227\}$ & $\{0.235\}$ & $\{0.268\}$ \\
$(\rho_{X},\rho_{e})$=(0.5,0.2) & $0.885$   & $0.896$   & $0.923$   & $0.875$   & $0.887$   & $0.920$   \\
                                & $\{0.230\}$ & $\{0.237\}$ & $\{0.253\}$ & $\{0.226\}$ & $\{0.234\}$ & $\{0.252\}$ \\
$(\rho_{X},\rho_{e})$=(0.5,0.5) & $0.849$   & $0.866$   & $0.920$   & $0.836$   & $0.852$   & $0.920$   \\
                                & $\{0.230\}$ & $\{0.238\}$ & $\{0.275\}$ & $\{0.226\}$ & $\{0.235\}$ & $\{0.277\}$ \\
\hline
\end{tabular}
}
\begin{tablenotes}
\vspace{-0.5cm}
\footnotesize
\item \hspace{-0.5cm} \textit{Note: Lengths of confidence intervals are shown in curly brackets.}
\end{tablenotes}\end{threeparttable}\end{center}
\end{table}

\begin{table}
\caption{Coverage and mean of length of 95\% CI for each standard error ($m_{\mathrm{LL}}$, Setup 2, $x=0.4$, undersmoothing)\label{tab:CI_LL2_0.4}}

\centering
\begin{center} 
\begin{threeparttable}
{\small

\begin{tabular}{l c c c c c c}
\hline
 & \multicolumn{3}{c}{$\max n_g=20$} & \multicolumn{3}{c}{$\max n_g=100$} \\
\cline{2-4} \cline{5-7}
& $CI$ & $CI_{\text{CR}}$ & $CI_{\lambda}$ & $CI$ & $CI_{\text{CR}}$ & $CI_{\lambda}$ \\
\hline
$(\rho_{X},\rho_{e})$=(0.2,0.2) & $0.996$   & $0.996$   & $1.000$   & $0.996$   & $0.999$   & $1.000$   \\
                                & $\{0.188\}$ & $\{0.192\}$ & $\{0.206\}$ & $\{0.185\}$ & $\{0.189\}$ & $\{0.205\}$ \\
$(\rho_{X},\rho_{e})$=(0.2,0.5) & $0.991$   & $0.993$   & $0.999$   & $0.991$   & $0.993$   & $0.998$   \\
                                & $\{0.188\}$ & $\{0.193\}$ & $\{0.226\}$ & $\{0.184\}$ & $\{0.189\}$ & $\{0.227\}$ \\
$(\rho_{X},\rho_{e})$=(0.5,0.2) & $0.995$   & $0.996$   & $0.998$   & $0.995$   & $0.996$   & $0.998$   \\
                                & $\{0.187\}$ & $\{0.191\}$ & $\{0.208\}$ & $\{0.184\}$ & $\{0.188\}$ & $\{0.208\}$ \\
$(\rho_{X},\rho_{e})$=(0.5,0.5) & $0.988$   & $0.992$   & $0.999$   & $0.986$   & $0.991$   & $0.998$   \\
                                & $\{0.187\}$ & $\{0.192\}$ & $\{0.231\}$ & $\{0.183\}$ & $\{0.188\}$ & $\{0.233\}$ \\
\hline
\end{tabular}
}
\begin{tablenotes}
\vspace{-0.5cm}
\footnotesize
\item \hspace{-0.5cm} \textit{Note: Lengths of confidence intervals are shown in curly brackets.}
\end{tablenotes}\end{threeparttable}\end{center}
\end{table}

In Setup 1 (Table $\text{\ref{tab:CI_LL1}}$, homoskedastic errors), $CI_{\mathrm{CR}}$ has slightly better coverages than $CI$ does although both confidence intervals have severe under-coverage values when $\rho_{e}=0.5$. These confidence intervals work more poorly for the case $\max_{g\leq G}n_{g}=100$. On the other hand, $CI_{\lambda}$ performs the best among the three methods. It has accurate coverage (94.5\%-96\%) for every data-generating process. 

For Setup 2 (heteroskedastic errors), we consider two different points (Table $\text{\ref{tab:CI_LL2_0.8}}$ for $x=0.8$ and Table $\text{\ref{tab:CI_LL2_0.4}}$ for $x=0.4$). Table $\text{\ref{tab:CI_LL2_0.8}}$ shows that $CI_{\lambda}$ improves the accuracy greatly, and it attains coverage ratios close to 95\%. $CI_{\mathrm{CR}}$ and $CI$ fail to reach even 90\% coverage ratios for almost all cases. However, Table $\text{\ref{tab:CI_LL2_0.4}}$ shows that all three methods have 95\% coverage ratios, and $CI_{\lambda}$ has over-coverage values at $x=0.4$. Differences between Table $\text{\ref{tab:CI_LL2_0.8}}$ and Table $\text{\ref{tab:CI_LL2_0.4}}$ come from the functional form of the error term $\sigma\left(X_{gj}\right)e_{gj}$. Since $\sigma\left(x\right)=\left(2+\cos\left(2\pi x\right)\right)/5$ takes a large value at $x=0.8$ and a small value at $x=0.4$, the conditional variance and covariance of error terms also do so. Overall, $CI_{\lambda}$ is the most conservative choice among the three methods. Our proposed confidence interval $CI_{\lambda}$ performs well even without undersmoothing (see Appendix $\text{\ref{sec:Add_sim}}$).

We recommend $CI_{\lambda}$ because it works the best for homoskedastic errors, and it provides a conservative interval for heteroskedastic errors in our simulation.

\section{\textbf{Empirical Illustration \label{sec:Empirical}}}

In this section, we will apply our methods to a dataset from \citet{alatas2012targeting},\footnote{Their replication package, including datasets, is available on the AEA website.} which ran an experiment in 640 Indonesian subvillages with heterogeneous cluster sizes from 17 to 72. The purpose is to investigate a good way to target people with low incomes. In their subvillage-level randomized assignments, they compare three different ways of targeting: using demographic characteristics as proxies of income, using the community knowledge on the ranking of wealth (community targeting), and using a hybrid of them. The wealth ranking for community targeting was measured as follows. In each subvillage, people were asked to rank everyone in the community from the richest to the poorest. A facilitator used randomly ordered index cards, each representing a household. Starting the first two cards, the facilitator asked the community which household was better off in terms of wealth. Based on the community\textquoteright s response, the cards were placed with the wealth order. By sequentially adding one more index card to the comparison, the facilitator continued the process until all the households had been ranked.

One concern for this ranking process is that human errors could happen since it took 1.68 hours on average. \citet{alatas2012targeting} investigated this concern by running a nonparametric regression of the mistarget rate ($Y_{gj}$) on the card order in the ranking process ($X_{gj}$). The mistarget rate is calculated based on the household\textquoteright s per capita consumption. The card orders in the ranking process are scaled from $0$ to $1$. In nonparametric regression, error terms may exhibit dependence within the same cluster due to unobserved subvillage-level shocks during the ranking process (e.g., instances where some individuals leave the room, distracting others) or strategic interactions (e.g., groups colluding to appear poorer than they actually are to secure future aid). These dependencies are captured by the subvillage-level cluster random effects.\footnote{\citet{alatas2012targeting} state that ``Since the targeting methods were assigned at the subvillage level, the standard errors are clustered to allow for arbitrary correlation within a subvillage.'' in a parametric regression analysis.} We revisit \citet{alatas2012targeting} with theoretically justified methods for cluster sampling. We will use the local linear regression with the Epachenikov kernel while \citet{alatas2012targeting} used the local linear regression with the quartic kernel (what they call nonparametric Fan regression). Since the regressor is an observed variable rather than a treatment assignment, the model-based approach is taken.\footnote{Although cluster sizes were randomly selected in their study, our theoretical results remain applicable when conditioning on the realized cluster sizes.}

By the random card order, it is reasonable to assume that the regressor $X_{gj}$ is independent within the cluster (subvillage). Since the distribution of $X_{gj}$ does not follow from $U[0,1]$ due to the lack of observations on the mistarget rate $Y_{gj}$, we also estimate it nonparametrically. Thanks to the independence of the regressor, we can estimate the joint density by the product of marginal densities. Other detailed calculations for the bandwidth selection and standard errors are done in the same way as in Section $\text{\ref{sec:Monte-Carlo-simulation}}$.

\begin{figure}
\includegraphics[scale=0.6]{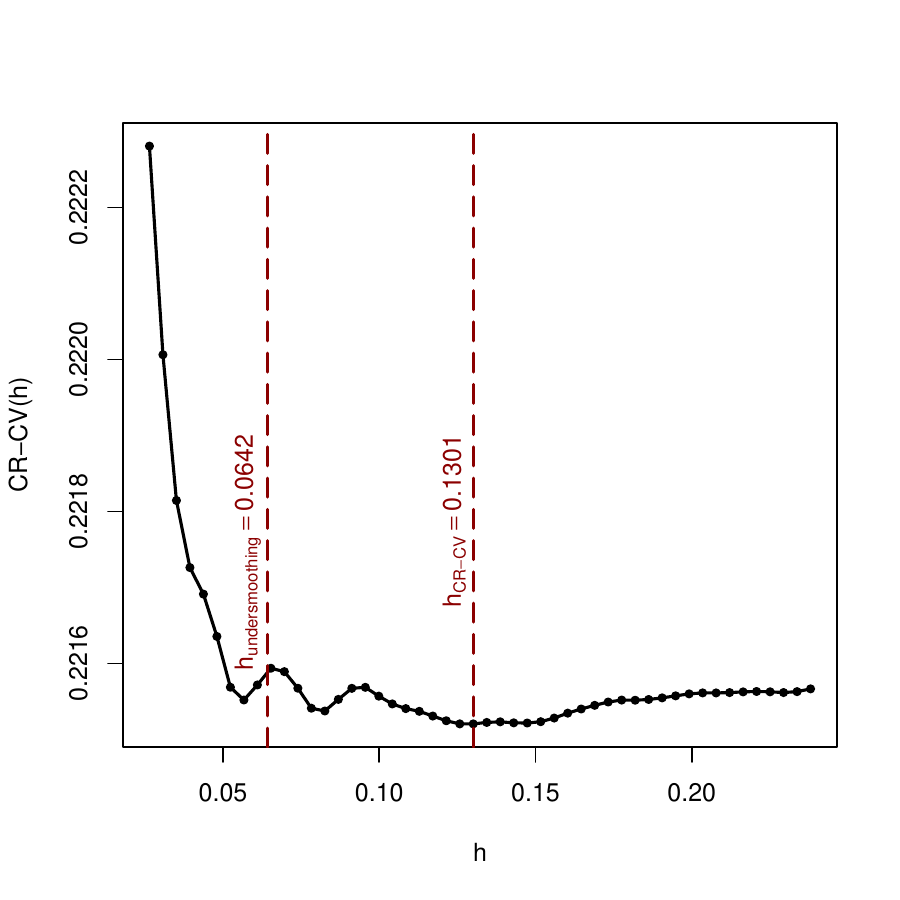} \caption{Cluster-robust cross-validation function $\mathrm{CV}(h)$\label{fig:cv_Alatas}}
\end{figure}

The sub-dataset for the above regression contains $n=3784$ observations, $G=431$ subvillages, and each subvillage has from 4 to 9 observations. Thus, $\max_{g\leq G}n_{g}$ is $9$. The bandwidth selected by CR-CV was $h_{\text{CR-CV}}=0.1301$, and the undersmoothing version, obtained by multiplying by the factor $n^{1/5}\times n^{-2/7}$ was $h_{\text{undersmoothing}}=0.0642$. In contrast, \citet{alatas2012targeting} heuristically choose a bandwidth of $(\max(X_{gj})-\min(X_{gj}))/5=0.1979$. We plot the cluster-robust cross-validation function in Figure $\text{\ref{fig:cv_Alatas}}$. 

\begin{figure}
\includegraphics[scale=0.8]{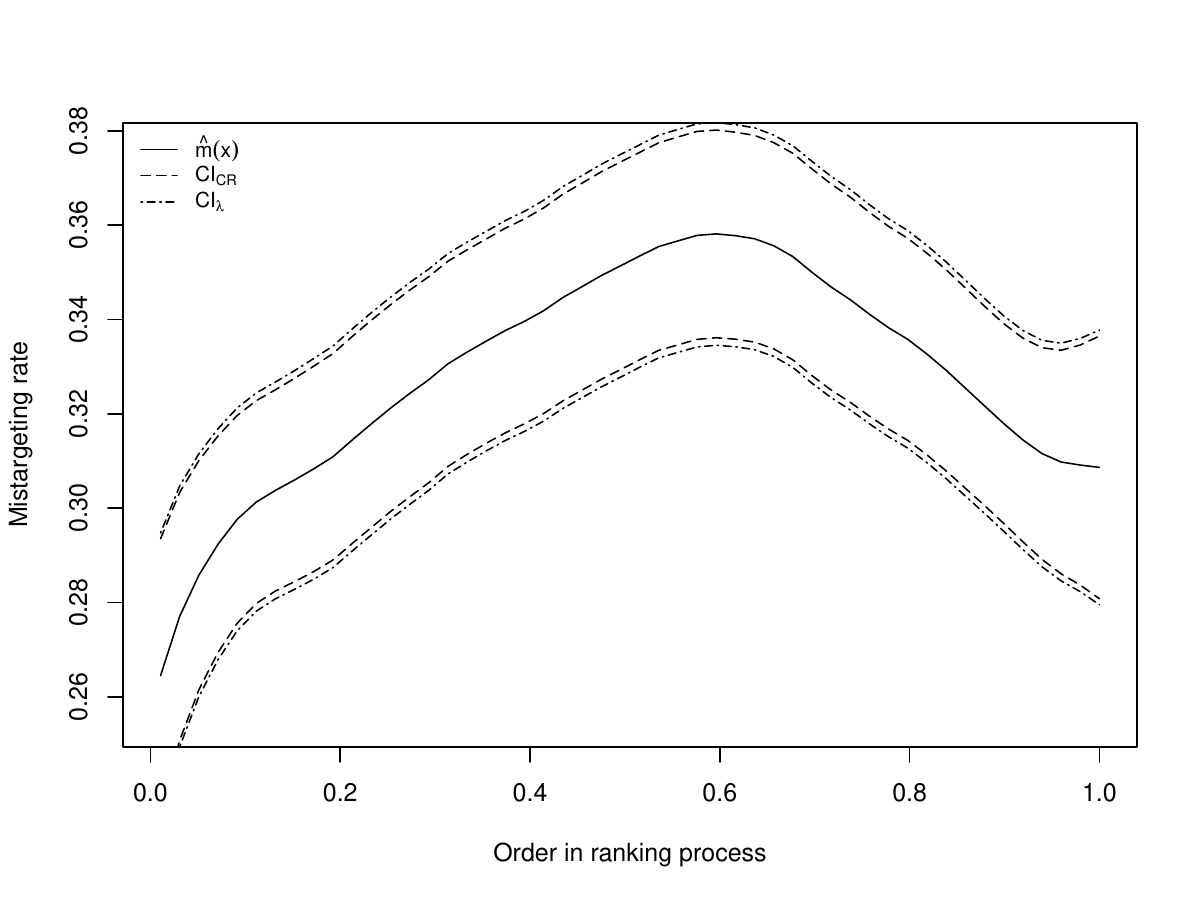}\caption{Local linear estimation and 95\% CIs on \citet{alatas2012targeting}'s dataset ($h_{m}=h_{\text{CR-CV}}$)\label{fig:Alatas}}
\end{figure}

\begin{figure}
\includegraphics[scale=0.8]{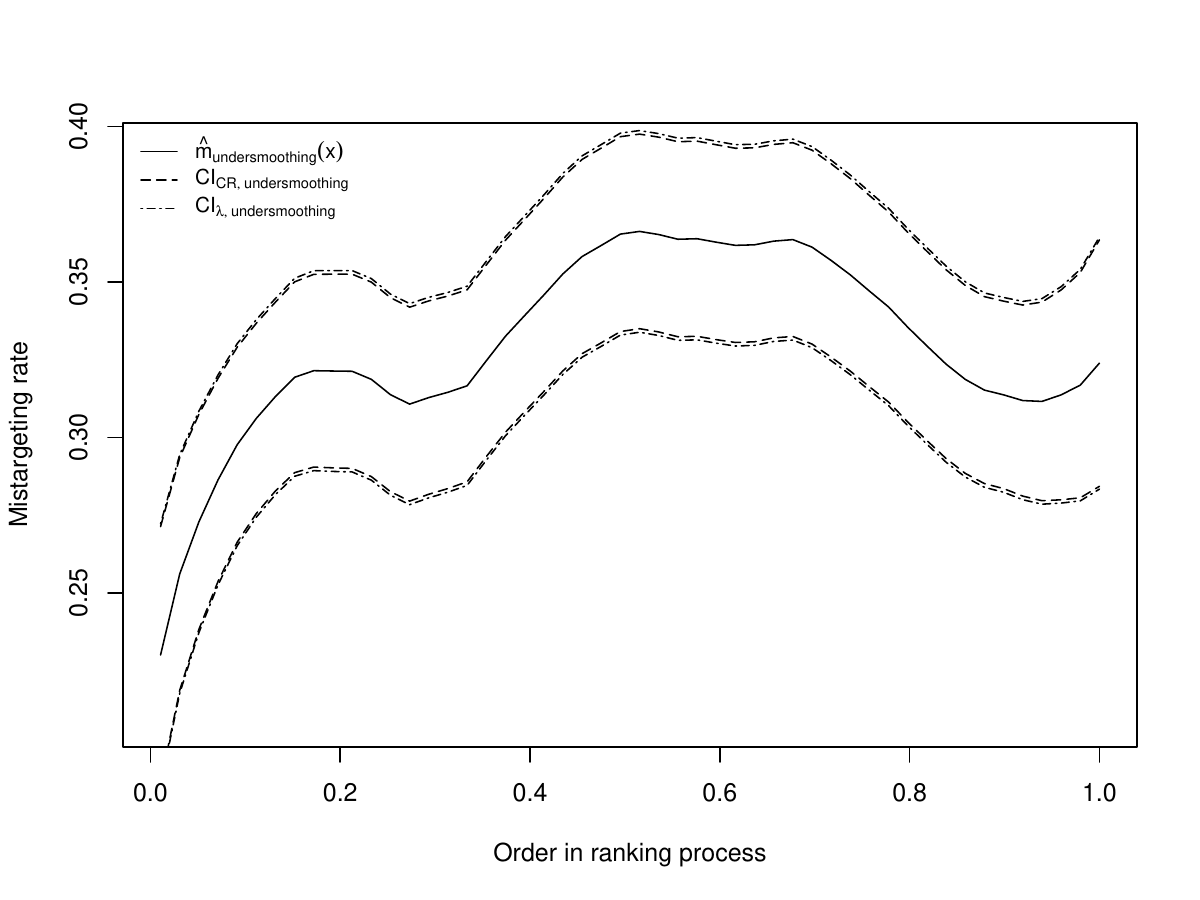}\caption{Local linear estimation and 95\% CIs on \citet{alatas2012targeting}'s dataset ($h_{m}=h_{\text{undersmoothing}}$)\label{fig:Alatas-undersmooth}}
\end{figure}

We calculated three 95\% confidence intervals: $CI$, $CI_{\mathrm{CR}}$, and $CI_{\lambda}$. We calculate $\widehat{\lambda}\approx1.148$. Since $CI$ and $CI_{\mathrm{CR}}$ are almost identical, we only draw $CI_{\mathrm{CR}}$ on the plot. Figures $\text{\ref{fig:Alatas}}$ and $\text{\ref{fig:Alatas-undersmooth}}$ show the estimated nonparametric regression values and estimated pointwise confidence intervals with $h_{m}=h_{\text{CR-CV}}$ and $h_{m}=h_{\text{undersmoothing}}$, respectively. The asymptotic bias exists under $h_{m}=h_{\text{CR-CV}}$, whereas it is asymptotically dominated under $h_{m}=h_{\text{undersmoothing}}$. The estimated nonparametric function appears more oscillatory and has wider CIs in Figure $\text{\ref{fig:Alatas-undersmooth}}$ due to the larger variance. We found that $CI_{\lambda}$ is slightly wider than $CI_{\mathrm{CR}}$ in both figures, however the difference is smaller in Figure $\text{\ref{fig:Alatas-undersmooth}}$, as the smaller bandwidth reduces within-cluster correlation in smaller neighborhoods. We recommend trying both bandwidth choices as a robustness check in practice. We still have significant pointwise differences between the first few households and the household in the middle of the ranking process (mistargeting rate rises 5-10\%) even under wider confidence intervals $CI_{\lambda}$. The conclusions are similar to \citet{alatas2012targeting}.

\section{\textbf{Conclusion \label{sec:Conclusion}}}

This article has developed a comprehensive theoretical framework for nonparametric regression analysis under cluster sampling. Our contributions are threefold, addressing critical aspects of cluster-dependent data analysis that have significant implications for econometric methodologies and applied research. First, we allow both growing and bounded size clusters. This extension is crucial, as growing cluster sizes introduce a non-negligible within-cluster dependence, necessitating the inclusion of an additional term in the asymptotic variance to capture this phenomenon accurately. Second, we cover the case where regressors contain common variables within the same clusters. These cluster-level regressors are the extreme case of cluster-dependent regressors, and they require the careful estimation of the joint density function. Third, our proposed inference is valid with heterogeneous and growing cluster sizes. The simulation studies illustrate the critical role of accounting for within-cluster dependence, affirming the practical relevance of our theoretical insights.

While this article establishes a foundation for nonparametric regression analysis under cluster sampling, several avenues for future research emerge. Theoretical work on other nonparametric estimators, such as local polynomial regressions and series regressions, would be an interesting extension. Investigating boundary analysis is crucial due to its impact on estimator bias. Additionally, developing cluster bootstrap inference methods for nonparametric regressions is important since it would provide more practical statistical inference for clustered data. Lastly, deriving honest and adaptive uniform confidence bands, as done by \citet{chernozhukov2014anti} and \citet{chen2024adaptive} in the i.i.d. case, represents an important extension. This direction would depend on future advancements in empirical process theory under cluster sampling.

\newpage{}

\appendix

\section{\textbf{Proofs for main results\label{app:math}}}

In this section, we will provide technical lemmas and proofs for the main results. The proofs for technical lemmas are in Appendix $\text{\ref{app:lemmas}}$.

Let $K_{h}\left(\cdot\right)=\frac{1}{h^{d}}K\left(\frac{\cdot}{h}\right)$.

\begin{lem}
\label{lem:Fr}Under Assumptions $\text{\ref{assu:dgp}}$ and $\text{\ref{assu:f}}$,
\begin{align*}
F_{0}(x) & \equiv\frac{1}{n}\sum_{g=1}^{G}\sum_{j=1}^{n_{g}}K_{h}\left(X_{gj}-x\right)=f\left(x\right)+o_{p}(1)\\
F_{1}(x) & \equiv\frac{1}{n}\sum_{g=1}^{G}\sum_{j=1}^{n_{g}}K_{h}\left(X_{gj}-x\right)\left(X_{gj}-x\right)=o_{p}\left(h\right)\mathbf{1}_{d},\\
F_{2}(x) & \equiv\frac{1}{n}\sum_{g=1}^{G}\sum_{j=1}^{n_{g}}K_{h}\left(X_{gj}-x\right)\left(X_{gj}-x\right)\left(X_{gj}-x\right)^{\top}=h^{2}f\left(x\right)\kappa_{2}\mathbf{I}_{d\times d}+o_{p}\left(h^{2}\right)\mathbf{1}_{d}\mathbf{1}_{d}^{\top}.
\end{align*}
\end{lem}
\begin{lem}
\label{lem:Jr} Under Assumptions $\text{\ref{assu:dgp}}$-$\text{\ref{assu:nw}}$,
\begin{align*}
J_{0}(x) & \equiv\frac{1}{n}\sum_{g=1}^{G}\sum_{j=1}^{n_{g}}K_{h}\left(X_{gj}-x\right)\left\{ m\left(X_{gj}\right)-m(x)\right\} \\
 & =\ensuremath{h^{2}\kappa_{2}\sum_{q=1}^{d}\left(\frac{1}{2}\partial_{qq}m(x)+f(x)^{-1}\partial_{q}f(x)\partial_{q}m(x)\right)}+o_{p}\left(h^{2}\right)+O_{p}\left(\sqrt{\frac{1}{nh^{d-2}}}\right),\\
J_{1}(x) & \equiv\frac{1}{n}\sum_{g=1}^{G}\sum_{j=1}^{n_{g}}K_{h}\left(X_{gj}-x\right)\left\{ m\left(X_{gj}\right)-m(x)\right\} \left(X_{gj}-x\right)\\
 & =h^{2}f(0)\kappa_{2}\nabla m\left(0\right)+o_{p}\left(h^{3}\right)\mathbf{1}_{d}+O_{p}\left(\sqrt{\frac{1}{nh^{d-4}}}\right)\mathbf{1}_{d}.
\end{align*}
\end{lem}
\begin{lem}
\label{lem:Hr} Under Assumptions $\text{\ref{assu:dgp}}$-$\text{\ref{assu:nw}}$,
\begin{align*}
H_{0}(x) & \equiv\frac{1}{n}\sum_{g=1}^{G}\sum_{j=1}^{n_{g}}K_{h}^{2}\left(X_{gj}-x\right)\sigma^{2}\left(X_{gj}\right)=\frac{1}{h^{d}}f\left(x\right)\sigma^{2}\left(x\right)R_{k}^{d}+o_{p}\left(h^{-d}\right),\\
H_{1}(x) & \equiv\frac{1}{n}\sum_{g=1}^{G}\sum_{j=1}^{n_{g}}K_{h}^{2}\left(X_{gj}-x\right)\sigma^{2}\left(X_{gj}\right)\left(X_{gj}-x\right)=o_{p}\left(h^{-d+1}\right)\mathbf{1}_{d},\\
H_{2}(x) & \equiv\frac{1}{n}\sum_{g=1}^{G}\sum_{j=1}^{n_{g}}K_{h}^{2}\left(X_{gj}-x\right)\sigma^{2}\left(X_{gj}\right)\left(X_{gj}-x\right)\left(X_{gj}-x\right)^{\top}\\
 & =\frac{1}{h^{d-2}}f\left(x\right)\sigma^{2}\left(x\right)\left\{ \int_{\mathbb{R}^{d}}K^{2}\left(T\right)TT^{\top}\mathrm{d}T\right\} +o_{p}\left(h^{-d+2}\right)\mathbf{1}_{d}\mathbf{1}_{d}^{\top}.
\end{align*}
\end{lem}
\begin{lem}
\label{lem:Ir} Under Assumptions $\text{\ref{assu:dgp}}$-$\text{\ref{assu:lambda}}$,
\begin{align*}
I_{0}(x) & \equiv\frac{1}{n}\sum_{g=1}^{G}\sum_{1\leq j<\ell\leq n_{g}}K_{h}\left(X_{gj}-x\right)K_{h}\left(X_{g\ell}-x\right)\sigma\left(X_{gj}^{(\mathrm{ind})},X_{g\ell}^{(\mathrm{ind})};X_{g}^{(\mathrm{cls})}\right)\\
 & =\frac{1}{2h^{d}}\lambda R_{k}^{d_{\mathrm{cls}}}f_{2}\left(x^{\mathrm{(ind)}},x^{\mathrm{(ind)}};x^{\mathrm{(cls)}}\right)\sigma\left(x^{\mathrm{(ind)}},x^{\mathrm{(ind)}};x^{\mathrm{(cls)}}\right)+o_{p}\left(h^{-d}\right),\\
I_{1}(x) & \equiv\frac{1}{n}\sum_{g=1}^{G}\sum_{1\leq j<\ell\leq n_{g}}K_{h}\left(X_{gj}-x\right)K_{h}\left(X_{g\ell}-x\right)\sigma\left(X_{gj}^{(\mathrm{ind})},X_{g\ell}^{(\mathrm{ind})};X_{g}^{(\mathrm{cls})}\right)\left(X_{gj}-x\right)\\
 & =o_{p}\left(h^{-d+1}\right)\mathbf{1}_{d},\\
I_{2}(x) & \equiv\frac{1}{n}\sum_{g=1}^{G}\sum_{1\leq j<\ell\leq n_{g}}K_{h}\left(X_{gj}-x\right)K_{h}\left(X_{g\ell}-x\right)\sigma\left(X_{gj}^{(\mathrm{ind})},X_{g\ell}^{(\mathrm{ind})};X_{g}^{(\mathrm{cls})}\right)\ensuremath{\left(X_{g\ell}-x\right)}\left(X_{gj}-x\right)^{\top}\\
 & =O_{p}\left(h^{-d+2}\right)\mathbf{1}_{d}\mathbf{1}_{d}^{\top}.
\end{align*}
\end{lem}
\begin{lem}
\label{lem:Er} Under Assumptions $\text{\ref{assu:dgp}}$-$\text{\ref{assu:lambda}}$,
\begin{align*}
\mathcal{E}_{0}(x) & \equiv\frac{1}{n}\sum_{g=1}^{G}\sum_{j=1}^{n_{g}}K_{h}\left(X_{gj}-x\right)e_{gj}=O_{p}\left(\sqrt{\frac{1}{nh^{d}}}\right),\\
\mathcal{E}_{1}(x) & \equiv\frac{1}{n}\sum_{g=1}^{G}\sum_{j=1}^{n_{g}}K_{h}\left(X_{gj}-x\right)e_{gj}\left(X_{gj}-x\right)=O_{p}\left(\sqrt{\frac{1}{nh^{d-2}}}\right)\mathbf{1}_{d}.
\end{align*}
\end{lem}
\begin{lem}
\label{lem:nabla_m_order}Under Assumptions \ref{assu:dgp}, \ref{assu:f} and \ref{assu:LL},
\begin{align*}
\frac{1}{n}\sum_{g=1}^{G}\sum_{j=1}^{n_{g}}K_{h}\left(X_{gj}-x\right)\left\{ \left(X_{gj}-x\right)^{\top}\nabla^{2}m(x)\left(X_{gj}-x\right)\right\}  & =h^{2}\kappa_{2}f(x)\sum_{q=1}^{d}\partial_{qq}m(x)+o_{p}\left(h^{2}\right),\\
\frac{1}{n}\sum_{g=1}^{G}\sum_{j=1}^{n_{g}}K_{h}\left(X_{gj}-x\right)\left(X_{gj}-x\right)\left\{ \left(X_{gj}-x\right)^{\top}\nabla^{2}m(x)\left(X_{gj}-x\right)\right\}  & =O_{p}\left(h^{3}\right)\mathbf{1}_{d}.
\end{align*}
\end{lem}
\bigskip{}

\subsection{Proof for Theorem $\text{\ref{thm:density_consistency}}$}
\begin{proof}
Lemma $\text{\ref{lem:Fr}}$ for $F_{0}(x)$ implies the result.
\end{proof}

\subsection{Proof for Theorem $\text{\ref{thm:nw_bias}}$}
\begin{proof}
Since observations belonging to different clusters are mutually independent and $\mathbb{E}\left[Y_{gj}\mid\mathbf{X}_{g}\right]=m\left(X_{gj}\right)$, 
\begin{align*}
\mathbb{E}\left[\hat{m}_{\mathrm{nw}}(x)\mid\mathbf{X}_{1},\cdots,\mathbf{X}_{G}\right] & =\frac{\sum_{g=1}^{G}\sum_{j=1}^{n_{g}}K\left(\frac{X_{gj}-x}{h}\right)m\left(X_{gj}\right)}{\sum_{g=1}^{G}\sum_{j=1}^{n_{g}}K\left(\frac{X_{gj}-x}{h}\right)}\\
 & =m(x)+\frac{J_{0}(x)}{\hat{f}(x)}.
\end{align*}
Theorem \ref{thm:density_consistency} implies $\widehat{f}(x)\overset{p}{\rightarrow}f(x)>0$. Thus, the continuous mapping theorem and Lemma $\text{\ref{lem:Jr}}$ imply the result.
\end{proof}

\subsection{Proof for Theorem $\text{\ref{thm:nw_var}}$}
\begin{proof}
Since $e_{gj}=Y_{gj}-m\left(X_{gj}\right)$,{\small{}
\begin{eqnarray*}
 &  & \operatorname{Var}\left[\hat{m}_{\mathrm{nw}}(x)\mid\mathbf{X}_{1},\cdots,\mathbf{X}_{G}\right]\\
 & = & \mathbb{E}\left[\left(\hat{m}_{\mathrm{nw}}(x)-\mathbb{E}\left[\left(\hat{m}_{\mathrm{nw}}(x)\right)\mid\mathbf{X}_{1},\cdots,\mathbf{X}_{G}\right]\right)^{2}\mid\mathbf{X}_{1},\cdots,\mathbf{X}_{G}\right]\\
 & = & \mathbb{E}\left[\left(\frac{\sum_{g=1}^{G}\sum_{j=1}^{n_{g}}K\left(\frac{X_{gj}-x}{h}\right)\left(Y_{gj}-m\left(X_{gj}\right)\right)}{\sum_{g=1}^{G}\sum_{j=1}^{n_{g}}K\left(\frac{X_{gj}-x}{h}\right)}\right)^{2}\mid\mathbf{X}_{1},\cdots,\mathbf{X}_{G}\right]\\
 & = & \frac{\mathbb{E}\left[\left(\sum_{g=1}^{G}\sum_{j=1}^{n_{g}}K\left(\frac{X_{gj}-x}{h}\right)e_{gj}\right)^{2}\mid\mathbf{X}_{1},\cdots,\mathbf{X}_{G}\right]}{\left(\sum_{g=1}^{G}\sum_{j=1}^{n_{g}}K\left(\frac{X_{gj}-x}{h}\right)\right)^{2}}\\
 & = & \frac{\sum_{g=1}^{G}\mathbb{E}\left[\left(\sum_{j=1}^{n_{g}}K\left(\frac{X_{gj}-x}{h}\right)e_{gj}\right)^{2}\mid\mathbf{X}_{g}\right]}{\left(\sum_{g=1}^{G}\sum_{j=1}^{n_{g}}K\left(\frac{X_{gj}-x}{h}\right)\right)^{2}}\\
 & = & \frac{\sum_{g=1}^{G}\left\{ \sum_{j=1}^{n_{g}}K\left(\frac{X_{gj}-x}{h}\right)^{2}\mathbb{E}\left[e_{gj}^{2}\mid\mathbf{X}_{g}\right]+2\sum_{1\leq j<\ell\leq n_{g}}K\left(\frac{X_{gj}-x}{h}\right)K\left(\frac{X_{g\ell}-x}{h}\right)\mathbb{E}\left[e_{gj}e_{g\ell}\mid\mathbf{X}_{g}\right]\right\} }{\left(\sum_{g=1}^{G}\sum_{j=1}^{n_{g}}K\left(\frac{X_{gj}-x}{h}\right)\right)^{2}}\\
 & = & \frac{\sum_{g=1}^{G}\left\{ \sum_{j=1}^{n_{g}}K\left(\frac{X_{gj}-x}{h}\right)^{2}\sigma^{2}\left(X_{gj}\right)+2\sum_{1\leq j<\ell\leq n_{g}}K\left(\frac{X_{gj}-x}{h}\right)K\left(\frac{X_{g\ell}-x}{h}\right)\sigma\left(X_{gj}^{(\mathrm{ind})},X_{g\ell}^{(\mathrm{ind})};X_{g}^{(\mathrm{cls})}\right)\right\} }{\left(\sum_{g=1}^{G}\sum_{j=1}^{n_{g}}K\left(\frac{X_{gj}-x}{h}\right)\right)^{2}}\\
 & = & \frac{h^{d}\left\{ H_{0}(x)+2I_{0}(x)\right\} }{nh^{d}\left(\hat{f}(x)\right)^{2}},
\end{eqnarray*}
}where the fourth equality follows from the mutual independence between clusters.

Theorem \ref{thm:density_consistency} implies $\widehat{f}(x)\overset{p}{\rightarrow}f(x)>0$. Lemmas $\text{\ref{lem:Hr}}$ and $\text{\ref{lem:Ir}}$ for $H_{0}(x)$ and $I_{0}(x)$ and the continuous mapping theorem together imply that
\begin{eqnarray*}
 &  & \operatorname{Var}\left[\hat{m}_{\mathrm{nw}}(x)\mid\mathbf{X}_{1},\cdots,\mathbf{X}_{G}\right]\\
 & = & \frac{1}{nh^{d}}\frac{f(x)\sigma^{2}(x)R_{k}^{d}+\lambda R_{k}^{d_{\mathrm{cls}}}f_{2}\left(x^{\mathrm{(ind)}},x^{\mathrm{(ind)}};x^{\mathrm{(cls)}}\right)\sigma\left(x^{\mathrm{(ind)}},x^{\mathrm{(ind)}};x^{\mathrm{(cls)}}\right)+o_{p}(1)}{f(x)^{2}+o_{p}(1)}\\
 & = & \frac{R_{k}^{d}\sigma^{2}(x)}{f(x)nh^{d}}+\frac{\lambda R_{k}^{d_{\mathrm{cls}}}f_{2}\left(x^{\mathrm{(ind)}},x^{\mathrm{(ind)}};x^{\mathrm{(cls)}}\right)\sigma\left(x^{\mathrm{(ind)}},x^{\mathrm{(ind)}};x^{\mathrm{(cls)}}\right)}{f(x)^{2}nh^{d}}+o_{p}\left(\frac{1}{nh^{d}}\right).
\end{eqnarray*}
\end{proof}

\subsection{Proof for Theorem $\text{\ref{thm:nw_cons}}$}
\begin{proof}
$ $
\begin{align}
\hat{m}_{\text{nw}}\left(x\right) & =\frac{\sum_{g=1}^{G}\sum_{j=1}^{n_{g}}K\left(\frac{X_{gj}-x}{h}\right)Y_{gj}}{\sum_{g=1}^{G}\sum_{j=1}^{n_{g}}K\left(\frac{X_{gj}-x}{h}\right)}\nonumber \\
 & =\frac{\frac{1}{n}\sum_{g=1}^{G}\sum_{j=1}^{n_{g}}K_{h}\left(X_{gj}-x\right)\left\{ m\left(x\right)+m\left(X_{gj}\right)-m\left(x\right)+e_{gj}\right\} }{\frac{1}{n}\sum_{g=1}^{G}\sum_{j=1}^{n_{g}}K_{h}\left(X_{gj}-x\right)}\nonumber \\
 & =m\left(x\right)+\frac{J_{0}(x)}{\widehat{f}(x)}+\frac{\mathcal{E}_{0}(x)}{\widehat{f}(x)}\label{eq:mJE}\\
 & \overset{p}{\rightarrow}0\nonumber 
\end{align}
by Theorem $\text{\ref{thm:density_consistency}}$ and Lemmas $\text{\ref{lem:Jr}}$ and $\text{\ref{lem:Er}}$.
\end{proof}

\subsection{Proof for Theorem $\ref{thm:nw_asy_dist}$}
\begin{proof}
Since we have $\text{\eqref{eq:mJE}},$ Theorem $\text{\ref{thm:density_consistency}}$, Lemma $\text{\ref{lem:Jr}}$, and $\text{\eqref{eq:nh^5}}$, 
\begin{align*}
\sqrt{nh^{d}}\left(\widehat{m}_{\mathrm{nw}}(x)-m(x)-h^{2}B_{\mathrm{nw}}(x)\right) & =\sqrt{nh^{d}}\left(\frac{\mathcal{E}_{0}(x)}{\widehat{f}(x)}\right)+\sqrt{nh^{d}}\left(\frac{J_{0}(x)}{\widehat{f}(x)}-h^{2}B_{\mathrm{nw}}(x)\right)\\
 & =\frac{\sqrt{nh^{d}}\mathcal{E}_{0}(x)}{\widehat{f}(x)}+\sqrt{nh^{d}}\left(o_{p}\left(h^{2}\right)+O_{p}\left(\sqrt{\frac{1}{nh^{d-2}}}\right)\right)\\
 & =\frac{\sqrt{nh^{d}}\mathcal{E}_{0}(x)}{\widehat{f}(x)}+\left(o_{p}\left(\sqrt{nh^{d+4}}\right)+O_{p}\left(h\right)\right)\\
 & =\frac{\sqrt{nh^{d}}\mathcal{E}_{0}(x)}{f(x)+o_{p}\left(1\right)}+o_{p}\left(1\right).
\end{align*}

Define $\widetilde{\mathbf{Z}}_{ng}=\sum_{j=1}^{n_{g}}K\left(\frac{X_{gj}-x}{h}\right)e_{gj}$. Note that $\left\{ \widetilde{\mathbf{Z}}_{ng}\right\} _{g=1}^{G}$ are independent and $\mathbb{E}\left[\widetilde{\mathbf{Z}}_{ng}\right]=0$. We can express $\sqrt{nh^{d}}\mathcal{E}_{0}(x)=\frac{1}{\sqrt{nh^{d}}}\sum_{g=1}^{G}\sum_{j=1}^{n_{g}}K\left(\frac{X_{gj}-x}{h}\right)e_{gj}=\frac{1}{\sqrt{nh^{d}}}\sum_{g=1}^{G}\widetilde{\mathbf{Z}}_{ng}$. Denote $s_{n}^{2}=\operatorname{Var}\left[\frac{1}{\sqrt{nh^{d}}}\sum_{g=1}^{G}\widetilde{\mathbf{Z}}_{ng}\right]$. By the proof for Lemma $\text{\ref{lem:Er}}$,
\begin{align*}
s_{n}^{2} & =nh^{d}\operatorname{Var}\left[\mathcal{E}_{0}(x)\right]=h^{d}\mathbb{E}\left[H_{0}(x)+2I_{0}(x)\right]\\
 & =f(x)\sigma^{2}(x)R_{k}^{d}+\lambda R_{k}^{d_{\mathrm{cls}}}f_{2}\left(x^{\mathrm{(ind)}},x^{\mathrm{(ind)}};x^{\mathrm{(cls)}}\right)\sigma\left(x^{\mathrm{(ind)}},x^{\mathrm{(ind)}};x^{\mathrm{(cls)}}\right)+o(1).
\end{align*}
By assumption, this implies that $s_{n}^{2}$ deterministically converges to some positive constant. The conclusion follows by applying the Lindeberg Central Limit Theorem and the Slutsky\textquoteright s Lemma. Thus, it is sufficient to verify the Lindeberg condition:
\begin{equation}
\frac{1}{nh^{d}s_{n}^{2}}\sum_{g=1}^{G}\mathbb{E}\left[\widetilde{\mathbf{Z}}_{ng}^{2}\mathbf{1}\left\{ \widetilde{\mathbf{Z}}_{ng}^{2}\geq nh^{d}s_{n}^{2}\varepsilon\right\} \right]=o(1)\label{eq:nw_Lindeberg}
\end{equation}
for all $\varepsilon>0$.

Pick any $\varepsilon>0$ and any $\delta>0$. 
\begin{eqnarray*}
 &  & \sup_{i}\mathbb{E}\left[\left|\frac{1}{n^{1/4}h^{d/2}}K\left(\frac{X_{i}-x}{h}\right)e_{i}\right|^{r}\mathbf{1}\left\{ \frac{1}{n^{1/4}h^{d/2}}\left|K\left(\frac{X_{i}-x}{h}\right)e_{i}\right|\geq B\right\} \right]\\
 & \leq & \sup_{i}\frac{1}{n^{r/4}h^{dr/2}}\mathbb{E}\left[K^{r}\left(\frac{X_{i}-x}{h}\right)\left|e_{i}\right|^{r}\mathbf{1}\left\{ \left|e_{i}\right|\geq\frac{n^{1/4}h^{d/2}B}{\bar{K}}\right\} \right]\\
 & = & \sup_{i}\frac{1}{n^{r/4}h^{dr/2}}\mathbb{E}\left[K^{r}\left(\frac{X_{i}-x}{h}\right)\mathbb{E}\left[\left|e_{i}\right|^{r}\mathbf{1}\left\{ \left|e_{i}\right|\geq\frac{n^{1/4}h^{d/2}B}{\bar{K}}\right\} \mid X_{i}\right]\right]\\
 & = & \sup_{i}\frac{1}{n^{r/4}h^{dr/2-d}}\int K^{r}\left(T_{i}\right)\mathbb{E}\left[\left|e_{i}\right|^{r}\mathbf{1}\left\{ \left|e_{i}\right|\geq\frac{n^{1/4}h^{d/2}B}{\bar{K}}\right\} \mid X_{i}=x+hT_{i}\right]f\left(x+hT_{i}\right)\mathrm{d}T_{i}\\
 & \leq & \int K^{r}\left(T\right)f\left(x+hT\right)\mathrm{d}T\frac{1}{n^{r/4}h^{dr/2-d}}\overline{v}^{2}\left|\frac{\bar{K}}{n^{1/4}h^{d/2}B}\right|^{r}\\
 & \leq & \bar{K}^{2r-2}R_{K}^{d}\left(f(x)+o(1)\right)\frac{1}{n^{r/2}h^{dr-d}}\overline{v}^{2}\frac{1}{\left|B\right|^{r}}\\
 & \leq & O(1)\cdot\frac{1}{\left|B\right|^{r}},
\end{eqnarray*}
where the first and third inequality follow from the definition of the kernel function (Definition $\text{\ref{def:kernel}}$) $K(u)\leq\bar{K}<\infty$, the first equality follows from the law of iterated expectations, the second equality follows from the change of variables $\left(X_{i}-x\right)/h=T_{i}$, the second inequality follows from 
\begin{align*}
\mathbb{E}\left[\left|e_{i}\right|^{r}\mathbf{1}\left\{ \left|e_{i}\right|\geq\frac{n^{1/4}h^{d/2}B}{\bar{K}}\right\} \mid X_{i}=x+hT_{i}\right] & =\mathbb{E}\left[\frac{\left|e_{i}\right|^{2r}}{\left|e_{i}\right|^{r}}\mathbf{1}\left\{ \left|e_{i}\right|\geq\frac{n^{1/4}h^{d/2}B}{\bar{K}}\right\} \mid X_{i}=x+hT_{i}\right]\\
 & \leq\mathbb{E}\left[\left|e_{i}\right|^{2r}\mid X_{i}=x+hT_{i}\right]\left|\frac{\bar{K}}{n^{1/4}h^{d/2}B}\right|^{r}\\
 & \leq\overline{v}^{2}\left|\frac{\bar{K}}{n^{1/4}h^{d/2}B}\right|^{r},\qquad\because\text{\eqref{eq:e^r}}
\end{align*}
and the fourth inequality follows by $\text{\eqref{eq:r_bound}}$. Thus, 
\[
\lim_{B\rightarrow\infty}\sup_{i}\mathbb{E}\left[\left|\frac{1}{n^{1/4}h^{d/2}}K\left(\frac{X_{i}-x}{h}\right)e_{i}\right|^{r}\mathbf{1}\left\{ \frac{1}{n^{1/4}h^{d/2}}\left|K\left(\frac{X_{i}-x}{h}\right)e_{i}\right|\geq B\right\} \right]=0
\]
holds. By Lemma 1 of \citet{hansen2019asymptotic}, this equation implies 
\[
\lim_{B\rightarrow\infty}\sup_{g}\mathbb{E}\left[\left|\frac{\widetilde{\mathbf{Z}}_{ng}}{n^{1/4}h^{d/2}n_{g}}\right|^{r}\mathbf{1}\left\{ \left|\frac{\widetilde{\mathbf{Z}}_{ng}}{n^{1/4}h^{d/2}n_{g}}\right|\geq B\right\} \right]=0.
\]

Hence, we can pick $B$ large enough so that 
\begin{equation}
\mathbb{E}\left[\left|\frac{\widetilde{\mathbf{Z}}_{ng}}{n^{1/4}h^{d/2}n_{g}}\right|^{r}\mathbf{1}\left\{ \left|\frac{\widetilde{\mathbf{Z}}_{ng}}{n^{1/4}h^{d/2}n_{g}}\right|\geq B\right\} \right]\leq\frac{s_{n}^{r}\varepsilon^{r/2-1}}{C^{r}}\delta\label{eq:nw_Lindeberg_bound}
\end{equation}
for large enough $n$. Now, let's verify the Lindeberg condition:
\begin{eqnarray*}
 &  & \frac{1}{nh^{d}s_{n}^{2}}\sum_{g=1}^{G}\mathbb{E}\left[\widetilde{\mathbf{Z}}_{ng}^{2}\mathbf{1}\left\{ \widetilde{\mathbf{Z}}_{ng}^{2}\geq nh^{d}s_{n}^{2}\varepsilon\right\} \right]\\
 & = & \frac{1}{nh^{d}s_{n}^{2}}\sum_{g=1}^{G}\mathbb{E}\left[\widetilde{\mathbf{Z}}_{ng}^{2}\mathbf{1}\left\{ \left|\widetilde{\mathbf{Z}}_{ng}\right|\geq\left(nh^{d}s_{n}^{2}\varepsilon\right)^{1/2}\right\} \right]\\
 & = & \frac{1}{nh^{d}s_{n}^{2}}\sum_{g=1}^{G}\mathbb{E}\left[\frac{\left|\widetilde{\mathbf{Z}}_{ng}\right|^{r}}{\left|\widetilde{\mathbf{Z}}_{ng}\right|^{r-2}}\mathbf{1}\left\{ \left|\widetilde{\mathbf{Z}}_{ng}\right|\geq\left(nh^{d}s_{n}^{2}\varepsilon\right)^{1/2}\right\} \right]\\
 & \leq & \frac{1}{nh^{d}s_{n}^{2}\left(\left(nh^{d}s_{n}^{2}\varepsilon\right)^{1/2}\right)^{r-2}}\sum_{g=1}^{G}\mathbb{E}\left[\left|\widetilde{\mathbf{Z}}_{ng}\right|^{r}\mathbf{1}\left\{ \left|\widetilde{\mathbf{Z}}_{ng}\right|\geq\left(nh^{d}s_{n}^{2}\varepsilon\right)^{1/2}\right\} \right]\\
 & = & \frac{1}{n^{r/4}s_{n}^{r}\varepsilon^{r/2-1}}\sum_{g=1}^{G}n_{g}^{r}\mathbb{E}\left[\left|\frac{\widetilde{\mathbf{Z}}_{ng}}{n^{1/4}h^{d/2}n_{g}}\right|^{r}\mathbf{1}\left\{ \left|\frac{\widetilde{\mathbf{Z}}_{ng}}{n^{1/4}h^{d/2}n_{g}}\right|\geq\frac{s_{n}n^{1/4}\varepsilon^{1/2}}{n_{g}}\right\} \right]\\
 & \leq & \frac{1}{n^{r/4}s_{n}^{r}\varepsilon^{r/2-1}}\sum_{g=1}^{G}n_{g}^{r}\mathbb{E}\left[\left|\frac{\widetilde{\mathbf{Z}}_{ng}}{n^{1/4}h^{d/2}n_{g}}\right|^{r}\mathbf{1}\left\{ \left|\frac{\widetilde{\mathbf{Z}}_{ng}}{n^{1/4}h^{d/2}n_{g}}\right|\geq B\right\} \right]\\
 & \leq & \frac{\sum_{g=1}^{G}n_{g}^{r}}{n^{r/4}C^{r}}\delta\\
 & \leq & \delta,
\end{eqnarray*}
where the second inequality holds for sufficiently large $n$ since $\text{\eqref{eq:n_g^4/(nhs)}}$ enables us to pick large enough $n^{*}$ to satisfy 
\[
\frac{1}{B}\geq\max_{g\leq G}\frac{n_{g}}{s_{n}n^{1/4}\varepsilon^{1/2}}\text{ for any }n\geq n^{*},
\]
the third inequality follows by $\text{\eqref{eq:nw_Lindeberg_bound}}$, and the fourth inequality follows by $\text{\eqref{eq:n_g/n_bound}}$.
\end{proof}

\subsection{Proof for Theorem $\text{\ref{thm:LL_bias}}$}
\begin{proof}
Define $\mathbf{M}=\left[m\left(X_{1}\right),\dots,m\left(X_{n}\right)\right]^{\top}$ and 
\[
\mathbf{D}_{h}=\left(\begin{array}{cc}
1 & 0\\
0 & h^{-2}\mathbf{I}_{d\times d}
\end{array}\right).
\]
Then, we can rewrite
\begin{align*}
\mathbb{E}\left[\hat{m}_{\mathrm{LL}}(x)\mid\mathbf{X}_{1},\cdots,\mathbf{X}_{G}\right] & =\mathbf{e}_{1}^{\top}\left(\mathbf{X}_{x}^{\top}\mathbf{W}_{x}\mathbf{X}_{x}\right)^{-1}\mathbf{X}_{x}^{\top}\mathbf{W}_{x}\mathbf{M}\\
 & =\mathbf{e}_{1}^{\top}\left(\mathbf{D}_{h}\mathbf{X}_{x}^{\top}\mathbf{W}_{x}\mathbf{X}_{x}\right)^{-1}\mathbf{D}_{h}\mathbf{X}_{x}^{\top}\mathbf{W}_{x}\mathbf{M}
\end{align*}
by $\mathbb{E}\left[e_{gj}\mid\mathbf{X}_{g}\right]=0$. Let $\mathbf{Q}_{m}(x)$ be a $n\times1$ vector
\[
\mathbf{Q}_{m}(x)=\left[\left(X_{1}-x\right)^{\top}\nabla^{2}m(x)\left(X_{1}-x\right),\dots,\left(X_{n}-x\right)^{\top}\nabla^{2}m(x)\left(X_{n}-x\right)\right]^{\top}.
\]
By Taylor expansion of $\mathbf{M}$ around $x$,
\[
\mathbf{M}=\mathbf{X}_{x}\left[m(x),\nabla m(x)^{\top}\right]^{\top}+\frac{1}{2}\mathbf{Q}_{m}(x)+\mathbf{R}_{m}(x),
\]
where $\mathbf{R}_{m}(x)$ is a $n\times1$ vector of remainder terms. Compact support of $K$ implies that there exists some constant $C>0$ such that we essentially use observations with $\left|X_{i}^{(q)}-x^{(q)}\right|\leq C\cdot h$ for any $i=1,\dots,n$ and any $q=1,\dots,d$. Thus, by the multivariate Taylor expansion, we can evaluate a scalar random variable as
\begin{equation}
\mathbf{e}_{1}^{\top}\left(\mathbf{X}_{x}^{\top}\mathbf{W}_{x}\mathbf{X}_{x}\right)^{-1}\mathbf{X}_{x}^{\top}\mathbf{W}_{x}\mathbf{R}_{m}(x)=o_{p}\left(h^{2}\right).\label{eq:Rem}
\end{equation}

By Lemma $\text{\ref{lem:Fr}}$, we can calculate
\begin{eqnarray}
 &  & \left(\frac{1}{n}\mathbf{D}_{h}\mathbf{X}_{x}^{\top}\mathbf{W}_{x}\mathbf{X}_{x}\right)^{-1}\nonumber \\
 & = & \left\{ \mathbf{D}_{h}\left[\begin{array}{cc}
\frac{1}{n}\sum_{g=1}^{G}\sum_{j=1}^{n_{g}}K_{h}\left(X_{gj}-x\right) & \frac{1}{n}\sum_{g=1}^{G}\sum_{j=1}^{n_{g}}K_{h}\left(X_{gj}-x\right)\left(X_{gj}-x\right)^{\top}\\
\frac{1}{n}\sum_{g=1}^{G}\sum_{j=1}^{n_{g}}K_{h}\left(X_{gj}-x\right)\left(X_{gj}-x\right) & \frac{1}{n}\sum_{g=1}^{G}\sum_{j=1}^{n_{g}}K_{h}\left(X_{gj}-x\right)\left(X_{gj}-x\right)\left(X_{gj}-x\right)^{\top}
\end{array}\right]\right\} ^{-1}\nonumber \\
 & = & \left[\begin{array}{cc}
f(x)+o_{p}(1) & o_{p}\left(h\right)\mathbf{1}_{d}^{\top}\\
o_{p}\left(h^{-1}\right)\mathbf{1}_{d} & f(x)\kappa_{2}I_{d\times d}+o_{p}\left(1\right)\mathbf{1}_{d}\mathbf{1}_{d}^{\top}
\end{array}\right]^{-1}\nonumber \\
 & = & \left[\begin{array}{cc}
f(x)^{-1}+o_{p}(1) & o_{p}\left(h\right)\mathbf{1}_{d}^{\top}\\
o_{p}\left(h^{-1}\right)\mathbf{1}_{d} & \left(f(x)\kappa_{2}I_{d\times d}\right)^{-1}+o_{p}\left(1\right)\mathbf{1}_{d}\mathbf{1}_{d}^{\top}
\end{array}\right]\label{eq:XWX}
\end{eqnarray}
 Also, by Lemma $\text{\ref{lem:nabla_m_order}}$, 
\begin{eqnarray}
 &  & \frac{1}{n}\mathbf{D}_{h}\mathbf{X}_{x}^{\top}\mathbf{W}_{x}\mathbf{Q}_{m}(x)\nonumber \\
 & = & \mathbf{D}_{h}\left[\begin{array}{c}
\frac{1}{n}\sum_{g=1}^{G}\sum_{j=1}^{n_{g}}K_{h}\left(X_{gj}-x\right)\left\{ \left(X_{gj}-x\right)^{\top}\nabla^{2}m(x)\left(X_{gj}-x\right)\right\} \\
\frac{1}{n}\sum_{g=1}^{G}\sum_{j=1}^{n_{g}}K_{h}\left(X_{gj}-x\right)\left(X_{gj}-x\right)\left\{ \left(X_{gj}-x\right)^{\top}\nabla^{2}m(x)\left(X_{gj}-x\right)\right\} 
\end{array}\right]\nonumber \\
 & = & \mathbf{D}_{h}\left[\begin{array}{c}
h^{2}\kappa_{2}f(x)\sum_{q=1}^{d}\partial_{qq}m(x)+o_{p}\left(h^{2}\right)\\
O_{p}\left(h^{3}\right)\mathbf{1}_{d}
\end{array}\right]\nonumber \\
 & = & \left[\begin{array}{c}
h^{2}\kappa_{2}f(x)\sum_{q=1}^{d}\partial_{qq}m(x)+o_{p}\left(h^{2}\right)\\
O_{p}\left(h\right)\mathbf{1}_{d}
\end{array}\right].\label{eq:XWQ}
\end{eqnarray}
Therefore, 
\begin{eqnarray*}
 &  & \mathbb{E}\left[\hat{m}_{\mathrm{LL}}(x)\mid\mathbf{X}_{1},\cdots,\mathbf{X}_{G}\right]-m(x)\\
 & = & \mathbf{e}_{1}^{\top}\left(\mathbf{D}_{h}\mathbf{X}_{x}^{\top}\mathbf{W}_{x}\mathbf{X}_{x}\right)^{-1}\mathbf{D}_{h}\mathbf{X}_{x}^{\top}\mathbf{W}_{x}\left(\frac{1}{2}\mathbf{Q}_{m}(x)+\mathbf{R}_{m}(x)\right)\\
 & = & \frac{1}{2}\mathbf{e}_{1}^{\top}\left(\frac{1}{n}\mathbf{D}_{h}\mathbf{X}_{x}^{\top}\mathbf{W}_{x}\mathbf{X}_{x}\right)^{-1}\left(\frac{1}{n}\mathbf{D}_{h}\mathbf{X}_{x}^{\top}\mathbf{W}_{x}\mathbf{Q}_{m}(x)\right)+o_{p}\left(h^{2}\right).\\
 & = & h^{2}\frac{\kappa_{2}}{2}\sum_{q=1}^{d}\partial_{qq}m(x)+o_{p}\left(h^{2}\right),
\end{eqnarray*}
where the first equality holds since $\mathbf{e}_{1}^{\top}\left(\mathbf{X}_{x}^{\top}\mathbf{W}_{x}\mathbf{X}_{x}\right)^{-1}\mathbf{X}_{x}^{\top}\mathbf{W}_{x}\mathbf{X}_{x}\left[m(x),\nabla m(x)^{\top}\right]^{\top}=m(x)$, the second equality follows from $\text{\eqref{eq:Rem}}$, and the third equality follows from $\text{\eqref{eq:XWX}}$ and $\text{\eqref{eq:XWQ}}$.
\end{proof}

\subsection{Proof for Theorem $\text{\ref{thm:LL_var}}$}
\begin{proof}
Let $\mathbf{Y}=\left[Y_{1},\dots,Y_{n}\right]^{\top}$. Then,

\begin{eqnarray*}
 &  & \operatorname{Var}\left[\hat{m}_{\mathrm{LL}}(x)\mid\mathbf{X}_{1},\cdots,\mathbf{X}_{G}\right]\\
 & = & \mathbf{e}_{1}^{\top}\left(\mathbf{X}_{x}^{\top}\mathbf{W}_{x}\mathbf{X}_{x}\right)^{-1}\mathbf{X}_{x}^{\top}\mathbf{W}_{x}\operatorname{Var}\left[\mathbf{Y}\mid\mathbf{X}_{1},\cdots,\mathbf{X}_{G}\right]\mathbf{W}_{x}\mathbf{X}_{x}\left(\mathbf{X}_{x}^{\top}\mathbf{W}_{x}\mathbf{X}_{x}\right)^{-1}\mathbf{e}_{1}.
\end{eqnarray*}
Here, $\operatorname{Var}\left[\mathbf{Y}\mid\mathbf{X}_{1},\cdots,\mathbf{X}_{G}\right]$ is a $n\times n$ matrix having the following structure. 
\[
\operatorname{Var}\left[\mathbf{Y}\mid\mathbf{X}_{1},\cdots,\mathbf{X}_{G}\right]=\left[\begin{array}{ccc}
\mathbf{V}_{1} &  & O\\
 & \ddots\\
O &  & \mathbf{V}_{G}
\end{array}\right],
\]
where $\mathbf{V}_{g}$ (for $g=1,\dots,G$) is a matrix with 
\[
\mathbf{V}_{g}=\left[\mathbb{E}\left[e_{gj}e_{g\ell}\mid\mathbf{X}_{g}\right]\right]_{n_{g}\times n_{g}}.
\]
The upper-left scalar element of 
\[
n^{-1}\mathbf{X}_{x}^{\top}\mathbf{W}_{x}\operatorname{Var}\left[\mathbf{Y}\mid\mathbf{X}_{1},\cdots,\mathbf{X}_{G}\right]\mathbf{W}_{x}\mathbf{X}_{x}\equiv\left[\begin{array}{cc}
\Omega_{11} & \boldsymbol{\Omega}_{12}\\
\boldsymbol{\Omega}_{21} & \boldsymbol{\Omega}_{22}
\end{array}\right]
\]
 is
\begin{eqnarray*}
\Omega_{11} & = & \frac{1}{n}\sum_{g=1}^{G}\left\{ \sum_{j=1}^{n_{g}}K_{h}^{2}\left(X_{gj}-x\right)\sigma^{2}\left(X_{gj}\right)\right.\\
 &  & \quad\left.+2\sum_{1\leq j<\ell\leq n_{g}}K_{h}\left(X_{gj}-x\right)K_{h}\left(X_{g\ell}-x\right)\sigma\left(X_{gj}^{(\mathrm{ind})},X_{g\ell}^{(\mathrm{ind})};X_{g}^{(\mathrm{cls})}\right)\right\} ,
\end{eqnarray*}
the lower-left $d\times1$ block is
\begin{eqnarray*}
\boldsymbol{\Omega}_{21} & = & \frac{1}{n}\sum_{g=1}^{G}\left\{ \sum_{j=1}^{n_{g}}K_{h}^{2}\left(X_{gj}-x\right)\sigma^{2}\left(X_{gj}\right)\left(X_{gj}-x\right)\right.\\
 &  & \quad\left.+2\sum_{1\leq j<\ell\leq n_{g}}K_{h}\left(X_{gj}-x\right)K_{h}\left(X_{g\ell}-x\right)\sigma\left(X_{gj}^{(\mathrm{ind})},X_{g\ell}^{(\mathrm{ind})};X_{g}^{(\mathrm{cls})}\right)\left(X_{gj}-x\right)\right\} ,
\end{eqnarray*}
and the lower-right $d\times d$ block is
\begin{eqnarray*}
\boldsymbol{\Omega}_{22} & = & \frac{1}{n}\sum_{g=1}^{G}\left\{ \sum_{j=1}^{n_{g}}K_{h}^{2}\left(X_{gj}-x\right)\sigma^{2}\left(X_{gj}\right)\left(X_{gj}-x\right)\left(X_{gj}-x\right)^{\top}\right.\\
 &  & \quad\left.+2\sum_{1\leq j<\ell\leq n_{g}}K_{h}\left(X_{gj}-x\right)K\left(X_{g\ell}-x\right)\sigma\left(X_{gj}^{(\mathrm{ind})},X_{g\ell}^{(\mathrm{ind})};X_{g}^{(\mathrm{cls})}\right)\left(X_{g\ell}-x\right)\left(X_{gj}-x\right)^{\top}\right\} .
\end{eqnarray*}
Here, we defined $\boldsymbol{\Omega}_{12}=\boldsymbol{\Omega}_{21}^{\top}$. By Lemma $\text{\ref{lem:Hr}}$ and $\text{\ref{lem:Ir}}$, 
\begin{align}
\Omega_{11} & =H_{0}(x)+2I_{0}(x)\nonumber \\
 & =\frac{1}{h^{d}}\left\{ f(x)\sigma^{2}(x)R_{k}^{d}+\lambda R_{k}^{d_{\mathrm{cls}}}f_{2}\left(x^{\mathrm{(ind)}},x^{\mathrm{(ind)}};x^{\mathrm{(cls)}}\right)\sigma\left(x^{\mathrm{(ind)}},x^{\mathrm{(ind)}};x^{\mathrm{(cls)}}\right)+o_{p}(1)\right\} ,\label{eq:Omega11}\\
\boldsymbol{\Omega}_{21} & =H_{1}(x)+2I_{1}(x)=\ensuremath{o_{p}\left(h^{-d+1}\right)}\mathbf{1}_{d},\label{eq:Omega21}\\
\boldsymbol{\Omega}_{22} & =H_{2}(x)+2I_{2}(x)=\ensuremath{O_{p}\left(h^{-d+2}\right)}\mathbf{1}_{d}\mathbf{1}_{d}^{\top}.\label{eq:Omega22}
\end{align}
 Therefore, 
\begin{eqnarray*}
 &  & \operatorname{Var}\left[\hat{m}_{\mathrm{LL}}(x)\mid\mathbf{X}_{1},\cdots,\mathbf{X}_{G}\right]\\
 & = & \frac{1}{n}\mathbf{e}_{1}^{\top}\left(\frac{1}{n}\mathbf{D}_{h}\mathbf{X}_{x}^{\top}\mathbf{W}_{x}\mathbf{X}_{x}\right)^{-1}\mathbf{D}_{h}\left[\begin{array}{cc}
\Omega_{11} & \boldsymbol{\Omega}_{12}\\
\boldsymbol{\Omega}_{21} & \boldsymbol{\Omega}_{22}
\end{array}\right]\mathbf{D}_{h}\left(\frac{1}{n}\mathbf{X}_{x}^{\top}\mathbf{W}_{x}\mathbf{X}_{x}\mathbf{D}_{h}\right)^{-1}\mathbf{e}_{1}\\
 & = & \frac{1}{n}\left[\begin{array}{c}
f(x)^{-1}+o_{p}(1)\\
o_{p}\left(h^{-1}\right)\mathbf{1}_{d}
\end{array}\right]^{\top}\left[\begin{array}{cc}
\Omega_{11} & \ensuremath{o_{p}\left(h^{-d+1}\right)}\mathbf{1}_{d}^{\top}\\
\ensuremath{o_{p}\left(h^{-d+1}\right)}\mathbf{1}_{d} & O_{p}\left(h^{-d+2}\right)\mathbf{1}_{d}\mathbf{1}_{d}^{\top}
\end{array}\right]\left[\begin{array}{c}
f(x)^{-1}+o_{p}(1)\\
o_{p}\left(h^{-1}\right)\mathbf{1}_{d}
\end{array}\right]\\
 & = & \frac{1}{n}\left\{ f(x)^{-1}+o_{p}(1)\right\} ^{2}\Omega_{11}+o_{p}\left(n^{-1}h^{-d+1}\right)\\
 & = & \frac{R_{k}^{d}\sigma^{2}(x)}{f(x)nh^{d}}+\frac{\lambda R_{k}^{d_{\mathrm{cls}}}f_{2}\left(x^{\mathrm{(ind)}},x^{\mathrm{(ind)}};x^{\mathrm{(cls)}}\right)\sigma\left(x^{\mathrm{(ind)}},x^{\mathrm{(ind)}};x^{\mathrm{(cls)}}\right)}{f(x)^{2}nh^{d}}+o_{p}\left(\frac{1}{nh^{d}}\right).
\end{eqnarray*}
where the second equality follows from $\text{\eqref{eq:XWX}}$, $\text{\eqref{eq:Omega21}}$, and $\text{\eqref{eq:Omega22}}$ and the last equality follows from $\text{\eqref{eq:Omega11}}$.
\end{proof}

\subsection{Proof for Theorem $\text{\ref{thm:LL_cons}}$}
\begin{proof}
Let $\boldsymbol{\mathcal{E}}=\left[e_{1},\dots,e_{n}\right]^{\top}$. Then, by Theorem $\text{\ref{thm:LL_bias}}$,
\begin{align*}
\hat{m}_{\mathrm{LL}}(x) & =\mathbf{e}_{1}^{\top}\left(\mathbf{X}_{x}^{\top}\mathbf{W}_{x}\mathbf{X}_{x}\right)^{-1}\mathbf{X}_{x}^{\top}\mathbf{W}_{x}\mathbf{Y}\\
 & =\mathbf{e}_{1}^{\top}\left(\mathbf{X}_{x}^{\top}\mathbf{W}_{x}\mathbf{X}_{x}\right)^{-1}\mathbf{X}_{x}^{\top}\mathbf{W}_{x}\mathbf{M}+\mathbf{e}_{1}^{\top}\left(\mathbf{X}_{x}^{\top}\mathbf{W}_{x}\mathbf{X}_{x}\right)^{-1}\mathbf{X}_{x}^{\top}\mathbf{W}_{x}\boldsymbol{\mathcal{E}}\\
 & =m(x)+o_{p}(1)+\mathbf{e}_{1}^{\top}\left(\mathbf{D}_{h}\mathbf{X}_{x}^{\top}\mathbf{W}_{x}\mathbf{X}_{x}\right)^{-1}\mathbf{D}_{h}\mathbf{X}_{x}^{\top}\mathbf{W}_{x}\boldsymbol{\mathcal{E}}.
\end{align*}
Here,
\begin{eqnarray}
\frac{1}{n}\mathbf{D}_{h}\mathbf{X}_{x}^{\top}\mathbf{W}_{x}\boldsymbol{\mathcal{E}} & = & \mathbf{D}_{h}\left[\begin{array}{c}
\frac{1}{n}\sum_{g=1}^{G}\sum_{j=1}^{n_{g}}K_{h}\left(X_{gj}-x\right)e_{gj}\\
\frac{1}{n}\sum_{g=1}^{G}\sum_{j=1}^{n_{g}}K_{h}\left(X_{gj}-x\right)\left(X_{gj}-x\right)e_{gj}
\end{array}\right]=\mathbf{D}_{h}\left[\begin{array}{c}
\mathcal{E}_{0}(x)\\
\mathcal{E}_{1}(x)
\end{array}\right]\nonumber \\
 & = & \left[\begin{array}{c}
O_{p}\left(\sqrt{\frac{1}{nh^{d}}}\right)\\
h^{-2}O_{p}\left(\sqrt{\frac{1}{nh^{d-2}}}\right)\mathbf{1}_{d}
\end{array}\right].\label{eq:XWE}
\end{eqnarray}
Thus, $\text{\eqref{eq:XWX}}$ and $\text{\eqref{eq:XWE}}$ together imply that
\begin{eqnarray*}
 &  & \mathbf{e}_{1}^{\top}\left(\mathbf{D}_{h}\mathbf{X}_{x}^{\top}\mathbf{W}_{x}\mathbf{X}_{x}\right)^{-1}\mathbf{D}_{h}\mathbf{X}_{x}^{\top}\mathbf{W}_{x}\boldsymbol{\mathcal{E}}\\
 & = & \mathbf{e}_{1}^{\top}\left[\begin{array}{cc}
f(x)^{-1}+o_{p}(1) & o_{p}\left(h\right)\mathbf{1}_{d}^{\top}\\
o_{p}\left(h^{-1}\right)\mathbf{1}_{d} & \left(f(x)\kappa_{2}I_{d\times d}\right)^{-1}+o_{p}\left(1\right)\mathbf{1}_{d}\mathbf{1}_{d}^{\top}
\end{array}\right]\left[\begin{array}{c}
O_{p}\left(\sqrt{\frac{1}{nh^{d}}}\right)\\
h^{-1}O_{p}\left(\sqrt{\frac{1}{nh^{d}}}\right)\mathbf{1}_{d}
\end{array}\right]\\
 & = & O_{p}\left(\sqrt{\frac{1}{nh^{d}}}\right)+o_{p}\left(\sqrt{\frac{1}{nh^{d}}}\right)=o_{p}(1).
\end{eqnarray*}
Hence, $\hat{m}_{\text{LL}}\left(x\right)\overset{p}{\rightarrow}m\left(x\right)$.
\end{proof}

\subsection{Proof for Theorem $\text{\ref{thm:LL_asy_dist}}$}
\begin{proof}
Theorem $\text{\ref{thm:LL_bias}}$ and $\hat{m}_{\mathrm{LL}}(x)=\mathbf{e}_{1}^{\top}\left(\mathbf{X}_{x}^{\top}\mathbf{W}_{x}\mathbf{X}_{x}\right)^{-1}\mathbf{X}_{x}^{\top}\mathbf{W}_{x}\left\{ \mathbf{M}+\boldsymbol{\mathcal{E}}\right\} $ together imply that
\begin{align*}
\sqrt{nh^{d}}\left(\hat{m}_{\mathrm{LL}}(x)-m(x)-h^{2}B_{\mathrm{LL}}(x)\right) & =\mathbf{e}_{1}^{\top}\left(\frac{1}{n}\mathbf{D}_{h}\mathbf{X}_{x}^{\top}\mathbf{W}_{x}\mathbf{X}_{x}\right)^{-1}\sqrt{nh^{d}}\frac{1}{n}\mathbf{D}_{h}\mathbf{X}_{x}^{\top}\mathbf{W}_{x}\boldsymbol{\mathcal{E}}+\sqrt{nh^{d}}o_{p}\left(h^{2}\right)\\
 & =\mathbf{e}_{1}^{\top}\left(\frac{1}{n}\mathbf{D}_{h}\mathbf{X}_{x}^{\top}\mathbf{W}_{x}\mathbf{X}_{x}\right)^{-1}\sqrt{nh^{d}}\frac{1}{n}\mathbf{D}_{h}\mathbf{X}_{x}^{\top}\mathbf{W}_{x}\boldsymbol{\mathcal{E}}+o_{p}\left(1\right),
\end{align*}
where the second equality follows from $nh^{d+4}=O(1)$.

Equations $\text{\eqref{eq:XWX}}$ and $\text{\eqref{eq:XWE}}$ together imply that the first term on the displayed equation will be
\begin{eqnarray*}
 &  & \mathbf{e}_{1}^{\top}\left(\frac{1}{n}\mathbf{D}_{h}\mathbf{X}_{x}^{\top}\mathbf{W}_{x}\mathbf{X}_{x}\right)^{-1}\sqrt{nh^{d}}\frac{1}{n}\mathbf{D}_{h}\mathbf{X}_{x}^{\top}\mathbf{W}_{x}\boldsymbol{\mathcal{E}}\\
 & = & \mathbf{e}_{1}^{\top}\left[\begin{array}{cc}
f(x)^{-1}+o_{p}(1) & o_{p}\left(h\right)\mathbf{1}_{d}^{\top}\\
o_{p}\left(h^{-1}\right)\mathbf{1}_{d} & \left(f(x)\kappa_{2}I_{d\times d}\right)^{-1}+o_{p}\left(1\right)\mathbf{1}_{d}\mathbf{1}_{d}^{\top}
\end{array}\right]\left[\begin{array}{c}
\sqrt{nh^{d}}\mathcal{E}_{0}(x)\\
\sqrt{nh^{d}}h^{-2}O_{p}\left(\sqrt{\frac{1}{nh^{d-2}}}\right)\mathbf{1}_{d}
\end{array}\right]\\
 & = & \left\{ f(x)^{-1}+o_{p}(1)\right\} \sqrt{nh^{d}}\mathcal{E}_{0}(x)+o_{p}\left(1\right)\\
 & = & \frac{\sqrt{nh^{d}}\mathcal{E}_{0}(x)}{f(x)}+o_{p}\left(1\right).
\end{eqnarray*}
We conclude with a similar argument to the proof of Theorem $\text{\ref{thm:nw_asy_dist}}$.
\end{proof}

\subsection{Proof for Theorem $\ref{thm:psi_unifconv}$}
\begin{proof}
We will show the theorem by the following three steps. The proof modifies time series results (Theorem 2 of \citet{hansen2008uniform}; Theorem 4.1 of \citet{vogt2012nonparametric}) to the cluster sampling case.

Let $\tau_{n}=C_{\tau}n^{1/s}$, where $C_{\tau}>0$ will be chosen in Step 1 below.\footnote{The choice of $\tau_{n}$ is different from \citet{hansen2008uniform}. For discussions on it, the reader can refer to the proof of Lemma B-1 in \citet{cattaneo2013generalized} and the proof of Theorem 4.1 in \citet{vogt2012nonparametric}.} Decompose $\hat{\psi}\left(x\right)$ into the tail $\hat{\psi}_{2}\left(x\right)$ and the truncated part $\hat{\psi}_{1}\left(x\right)$. 

\begin{eqnarray*}
\hat{\psi}\left(x\right) & = & \frac{1}{n}\sum_{g=1}^{G}\sum_{j=1}^{n_{g}}K_{h}\left(X_{gj}-x\right)W_{gj}\mathbf{1}\left\{ \left|W_{gj}\right|\leq\tau_{n}\right\} \\
 &  & +\frac{1}{n}\sum_{g=1}^{G}\sum_{j=1}^{n_{g}}K_{h}\left(X_{gj}-x\right)W_{gj}\mathbf{1}\left\{ \left|W_{gj}\right|>\tau_{n}\right\} \\
 & \equiv & \hat{\psi}_{1}\left(x\right)+\hat{\psi}_{2}\left(x\right).
\end{eqnarray*}
Then,
\[
\hat{\psi}\left(x\right)-\mathbb{E}\left[\hat{\psi}\left(x\right)\right]=\hat{\psi}_{1}\left(x\right)-\mathbb{E}\left[\hat{\psi}_{1}\left(x\right)\right]+\hat{\psi}_{2}\left(x\right)-\mathbb{E}\left[\hat{\psi}_{2}\left(x\right)\right].
\]

\textbf{Step 1}: Evaluate the tail part $\hat{\psi}_{2}\left(x\right)-\mathbb{E}\left[\hat{\psi}_{2}\left(x\right)\right]$

The tail part has the following bounds. Pick any $\varepsilon>0$.
\begin{align*}
\mathbb{P}\left(\sup_{x}\left|\hat{\psi}_{2}\left(x\right)\right|>a_{n}\right) & \leq\mathbb{P}\left(\left|W_{i}\right|>\tau_{n}\text{ for some }i\right)\\
 & \leq n\mathbb{P}\left(\left|W\right|>\tau_{n}\right)\\
 & \leq n\mathbb{E}\left[\left|W\right|^{s}\right]\tau_{n}^{-s}\\
 & \leq nB_{1}\tau_{n}^{-s}\leq B_{1}/C_{\tau}^{s},
\end{align*}
where the first inequality follows from the construction of $\hat{\psi}_{2}\left(x\right)$, the second inequality follows from the union bound, the third inequality follows from Markov's inequality, the fourth inequality follows from $\text{\eqref{eq:Ws_bound}}$, and the last equality follows from the definition of $\tau_{n}$. Then, we can choose a large enough number $C_{\tau}$ such that $\mathbb{P}\left(\sup_{x}\left|\hat{\psi}_{2}\left(x\right)\right|>a_{n}\right)\leq\varepsilon$. Hence, $\left|\hat{\psi}_{2}\left(x\right)\right|=O_{p}\left(a_{n}\right)$ uniformly. Note that $C_{\tau}$ depends on $\varepsilon$, but does not on $n$.

Also, 
\begin{eqnarray*}
 &  & \mathbb{E}\left[\left|\hat{\psi}_{2}\left(x\right)\right|\right]\\
 & \leq & \frac{1}{h^{d}}\int_{\mathbb{R}^{d}}K\left(\frac{X-x}{h}\right)\mathbb{E}\left[\left|W\right|\mathbf{1}\left\{ \left|W\right|>\tau_{n}\right\} \mid X\right]f\left(X\right)\mathrm{d}X\\
 & \leq & \int_{\mathbb{R}^{d}}K\left(T\right)\mathbb{E}\left[\left|W\right|\mathbf{1}\left\{ \left|W\right|>\tau_{n}\right\} \mid X=x+hT\right]f\left(x+hT\right)\mathrm{d}T\\
 & \leq & \frac{1}{\tau_{n}^{s-1}}\int_{\mathbb{R}^{d}}K\left(T\right)\mathbb{E}\left[\left|W\right|^{s}\mathbf{1}\left\{ \left|W\right|>\tau_{n}\right\} \mid X=x+hT\right]f\left(x+hT\right)\mathrm{d}T\\
 & \leq & \frac{1}{\tau_{n}^{s-1}}B_{2}=O\left(a_{n}\right),
\end{eqnarray*}
where the fourth inequality follows uniformly from $\text{\eqref{eq:Ws_cond}}$, and the last equality follows from
\begin{align*}
\frac{1}{\tau_{n}^{s-1}} & =O\left(n^{1/s-1}\right)\\
 & \leq O\left(a_{n}\right).\qquad\because\text{\eqref{eq:theta_logn}}
\end{align*}

\bigskip{}

In the next two steps, we evaluate the truncated part $\hat{\psi}_{1}\left(x\right)-\mathbb{E}\left[\hat{\psi}_{1}\left(x\right)\right]$ .\bigskip{}

\textbf{Step 2}: Bound the supremum over $\left\Vert x\right\Vert \leq c_{n}$ with the maximum over a finite grid

We can cover the region $\left\{ x\in\mathbb{R}^{d}:\left\Vert x\right\Vert \leq c_{n}\right\} $ with $N_{\mathrm{ball}}\leq c_{n}^{d}h^{-d}a_{n}^{-d}$ balls 
\[
B_{a_{n}h}\left(x_{k}\right)=\left\{ x\in\mathbb{R}^{d}:\left\Vert x-x_{k}\right\Vert \leq a_{n}h\right\} ,
\]
 where $x_{k}$ is the midpoint of $B_{a_{n}h}\left(x_{k}\right)$. Assumption $\text{\ref{assu:kernel}}$ implies that for all $\left\Vert x-x^{\prime}\right\Vert \leq a\leq L$, there exists some function $K^{*}(\cdot)$ and some constant $A>0$ such that 
\begin{equation}
\left|K(x)-K(x^{\prime})\right|\leq aAK^{*}(x^{\prime})\label{eq:liplike}
\end{equation}
where $K^{*}(u)=\prod_{q=1}^{d}k^{*}\left(u^{(q)}\right)$ and $k^{*}\left(\cdot\right)$ satisfies the definition of the kernel function (Definition $\text{\ref{def:kernel}}$). To construct such functions, we can define 
\begin{align*}
K^{*}\left(u\right) & \equiv\prod_{q=1}^{d}1/(4L)\mathbf{1}\left\{ \left|u^{(q)}\right|\leq2L\right\} \\
 & \equiv\prod_{q=1}^{d}k^{*}\left(u^{(q)}\right)
\end{align*}
 and set $A=4^{d}L^{d}\Lambda$. Also, let $K_{h}^{*}\left(\cdot\right)=\frac{1}{h^{d}}K^{*}\left(\frac{\cdot}{h}\right)$.\footnote{Under Assumption $\text{\ref{assu:kernel}}$,
\begin{align*}
\left|K(x)-K(x^{\prime})\right| & \leq\Lambda\left\Vert x-x^{\prime}\right\Vert \mathbf{1}\left\{ \left\Vert x^{\prime}\right\Vert \leq2L\right\} \leq a\Lambda\mathbf{1}\left\{ \left\Vert x^{\prime}\right\Vert \leq2L\right\} \\
 & \leq a\Lambda\prod_{q=1}^{d}\mathbf{1}\left\{ \left|x^{\prime}{}^{(q)}\right|\leq2L\right\} =a\Lambda4^{d}L^{d}K^{*}\left(x^{\prime}\right),
\end{align*}
where the first inequality follows from the support of $K$, the second inequality follows from $\left\Vert x-x^{\prime}\right\Vert \leq a$, and the third inequality follows from $\left|x^{\prime}{}^{(q)}\right|\leq\sqrt{\sum_{p=1}^{d}\left(x^{\prime}{}^{(p)}\right)^{2}}=\left\Vert x^{\prime}\right\Vert $.

Since $k^{*}$ is bounded, symmetric, and has finite moments, it satisfies the definition of the kernel function.} 

Then, for any $x\in B_{a_{n}h}\left(x_{k}\right)$ equation $\text{\eqref{eq:liplike}}$ implies 
\begin{equation}
\left|K\left(\frac{X_{gj}-x}{h}\right)-K\left(\frac{X_{gj}-x_{k}}{h}\right)\right|\leq a_{n}AK^{*}\left(\frac{X_{gj}-x_{k}}{h}\right)\label{eq:K*}
\end{equation}
since 
\[
\left\Vert \frac{X_{gj}-x}{h}-\frac{X_{gj}-x_{k}}{h}\right\Vert =\frac{\left\Vert x-x_{k}\right\Vert }{h}\leq a_{n},
\]
and $a_{n}\leq L$ for large enough $n$.

Define $\widetilde{\psi}_{1}\left(x\right)$ by replacing $K_{h}(\cdot)$ on $\hat{\psi}_{1}\left(x\right)$ with $K_{h}^{*}(\cdot)$,
\[
\widetilde{\psi}_{1}\left(x\right)\equiv\frac{1}{n}\sum_{g=1}^{G}\sum_{j=1}^{n_{g}}K_{h}^{*}\left(X_{gj}-x\right)W_{gj}\mathbf{1}\left\{ \left|W_{gj}\right|\leq\tau_{n}\right\} .
\]
Then, $\mathbb{E}\left[\widetilde{\psi}_{1}\left(x\right)\right]$ is bounded since 
\begin{align*}
\mathbb{E}\left[\widetilde{\psi}_{1}\left(x\right)\right] & =\frac{1}{h^{d}}\mathbb{E}\left[K^{*}\left(\frac{X_{gj}-x}{h}\right)\mathbb{E}\left[\left|W_{gj}\right|\mathbf{1}\left\{ \left|W_{gj}\right|\leq\tau_{n}\right\} \mid X_{gj}\right]\right]\\
 & \leq\int_{\mathbb{R}^{d}}K^{*}\left(u_{gj}\right)\mathbb{E}\left[\left|W_{gj}\right|\mid X_{gj}=x+hu_{gj}\right]f\left(x+hu_{gj}\right)\mathrm{d}u_{gj}\\
 & \leq B_{3}\int_{\mathbb{R}^{d}}K^{*}\left(u_{gj}\right)\mathrm{d}u_{gj}=B_{3}<\infty,
\end{align*}
where $B_{3}$ exists by $\text{\eqref{eq:Ws_cond}}$. Thus, $A\mathbb{E}\left[\widetilde{\psi}_{1}\left(x\right)\right]<M$ for large enough $M$, and within each ball $B_{a_{n}h}\left(x_{k}\right)$,
\begin{eqnarray*}
 &  & \sup_{x\in B_{a_{n}h}\left(x_{k}\right)}\left|\hat{\psi}_{1}\left(x\right)-\mathbb{E}\left[\hat{\psi}_{1}\left(x\right)\right]\right|\\
 & = & \sup_{x\in B_{a_{n}h}\left(x_{k}\right)}\left|\hat{\psi}_{1}\left(x\right)-\hat{\psi}_{1}\left(x_{k}\right)+\hat{\psi}_{1}\left(x_{k}\right)-\mathbb{E}\left[\hat{\psi}_{1}\left(x_{k}\right)\right]+\mathbb{E}\left[\hat{\psi}_{1}\left(x_{k}\right)\right]-\mathbb{E}\left[\hat{\psi}_{1}\left(x\right)\right]\right|\\
 & \leq & \left|\hat{\psi}_{1}\left(x_{k}\right)-\mathbb{E}\left[\hat{\psi}_{1}\left(x_{k}\right)\right]\right|+\sup_{x\in B_{a_{n}h}\left(x_{k}\right)}\left|\hat{\psi}_{1}\left(x\right)-\hat{\psi}_{1}\left(x_{k}\right)\right|+\sup_{x\in B_{a_{n}h}\left(x_{k}\right)}\left|\mathbb{E}\left[\hat{\psi}_{1}\left(x_{k}\right)\right]-\mathbb{E}\left[\hat{\psi}_{1}\left(x\right)\right]\right|\\
 & \leq & \left|\hat{\psi}_{1}\left(x_{k}\right)-\mathbb{E}\left[\hat{\psi}_{1}\left(x_{k}\right)\right]\right|+a_{n}A\left\{ \left|\widetilde{\psi}_{1}\left(x_{k}\right)\right|+\mathbb{E}\left[\left|\widetilde{\psi}_{1}\left(x_{k}\right)\right|\right]\right\} \\
 & \leq & \left|\hat{\psi}_{1}\left(x_{k}\right)-\mathbb{E}\left[\hat{\psi}_{1}\left(x_{k}\right)\right]\right|+a_{n}A\left|\widetilde{\psi}_{1}\left(x_{k}\right)-\mathbb{E}\left[\widetilde{\psi}_{1}\left(x_{k}\right)\right]\right|+2a_{n}A\mathbb{E}\left[\left|\widetilde{\psi}_{1}\left(x_{k}\right)\right|\right]\\
 & \leq & \left|\hat{\psi}_{1}\left(x_{k}\right)-\mathbb{E}\left[\hat{\psi}_{1}\left(x_{k}\right)\right]\right|+\left|\widetilde{\psi}_{1}\left(x_{k}\right)-\mathbb{E}\left[\widetilde{\psi}_{1}\left(x_{k}\right)\right]\right|+2a_{n}M,
\end{eqnarray*}
where the first and third inequalities follow from the triangle inequality, the second inequality follows from $\text{\eqref{eq:K*}}$, and the last inequality comes from $a_{n}\leq A^{-1}$ for large enough $n$ and $A\mathbb{E}\left[\widetilde{\psi}_{1}\left(x\right)\right]<M$.

As a consequence,
\begin{eqnarray*}
 &  & \mathbb{P}\left[\sup_{\left\Vert x\right\Vert \leq c_{n}}\left|\hat{\psi}_{1}\left(x\right)-\mathbb{E}\left[\hat{\psi}_{1}\left(x\right)\right]\right|>4Ma_{n}\right]\\
 & \leq & N_{\mathrm{ball}}\max_{1\leq k\leq N_{\mathrm{ball}}}\mathbb{P}\left[\sup_{x\in B_{a_{n}h}\left(x_{k}\right)}\left|\hat{\psi}_{1}\left(x\right)-\mathbb{E}\left[\hat{\psi}_{1}\left(x\right)\right]\right|>4Ma_{n}\right]\\
 & \leq & N_{\mathrm{ball}}\max_{1\leq k\leq N_{\mathrm{ball}}}\left\{ \mathbb{P}\left[\left|\hat{\psi}_{1}\left(x_{k}\right)-\mathbb{E}\left[\hat{\psi}_{1}\left(x_{k}\right)\right]\right|>Ma_{n}\right]+\mathbb{P}\left[\left|\widetilde{\psi}_{1}\left(x_{k}\right)-\mathbb{E}\left[\widetilde{\psi}_{1}\left(x_{k}\right)\right]\right|>Ma_{n}\right]\right\} .
\end{eqnarray*}
Since we can evaluate both of $\mathbb{P}\left[\left|\hat{\psi}_{1}\left(x_{k}\right)-\mathbb{E}\left[\hat{\psi}_{1}\left(x_{k}\right)\right]\right|>Ma_{n}\right]$ and $\mathbb{P}\left[\left|\widetilde{\psi}_{1}\left(x_{k}\right)-\mathbb{E}\left[\widetilde{\psi}_{1}\left(x_{k}\right)\right]\right|>Ma_{n}\right]$ in the same way, we will focus on $\mathbb{P}\left[\left|\hat{\psi}_{1}\left(x_{k}\right)-\mathbb{E}\left[\hat{\psi}_{1}\left(x_{k}\right)\right]\right|>Ma_{n}\right]$ in the next step.

\bigskip{}

\textbf{Step 3}: Apply the Bernstein's inequality.

Define
\[
\widetilde{\mathbf{U}}_{g}=\sum_{j=1}^{n_{g}}\left\{ K\left(\frac{X_{gj}-x}{h}\right)W_{gj}\mathbf{1}\left\{ \left|W_{gj}\right|\leq\tau_{n}\right\} -\mathbb{E}\left[K\left(\frac{X_{gj}-x}{h}\right)W_{gj}\mathbf{1}\left\{ \left|W_{gj}\right|\leq\tau_{n}\right\} \right]\right\} .
\]
Then, 
\[
\hat{\psi}_{1}\left(x_{k}\right)-\mathbb{E}\left[\hat{\psi}_{1}\left(x_{k}\right)\right]=\frac{1}{nh^{d}}\sum_{g=1}^{G}\widetilde{\mathbf{U}}_{g}.
\]
 Since 
\[
\left|K\left(\frac{X_{gj}-x}{h}\right)W_{gj}\mathbf{1}\left\{ \left|W_{gj}\right|\leq\tau_{n}\right\} -\mathbb{E}\left[K\left(\frac{X_{gj}-x}{h}\right)W_{gj}\mathbf{1}\left\{ \left|W_{gj}\right|\leq\tau_{n}\right\} \right]\right|\leq2\bar{K}\tau_{n}
\]
and
\begin{align*}
\operatorname{Var}\left(\sum_{g=1}^{G}\widetilde{\mathbf{U}}_{g}\right) & =n^{2}h^{2d}\operatorname{Var}\left(\hat{\psi}\left(x\right)\right)\leq nh^{d}\overline{V},\qquad\because\text{Assumption \text{\ref{assu:psi_var}}}
\end{align*}
the Bernstein's inequality for cluster sampling (Lemma $\text{\ref{lem:Bernstein}}$) implies
\begin{align*}
\mathbb{P}\left[\left|\hat{\psi}_{1}\left(x_{k}\right)-\mathbb{E}\left[\hat{\psi}_{1}\left(x_{k}\right)\right]\right|>Ma_{n}\right] & =\mathbb{P}\left[\left|\sum_{g=1}^{G}\widetilde{\mathbf{U}}_{g}\right|>Ma_{n}nh^{d}\right]\\
 & \leq2\exp\left\{ -\frac{1}{2}\frac{M^{2}a_{n}^{2}n^{2}h^{2d}}{nh^{d}\overline{V}+2\left(\max_{g\leq G}n_{g}\right)\bar{K}\tau_{n}Ma_{n}nh^{d}/3}\right\} \\
 & \leq2\exp\left\{ -\frac{1}{2}\frac{M^{2}a_{n}^{2}nh^{d}}{\overline{V}+2C_{a}\bar{K}M/3}\right\} \\
 & =2\exp\left\{ -\frac{1}{2}\frac{M^{2}\log n}{\overline{V}+2C_{a}\bar{K}M/3}\right\} \\
 & \leq2\exp\left\{ -\frac{6M\log n}{3+2C_{a}\bar{K}}\right\} \\
 & =2n^{-6M/\left(3+2C_{a}\bar{K}\right)},
\end{align*}
where the second inequality with some $C_{a}>0$ follows from $\left(\max_{g\leq G}n_{g}\right)\tau_{n}a_{n}=O(1)$ by $\eqref{eq:theta_logn}$, the second equality follows from $a_{n}^{2}=\log n/(nh^{d})$, the third inequality follows by choosing $M>\overline{V}$. Thus, 
\begin{align}
\mathbb{P}\left[\sup_{\left\Vert x\right\Vert \leq c_{n}}\left|\hat{\psi}_{1}\left(x\right)-\mathbb{E}\left[\hat{\psi}_{1}\left(x\right)\right]\right|>4Ma_{n}\right] & \leq4N_{\mathrm{ball}}n^{-6M/\left(3+2C_{a}\bar{K}\right)}\nonumber \\
 & \leq O\left(T_{n}\right),\label{eq:bound_psi1}
\end{align}
where $T_{n}=c_{n}^{d}h^{-d}a_{n}^{-d}n^{-6M/\left(3+2C_{a}\bar{K}\right)}$. We can evaluate
\begin{align*}
c_{n}^{d}h^{-d} & =O\left(\frac{\left(\max_{g\leq G}n_{g}\right)^{2}\log n}{h^{d}}\right)\qquad\because\text{\eqref{eq:cn}}\\
 & =O\left(n^{1-(2/s)}\right)\qquad\because\text{\eqref{eq:theta_logn}}
\end{align*}
and
\begin{align*}
a_{n}^{-d} & =\left(\frac{nh^{d}}{\log n}\right)^{d/2}=o\left(n^{d/2}\right).
\end{align*}
Thus,
\begin{align*}
T_{n} & =o\left(n^{1-(2/s)+(d/2)-(6M/(3+2C_{a}\bar{K}))}\right)\\
 & \leq o(1),
\end{align*}
where the inequality holds for large enough $M$. Therefore, $\text{\eqref{eq:bound_psi1}}$ implies $\sup_{\left\Vert x\right\Vert \leq c_{n}}\left|\hat{\psi}_{1}\left(x\right)-\mathbb{E}\left[\hat{\psi}_{1}\left(x\right)\right]\right|=O_{p}\left(a_{n}\right)$.
\end{proof}

\subsection{Proof for Lemma $\ref{lem:Bernstein}$}
\begin{proof}
By the triangle inequality, $\left|\widetilde{\mathbf{Y}}_{g}\right|=\left|\sum_{j=1}^{n_{g}}Y_{gj}\right|\leq n_{g}B$. Thus, 
\[
\max_{g\leq G}\left|\widetilde{\mathbf{Y}}_{g}\right|\leq\left(\max_{g\leq G}n_{g}\right)B.
\]
The result follows from the standard Bernstein's inequality for the independent and zero mean random variables $\widetilde{\mathbf{Y}}_{1},\dots,\widetilde{\mathbf{Y}}_{G}$.
\end{proof}

\subsection{Proof for Theorem $\text{\ref{thm:density_unifconv}}$}
\begin{proof}
As the proof for Lemma $\text{\ref{lem:Fr}}$, we can prove that
\begin{align*}
\operatorname{Var}\left[\hat{f}\left(x\right)\right] & =\operatorname{Var}\left[F_{0}\left(x\right)\right]\\
 & \le O\left(n^{-1}h^{-d}\right)+O\left(\frac{1}{n}\left(\max_{g}n_{g}\right)\right)=O\left(\frac{1}{nh^{d}}\right).
\end{align*}
 Under Assumption \ref{assu:nonpara_unif}, this bound holds uniformly for any $x\in\mathbb{R}^{d}$. Thus, Assumption $\text{\ref{assu:psi_var}}$ for $\hat{\psi}\left(x\right)=\hat{f}\left(x\right)$ with $W_{gj}=1$ is satisfied. Since we also have Assumption $\text{\ref{assu:Wgj}}$ with $s=\infty$, all assumptions for Theorem $\text{\ref{thm:psi_unifconv}}$ are satisfied. Hence,
\begin{equation}
\sup_{\left\Vert x\right\Vert \leq c_{n}}\left|\hat{f}\left(x\right)-\mathbb{E}\left[\hat{f}\left(x\right)\right]\right|=O_{p}\left(a_{n}\right).\label{eq:fhat-Ef}
\end{equation}
As the proof for Lemma $\text{\ref{lem:Fr}}$, we can also show
\begin{equation}
\sup_{x\in\mathbb{R}^{d}}\left|\mathbb{E}\left[\hat{f}\left(x\right)\right]-f\left(x\right)\right|=O\left(h^{2}\right),\label{eq:Ef-f}
\end{equation}
where we have the sup bound under Assumption \ref{assu:nonpara_unif}. The triangle inequality, $\text{\eqref{eq:fhat-Ef}}$, and $\text{\eqref{eq:Ef-f}}$ together imply the result.
\end{proof}

\subsection{Proof for Theorem $\text{\ref{thm:nw_unifconv}}$}
\begin{proof}
\textbf{For the case $\widehat{m}_{*}(x)=\widehat{m}_{\mathrm{nw}}(x)$.}

First, Theorem $\text{\ref{thm:density_unifconv}}$ implies
\begin{equation}
\sup_{\left\Vert x\right\Vert \leq c_{n}}\left|\frac{\hat{f}\left(x\right)}{f\left(x\right)}-1\right|\leq\frac{\sup_{\left\Vert x\right\Vert \leq c_{n}}\left|\hat{f}\left(x\right)-f\left(x\right)\right|}{\inf_{\left\Vert x\right\Vert \leq c_{n}}f(x)}=O_{p}\left(\delta_{n}^{-1}\left(a_{n}+h^{2}\right)\right).\label{eq:fhat_ratio_unif}
\end{equation}

Next, define
\[
\widehat{\phi}\left(x\right)=\frac{1}{n}\sum_{g=1}^{G}\sum_{j=1}^{n_{g}}K_{h}\left(X_{gj}-x\right)Y_{gj}.
\]
Then,
\begin{eqnarray*}
 &  & \operatorname{Var}\left[\widehat{\phi}\left(x\right)\right]\\
 & = & \operatorname{Var}\left[\frac{1}{n}\sum_{g=1}^{G}\sum_{j=1}^{n_{g}}K\left(X_{gj}-x\right)\left\{ m\left(X_{gj}\right)-m(x)+e_{gj}+m(x)\right\} \right]\\
 & = & \operatorname{Var}\left[J_{0}(x)+m(x)F_{0}(x)+\mathcal{E}_{0}(x)\right]\\
 & \leq & \left(\sqrt{\operatorname{Var}\left[J_{0}(x)\right]}+m(x)\sqrt{\operatorname{Var}\left[F_{0}(x)\right]}+\sqrt{\operatorname{Var}\left[\mathcal{E}_{0}(x)\right]}\right)^{2},
\end{eqnarray*}
where the inequality follows since the absolute value of covariance is bonded by the product of the square root of variances. By the similar way as in the proof of Lemmas $\text{\ref{lem:Fr}}$, $\text{\ref{lem:Jr}}$, and $\text{\ref{lem:Er}}$, we can evaluate
\begin{align*}
\operatorname{Var}\left[F_{0}(x)\right] & \leq O\left(n^{-1}h^{-d}\right)+O\left(n^{-1}\left(\max_{g}n_{g}\right)\right)\leq O\left(\frac{1}{nh^{d}}\right),\\
\operatorname{Var}\left[J_{0}(x)\right] & \leq O\left(\frac{h^{2}}{nh^{d}}\right),\\
\operatorname{Var}\left[\mathcal{E}_{0}(x)\right] & \leq O\left(\frac{1}{nh^{d}}\right).
\end{align*}
Under Assumption \ref{assu:nonpara_unif}, these bounds hold uniformly for any $x\in\mathbb{R}$. Combining these equations and the uniform boundedness of $m(x)$, we have
\begin{align*}
\operatorname{Var}\left[\widehat{\phi}\left(x\right)\right] & \leq O\left(\frac{1}{nh^{d}}\right)
\end{align*}
 uniformly for any $x\in\mathbb{R}^{d}$. Then, all assumptions for Theorem $\text{\ref{thm:psi_unifconv}}$ are satisfied. Hence,
\begin{equation}
\sup_{\left\Vert x\right\Vert \leq c_{n}}\left|\widehat{\phi}\left(x\right)-\mathbb{E}\left[\widehat{\phi}\left(x\right)\right]\right|=O_{p}\left(a_{n}\right).\label{eq:nw_phi_unif}
\end{equation}
Also,
\begin{eqnarray}
 &  & \sup_{\left\Vert x\right\Vert \leq c_{n}}\left|\mathbb{E}\left[\widehat{\phi}\left(x\right)\right]-m(x)f(x)\right|\nonumber \\
 & = & \sup_{\left\Vert x\right\Vert \leq c_{n}}\left|\mathbb{E}\left[J_{0}(x)\right]+\mathbb{E}\left[\mathcal{E}_{0}(x)\right]+m(x)\mathbb{E}\left[F_{0}(x)\right]-m(x)f(x)\right|\nonumber \\
 & \leq & \sup_{\left\Vert x\right\Vert \leq c_{n}}\left|\mathbb{E}\left[J_{0}(x)\right]\right|+\sup_{\left\Vert x\right\Vert \leq c_{n}}\left|\mathbb{E}\left[\mathcal{E}_{0}(x)\right]\right|+\sup_{\left\Vert x\right\Vert \leq c_{n}}\left|m(x)\right|\sup_{\left\Vert x\right\Vert \leq c_{n}}\left|\mathbb{E}\left[F_{0}(x)\right]-f(x)\right|\nonumber \\
 & \leq & O\left(h^{2}\right)+0+O\left(1\right)O\left(h^{2}\right)\nonumber \\
 & = & O\left(h^{2}\right),\label{eq:phi_hat_e}
\end{eqnarray}
 where the first inequality follows from the triangle inequality, the second inequality can be shown as in the proof of Lemmas $\text{\ref{lem:Fr}}$, $\text{\ref{lem:Jr}}$, and $\text{\ref{lem:Er}}$, and Assumption \ref{assu:nonpara_unif} implies these bounds hold uniformly for any $x\in\mathbb{R}^{d}$. Hence,
\begin{eqnarray*}
 &  & \sup_{\left\Vert x\right\Vert \leq c_{n}}\left|\hat{m}_{\mathrm{nw}}\left(x\right)-m\left(x\right)\right|\\
 & = & \sup_{\left\Vert x\right\Vert \leq c_{n}}\left|\frac{\widehat{\phi}\left(x\right)}{f\left(x\right)}\cdot\frac{f\left(x\right)}{\widehat{f}\left(x\right)}-m\left(x\right)\right|\\
 & = & \sup_{\left\Vert x\right\Vert \leq c_{n}}\left|\left(\frac{\widehat{\phi}\left(x\right)}{f\left(x\right)}-\frac{m\left(x\right)\widehat{f}\left(x\right)}{f\left(x\right)}\right)\frac{f\left(x\right)}{\widehat{f}\left(x\right)}\right|\\
 & \leq & \sup_{\left\Vert x\right\Vert \leq c_{n}}\left|\frac{\widehat{\phi}\left(x\right)}{f\left(x\right)}-\frac{m\left(x\right)\widehat{f}\left(x\right)}{f\left(x\right)}\right|\sup_{\left\Vert x\right\Vert \leq c_{n}}\left|\frac{f\left(x\right)}{\widehat{f}\left(x\right)}\right|\\
 & \leq & \sup_{\left\Vert x\right\Vert \leq c_{n}}\left|\widehat{\phi}\left(x\right)-m\left(x\right)\widehat{f}\left(x\right)\right|\delta_{n}^{-1}\left\{ 1+O_{p}\left(\delta_{n}^{-1}\left(a_{n}+h^{2}\right)\right)\right\} \\
 & \leq & \left\{ \sup_{\left\Vert x\right\Vert \leq c_{n}}\left|\widehat{\phi}\left(x\right)-\mathbb{E}\left[\widehat{\phi}\left(x\right)\right]\right|+\sup_{\left\Vert x\right\Vert \leq c_{n}}\left|\mathbb{E}\left[\widehat{\phi}\left(x\right)\right]-m\left(x\right)f\left(x\right)\right|+\sup_{\left\Vert x\right\Vert \leq c_{n}}\left|m\left(x\right)f\left(x\right)-m\left(x\right)\widehat{f}\left(x\right)\right|\right\} \\
 &  & \times\delta_{n}^{-1}\left\{ 1+O_{p}\left(\delta_{n}^{-1}\left(a_{n}+h^{2}\right)\right)\right\} \\
 & \leq & \left\{ O_{p}\left(a_{n}\right)+O\left(h^{2}\right)+O_{p}\left(a_{n}\right)\right\} \delta_{n}^{-1}\left\{ 1+O_{p}\left(\delta_{n}^{-1}\left(a_{n}+h^{2}\right)\right)\right\} \\
 & \leq & O_{p}\left(\delta_{n}^{-1}\left(a_{n}+h^{2}\right)\right)\left\{ 1+O_{p}\left(\delta_{n}^{-1}\left(a_{n}+h^{2}\right)\right)\right\} \\
 & = & O_{p}\left(\delta_{n}^{-1}\left(a_{n}+h^{2}\right)\right),
\end{eqnarray*}
where the second inequality follows from $\text{\eqref{eq:delta_n}}$ and $\text{\eqref{eq:fhat_ratio_unif}}$, the third inequality follows from the triangle inequality, and the fourth inequality follows from $\text{\eqref{eq:nw_phi_unif}}$, the uniform boundedness of $m\left(x\right)$, the result of Theorem $\text{\ref{thm:density_unifconv}}$, and $\text{\eqref{eq:phi_hat_e}}$. 

\textbf{For the case $\widehat{m}_{*}(x)=\widehat{m}_{\mathrm{LL}}(x)$.}

Using the partition matrix inversion, we can rewrite
\begin{align}
\widehat{m}_{\mathrm{LL}}(x) & =\mathbf{e}_{1}^{\top}\left(\mathbf{X}_{x}^{\top}\mathbf{W}_{x}\mathbf{X}_{x}\right)^{-1}\mathbf{X}_{x}^{\top}\mathbf{W}_{x}\mathbf{Y}\nonumber \\
 & =\frac{\widehat{f}(x)\widehat{m}_{\text{nw}}(x)-S(x)^{\top}M(x)^{-1}N(x)}{\widehat{f}(x)-S(x)^{\top}M(x)^{-1}S(x)},\label{eq:LL_nw}
\end{align}
where
\begin{align*}
S(x) & \equiv\frac{1}{n}\sum_{g=1}^{G}\sum_{j=1}^{n_{g}}K_{h}\left(X_{gj}-x\right)\left(\frac{X_{gj}-x}{h}\right),\\
M(x) & \equiv\frac{1}{n}\sum_{g=1}^{G}\sum_{j=1}^{n_{g}}K_{h}\left(X_{gj}-x\right)\left(\frac{X_{gj}-x}{h}\right)\left(\frac{X_{gj}-x}{h}\right)^{\top},\\
N(x) & \equiv\frac{1}{n}\sum_{g=1}^{G}\sum_{j=1}^{n_{g}}K_{h}\left(X_{gj}-x\right)\left(\frac{X_{gj}-x}{h}\right)Y_{gj}.
\end{align*}
Define
\begin{align*}
S^{(q)}(x) & \equiv\frac{1}{n}\sum_{g=1}^{G}\sum_{j=1}^{n_{g}}K_{h}\left(X_{gj}-x\right)\left(\frac{X_{gj}^{(q)}-x^{(q)}}{h}\right)=h^{-1}F_{1}^{(q)}(x),\\
M^{(p,q)}(x) & \equiv\frac{1}{n}\sum_{g=1}^{G}\sum_{j=1}^{n_{g}}K_{h}\left(X_{gj}-x\right)\left(\frac{X_{gj}^{(p)}-x^{(p)}}{h}\right)\left(\frac{X_{gj}^{(q)}-x^{(q)}}{h}\right)^{\top}\\
 & =\begin{cases}
h^{-2}F_{2}^{(q)}(x) & \text{ if }p=q\\
h^{-2}F^{(p,q)}(x) & \text{ if }p\neq q
\end{cases},
\end{align*}
\begin{align*}
N^{(q)}(x) & \equiv\frac{1}{n}\sum_{g=1}^{G}\sum_{j=1}^{n_{g}}K_{h}\left(X_{gj}-x\right)\left(\frac{X_{gj}^{(q)}-x^{(q)}}{h}\right)\left\{ m\left(X_{gj}\right)-m(x)+m(x)+e_{gj}\right\} \\
 & =h^{-1}J_{1}^{(q)}(x)+h^{-1}m(x)F_{1}^{(q)}(x)+h^{-1}\mathcal{E}_{1}^{(q)}(x).
\end{align*}
Similar way as in the proof of Lemmas $\text{\ref{lem:Fr}}$, $\text{\ref{lem:Jr}}$, and $\text{\ref{lem:Er}}$, we can evaluate
\begin{align*}
\operatorname{Var}\left[h^{-1}F_{1}^{(q)}\right] & \leq O\left(n^{-1}h^{-d}\right)+O\left(n^{-1}\left(\max_{g}n_{g}\right)\right)=O\left(n^{-1}h^{-d}\right),
\end{align*}
\begin{align*}
\operatorname{Var}\left[h^{-2}F_{2}^{(q)}\right] & \leq O\left(n^{-1}h^{-d}\right)+O\left(n^{-1}\left(\max_{g}n_{g}\right)\right)=O\left(n^{-1}h^{-d}\right),
\end{align*}
\begin{align*}
\operatorname{Var}\left[h^{-2}F^{(p,q)}(x)\right] & \leq O\left(n^{-1}h^{-d}\right)+O\left(n^{-1}\left(\max_{g}n_{g}\right)\right)=O\left(n^{-1}h^{-d}\right),
\end{align*}
\begin{align*}
\operatorname{Var}\left[h^{-1}J_{1}^{(q)}(x)\right] & \leq O\left(n^{-1}h^{2-d}\right)+O\left(n^{-1}\left(\max_{g}n_{g}\right)h^{2}\right)=O\left(n^{-1}h^{-d}\right),
\end{align*}
and
\begin{align*}
\operatorname{Var}\left[h^{-1}\mathcal{E}_{1}^{(q)}(x)\right] & =O\left(n^{-1}h^{-d}\right).
\end{align*}
Since these bounds are uniform for any $x\in\mathbb{R}$ under Assumption $\ref{assu:nonpara_unif}$ and the compact kernel function enables us to treat $(X_{gj}^{(q)}-x^{(q)})/h$ as bounded in $S^{(q)}(x)$, $M^{(p,q)}(x)$, and $N^{(q)}(x)$, we can apply Theorem $\text{\ref{thm:psi_unifconv}}$:
\begin{align*}
\sup_{\|x\|\leq c_{n}}\left|S^{(q)}(x)-\mathbb{E}\left[h^{-1}F_{1}^{(q)}(x)\right]\right| & =\sup_{\|x\|\leq c_{n}}\left|S^{(q)}(x)-h\partial_{q}f(x)\kappa_{2}+O\left(h^{2}\right)\right|=O_{p}\left(a_{n}\right),\\
\sup_{\|x\|\leq c_{n}}\left|M^{(q,q)}(x)-\mathbb{E}\left[h^{-2}F_{2}^{(q)}(x)\right]\right| & =\sup_{\|x\|\leq c_{n}}\left|M^{(q,q)}(x)-f(x)\kappa_{2}+O\left(h^{2}\right)\right|=O_{p}\left(a_{n}\right),\\
\sup_{\|x\|\leq c_{n}}\left|M^{(p,q)}(x)-\mathbb{E}\left[h^{-2}F^{(p,q)}(x)\right]\right| & =\sup_{\|x\|\leq c_{n}}\left|M^{(p,q)}(x)+O\left(h^{2}\right)\right|=O_{p}\left(a_{n}\right),
\end{align*}
and
\begin{eqnarray*}
 &  & \sup_{\|x\|\leq c_{n}}\left|N^{(q)}(x)-\mathbb{E}\left[N^{(q)}(x)\right]\right|\\
 & = & \sup_{\|x\|\leq c_{n}}\left|N^{(q)}(x)-hf(x)\partial_{q}m(x)\kappa_{2}-hm(x)\partial_{q}f(x)\kappa_{2}-0+O\left(h^{2}\right)\right|=O_{p}\left(a_{n}\right).
\end{eqnarray*}
By element-wise comparisons, we obtain 
\begin{align*}
S(x) & =h\kappa_{2}\nabla f\left(x\right)+O_{p}\left(a_{n}+h^{2}\right)\mathbf{1}_{d},\\
M(x) & =f(x)\kappa_{2}\mathbf{I}_{d\times d}+O_{p}\left(a_{n}+h^{2}\right)\mathbf{1}_{d}\mathbf{1}_{d}^{\top},\\
N(x) & =h\kappa_{2}\nabla\left\{ f\left(x\right)m\left(x\right)\right\} +O_{p}\left(a_{n}+h^{2}\right)\mathbf{1}_{d},
\end{align*}
where asymptotic orders are uniform over $\left\Vert x\right\Vert \leq c_{n}$. 

Therefore, by the same matrix calculations as \citet{hansen2008uniform}, we obtain 
\[
\widehat{m}_{\mathrm{LL}}(x)=m(x)+O_{p}\left(\delta_{n}^{-1}\left(a_{n}+h^{2}\right)\right)
\]
uniform over $\left\Vert x\right\Vert \leq c_{n}$. 
\end{proof}

\subsection{Proof for Theorem $\text{\ref{thm:bw}}$}

\begin{proof}
Necessary condition is
\[
\frac{\partial}{\partial h}\operatorname{AIMSE}=4h^{3}\bar{B}-\frac{dR_{k}^{d}\bar{\sigma}^{2}}{nh^{d+1}}=0.
\]
We obtain $h_{0}$ by solving this equation since by
\[
\frac{\partial}{\partial h^{2}}\operatorname{AIMSE}=12h^{2}\bar{B}+\frac{d(d+1)R_{k}^{d}\bar{\sigma}^{2}}{nh^{d+2}}>0,
\]
the first-order condition is sufficient.
\end{proof}

\subsection{Proof for Theorem $\text{\ref{thm:cr-cv}}$}
\begin{proof}
For any $g$ and $j$,
\begin{align*}
\mathbb{E}\left[\tilde{e}_{gj}\left(h\right)^{2}w\left(X_{gj}\right)\right] & =\mathbb{E}\left[e_{gj}^{2}w\left(X_{gj}\right)\right]+\mathbb{E}\left[\left\{ m\left(X_{gj}\right)-\widetilde{m}_{-g}\left(X_{gj},h\right)\right\} ^{2}w\left(X_{gj}\right)\right]\\
 & \qquad+2\mathbb{E}\left[\left\{ m\left(X_{gj}\right)-\widetilde{m}_{-g}\left(X_{gj},h\right)\right\} e_{gj}w\left(X_{gj}\right)\right]\\
 & \overset{\mathrm{(i)}}{=}\overline{\sigma}_{w}^{2}+\mathbb{E}\left[\left\{ m\left(X_{gj}\right)-\widetilde{m}_{-g}\left(X_{gj},h\right)\right\} ^{2}w\left(X_{gj}\right)\right]\\
 & \overset{\mathrm{(ii)}}{=}\overline{\sigma}_{w}^{2}+\mathbb{E}_{-g}\left[\int_{\mathbb{R}^{d}}\left\{ m\left(x\right)-\widetilde{m}_{-g}\left(x,h\right)\right\} ^{2}f\left(x\right)w\left(x\right)\mathrm{d}x\right]
\end{align*}
where (i) follows from the definition of $\overline{\sigma}_{w}^{2}$ and
\begin{align*}
\mathbb{E}\left[\left\{ m\left(X_{gj}\right)-\widetilde{m}_{-g}\left(X_{gj},h\right)\right\} e_{gj}w\left(X_{gj}\right)\right] & =\mathbb{E}\left[\mathbb{E}\left[\left\{ m\left(X_{gj}\right)-\widetilde{m}_{-g}\left(X_{gj},h\right)\right\} e_{gj}w\left(X_{gj}\right)\mid\mathbf{X}_{g}\right]\right]\\
 & =\mathbb{E}\left[\mathbb{E}\left[\left\{ m\left(X_{gj}\right)-\widetilde{m}_{-g}\left(X_{gj},h\right)\right\} w\left(X_{gj}\right)\mid\mathbf{X}_{g}\right]\mathbb{E}\left[e_{gj}\mid\mathbf{X}_{g}\right]\right]\\
 & =0
\end{align*}
since $\widetilde{m}_{-g}\left(X_{gj},h\right)$ is independent of $e_{gj}$ after conditioning $X_{g}$, and (ii) follows from
\begin{align*}
\mathbb{E}\left[\left\{ m\left(X_{gj}\right)-\widetilde{m}_{-g}\left(X_{gj},h\right)\right\} ^{2}w\left(X_{gj}\right)\right] & =\mathbb{E}_{-g}\left[\mathbb{E}\left[\left\{ m\left(X_{gj}\right)-\widetilde{m}_{-g}\left(X_{gj},h\right)\right\} ^{2}w\left(X_{gj}\right)\mid\mathbf{Y}_{-g},\mathbf{X}_{-g}\right]\right]\\
 & =\mathbb{E}_{-g}\left[\int_{\mathbb{R}^{d}}\left\{ m\left(x\right)-\widetilde{m}_{-g}\left(x,h\right)\right\} ^{2}f\left(x\right)w\left(x\right)\mathrm{d}x\right].
\end{align*}
Thus,
\begin{eqnarray*}
\mathbb{E}\left[\mathrm{CV}(h)\right] & = & \frac{1}{n}\sum_{g=1}^{G}\sum_{j=1}^{n_{g}}\mathbb{E}\left[\tilde{e}_{gj}\left(h\right)^{2}w\left(X_{gj}\right)\right]\\
 & = & \overline{\sigma}_{w}^{2}+\sum_{g=1}^{G}\frac{n_{g}}{n}\mathbb{E}_{-g}\left[\int_{\mathbb{R}^{d}}\left\{ m\left(x\right)-\widetilde{m}_{-g}\left(x,h\right)\right\} ^{2}f\left(x\right)w\left(x\right)\mathrm{d}x\right]\\
 & = & \overline{\sigma}_{w}^{2}+\operatorname{IMSE}_{G-1}(h).
\end{eqnarray*}
\end{proof}

\subsection{Proof for Theorem $\text{\ref{thm:joint_density_consistency}}$}
\begin{proof}
We can interpret $\text{\eqref{eq:joint_density_estimator}}$ as a standard nonparametric density estimator. Under Assumption $\text{\ref{assu:condvar}}$, Theorem $\text{\ref{thm:density_consistency}}$ is applicable for $\left(2d_{\mathrm{ind}}+d_{\mathrm{cls}}\right)$-dimensional regressors and $n_{g}(n_{g}-1)/2$ size clusters.
\end{proof}

\subsection{Proof for Theorem $\text{\ref{thm:nw_condvar_cons}}$}
\begin{proof}
First, we show that the feasible estimator of conditional variance can be asymptotically replaced with the infeasible estimator, i.e., 
\begin{equation}
\left|\widehat{\sigma}_{\mathrm{nw}}^{2}\left(x\right)-\widehat{\sigma}_{\mathrm{nw}}^{2*}\left(x\right)\right|=o_{p}(1).\label{eq:nw_condvar_star}
\end{equation}
Here,
\begin{align*}
\left|\widehat{\sigma}_{\mathrm{nw}}^{2}\left(x\right)-\widehat{\sigma}_{\mathrm{nw}}^{2*}\left(x\right)\right| & \leq\frac{\left|\frac{1}{n}\sum_{g=1}^{G}\sum_{j=1}^{n_{g}}K_{h}\left(X_{gj}-x\right)\left(\widehat{e}_{gj}^{2}-e_{gj}^{2}\right)\right|}{\frac{1}{n}\sum_{g=1}^{G}\sum_{j=1}^{n_{g}}K_{h}\left(X_{gj}-x\right)}\\
 & \leq\max_{g}\max_{j}\left|\widehat{e}_{gj}^{2}-e_{gj}^{2}\right|.
\end{align*}
and
\begin{eqnarray*}
 &  & \max_{g}\max_{j}\left|\widehat{e}_{gj}^{2}-e_{gj}^{2}\right|\\
 & \leq & \max_{g}\max_{j}\left|\left\{ e_{gj}+m\left(X_{gj}\right)-\widehat{m}_{*}\left(X_{gj}\right)\right\} ^{2}-e_{gj}^{2}\right|\\
 & = & \max_{g}\max_{j}\left|2e_{gj}\left\{ m\left(X_{gj}\right)-\widehat{m}_{*}\left(X_{gj}\right)\right\} +\left\{ m\left(X_{gj}\right)-\widehat{m}_{*}\left(X_{gj}\right)\right\} ^{2}\right|\\
 & \leq & 2\max_{g}\max_{j}\left|e_{gj}\right|\cdot\max_{g}\max_{j}\left|\left\{ m\left(X_{gj}\right)-\widehat{m}_{*}\left(X_{gj}\right)\right\} \right|\\
 &  & +\left\{ \max_{g}\max_{j}\left|m\left(X_{gj}\right)-\widehat{m}_{*}\left(X_{gj}\right)\right|\right\} ^{2}.
\end{eqnarray*}
Pick any $\varepsilon>0$. By Theorem $\text{\ref{thm:nw_unifconv}}$ and Assumption $\text{\ref{assu:condvar}}$ (v),
\begin{eqnarray*}
 &  & \Pr\left(\max_{g}\max_{j}\left|\left\{ m\left(X_{gj}\right)-\widehat{m}_{*}\left(X_{gj}\right)\right\} \right|>\varepsilon\right)\\
 & \leq & \Pr\left(\max_{g}\max_{j}\left|\left\{ m\left(X_{gj}\right)-\widehat{m}_{*}\left(X_{gj}\right)\right\} \right|>\varepsilon\mid\left\Vert X_{gj}\right\Vert \leq c_{n}\right)\Pr\left(\left\Vert X_{gj}\right\Vert \leq c_{n}\right)\\
 &  & +o(1)\\
 & \leq & \Pr\left(\sup_{\left\Vert x\right\Vert \leq c_{n}}\left|m\left(x\right)-\widehat{m}_{*}\left(x\right)\right|>\varepsilon\right)+o(1)\\
 & \leq & o(1).
\end{eqnarray*}
We also know that assumptions for Theorem $\text{\ref{thm:nw_unifconv}}$ imply
\[
\max_{g}\max_{j}\left|Y_{gj}\right|=o_{p}\left(n^{-1/s}\right),
\]
and
\[
\max_{g}\max_{j}\left|m\left(X_{gj}\right)\right|=O(1),
\]
thus
\begin{align*}
\max_{g}\max_{j}\left|e_{gj}\right| & =\max_{g}\max_{j}\left|Y_{gj}\right|+\max_{g}\max_{j}\left|m\left(X_{gj}\right)\right|\\
 & \leq O_{p}(1).
\end{align*}
Hence,
\[
\max_{g}\max_{j}\left|\widehat{e}_{gj}^{2}-e_{gj}^{2}\right|=o_{p}(1).
\]

Thus, it is sufficient to show that
\begin{equation}
\widehat{\sigma}_{\mathrm{nw}}^{2*}\left(x\right)\overset{p}{\rightarrow}\sigma^{2}\left(x\right).\label{eq:nw_sigma_cons}
\end{equation}
Let $v_{gj}=e_{gj}^{2}-\sigma^{2}\left(X_{gj}\right)$. Since $\sigma^{2}(x)=\mathbb{E}\left[e^{2}\mid X=x\right]$, we have
\begin{align*}
\mathbb{E}\left[v_{gj}\mid\mathbf{X}_{g}\right] & =0,\\
\mathbb{E}\left[v_{gj}^{2}\mid\mathbf{X}_{g}\right] & =\mathbb{E}\left[v_{gj}^{2}\mid X_{gj}\right]=\mathbb{E}\left[\left\{ e_{gj}^{2}-\sigma^{2}\left(X_{gj}\right)\right\} ^{2}\mid X_{gj}\right]=\mathbb{E}\left[e_{gj}^{4}\mid X_{gj}\right]-\left\{ \sigma^{2}\left(X_{gj}\right)\right\} ^{2}\\
 & =\varsigma^{2}\left(X_{gj}\right)-\left\{ \sigma^{2}\left(X_{gj}\right)\right\} ^{2},\\
\mathbb{E}\left[v_{gj}v_{g\ell}\mid\mathbf{X}_{g}\right] & =\mathbb{E}\left[v_{gj}v_{g\ell}\mid X_{gj}^{(\mathrm{ind})},X_{g\ell}^{(\mathrm{ind})};X_{g}^{(\mathrm{cls})}\right]\\
 & =\mathbb{E}\left[\left\{ e_{gj}^{2}-\sigma^{2}\left(X_{gj}\right)\right\} \left\{ e_{g\ell}^{2}-\sigma^{2}\left(X_{g\ell}\right)\right\} \mid X_{gj}^{(\mathrm{ind})},X_{g\ell}^{(\mathrm{ind})};X_{g}^{(\mathrm{cls})}\right]\\
 & =\mathbb{E}\left[e_{gj}^{2}e_{g\ell}^{2}\mid X_{gj}^{(\mathrm{ind})},X_{g\ell}^{(\mathrm{ind})};X_{g}^{(\mathrm{cls})}\right]-\sigma^{2}\left(X_{g\ell}\right)\sigma^{2}\left(X_{g\ell}\right)\\
 & =\varsigma\left(X_{gj}^{(\mathrm{ind})},X_{gj}^{(\mathrm{ind})},X_{g\ell}^{(\mathrm{ind})},X_{g\ell}^{(\mathrm{ind})};X_{g}^{(\mathrm{cls})}\right)-\sigma^{2}\left(X_{g\ell}\right)\sigma^{2}\left(X_{g\ell}\right).
\end{align*}
Under Assumption $\text{\ref{assu:condvar}}$, we can apply Theorem $\text{\ref{thm:nw_cons}}$ after replacing $m(x)$ with $\sigma^{2}\left(x\right)$ and obtain $\text{\eqref{eq:nw_sigma_cons}}$.
\end{proof}

\subsection{Proof for Theorem $\text{\ref{thm:nw_condcov_cons}}$}
\begin{proof}
We will first show 
\begin{equation}
\left|\widehat{\sigma}_{\mathrm{nw}}\left(x^{\mathrm{(ind)}},x^{\mathrm{(ind)}};x^{\mathrm{(cls)}}\right)-\widehat{\sigma}_{\mathrm{nw}}^{2}\left(x^{\mathrm{(ind)}},x^{\mathrm{(ind)}};x^{\mathrm{(cls)}}\right)\right|=o_{p}(1),\label{eq:nw_condcovar_star}
\end{equation}
and then show that 
\begin{equation}
\widehat{\sigma}_{\mathrm{nw}}\left(x^{\mathrm{(ind)}},x^{\mathrm{(ind)}};x^{\mathrm{(cls)}}\right)\overset{p}{\rightarrow}\sigma\left(x^{\mathrm{(ind)}},x^{\mathrm{(ind)}};x^{\mathrm{(cls)}}\right).\label{eq:nw_sigmacov_cons}
\end{equation}

For the first step,
\begin{eqnarray*}
 &  & \left|\widehat{\sigma}_{\mathrm{nw}}\left(x^{\mathrm{(ind)}},x^{\mathrm{(ind)}};x^{\mathrm{(cls)}}\right)-\widehat{\sigma}_{\mathrm{nw}}^{2}\left(x^{\mathrm{(ind)}},x^{\mathrm{(ind)}};x^{\mathrm{(cls)}}\right)\right|\\
 & \leq & \frac{\left|\sum_{g:n_{g}\geq2}\sum_{1\leq j<\ell\leq n_{g}}K\left(\frac{\left(X_{gj}^{(\mathrm{ind})\top},X_{g\ell}^{(\mathrm{ind})\top},X_{g}^{\mathrm{(cls)}\top}\right)^{\top}-\left(x^{(\mathrm{ind})\top},x^{(\mathrm{ind})\top},x^{\mathrm{(cls)}\top}\right)^{\top}}{b}\right)\left(\widehat{e}_{gj}\widehat{e}_{g\ell}-e_{gj}e_{g\ell}\right)\right|}{\sum_{g:n_{g}\geq2}\sum_{1\leq j<\ell\leq n_{g}}K\left(\frac{\left(X_{gj}^{(\mathrm{ind})\top},X_{g\ell}^{(\mathrm{ind})\top},X_{g}^{\mathrm{(cls)}\top}\right)^{\top}-\left(x^{(\mathrm{ind})\top},x^{(\mathrm{ind})\top},x^{\mathrm{(cls)}\top}\right)^{\top}}{b}\right)}\\
 & \leq & \max_{g}\max_{j,\ell}\left|\widehat{e}_{gj}\widehat{e}_{g\ell}-e_{gj}e_{g\ell}\right|,
\end{eqnarray*}
and
\begin{eqnarray*}
 &  & \max_{g}\max_{j,\ell}\left|\widehat{e}_{gj}\widehat{e}_{g\ell}-e_{gj}e_{g\ell}\right|\\
 & \leq & \max_{g}\max_{j,\ell}\left|\left\{ e_{gj}+m\left(X_{gj}\right)-\widehat{m}_{*}\left(X_{gj}\right)\right\} \left\{ e_{g\ell}+m\left(X_{g\ell}\right)-\widehat{m}_{*}\left(X_{g\ell}\right)\right\} -e_{gj}e_{g\ell}\right|\\
 & \leq & \max_{g}\max_{j,\ell}\left|e_{gj}\left\{ m\left(X_{g\ell}\right)-\widehat{m}_{*}\left(X_{g\ell}\right)\right\} \right|\\
 &  & +\max_{g}\max_{j,\ell}\left|e_{g\ell}\left\{ m\left(X_{gj}\right)-\widehat{m}_{*}\left(X_{gj}\right)\right\} \right|\\
 &  & +\max_{g}\max_{j,\ell}\left|\left\{ m\left(X_{gj}\right)-\widehat{m}_{*}\left(X_{gj}\right)\right\} \left\{ m\left(X_{g\ell}\right)-\widehat{m}_{*}\left(X_{g\ell}\right)\right\} \right|\\
 & \leq & 2\max_{g}\max_{j}\left|e_{gj}\right|\cdot\max_{g}\max_{j}\left|\left\{ m\left(X_{gj}\right)-\widehat{m}_{*}\left(X_{gj}\right)\right\} \right|\\
 &  & +\max_{g}\max_{j}\left|m\left(X_{gj}\right)-\widehat{m}_{*}\left(X_{gj}\right)\right|^{2}
\end{eqnarray*}
Thus, similarly to the proof of Theorem $\text{\ref{thm:nw_condvar_cons}}$, we can show that
\[
\max_{g}\max_{j,\ell}\left|\widehat{e}_{gj}\widehat{e}_{g\ell}-e_{gj}e_{g\ell}\right|=o_{p}(1),
\]
and $\text{\eqref{eq:nw_condcovar_star}}$ is shown. 

Next, let's prove $\text{\eqref{eq:nw_sigmacov_cons}}$. Let $u_{gj\ell}=e_{gj}e_{g\ell}-\sigma\left(X_{gj}^{(\mathrm{ind})},X_{g\ell}^{(\mathrm{ind})};X_{g}^{(\mathrm{cls})}\right)$. Since $\sigma\left(X_{gj}^{(\mathrm{ind})},X_{g\ell}^{(\mathrm{ind})};X_{g}^{(\mathrm{cls})}\right)=\mathbb{E}\left[e_{gj}e_{g\ell}\mid X_{gj}^{(\mathrm{ind})},X_{g\ell}^{(\mathrm{ind})};X_{g}^{(\mathrm{cls})}\right]$, we have
\begin{align*}
\mathbb{E}\left[u_{gj\ell}\mid\mathbf{X}_{g}\right] & =0,\\
\mathbb{E}\left[u_{gj\ell}^{2}\mid\mathbf{X}_{g}\right] & =\mathbb{E}\left[u_{gj\ell}^{2}\mid X_{gj}^{(\mathrm{ind})},X_{g\ell}^{(\mathrm{ind})};X_{g}^{(\mathrm{cls})}\right]\\
 & =\mathbb{E}\left[\left\{ e_{gj}e_{g\ell}-\sigma\left(X_{gj}^{(\mathrm{ind})},X_{g\ell}^{(\mathrm{ind})};X_{g}^{(\mathrm{cls})}\right)\right\} ^{2}\mid X_{gj}^{(\mathrm{ind})},X_{g\ell}^{(\mathrm{ind})};X_{g}^{(\mathrm{cls})}\right]\\
 & =\mathbb{E}\left[e_{gj}^{2}e_{g\ell}^{2}\mid X_{gj}^{(\mathrm{ind})},X_{g\ell}^{(\mathrm{ind})};X_{g}^{(\mathrm{cls})}\right]-\sigma^{2}\left(X_{gj}^{(\mathrm{ind})},X_{g\ell}^{(\mathrm{ind})};X_{g}^{(\mathrm{cls})}\right)\\
 & =\varsigma\left(X_{gj}^{(\mathrm{ind})},X_{gj}^{(\mathrm{ind})},X_{g\ell}^{(\mathrm{ind})},X_{g\ell}^{(\mathrm{ind})};X_{g}^{(\mathrm{cls})}\right)-\sigma^{2}\left(X_{gj}^{(\mathrm{ind})},X_{g\ell}^{(\mathrm{ind})};X_{g}^{(\mathrm{cls})}\right),
\end{align*}
and for $\left(j,\ell\right)\neq\left(t,s\right)$,
\begin{eqnarray*}
 &  & \mathbb{E}\left[u_{gj\ell}u_{gts}\mid\mathbf{X}_{g}\right]\\
 & = & \mathbb{E}\left[u_{gj\ell}u_{gts}\mid X_{gj}^{(\mathrm{ind})},X_{g\ell}^{(\mathrm{ind})},X_{gt}^{(\mathrm{ind})},X_{gs}^{(\mathrm{ind})};X_{g}^{(\mathrm{cls})}\right]\\
 & = & \mathbb{E}\left[\left\{ e_{gj}e_{g\ell}-\sigma\left(X_{gj}^{(\mathrm{ind})},X_{g\ell}^{(\mathrm{ind})};X_{g}^{(\mathrm{cls})}\right)\right\} \right.\\
 &  & \qquad\left.\times\left\{ e_{gt}e_{gs}-\sigma\left(X_{gt}^{(\mathrm{ind})},X_{gs}^{(\mathrm{ind})};X_{g}^{(\mathrm{cls})}\right)\right\} \mid X_{gj}^{(\mathrm{ind})},X_{g\ell}^{(\mathrm{ind})},X_{gt}^{(\mathrm{ind})},X_{gs}^{(\mathrm{ind})};X_{g}^{(\mathrm{cls})}\right]\\
 & = & \varsigma\left(X_{gj}^{(\mathrm{ind})},X_{g\ell}^{(\mathrm{ind})},X_{gt}^{(\mathrm{ind})},X_{gs}^{(\mathrm{ind})};X_{g}^{(\mathrm{cls})}\right)\\
 &  & -\sigma\left(X_{gj}^{(\mathrm{ind})},X_{g\ell}^{(\mathrm{ind})};X_{g}^{(\mathrm{cls})}\right)\sigma\left(X_{gt}^{(\mathrm{ind})},X_{gs}^{(\mathrm{ind})};X_{g}^{(\mathrm{cls})}\right).
\end{eqnarray*}
Under Assumption $\text{\ref{assu:condvar}}$, we can apply Theorem $\text{\ref{thm:nw_cons}}$ for $\left(2d_{\mathrm{ind}}+d_{\mathrm{cls}}\right)$-dimensional regressors and $n_{g}(n_{g}-1)/2$ size clusters.
\end{proof}

\subsection{Proof for Corollary $\text{\ref{cor:nw_asydist_condvar}}$}
\begin{proof}
Apply the Slutsky's Lemma.
\end{proof}
\newpage{}

\section{\textbf{Proofs for technical lemmas\label{app:lemmas}}}

For the following proofs, we focus on the case $x=\left(x^{(\mathrm{ind})\top},x^{\mathrm{(cls)}\top}\right)^{\top}=0$ to make notation lighter. The notation $X_{gj}^{(q)}$ denotes the $q$-th element of the vector $X_{gj}$. We also suppress subscripts such as $g$ and $j$ if the meaning is implied from the context.

\subsection{Proof for Lemma $\text{\ref{lem:Fr}}$}
\begin{proof}
Define $F_{r}^{(q)}=\frac{1}{n}\sum_{g=1}^{G}\sum_{j=1}^{n_{g}}K_{h}\left(X_{gj}\right)\left(X_{gj}^{(q)}\right)^{r}$, $F^{(p,q)}=\frac{1}{n}\sum_{g=1}^{G}\sum_{j=1}^{n_{g}}K_{h}\left(X_{gj}\right)X_{gj}^{(p)}X_{gj}^{(q)}$ for $p\neq q$, and $\nu_{r}^{(q)}=\int_{\mathbb{R}^{d}}K\left(T\right)\left(T^{(q)}\right)^{r}\mathrm{d}T=\int_{-\infty}^{\infty}k\left(T^{(q)}\right)\left(T^{(q)}\right)^{r}\mathrm{d}T^{(q)}$ for $r=0,1,2$. \footnote{When $r=0$, $F_{r}^{(q)}$ does not depend on $q$ because $\left(X_{gj}^{(q)}\right)^{r}=1$.} Note that 
\[
\nu_{r}^{(q)}=\begin{cases}
1 & \text{ if }r=0\\
0 & \text{ if }r=1\\
\kappa_{2} & \text{ if }r=2
\end{cases}.
\]
We will evaluate expectations and variances of $F^{(p,q)}$ and $F_{r}^{(q)}$ and obtain a conclusion by Markov's inequality. For expectations, we have
\begin{align*}
\mathbb{E}\left[F_{r}^{(q)}\right] & =h^{r}\int_{\mathbb{R}^{d}}K\left(T\right)\left(T^{(q)}\right)^{r}f\left(hT\right)\mathrm{d}T\\
 & =h^{r}\int_{\mathbb{R}^{d}}K\left(T\right)\left(T^{(q)}\right)^{r}\left\{ f\left(0\right)+hT^{\top}\nabla f\left(0\right)+\frac{h^{2}}{2}T^{\top}\nabla^{2}f\left(h\tilde{T}\right)T\right\} \mathrm{d}T\\
 & =h^{r}\int_{\mathbb{R}^{d}}K\left(T\right)\left(T^{(q)}\right)^{r}\left\{ f\left(0\right)+hT^{\top}\nabla f\left(0\right)\right\} \mathrm{d}T+O\left(h^{r+2}\right)\\
 & =\begin{cases}
h^{r}f\left(0\right)\nu_{r}^{(q)}+O\left(h^{r+2}\right) & \text{ if }r\text{ is even}\\
h^{r+1}\partial_{q}f\left(0\right)\int_{-\infty}^{\infty}k\left(T^{(q)}\right)\left(T^{(q)}\right)^{r+1}\mathrm{d}T^{(q)}+O\left(h^{r+1}\right) & \text{ if }r\text{ is odd}
\end{cases},
\end{align*}
by the identical marginal distribution, the change of variables $T=X/h$, the Taylor expansion ($\tilde{T}$ is between $0$ and $T$), the dominated convergence theorem, and the symmetry of the kernel function.\footnote{We use the continuity of $\nabla^{2}f\left(x\right)$ in some neighborhood $\mathcal{N}$ of $x=0$. The continuity implies $\nabla^{2}f\left(h\tilde{T}\right)\rightarrow\nabla^{2}f\left(0\right)$ as $h\rightarrow0$. Since $\nabla^{2}f\left(0\right)$ exists, it is bounded. Thus, we can apply the dominated convergence theorem.} Thus,
\[
\mathbb{E}\left[F_{r}^{(q)}\right]=\begin{cases}
f\left(0\right)+o\left(1\right) & \text{ if }r=0\\
o\left(h\right) & \text{ if }r=1\\
h^{2}f\left(0\right)\kappa_{2}+o\left(h^{2}\right) & \text{ if }r=2
\end{cases}.
\]

Similarly, for $p\neq q$,
\begin{align*}
\mathbb{E}\left[F^{(p,q)}\right] & =h^{2}\int_{\mathbb{R}^{d}}K\left(T\right)T^{(p)}T^{(q)}f\left(hT\right)\mathrm{d}T\\
 & =h^{2}\int_{\mathbb{R}^{d}}K\left(T\right)T^{(p)}T^{(q)}\left\{ f\left(0\right)+hT^{\top}\nabla f\left(0\right)+\frac{h^{2}}{2}T^{\top}\nabla^{2}f\left(0\right)T\right\} \mathrm{d}T+o\left(h^{4}\right)\\
 & =h^{2}f\left(0\right)\int_{\mathbb{R}^{d}}K\left(T\right)T^{(p)}T^{(q)}\mathrm{d}T+h^{3}\int_{\mathbb{R}^{d}}K\left(T\right)T^{(p)}T^{(q)}T^{\top}\nabla f\left(0\right)\mathrm{d}T+O\left(h^{4}\right)\\
 & =O\left(h^{4}\right).
\end{align*}
For variances,
\begin{eqnarray*}
 &  & \operatorname{Var}\left[F_{r}^{(q)}\right]\\
 & = & \operatorname{Var}\left[\frac{1}{n}\sum_{g=1}^{G}\sum_{j=1}^{n_{g}}K_{h}\left(X_{gj}\right)\left(X_{gj}^{(q)}\right)^{r}\right]=\frac{1}{n^{2}}\sum_{g=1}^{G}\operatorname{Var}\left[\sum_{j=1}^{n_{g}}K_{h}\left(X_{gj}\right)\left(X_{gj}^{(q)}\right)^{r}\right]\\
 & = & \frac{1}{n^{2}}\sum_{g=1}^{G}\left\{ \sum_{j=1}^{n_{g}}\operatorname{Var}\left[K_{h}\left(X_{gj}\right)\left(X_{gj}^{(q)}\right)^{r}\right]+2\ensuremath{\sum_{1\leq j<\ell\leq n_{g}}}\operatorname{Cov}\left[K_{h}\left(X_{gj}\right)\left(X_{gj}^{(q)}\right)^{r},K_{h}\left(X_{g\ell}\right)\left(X_{g\ell}^{(q)}\right)^{r}\right]\right\} \\
 & \leq & \frac{1}{n^{2}}\sum_{g=1}^{G}\sum_{j=1}^{n_{g}}\mathbb{E}\left[K_{h}^{2}\left(X_{gj}\right)\left(X_{gj}^{(q)}\right)^{2r}\right]\\
 &  & +\ensuremath{\frac{2}{n^{2}}\sum_{g=1}^{G}\sum_{1\leq j<\ell\leq n_{g}}}\left(\mathbb{E}\left[K_{h}\left(X_{gj}\right)\left(X_{gj}^{(q)}\right)^{r}K_{h}\left(X_{g\ell}\right)\left(X_{g\ell}^{(q)}\right)^{r}\right]\right.\\
 &  & \qquad\qquad\qquad\qquad\qquad\qquad\qquad\qquad\left.-\underbrace{\mathbb{E}\left[K_{h}\left(X_{gj}\right)\left(X_{gj}^{(q)}\right)^{r}\right]\mathbb{E}\left[K_{h}\left(X_{g\ell}\right)\left(X_{g\ell}^{(q)}\right)^{r}\right]}_{=\mathbb{E}\left[F_{r}^{(q)}\right]^{2}}\right),
\end{eqnarray*}
where the second equality follows from the independence between clusters and the inequality follows from $\operatorname{Var}\left[K_{h}\left(X_{gj}\right)\left(X_{gj}^{(q)}\right)^{r}\right]\leq\mathbb{E}\left[K_{h}^{2}\left(X_{gj}\right)\left(X_{gj}^{(q)}\right)^{2r}\right]$. We will bound the following two expectations 
\begin{align}
 & \mathbb{E}\left[K_{h}^{2}\left(X\right)\left(X^{(q)}\right)^{2r}\right],\label{eq:var_Fr_e1}\\
 & \mathbb{E}\left[K_{h}\left(X_{j}\right)\left(X_{j}^{(q)}\right)^{r}K_{h}\left(X_{\ell}\right)\left(X_{\ell}^{(q)}\right)^{r}\right].\label{eq:var_Fr_e2}
\end{align}

\begin{eqnarray*}
 & \eqref{eq:var_Fr_e1}: & \mathbb{E}\left[K_{h}^{2}\left(X\right)\left(X^{(q)}\right)^{2r}\right]=\frac{1}{h^{2d}}\int_{\mathbb{R}^{d}}K\left(\frac{X}{h}\right)^{2}\left(X^{(q)}\right)^{2r}f\left(X\right)\mathrm{d}X\\
 & = & \frac{1}{h^{d-2r}}\int_{\mathbb{R}^{d}}K\left(T\right)^{2}\left(T^{(q)}\right)^{2r}f\left(Th\right)\mathrm{d}T\\
 & = & \frac{1}{h^{d-2r}}\int_{\mathbb{R}^{d}}K\left(T\right)^{2}\left(T^{(q)}\right)^{2r}f(0)\mathrm{d}T+o\left(h^{2r-d}\right)\\
 & = & O\left(h^{2r-d}\right),
\end{eqnarray*}
where the second equality follows from the change of variables $T=X/h$ and the third equality follows from the continuity.

Also,
\begin{eqnarray*}
 & \eqref{eq:var_Fr_e2}: & \mathbb{E}\left[K_{h}\left(X_{j}\right)\left(X_{j}^{(q)}\right)^{r}K_{h}\left(X_{\ell}\right)\left(X_{\ell}^{(q)}\right)^{r}\right]\\
 & = & \frac{1}{h^{2d}}\int_{\mathbb{R}^{d_{\mathrm{ind}}}}\int_{\mathbb{R}^{d_{\mathrm{ind}}}}\int_{\mathbb{R}^{d_{\mathrm{cls}}}}K\left(\frac{X_{j}}{h}\right)K\left(\frac{X_{\ell}}{h}\right)\left(X_{j}^{(q)}\right)^{r}\left(X_{\ell}^{(q)}\right)^{r}\\
 &  & \qquad\qquad\times f_{2}\left(X_{j}^{(\mathrm{ind})},X_{\ell}^{(\mathrm{ind})};X^{(\mathrm{cls})}\right)\mathrm{d}X_{j}^{(\mathrm{ind})}\mathrm{d}X_{\ell}^{(\mathrm{ind})}\mathrm{d}X^{(\mathrm{cls})}\\
 & = & h^{2r-d_{\mathrm{cls}}}\int_{\mathbb{R}^{d_{\mathrm{ind}}}}\int_{\mathbb{R}^{d_{\mathrm{ind}}}}\int_{\mathbb{R}^{d_{\mathrm{cls}}}}K\left(T_{j}\right)K\left(T_{\ell}\right)\left(T_{j}^{(q)}\right)^{r}\left(T_{\ell}^{(q)}\right)^{r}\\
 &  & \qquad\qquad\times f_{2}\left(hT_{j}^{(\mathrm{ind})},hT_{\ell}^{(\mathrm{ind})};hT^{(\mathrm{cls})}\right)\mathrm{d}T_{j}^{(\mathrm{ind})}\mathrm{d}T_{\ell}^{(\mathrm{ind})}\mathrm{d}T^{(\mathrm{cls})}\\
 & = & h^{2r-d_{\mathrm{cls}}}\int_{\mathbb{R}^{d_{\mathrm{ind}}}}\int_{\mathbb{R}^{d_{\mathrm{ind}}}}\int_{\mathbb{R}^{d_{\mathrm{cls}}}}K\left(T_{j}\right)K\left(T_{\ell}\right)\left(T_{j}^{(q)}\right)^{r}\left(T_{\ell}^{(q)}\right)^{r}\\
 &  & \qquad\qquad\times f_{2}\left(0,0;0\right)\mathrm{d}T_{j}^{(\mathrm{ind})}\mathrm{d}T_{\ell}^{(\mathrm{ind})}\mathrm{d}T^{(\mathrm{cls})}+o\left(h^{2r-d_{\mathrm{cls}}}\right)\\
 & = & O\left(h^{2r-d_{\mathrm{cls}}}\right),
\end{eqnarray*}
where the second equality follows from the change of variables $T_{j}^{(\mathrm{ind})}=X_{j}^{(\mathrm{ind})}/h$, $T_{\ell}^{(\mathrm{ind})}=X_{\ell}^{(\mathrm{ind})}/h$, and $T^{(\mathrm{cls})}=X^{(\mathrm{cls})}/h$ (we define $T_{j}=\left(T_{j}^{(\mathrm{ind})\top},T^{(\mathrm{cls})\top}\right)^{\top}$, $T_{\ell}=\left(T_{\ell}^{(\mathrm{ind})\top},T^{(\mathrm{cls})\top}\right)^{\top}$), and the third equality follows from the continuity.

Thus, since $\sum_{g=1}^{G}n_{g}=n$ and $\left(\max_{g}n_{g}\right)/\left(nh^{d_{\mathrm{cls}}}\right)=\left(\max_{g}n_{g}h^{d_{\mathrm{ind}}}\right)/\left(nh^{d}\right)=o(1)$,
\begin{eqnarray*}
\operatorname{Var}\left[F_{r}^{(q)}\right] & \leq & \frac{1}{n^{2}}\sum_{g=1}^{G}\sum_{j=1}^{n_{g}}O\left(h^{2r-d}\right)+\ensuremath{\frac{1}{n^{2}}\sum_{g=1}^{G}\sum_{1\leq j<\ell\leq n_{g}}}O\left(h^{2r-d_{\mathrm{cls}}}\right)\\
 & \leq & O\left(n^{-1}h^{2r-d}\right)+\frac{1}{n}\left(\max_{g}n_{g}\right)O\left(h^{2r-d_{\mathrm{cls}}}\right)=o\left(h^{2r}\right).
\end{eqnarray*}

Similarly,
\begin{eqnarray*}
 &  & \operatorname{Var}\left[F^{(p,q)}\right]\\
 & \leq & \frac{1}{n^{2}}\sum_{g=1}^{G}\sum_{j=1}^{n_{g}}\mathbb{E}\left[K_{h}^{2}\left(X_{gj}\right)\left(X_{gj}^{(p)}\right)^{2}\left(X_{gj}^{(q)}\right)^{2}\right]\\
 &  & +\ensuremath{\frac{2}{n^{2}}\sum_{g=1}^{G}\sum_{1\leq j<\ell\leq n_{g}}}\left(\mathbb{E}\left[K_{h}\left(X_{gj}\right)X_{gj}^{(p)}X_{gj}^{(q)}K_{h}\left(X_{g\ell}\right)X_{g\ell}^{(p)}X_{g\ell}^{(q)}\right]\right.\\
 &  & \qquad\qquad\qquad\qquad\qquad\left.-\underbrace{\mathbb{E}\left[K_{h}\left(X_{gj}\right)X_{gj}^{(p)}X_{gj}^{(q)}\right]\mathbb{E}\left[K_{h}\left(X_{g\ell}\right)X_{g\ell}^{(p)}X_{g\ell}^{(q)}\right]}_{=\mathbb{E}\left[F^{(p,q)}\right]^{2}}\right)\\
 & \leq & \frac{1}{n^{2}}\sum_{g=1}^{G}\sum_{j=1}^{n_{g}}O\left(h^{4-d}\right)+\ensuremath{\frac{1}{n^{2}}\sum_{g=1}^{G}\sum_{1\leq j<\ell\leq n_{g}}}O\left(h^{4-d_{\mathrm{cls}}}\right)\\
 & \leq & O\left(n^{-1}h^{4-d}\right)+O\left(n^{-1}\left(\max_{g}n_{g}\right)h^{4-d_{\mathrm{cls}}}\right)=o\left(h^{4}\right).
\end{eqnarray*}

Therefore, by Markov's inequality and Jensen's inequality,
\begin{align*}
\mathbb{P}\left[\left|h^{2}\left(F_{2}^{(q)}-h^{2}f\left(0\right)\kappa_{2}\right)\right|>\delta\right] & \leq\frac{\mathbb{E}\left[\left|F_{2}^{(q)}-h^{2}f\left(0\right)\kappa_{2}\right|\right]}{h^{2}\delta}\leq\frac{\mathbb{E}\left[\left(F_{2}^{(q)}-h^{2}f\left(0\right)\kappa_{2}\right)^{2}\right]^{1/2}}{h^{2}\delta}\\
 & =\frac{\left|\mathbb{E}\left[F_{2}^{(q)}\right]-h^{2}f\left(0\right)\kappa_{2}\right|+\sqrt{\operatorname{Var}\left[F_{2}^{(q)}\right]}}{h^{2}\delta}\\
 & \leq o\left(1\right)\qquad\text{for any }\delta,
\end{align*}
which implies that $F_{2}^{(q)}=h^{2}f\left(0\right)\kappa_{2}+o_{p}\left(h^{2}\right)$. Similarly, we have $F_{0}^{(q)}=f\left(0\right)+o_{p}\left(1\right)$, $F_{1}^{(q)}=o_{p}\left(h\right)$, and $F^{(p,q)}=o_{p}\left(h^{2}\right)$. We conclude by element-wise comparisons.
\end{proof}

\subsection{Proof for Lemma $\text{\ref{lem:Jr}}$}
\begin{proof}
Define $J_{r}^{(q)}=\frac{1}{n}\sum_{g=1}^{G}\sum_{j=1}^{n_{g}}K_{h}\left(X_{gj}\right)\left\{ m\left(X_{gj}\right)-m(0)\right\} \left(X_{gj}^{(q)}\right)^{r}$ for $r=0,1$ . For expectations,
\begin{eqnarray*}
 &  & \mathbb{E}\left[J_{r}^{(q)}\right]\\
 & = & \mathbb{E}\left[\frac{1}{n}\sum_{g=1}^{G}\sum_{j=1}^{n_{g}}K_{h}\left(X_{gj}\right)\left\{ m\left(X_{gj}\right)-m(0)\right\} \left(X_{gj}^{(q)}\right)^{r}\right]\\
 & = & \frac{1}{h^{d}}\int_{\mathbb{R}^{d}}K\left(\frac{X}{h}\right)\left\{ m\left(X\right)-m(0)\right\} \left(X^{(q)}\right)^{r}f\left(X\right)\mathrm{d}X\\
 & = & h^{r}\int_{\mathbb{R}^{d}}K\left(T\right)\left\{ m\left(hT\right)-m(0)\right\} \left(T^{(q)}\right)^{r}f\left(hT\right)\mathrm{d}T\\
 & = & h^{r}\int_{\mathbb{R}^{d}}K\left(T\right)\left(T^{(q)}\right)^{r}\left\{ hT^{\top}\nabla m\left(0\right)+\frac{h^{2}}{2}T^{\top}\nabla^{2}m\left(h\tilde{T}\right)T\right\} \left\{ f\left(0\right)+hT^{\top}\nabla f\left(h\grave{T}\right)\right\} \mathrm{d}T\\
 & = & h^{r+1}f(0)\int_{\mathbb{R}^{d}}\left(T^{(q)}\right)^{r}T^{\top}\nabla m\left(0\right)K\left(T\right)\mathrm{d}T\\
 &  & +\frac{h^{r+2}}{2}f(0)\int_{\mathbb{R}^{d}}\left(T^{(q)}\right)^{r}T^{\top}\nabla^{2}m\left(0\right)TK\left(T\right)\mathrm{d}T\\
 &  & +h^{r+2}\int_{\mathbb{R}^{d}}\left(T^{(q)}\right)^{r}T^{\top}\nabla m\left(0\right)T^{\top}\nabla f\left(0\right)K\left(T\right)\mathrm{d}T+O\left(h^{r+3}\right)+o\left(h^{r+2}\right)\\
 & = & \begin{cases}
h^{2}\sum_{q=1}^{d}\left\{ \frac{1}{2}f(0)\partial_{qq}m\left(0\right)+\partial_{q}m\left(0\right)\partial_{q}f\left(0\right)\right\} \kappa_{2}+o\left(h^{2}\right) & \text{ if }r=0\\
h^{2}f(0)\partial_{q}m\left(0\right)\kappa_{2}+o\left(h^{3}\right) & \text{ if }r=1
\end{cases},
\end{eqnarray*}
where the second equality follows from the linearity of the expectation and the identical marginal distribution, the third equality follows from the change of variables $T=X/h$, the fourth equality follows from the Taylor expansion ($\tilde{T}$ and $\grave{T}$ are between 0 and $T$), the fifth equality follows from the dominated convergence theorem, and the sixth equality follows from the symmetry of the kernel function. 

For variances,
\begin{eqnarray*}
 &  & \operatorname{Var}\left[J_{r}^{(q)}\right]\\
 & = & \operatorname{Var}\left[\frac{1}{n}\sum_{g=1}^{G}\sum_{j=1}^{n_{g}}K_{h}\left(X_{gj}\right)\left\{ m\left(X_{gj}\right)-m(0)\right\} \left(X_{gj}^{(q)}\right)^{r}\right]\\
 & = & \frac{1}{n^{2}}\sum_{g=1}^{G}\operatorname{Var}\left[\sum_{j=1}^{n_{g}}K_{h}\left(X_{gj}\right)\left\{ m\left(X_{gj}\right)-m(0)\right\} \left(X_{gj}^{(q)}\right)^{r}\right]\\
 & = & \frac{1}{n^{2}}\sum_{g=1}^{G}\sum_{j=1}^{n_{g}}\operatorname{Var}\left[K_{h}\left(X_{gj}\right)\left\{ m\left(X_{gj}\right)-m(0)\right\} \left(X_{gj}^{(q)}\right)^{r}\right]\\
 &  & +2\frac{1}{n^{2}}\sum_{g=1}^{G}\ensuremath{\sum_{1\leq j<\ell\leq n_{g}}}\operatorname{Cov}\left[K_{h}\left(X_{gj}\right)\left\{ m\left(X_{gj}\right)-m(0)\right\} \left(X_{gj}^{(q)}\right)^{r},K_{h}\left(X_{g\ell}\right)\left\{ m\left(X_{g\ell}\right)-m(0)\right\} \left(X_{g\ell}^{(q)}\right)^{r}\right]\\
 & \leq & \frac{1}{n^{2}}\sum_{g=1}^{G}\sum_{j=1}^{n_{g}}\mathbb{E}\left[K_{h}^{2}\left(X_{gj}\right)\left\{ m\left(X_{gj}\right)-m(0)\right\} ^{2}\left(X_{gj}^{(q)}\right)^{2r}\right]\\
 &  & +\ensuremath{\frac{2}{n^{2}}\sum_{g=1}^{G}\sum_{1\leq j<\ell\leq n_{g}}}\mathbb{E}\left[K_{h}\left(X_{gj}\right)\left\{ m\left(X_{gj}\right)-m(0)\right\} \left(X_{gj}^{(q)}\right)^{r}K_{h}\left(X_{g\ell}\right)\left\{ m\left(X_{g\ell}\right)-m(0)\right\} \left(X_{g\ell}^{(q)}\right)^{r}\right]\\
 &  & -\ensuremath{\frac{2}{n^{2}}\sum_{g=1}^{G}\sum_{1\leq j<\ell\leq n_{g}}}\underbrace{\mathbb{E}\left[K_{h}\left(X_{gj}\right)\left\{ m\left(X_{gj}\right)-m(0)\right\} \left(X_{gj}^{(q)}\right)^{r}\right]\mathbb{E}\left[K_{h}\left(X_{g\ell}\right)\left\{ m\left(X_{g\ell}\right)-m(0)\right\} \left(X_{g\ell}^{(q)}\right)^{r}\right]}_{=\mathbb{E}\left[J_{r}^{(q)}\right]^{2}},
\end{eqnarray*}
where the second equality follows from the independence between clusters. We will bound the following two expectations
\begin{align}
 & \mathbb{E}\left[K_{h}^{2}\left(X\right)\left\{ m\left(X\right)-m(0)\right\} ^{2}\left(X^{(q)}\right)^{2r}\right],\label{eq:var_Jr_e1}\\
 & \mathbb{E}\left[K_{h}\left(X_{j}\right)\left\{ m\left(X_{j}\right)-m(0)\right\} \left(X_{j}^{(q)}\right)^{r}K_{h}\left(X_{\ell}\right)\left\{ m\left(X_{\ell}\right)-m(0)\right\} \left(X_{\ell}^{(q)}\right)^{r}\right].\label{eq:var_Jr_e2}
\end{align}
\begin{eqnarray*}
 & \eqref{eq:var_Jr_e1}: & \mathbb{E}\left[K_{h}^{2}\left(X\right)\left\{ m\left(X\right)-m(0)\right\} ^{2}\left(X^{(q)}\right)^{2r}\right]\\
 & = & \frac{1}{h^{2d}}\int_{\mathbb{R}^{d}}K\left(\frac{X}{h}\right)^{2}\left\{ m\left(X\right)-m(0)\right\} ^{2}\left(X^{(q)}\right)^{2r}f\left(X\right)\mathrm{d}X\\
 & = & \frac{1}{h^{d-2r}}\int_{\mathbb{R}^{d}}K\left(T\right)^{2}\left(T^{(q)}\right)^{2r}\left\{ m\left(hT\right)-m(0)\right\} ^{2}f\left(hT\right)\mathrm{d}T\\
 & = & \frac{1}{h^{d-2(r+1)}}\int_{\mathbb{R}^{d}}K\left(T\right)^{2}\left(T^{(q)}\right)^{2r}\left\{ T^{\top}\nabla m\left(0\right)\right\} ^{2}f\left(0\right)\mathrm{d}T+o\left(h^{2(r+1)-d}\right)\\
 & = & O\left(h^{2(r+1)-d}\right).
\end{eqnarray*}
where the second equality follows from the change of variables $T=X/h$, and the third equality follows from the Taylor expansion and the dominated convergence theorem. 

Also,
\begin{eqnarray*}
 & \eqref{eq:var_Jr_e2}: & \mathbb{E}\left[K_{h}\left(X_{j}\right)\left\{ m\left(X_{j}\right)-m(0)\right\} \left(X_{j}^{(q)}\right)^{r}K_{h}\left(X_{\ell}\right)\left\{ m\left(X_{\ell}\right)-m(0)\right\} \left(X_{\ell}^{(q)}\right)^{r}\right]\\
 & = & \frac{1}{h^{2d}}\int_{\mathbb{R}^{d_{\mathrm{ind}}}}\int_{\mathbb{R}^{d_{\mathrm{ind}}}}\int_{\mathbb{R}^{d_{\mathrm{cls}}}}K\left(\frac{X_{j}}{h}\right)K\left(\frac{X_{\ell}}{h}\right)\left(X_{j}^{(q)}\right)^{r}\left(X_{\ell}^{(q)}\right)^{r}\\
 &  & \qquad\qquad\times\left\{ m\left(X_{j}\right)-m(0)\right\} \left\{ m\left(X_{g\ell}\right)-m(0)\right\} \\
 &  & \qquad\qquad\times f_{2}\left(X_{j}^{(\mathrm{ind})},X_{\ell}^{(\mathrm{ind})};X^{(\mathrm{cls})}\right)\mathrm{d}X_{j}^{(\mathrm{ind})}\mathrm{d}X_{\ell}^{(\mathrm{ind})}\mathrm{d}X^{(\mathrm{cls})}\\
 & = & h^{2r-d_{\mathrm{cls}}}\int_{\mathbb{R}^{d_{\mathrm{ind}}}}\int_{\mathbb{R}^{d_{\mathrm{ind}}}}\int_{\mathbb{R}^{d_{\mathrm{cls}}}}K\left(T_{j}\right)K\left(T_{\ell}\right)\left(T_{j}^{(q)}\right)^{r}\left(T_{\ell}^{(q)}\right)^{r}\\
 &  & \qquad\qquad\times\left\{ m\left(hT_{j}\right)-m(0)\right\} \left\{ m\left(hT_{\ell}\right)-m(0)\right\} \\
 &  & \qquad\qquad\times f_{2}\left(hT_{j}^{(\mathrm{ind})},hT_{\ell}^{(\mathrm{ind})};hT^{(\mathrm{cls})}\right)\mathrm{d}T_{j}^{(\mathrm{ind})}\mathrm{d}T_{\ell}^{(\mathrm{ind})}\mathrm{d}T^{(\mathrm{cls})}\\
 & = & h^{2r+2-d_{\mathrm{cls}}}f_{2}\left(0,0;0\right)\int_{\mathbb{R}^{d_{\mathrm{ind}}}}\int_{\mathbb{R}^{d_{\mathrm{ind}}}}\int_{\mathbb{R}^{d_{\mathrm{cls}}}}K\left(T_{j}\right)K\left(T_{\ell}\right)\left(T_{j}^{(q)}\right)^{r}\left(T_{\ell}^{(q)}\right)^{r}\\
 &  & \qquad\qquad\times\left\{ T_{j}^{\top}\nabla m\left(0\right)\right\} \left\{ T_{\ell}^{\top}\nabla m\left(0\right)\right\} \mathrm{d}T_{j}^{(\mathrm{ind})}\mathrm{d}T_{\ell}^{(\mathrm{ind})}\mathrm{d}T^{(\mathrm{cls})}\\
 &  & +o\left(h^{2r+2-d_{\mathrm{cls}}}\right)\\
 & = & \begin{cases}
O\left(h^{2-d_{\mathrm{cls}}}\right) & \text{ if }r=0\\
O\left(h^{4-d_{\mathrm{cls}}}\right) & \text{ if }r=1
\end{cases},
\end{eqnarray*}
 where the second equality follows from the change of variables $T_{j}^{(\mathrm{ind})}=X_{j}^{(\mathrm{ind})}/h$, $T_{\ell}^{(\mathrm{ind})}=X_{\ell}^{(\mathrm{ind})}/h$, and $T^{(\mathrm{cls})}=X^{(\mathrm{cls})}/h$ (we define $T_{j}=\left(T_{j}^{(\mathrm{ind})\top},T^{(\mathrm{cls})\top}\right)^{\top}$, $T_{\ell}=\left(T_{\ell}^{(\mathrm{ind})\top},T^{(\mathrm{cls})\top}\right)^{\top}$), and the third equality follows from the Taylor expansion and the dominated convergence theorem. Thus, for $r=0$,
\begin{align*}
\operatorname{Var}\left[J_{0}^{(q)}\right] & =\frac{1}{n^{2}}\sum_{g=1}^{G}\left[\sum_{j=1}^{n_{g}}O\left(h^{2-d}\right)+2\sum_{1\leq j<\ell\leq n_{g}}O\left(h^{2-d_{\mathrm{cls}}}\right)\right]\\
 & \leq O\left(n^{-1}h^{2-d}\right)+O\left(n^{-1}\left(\max_{g}n_{g}\right)h^{2-d_{\mathrm{cls}}}\right)\\
 & =O\left(\frac{h^{2}}{nh^{d}}\right)+\left\{ \left(\max_{g\leq G}n_{g}\right)h^{d_{\mathrm{ind}}}\right\} O\left(\frac{h^{2}}{nh^{d}}\right)\\
 & =O\left(\frac{h^{2}}{nh^{d}}\right),
\end{align*}
and for $r=1$,
\begin{eqnarray*}
\operatorname{Var}\left[J_{1}^{(q)}\right] & \leq & \frac{1}{n^{2}}\sum_{g=1}^{G}\left[\sum_{j=1}^{n_{g}}O\left(h^{4-d}\right)+\ensuremath{2\sum_{1\leq j<\ell\leq n_{g}}}O\left(h^{4-d_{\mathrm{cls}}}\right)\right]\\
 & \leq & O\left(\frac{h^{4}}{nh^{d}}\right).
\end{eqnarray*}

Therefore, by Markov's inequality, $J_{0}^{(q)}=h^{2}\kappa_{2}\sum_{q=1}^{d}\left\{ \frac{1}{2}f(0)\partial_{qq}m\left(0\right)+\partial_{q}m\left(0\right)\partial_{q}f\left(0\right)\right\} +o_{p}\left(h^{2}\right)+O_{p}\left(\sqrt{\frac{1}{nh^{d-2}}}\right)$, $J_{1}^{(q)}=h^{2}\kappa_{2}f(0)\partial_{q}m\left(0\right)+o_{p}\left(h^{3}\right)+O_{p}\left(\sqrt{\frac{1}{nh^{d-4}}}\right)$. We conclude by element-wise comparisons.
\end{proof}

\subsection{Proof for Lemma $\text{\ref{lem:Hr}}$}
\begin{proof}
Define $H_{r}^{(q)}=\frac{1}{n}\sum_{g=1}^{G}\sum_{j=1}^{n_{g}}K_{h}^{2}\left(X_{gj}\right)\sigma^{2}\left(X_{gj}\right)\left(X_{gj}^{(q)}\right)^{r}$, $H^{(p,q)}=\frac{1}{n}\sum_{g=1}^{G}\sum_{j=1}^{n_{g}}K_{h}^{2}\left(X_{gj}\right)\allowbreak\sigma^{2}\left(X_{gj}\right)X_{gj}^{(p)}X_{gj}^{(q)}$ for $p\neq q$, and $\pi_{r}^{(q)}=\int_{\mathbb{R}^{d}}K^{2}\left(T\right)\left(T^{(q)}\right)^{r}\mathrm{d}T$ for $r=0,1,2$. For expectations,
\begin{eqnarray*}
\mathbb{E}\left[H_{r}^{(q)}\right] & = & \frac{1}{h^{2d}}\int_{\mathbb{R}^{d}}K^{2}\left(\frac{X}{h}\right)\sigma^{2}\left(X\right)\left(X^{(q)}\right)^{r}f\left(X\right)\mathrm{d}X\\
 & = & \frac{1}{h^{d-r}}\int_{\mathbb{R}^{d}}K^{2}\left(T\right)\sigma^{2}\left(hT\right)\left(T^{(q)}\right)^{r}f\left(hT\right)\mathrm{d}T\\
 & = & \frac{1}{h^{d-r}}\int_{\mathbb{R}^{d}}K^{2}\left(T\right)\left(T^{(q)}\right)^{r}\sigma^{2}\left(0\right)\left\{ f\left(0\right)+hT^{\top}\nabla f\left(0\right)\right\} \mathrm{d}T\\
 &  & +o\left(h^{r+1-d}\right)\\
 & = & \begin{cases}
\frac{1}{h^{d-r}}f\left(0\right)\sigma^{2}\left(0\right)\pi_{r}^{(q)}+o\left(h^{r+1-d}\right) & \text{ if }r\text{ is even }\\
O\left(h^{r+1-d}\right) & \text{ if }r\text{ is odd }
\end{cases},
\end{eqnarray*}
where the second equality follows from the change of variables $T=X/h$, and the third equality follows from the Taylor expansion. Since $\pi_{0}^{(q)}=R_{k}^{d}$,
\[
\mathbb{E}\left[H_{r}^{(q)}\right]=\begin{cases}
\frac{1}{h^{d}}\left\{ f\left(0\right)\sigma^{2}\left(0\right)R_{k}^{d}+o\left(1\right)\right\}  & \text{ if }r=0\\
O\left(h^{-d+2}\right) & \text{ if }r=1\\
\frac{1}{h^{d-2}}f\left(0\right)\sigma^{2}\left(0\right)\left\{ \int_{\mathbb{R}^{d}}K^{2}\left(T\right)\left(T^{(q)}\right)^{2}\mathrm{d}T\right\} +o\left(h^{-d+2}\right) & \text{ if }r=2
\end{cases}.
\]

Similarly, for $p\neq q$, 
\begin{align*}
\mathbb{E}\left[H^{(p,q)}\right] & =\mathbb{E}\left[\frac{1}{n}\sum_{g=1}^{G}\sum_{j=1}^{n_{g}}K_{h}^{2}\left(X_{gj}\right)\sigma^{2}\left(X_{gj}\right)X_{gj}^{(p)}X_{gj}^{(q)}\right]=\frac{1}{h^{2d}}\int_{\mathbb{R}^{d}}K^{2}\left(\frac{X}{h}\right)\sigma^{2}\left(X\right)X^{(p)}X^{(q)}f\left(X\right)\mathrm{d}X\\
 & =\frac{1}{h^{d-2}}\int_{\mathbb{R}^{d}}K^{2}\left(T\right)\sigma^{2}\left(hT\right)T^{(p)}T^{(q)}f\left(hT\right)\mathrm{d}T\\
 & =\frac{1}{h^{d-2}}f\left(0\right)\sigma^{2}\left(0\right)\int_{\mathbb{R}^{d}}K^{2}\left(T\right)T^{(p)}T^{(q)}\mathrm{d}T+o\left(\frac{1}{h^{d-2}}\right)\\
 & =o\left(h^{-d+2}\right).
\end{align*}

For variances,
\begin{eqnarray*}
 &  & \operatorname{Var}\left[H_{r}^{(q)}\right]\\
 & = & \operatorname{Var}\left[\frac{1}{n}\sum_{g=1}^{G}\sum_{j=1}^{n_{g}}K_{h}^{2}\left(X_{gj}\right)\sigma^{2}\left(X_{gj}\right)\left(X_{gj}^{(q)}\right)^{r}\right]=\frac{1}{n^{2}}\sum_{g=1}^{G}\operatorname{Var}\left[\sum_{j=1}^{n_{g}}K_{h}^{2}\left(X_{gj}\right)\sigma^{2}\left(X_{gj}\right)\left(X_{gj}^{(q)}\right)^{r}\right]\\
 & = & \frac{1}{n^{2}}\sum_{g=1}^{G}\sum_{j=1}^{n_{g}}\operatorname{Var}\left[K_{h}^{2}\left(X_{gj}\right)\sigma^{2}\left(X_{gj}\right)\left(X_{gj}^{(q)}\right)^{r}\right]\\
 &  & +2\frac{1}{n^{2}}\sum_{g=1}^{G}\ensuremath{\sum_{1\leq j<\ell\leq n_{g}}}\operatorname{Cov}\left[K_{h}^{2}\left(X_{gj}\right)\sigma^{2}\left(X_{gj}\right)\left(X_{gj}^{(q)}\right)^{r},K_{h}^{2}\left(X_{g\ell}\right)\sigma^{2}\left(X_{g\ell}\right)\left(X_{g\ell}^{(q)}\right)^{r}\right]\\
 & \leq & \frac{1}{n^{2}}\sum_{g=1}^{G}\sum_{j=1}^{n_{g}}\mathbb{E}\left[K_{h}^{4}\left(X_{gj}\right)\left(\sigma^{2}\left(X_{gj}\right)\right)^{2}\left(X_{gj}^{(q)}\right)^{2r}\right]\\
 &  & +\ensuremath{\frac{2}{n^{2}}\sum_{g=1}^{G}\sum_{1\leq j<\ell\leq n_{g}}}\mathbb{E}\left[K_{h}^{2}\left(X_{gj}\right)\sigma^{2}\left(X_{gj}\right)\left(X_{gj}^{(q)}\right)^{r}K_{h}^{2}\left(X_{g\ell}\right)\sigma^{2}\left(X_{g\ell}\right)\left(X_{g\ell}^{(q)}\right)^{r}\right]\\
 &  & -\ensuremath{\frac{2}{n^{2}}\sum_{g=1}^{G}\sum_{1\leq j<\ell\leq n_{g}}}\underbrace{\mathbb{E}\left[K_{h}^{2}\left(X_{gj}\right)\sigma^{2}\left(X_{gj}\right)\left(X_{gj}^{(q)}\right)^{r}\right]\mathbb{E}\left[K_{h}^{2}\left(X_{g\ell}\right)\sigma^{2}\left(X_{g\ell}\right)\left(X_{g\ell}^{(q)}\right)^{r}\right]}_{=\mathbb{E}\left[H_{r}^{(q)}\right]^{2}},
\end{eqnarray*}
where the second equality follows from the independence between clusters. We will bound the following two expectations 
\begin{align}
 & \mathbb{E}\left[K_{h}^{4}\left(X\right)\left(\sigma^{2}\left(X\right)\right)^{2}\left(X^{(q)}\right)^{2r}\right],\label{eq:var_Hr_e1}\\
 & \mathbb{E}\left[K_{h}^{2}\left(X_{j}\right)\sigma^{2}\left(X_{j}\right)\left(X_{j}^{(q)}\right)^{r}K_{h}^{2}\left(X_{\ell}\right)\sigma^{2}\left(X_{\ell}\right)\left(X_{\ell}^{(q)}\right)^{r}\right].\label{eq:var_Hr_e2}
\end{align}

\begin{eqnarray*}
 & \eqref{eq:var_Hr_e1}: & \mathbb{E}\left[K_{h}^{4}\left(X\right)\left(\sigma^{2}\left(X\right)\right)^{2}\left(X^{(q)}\right)^{2r}\right]=\frac{1}{h^{4d}}\int_{\mathbb{R}^{d}}K\left(\frac{X}{h}\right)^{4}\left(\sigma^{2}\left(X\right)\right)^{2}\left(X^{(q)}\right)^{2r}f\left(X\right)\mathrm{d}X\\
 & = & \frac{1}{h^{3d-2r}}\int_{\mathbb{R}^{d}}K\left(T\right)^{4}\left(T^{(q)}\right)^{2r}\left(\sigma^{2}\left(hT\right)\right)^{2}f\left(hT\right)\mathrm{d}T\\
 & = & \frac{1}{h^{3d-2r}}\int_{\mathbb{R}^{d}}K\left(T\right)^{4}\left(T^{(q)}\right)^{2r}\left(\sigma^{2}\left(0\right)\right)^{2}\left\{ f(0)+hT^{\top}\nabla f(0)\right\} \mathrm{d}T+o\left(h^{2r-3d+1}\right)\\
 & = & O\left(h^{2r-3d}\right),
\end{eqnarray*}
where the second equality follows from the change of variables $T=X/h$, and the third equality follows from the Taylor expansion. 

Also,
\begin{eqnarray*}
 & \eqref{eq:var_Hr_e2}: & \mathbb{E}\left[K_{h}^{2}\left(X_{j}\right)\sigma^{2}\left(X_{j}\right)\left(X_{j}^{(q)}\right)^{r}K_{h}^{2}\left(X_{\ell}\right)\sigma^{2}\left(X_{\ell}\right)\left(X_{\ell}^{(q)}\right)^{r}\right]\\
 & = & \frac{1}{h^{4d}}\int_{\mathbb{R}^{d_{\mathrm{ind}}}}\int_{\mathbb{R}^{d_{\mathrm{ind}}}}\int_{\mathbb{R}^{d_{\mathrm{cls}}}}K^{2}\left(\frac{X_{j}}{h}\right)\sigma^{2}\left(X_{j}\right)\left(X_{j}^{(q)}\right)^{r}K^{2}\left(\frac{X_{\ell}}{h}\right)\sigma^{2}\left(X_{\ell}\right)\left(X_{\ell}^{(q)}\right)^{r}\\
 &  & \qquad\qquad\times f_{2}\left(X_{j}^{(\mathrm{ind})},X_{\ell}^{(\mathrm{ind})};X^{(\mathrm{cls})}\right)\mathrm{d}X_{j}^{(\mathrm{ind})}\mathrm{d}X_{\ell}^{(\mathrm{ind})}\mathrm{d}X^{(\mathrm{cls})}\\
 & = & \frac{1}{h^{2d-2r+d_{\mathrm{cls}}}}\int_{\mathbb{R}^{d_{\mathrm{ind}}}}\int_{\mathbb{R}^{d_{\mathrm{ind}}}}\int_{\mathbb{R}^{d_{\mathrm{cls}}}}K^{2}\left(T_{j}\right)\sigma^{2}\left(hT_{j}\right)\left(T_{j}^{(q)}\right)^{r}K^{2}\left(T_{\ell}\right)\left(T_{\ell}^{(q)}\right)^{r}\sigma^{2}\left(hT_{\ell}\right)\\
 &  & \qquad\qquad\times f_{2}\left(hT_{j}^{(\mathrm{ind})},hT_{\ell}^{(\mathrm{ind})};hT^{(\mathrm{cls})}\right)\mathrm{d}T_{j}^{(\mathrm{ind})}\mathrm{d}T_{\ell}^{(\mathrm{ind})}\mathrm{d}T^{(\mathrm{cls})}\\
 & = & O\left(h^{2r-2d-d_{\mathrm{cls}}}\right).
\end{eqnarray*}

Thus,
\begin{eqnarray*}
\operatorname{Var}\left[F_{r}^{(q)}\right] & \leq & \frac{1}{n^{2}}\sum_{g=1}^{G}\sum_{j=1}^{n_{g}}O\left(h^{2r-3d}\right)+\ensuremath{\frac{1}{n^{2}}\sum_{g=1}^{G}\sum_{1\leq j<\ell\leq n_{g}}}O\left(h^{2r-2d-d_{\mathrm{cls}}}\right)\\
 & \leq & O\left(n^{-1}h^{2r-3d}\right)+O\left(n^{-1}\left(\max_{g}n_{g}\right)h^{2r-2d-d_{\mathrm{cls}}}\right)=o\left(h^{2r-2d}\right).
\end{eqnarray*}
Similarly,
\begin{eqnarray*}
 &  & \operatorname{Var}\left[F^{(p,q)}\right]\\
 & \leq & \frac{1}{n^{2}}\sum_{g=1}^{G}\sum_{j=1}^{n_{g}}\mathbb{E}\left[K_{h}^{4}\left(X_{gj}\right)\left(\sigma^{2}\left(X_{gj}\right)\right)^{2}\left(X_{gj}^{(q)}\right)^{2}\right]\\
 &  & +\ensuremath{\frac{2}{n^{2}}\sum_{g=1}^{G}\sum_{1\leq j<\ell\leq n_{g}}}\mathbb{E}\left[K_{h}^{2}\left(X_{gj}\right)\sigma^{2}\left(X_{gj}\right)X_{gj}^{(p)}X_{gj}^{(q)}K_{h}^{2}\left(X_{g\ell}\right)\sigma^{2}\left(X_{g\ell}\right)X_{g\ell}^{(p)}X_{g\ell}^{(q)}\right]\\
 &  & -\ensuremath{\frac{2}{n^{2}}\sum_{g=1}^{G}\sum_{1\leq j<\ell\leq n_{g}}}\underbrace{\mathbb{E}\left[K_{h}^{2}\left(X_{gj}\right)\sigma^{2}\left(X_{gj}\right)X_{gj}^{(p)}X_{gj}^{(q)}\right]\mathbb{E}\left[K_{h}^{2}\left(X_{gj}\right)\sigma^{2}\left(X_{g\ell}\right)X_{g\ell}^{(p)}X_{g\ell}^{(q)}\right]}_{=\mathbb{E}\left[H^{(p,q)}\right]^{2}}\\
 & \leq & \frac{1}{n^{2}}\sum_{g=1}^{G}\sum_{j=1}^{n_{g}}O\left(h^{4-3d}\right)+\ensuremath{\frac{1}{n^{2}}\sum_{g=1}^{G}\sum_{1\leq j<\ell\leq n_{g}}}O\left(h^{4-2d-d_{\mathrm{cls}}}\right)\\
 & \leq & O\left(n^{-1}h^{4-3d}\right)+O\left(n^{-1}\left(\max_{g}n_{g}\right)h^{4-2d-d_{\mathrm{cls}}}\right)=o\left(h^{4-2d}\right).
\end{eqnarray*}
Therefore, by Markov's inequality, $H_{0}^{(q)}=\frac{1}{h^{d}}f\left(0\right)\sigma^{2}\left(0\right)R_{k}^{d}+o_{p}\left(h^{-d}\right)$, $H_{1}^{(q)}=o_{p}\left(h^{-d+1}\right)$, $H_{2}^{(q)}=\frac{1}{h^{d-2}}f\left(0\right)\sigma^{2}\left(0\right)\left\{ \int_{\mathbb{R}^{d}}K^{2}\left(T\right)\left(T^{(q)}\right)^{2}\mathrm{d}T\right\} +o_{p}\left(h^{-d+2}\right)$, and $H^{(p,q)}=o_{p}\left(h^{-d+2}\right)$. We conclude by element-wise comparisons.
\end{proof}

\subsection{Proof for Lemma $\text{\ref{lem:Ir}}$}
\begin{proof}
Define $I_{r}^{(q)}=\frac{1}{n}\sum_{g=1}^{G}\sum_{1\leq j<\ell\leq n_{g}}K_{h}\left(X_{gj}\right)K_{h}\left(X_{g\ell}\right)\sigma\left(X_{gj}^{(\mathrm{ind})},X_{g\ell}^{(\mathrm{ind})};X_{g}^{(\mathrm{cls})}\right)\left(X_{gj}^{(q)}\right)^{r}$ for $r=0,1$, and $I^{(p,q)}=\frac{1}{n}\sum_{g=1}^{G}\sum_{1\leq j<\ell\leq n_{g}}K_{h}\left(X_{gj}\right)K_{h}\left(X_{g\ell}\right)\sigma\left(X_{gj}^{(\mathrm{ind})},X_{g\ell}^{(\mathrm{ind})};X_{g}^{(\mathrm{cls})}\right)X_{gj}^{(p)}X_{g\ell}^{(q)}$ for any $p$ and $q$ (allow $p=q$ here). For expectations,
\begin{align*}
\mathbb{E}\left[I_{r}^{(q)}\right] & =\mathbb{E}\left[\frac{1}{n}\sum_{g=1}^{G}\sum_{1\leq j<\ell\leq n_{g}}K_{h}\left(X_{gj}\right)K_{h}\left(X_{g\ell}\right)\sigma\left(X_{gj}^{(\mathrm{ind})},X_{g\ell}^{(\mathrm{ind})};X_{g}^{(\mathrm{cls})}\right)\left(X_{gj}^{(q)}\right)^{r}\right]\\
 & =\frac{1}{n}\sum_{g=1}^{G}\sum_{1\leq j<\ell\leq n_{g}}\mathbb{E}\left[K_{h}\left(X_{gj}\right)K_{h}\left(X_{g\ell}\right)\sigma\left(X_{gj}^{(\mathrm{ind})},X_{g\ell}^{(\mathrm{ind})};X_{g}^{(\mathrm{cls})}\right)\left(X_{gj}^{(q)}\right)^{r}\right],
\end{align*}
and
\begin{align*}
\mathbb{E}\left[I^{(p,q)}\right] & =\mathbb{E}\left[\frac{1}{n}\sum_{g=1}^{G}\sum_{1\leq j<\ell\leq n_{g}}K_{h}\left(X_{gj}\right)K_{h}\left(X_{g\ell}\right)\sigma\left(X_{gj}^{(\mathrm{ind})},X_{g\ell}^{(\mathrm{ind})};X_{g}^{(\mathrm{cls})}\right)X_{gj}^{(p)}X_{g\ell}^{(q)}\right]\\
 & =\frac{1}{n}\sum_{g=1}^{G}\sum_{1\leq j<\ell\leq n_{g}}\mathbb{E}\left[K_{h}\left(X_{gj}\right)K_{h}\left(X_{g\ell}\right)\sigma\left(X_{gj}^{(\mathrm{ind})},X_{g\ell}^{(\mathrm{ind})};X_{g}^{(\mathrm{cls})}\right)X_{gj}^{(p)}X_{g\ell}^{(q)}\right].
\end{align*}
We will evaluate 
\begin{align}
 & \mathbb{E}\left[K_{h}\left(X_{j}\right)K_{h}\left(X_{\ell}\right)\sigma\left(X_{j}^{(\mathrm{ind})},X_{\ell}^{(\mathrm{ind})};X^{(\mathrm{cls})}\right)\left(X_{j}^{(q)}\right)^{r}\right],\label{eq:e_Ir_e1}\\
 & \mathbb{E}\left[K_{h}\left(X_{j}\right)K_{h}\left(X_{\ell}\right)\sigma\left(X_{j}^{(\mathrm{ind})},X_{\ell}^{(\mathrm{ind})};X^{(\mathrm{cls})}\right)X_{j}^{(p)}X_{\ell}^{(q)}\right].\label{eq:e_Ir_e2}
\end{align}
Denote 
\begin{align*}
\nabla_{1}f_{2}\left(0,0;0\right) & =\left.\frac{\partial f_{2}\left(X_{j}^{(\mathrm{ind})},X_{\ell}^{(\mathrm{ind})};X^{(\mathrm{cls})}\right)}{\partial X_{j}^{(\mathrm{ind})}}\right|_{\left(X_{j}^{(\mathrm{ind})},X_{\ell}^{(\mathrm{ind})};X^{(\mathrm{cls})}\right)=\left(0,0;0\right)},\\
\nabla_{2}f_{2}\left(0,0;0\right) & =\left.\frac{\partial f_{2}\left(X_{j}^{(\mathrm{ind})},X_{\ell}^{(\mathrm{ind})};X^{(\mathrm{cls})}\right)}{\partial X_{\ell}^{(\mathrm{ind})}}\right|_{\left(X_{j}^{(\mathrm{ind})},X_{\ell}^{(\mathrm{ind})};X^{(\mathrm{cls})}\right)=\left(0,0;0\right)},\\
\nabla_{c}f_{2}\left(0,0;0\right) & =\left.\frac{\partial f_{2}\left(X_{j}^{(\mathrm{ind})},X_{\ell}^{(\mathrm{ind})};X^{(\mathrm{cls})}\right)}{\partial X^{(\mathrm{cls})}}\right|_{\left(X_{j}^{(\mathrm{ind})},X_{\ell}^{(\mathrm{ind})};X^{(\mathrm{cls})}\right)=\left(0,0;0\right)}.
\end{align*}
\begin{eqnarray*}
 & \eqref{eq:e_Ir_e1}: & \mathbb{E}\left[K_{h}\left(X_{j}\right)K_{h}\left(X_{\ell}\right)\sigma\left(X_{j}^{(\mathrm{ind})},X_{\ell}^{(\mathrm{ind})};X^{(\mathrm{cls})}\right)\left(X_{j}^{(q)}\right)^{r}\right]\\
 & = & \frac{1}{h^{2d}}\int_{\mathbb{R}^{d_{\mathrm{ind}}}}\int_{\mathbb{R}^{d_{\mathrm{ind}}}}\int_{\mathbb{R}^{d_{\mathrm{cls}}}}K\left(\frac{X_{j}}{h}\right)\left(X_{j}^{(q)}\right)^{r}K\left(\frac{X_{\ell}}{h}\right)\sigma\left(X_{j}^{(\mathrm{ind})},X_{\ell}^{(\mathrm{ind})};X^{(\mathrm{cls})}\right)\\
 &  & \qquad\times f_{2}\left(X_{j}^{(\mathrm{ind})},X_{\ell}^{(\mathrm{ind})};X^{(\mathrm{cls})}\right)\mathrm{d}X_{j}^{(\mathrm{ind})}\mathrm{d}X_{\ell}^{(\mathrm{ind})}\mathrm{d}X^{(\mathrm{cls})}\\
 & = & h^{r-d_{\mathrm{cls}}}\int_{\mathbb{R}^{d_{\mathrm{ind}}}}\int_{\mathbb{R}^{d_{\mathrm{ind}}}}\int_{\mathbb{R}^{d_{\mathrm{cls}}}}K\left(T_{j}\right)\left(T_{j}^{(q)}\right)^{r}K\left(T_{\ell}\right)\sigma\left(hT_{j}^{(\mathrm{ind})},hT_{\ell}^{(\mathrm{ind})};hT^{(\mathrm{cls})}\right)\\
 &  & \qquad\times f_{2}\left(hT_{j}^{(\mathrm{ind})},hT_{\ell}^{(\mathrm{ind})};hT^{(\mathrm{cls})}\right)\mathrm{d}T_{j}^{(\mathrm{ind})}\mathrm{d}T_{\ell}^{(\mathrm{ind})}\mathrm{d}T^{(\mathrm{cls})}\\
 & = & h^{r-d_{\mathrm{cls}}}\int_{\mathbb{R}^{d_{\mathrm{ind}}}}\int_{\mathbb{R}^{d_{\mathrm{ind}}}}\int_{\mathbb{R}^{d_{\mathrm{cls}}}}K\left(T_{j}\right)\left(T_{j}^{(q)}\right)^{r}K\left(T_{\ell}\right)\sigma\left(0,0;0\right)\\
 &  & \qquad\times\left\{ f_{2}\left(0,0;0\right)+hT_{j}^{(\mathrm{ind})\top}\nabla_{1}f_{2}\left(0,0;0\right)+hT_{\ell}^{(\mathrm{ind})\top}\nabla_{2}f_{2}\left(0,0;0\right)+hT^{(\mathrm{cls})\top}\nabla_{c}f_{2}\left(0,0;0\right)\right\} \\
 &  & \qquad\times\mathrm{d}T_{j}^{(\mathrm{ind})}\mathrm{d}T_{\ell}^{(\mathrm{ind})}\mathrm{d}T^{(\mathrm{cls})}+o\left(h^{r+1-d_{\mathrm{cls}}}\right)\\
 & = & \begin{cases}
h^{-d_{\mathrm{cls}}}R_{k}^{d_{\mathrm{cls}}}\sigma\left(0,0;0\right)f_{2}\left(0,0;0\right)+O\left(h\right) & \text{ if }r=0\\
O\left(h^{2-d_{\mathrm{cls}}}\right) & \text{ if }r=1
\end{cases},
\end{eqnarray*}
where the second equality follows from the change of variables $T_{j}^{(\mathrm{ind})}=X_{j}^{(\mathrm{ind})}/h$, $T_{\ell}^{(\mathrm{ind})}=X_{\ell}^{(\mathrm{ind})}/h$, and $T^{(\mathrm{cls})}=X^{(\mathrm{cls})}/h$ (we define $T_{j}=\left(T_{j}^{(\mathrm{ind})\top},T^{(\mathrm{cls})\top}\right)^{\top}$, $T_{\ell}=\left(T_{\ell}^{(\mathrm{ind})\top},T^{(\mathrm{cls})\top}\right)^{\top}$), and the third equality follows from the Taylor expansion.

Similarly,
\begin{eqnarray*}
 & \eqref{eq:e_Ir_e2}: & \mathbb{E}\left[K_{h}\left(X_{j}\right)K_{h}\left(X_{\ell}\right)\sigma\left(X_{j}^{(\mathrm{ind})},X_{\ell}^{(\mathrm{ind})};X^{(\mathrm{cls})}\right)X_{j}^{(p)}X_{\ell}^{(q)}\right]\\
 & = & \frac{1}{h^{2d}}\int_{\mathbb{R}^{d_{\mathrm{ind}}}}\int_{\mathbb{R}^{d_{\mathrm{ind}}}}\int_{\mathbb{R}^{d_{\mathrm{cls}}}}K\left(\frac{X_{j}}{h}\right)X_{j}^{(p)}K\left(\frac{X_{\ell}}{h}\right)X_{\ell}^{(q)}\sigma\left(X_{j}^{(\mathrm{ind})},X_{\ell}^{(\mathrm{ind})};X^{(\mathrm{cls})}\right)\\
 &  & \qquad\qquad\times f_{2}\left(X_{j}^{(\mathrm{ind})},X_{\ell}^{(\mathrm{ind})};X^{(\mathrm{cls})}\right)\mathrm{d}X_{j}^{(\mathrm{ind})}\mathrm{d}X_{\ell}^{(\mathrm{ind})}\mathrm{d}X^{(\mathrm{cls})}\\
 & = & h^{2-d_{\mathrm{cls}}}\int_{\mathbb{R}^{d_{\mathrm{ind}}}}\int_{\mathbb{R}^{d_{\mathrm{ind}}}}\int_{\mathbb{R}^{d_{\mathrm{cls}}}}K\left(T_{j}\right)T_{j}^{(p)}K\left(T_{\ell}\right)T_{\ell}^{(q)}\sigma\left(hT_{j}^{(\mathrm{ind})},hT_{\ell}^{(\mathrm{ind})};hT^{(\mathrm{cls})}\right)\\
 &  & \qquad\qquad\times f_{2}\left(hT_{j}^{(\mathrm{ind})},hT_{\ell}^{(\mathrm{ind})};hT^{(\mathrm{cls})}\right)\mathrm{d}T_{j}^{(\mathrm{ind})}\mathrm{d}T_{\ell}^{(\mathrm{ind})}\mathrm{d}T^{(\mathrm{cls})}\\
 & = & h^{2-d_{\mathrm{cls}}}\sigma\left(0,0;0\right)f_{2}\left(0,0;0\right)\int_{\mathbb{R}^{d_{\mathrm{ind}}}}\int_{\mathbb{R}^{d_{\mathrm{ind}}}}\int_{\mathbb{R}^{d_{\mathrm{cls}}}}K\left(T_{j}\right)T_{j}^{(p)}K\left(T_{\ell}\right)T_{\ell}^{(q)}\mathrm{d}T_{j}\mathrm{d}T_{\ell}+o\left(h^{2-d_{\mathrm{cls}}}\right)\\
 & = & O\left(h^{2-d_{\mathrm{cls}}}\right).
\end{eqnarray*}

Thus, since 
\[
\frac{1}{n}\sum_{g=1}^{G}\sum_{1\leq j<\ell\leq n_{g}}1=\frac{1}{n}\sum_{g=1}^{G}\left(1+2+\cdots+\left(n_{g}-1\right)\right)=\frac{1}{n}\sum_{g=1}^{G}\frac{n_{g}(n_{g}-1)}{2}=\left(\frac{1}{n}\sum_{g=1}^{G}\frac{n_{g}^{2}}{2}\right)-\frac{1}{2}
\]
and
\[
\left(\frac{1}{n}\sum_{g=1}^{G}n_{g}^{2}-1\right)h^{d_{\mathrm{ind}}}=\lambda+o(1),
\]
 we have for $r=0$,
\begin{align*}
\mathbb{E}\left[I_{0}^{(q)}\right] & =\frac{1}{n}\sum_{g=1}^{G}\sum_{1\leq j<\ell\leq n_{g}}\left\{ h^{-d_{\mathrm{cls}}}R_{k}^{d_{\mathrm{cls}}}\sigma\left(0,0;0\right)f_{2}\left(0,0;0\right)+O\left(h^{1-d_{\mathrm{cls}}}\right)\right\} \\
 & =\ensuremath{\frac{1}{2}\left(\frac{1}{n}\sum_{g=1}^{G}n_{g}^{2}-1\right)h^{-d_{\mathrm{cls}}}R_{k}^{d_{\mathrm{cls}}}\sigma\left(0,0;0\right)f_{2}\left(0,0;0\right)+o\left(\left(\max_{g}n_{g}\right)h^{-d_{\mathrm{cls}}}\right)}\\
 & =h^{-d}\left\{ \frac{\lambda}{2}R_{k}^{d_{\mathrm{cls}}}\sigma\left(0,0;0\right)f_{2}\left(0,0;0\right)\ensuremath{+o\left(1\right)}\right\} ,
\end{align*}
and for $r=1$,
\begin{align*}
\mathbb{E}\left[I_{1}^{(q)}\right] & =\frac{1}{n}\sum_{g=1}^{G}\sum_{1\leq j<\ell\leq n_{g}}O\left(h^{2-d_{\mathrm{cls}}}\right)=\ensuremath{O\left(\left(\max_{g}n_{g}\right)h^{2-d_{\mathrm{cls}}}\right)}=O\left(h^{-d+2}\right).
\end{align*}
Also,
\begin{align*}
\mathbb{E}\left[I^{(p,q)}\right] & =\frac{1}{n}\sum_{g=1}^{G}\sum_{1\leq j<\ell\leq n_{g}}O\left(h^{2-d_{\mathrm{cls}}}\right)=O\left(h^{-d+2}\right).
\end{align*}

For variances, by the mutual independence between clusters,
\begin{eqnarray*}
 &  & \operatorname{Var}\left[I_{r}^{(q)}\right]\\
 & = & \operatorname{Var}\left[\frac{1}{n}\sum_{g=1}^{G}\sum_{1\leq j<\ell\leq n_{g}}K_{h}\left(X_{gj}\right)K_{h}\left(X_{g\ell}\right)\sigma\left(X_{gj}^{(\mathrm{ind})},X_{g\ell}^{(\mathrm{ind})};X_{g}^{(\mathrm{cls})}\right)\left(X_{gj}^{(q)}\right)^{r}\right]\\
 & = & \frac{1}{n^{2}}\sum_{g=1}^{G}\operatorname{Var}\left[\sum_{1\leq j<\ell\leq n_{g}}K_{h}\left(X_{gj}\right)K_{h}\left(X_{g\ell}\right)\sigma\left(X_{gj}^{(\mathrm{ind})},X_{g\ell}^{(\mathrm{ind})};X_{g}^{(\mathrm{cls})}\right)\left(X_{gj}^{(q)}\right)^{r}\right].
\end{eqnarray*}
Here, 
\begin{eqnarray*}
 &  & \operatorname{Var}\left[\sum_{1\leq j<\ell\leq n_{g}}K_{h}\left(X_{gj}\right)K_{h}\left(X_{g\ell}\right)\sigma\left(X_{gj}^{(\mathrm{ind})},X_{g\ell}^{(\mathrm{ind})};X_{g}^{(\mathrm{cls})}\right)\left(X_{gj}^{(q)}\right)^{r}\right]\\
 & = & \sum_{1\leq j<\ell\leq n_{g}}\sum_{1\leq t<s\leq n_{g}}\operatorname{Cov}\left[K_{h}\left(X_{gj}\right)K_{h}\left(X_{g\ell}\right)\sigma\left(X_{gj}^{(\mathrm{ind})},X_{g\ell}^{(\mathrm{ind})};X_{g}^{(\mathrm{cls})}\right)\left(X_{gj}^{(q)}\right)^{r},\right.\\
 &  & \qquad\qquad\qquad\qquad\qquad\qquad\qquad\qquad\left.K_{h}\left(X_{gt}\right)K_{h}\left(X_{gs}\right)\sigma\left(X_{gj}^{(\mathrm{ind})},X_{g\ell}^{(\mathrm{ind})};X_{g}^{(\mathrm{cls})}\right)\left(X_{gt}^{(q)}\right)^{r}\right],
\end{eqnarray*}
and there are the following three cases (i) $j=t$ and $\ell=s$, (ii) $j=t,\ell\neq s$, (iii) $j\neq t,\ell=s$, (iv) $j\neq t$ and $\ell\neq s$. 

{\small{}(i) When $j=t$ and $\ell=s$,
\begin{eqnarray*}
 &  & \operatorname{Cov}\left[K_{h}\left(X_{j}\right)K_{h}\left(X_{\ell}\right)\sigma\left(X_{j}^{(\mathrm{ind})},X_{\ell}^{(\mathrm{ind})};X^{(\mathrm{cls})}\right)\left(X_{j}^{(q)}\right)^{r},\right.\\
 &  & \qquad\qquad\qquad\qquad\qquad\qquad\left.K_{h}\left(X_{j}\right)K_{h}\left(X_{\ell}\right)\sigma\left(X_{j}^{(\mathrm{ind})},X_{\ell}^{(\mathrm{ind})};X^{(\mathrm{cls})}\right)\left(X_{j}^{(q)}\right)^{r}\right]\\
 & = & \operatorname{Var}\left[K_{h}\left(X_{j}\right)K_{h}\left(X_{\ell}\right)\sigma\left(X_{j}^{(\mathrm{ind})},X_{\ell}^{(\mathrm{ind})};X^{(\mathrm{cls})}\right)\left(X_{j}^{(q)}\right)^{r}\right]\\
 & \leq & \mathbb{E}\left[K_{h}^{2}\left(X_{j}\right)K_{h}^{2}\left(X_{\ell}\right)\sigma^{2}\left(X_{j}^{(\mathrm{ind})},X_{\ell}^{(\mathrm{ind})};X^{(\mathrm{cls})}\right)\left(X_{j}^{(q)}\right)^{2r}\right]\\
 & = & \frac{1}{h^{4d}}\int_{\mathbb{R}^{d_{\mathrm{ind}}}}\int_{\mathbb{R}^{d_{\mathrm{ind}}}}\int_{\mathbb{R}^{d_{\mathrm{cls}}}}K^{2}\left(\frac{X_{j}}{h}\right)K^{2}\left(\frac{X_{\ell}}{h}\right)\sigma^{2}\left(X_{j}^{(\mathrm{ind})},X_{\ell}^{(\mathrm{ind})};X^{(\mathrm{cls})}\right)\left(X_{j}^{(q)}\right)^{2r}\\
 &  & \qquad\qquad\times f_{2}\left(X_{j}^{(\mathrm{ind})},X_{\ell}^{(\mathrm{ind})};X^{(\mathrm{cls})}\right)\mathrm{d}X_{j}^{(\mathrm{ind})}\mathrm{d}X_{\ell}^{(\mathrm{ind})}\mathrm{d}X^{(\mathrm{cls})}\\
 & = & \frac{1}{h^{2d-2r+d_{\mathrm{cls}}}}\int_{\mathbb{R}^{d_{\mathrm{ind}}}}\int_{\mathbb{R}^{d_{\mathrm{ind}}}}\int_{\mathbb{R}^{d_{\mathrm{cls}}}}K^{2}\left(T_{j}\right)K^{2}\left(T_{\ell}\right)\sigma^{2}\left(hT_{j}^{(\mathrm{ind})},hT_{\ell}^{(\mathrm{ind})};hT^{(\mathrm{cls})}\right)\left(T_{j}^{(q)}\right)^{2r}\\
 &  & \qquad\qquad\times f_{2}\left(hT_{j}^{(\mathrm{ind})},hT_{\ell}^{(\mathrm{ind})};hT^{(\mathrm{cls})}\right)\mathrm{d}T_{j}^{(\mathrm{ind})}\mathrm{d}T_{\ell}^{(\mathrm{ind})}\mathrm{d}T^{(\mathrm{cls})}\\
 & = & O\left(h^{2r-2d-d_{\mathrm{cls}}}\right),
\end{eqnarray*}
where the third equality follows from the change of variables $T_{j}^{(\mathrm{ind})}=X_{j}^{(\mathrm{ind})}/h$, $T_{\ell}^{(\mathrm{ind})}=X_{\ell}^{(\mathrm{ind})}/h$, and $T^{(\mathrm{cls})}=X^{(\mathrm{cls})}/h$ (we define $T_{j}=\left(T_{j}^{(\mathrm{ind})\top},T^{(\mathrm{cls})\top}\right)^{\top}$, $T_{\ell}=\left(T_{\ell}^{(\mathrm{ind})\top},T^{(\mathrm{cls})\top}\right)^{\top}$).}{\small\par}

{\small{}(ii) When $j=t,\ell\neq s$,
\begin{eqnarray}
 &  & \operatorname{Cov}\left[K_{h}\left(X_{j}\right)K_{h}\left(X_{\ell}\right)\sigma\left(X_{j}^{(\mathrm{ind})},X_{\ell}^{(\mathrm{ind})};X^{(\mathrm{cls})}\right)\left(X_{j}^{(q)}\right)^{r},\right.\nonumber \\
 &  & \qquad\qquad\qquad\qquad\qquad\qquad\left.K_{h}\left(X_{j}\right)K_{h}\left(X_{s}\right)\sigma\left(X_{j}^{(\mathrm{ind})},X_{s}^{(\mathrm{ind})};X^{(\mathrm{cls})}\right)\left(X_{j}^{(q)}\right)^{r}\right]\nonumber \\
 & = & \mathbb{E}\left[K_{h}^{2}\left(X_{j}\right)K_{h}\left(X_{\ell}\right)\sigma\left(X_{j}^{(\mathrm{ind})},X_{\ell}^{(\mathrm{ind})};X^{(\mathrm{cls})}\right)\left(X_{j}^{(q)}\right)^{2r}K_{h}\left(X_{s}\right)\sigma\left(X_{j}^{(\mathrm{ind})},X_{s}^{(\mathrm{ind})};X^{(\mathrm{cls})}\right)\right]\label{eq:var_Ir_e1}\\
 &  & -\underbrace{\mathbb{E}\left[K_{h}\left(X_{j}\right)K_{h}\left(X_{\ell}\right)\sigma\left(X_{j}^{(\mathrm{ind})},X_{\ell}^{(\mathrm{ind})};X^{(\mathrm{cls})}\right)\left(X_{j}^{(q)}\right)^{r}\right]^{2}}_{=\left\{ \eqref{eq:e_Ir_e1}\right\} ^{2}}.\nonumber 
\end{eqnarray}
We can evaluate an expectation as
\begin{eqnarray*}
 & \eqref{eq:var_Ir_e1}: & \mathbb{E}\left[K_{h}^{2}\left(X_{j}\right)K_{h}\left(X_{\ell}\right)\sigma\left(X_{j}^{(\mathrm{ind})},X_{\ell}^{(\mathrm{ind})};X^{(\mathrm{cls})}\right)\left(X_{j}^{(q)}\right)^{2r}K_{h}\left(X_{s}\right)\sigma\left(X_{j}^{(\mathrm{ind})},X_{s}^{(\mathrm{ind})};X^{(\mathrm{cls})}\right)\right]\\
 & = & \frac{1}{h^{4d}}\int_{\mathbb{R}^{d_{\mathrm{ind}}}}\int_{\mathbb{R}^{d_{\mathrm{ind}}}}\int_{\mathbb{R}^{d_{\mathrm{ind}}}}\int_{\mathbb{R}^{d_{\mathrm{cls}}}}K^{2}\left(\frac{X_{j}}{h}\right)K\left(\frac{X_{\ell}}{h}\right)\sigma\left(X_{j}^{(\mathrm{ind})},X_{\ell}^{(\mathrm{ind})};X^{(\mathrm{cls})}\right)\left(X_{j}^{(q)}\right)^{2r}\\
 &  & \qquad\qquad\times K_{h}\left(\frac{X_{s}}{h}\right)\sigma\left(X_{j}^{(\mathrm{ind})},X_{s}^{(\mathrm{ind})};X^{(\mathrm{cls})}\right)\\
 &  & \qquad\qquad\times f_{3}\left(X_{j}^{(\mathrm{ind})},X_{\ell}^{(\mathrm{ind})},X_{s}^{(\mathrm{ind})};X^{(\mathrm{cls})}\right)\mathrm{d}X_{j}^{(\mathrm{ind})}\mathrm{d}X_{\ell}^{(\mathrm{ind})}\mathrm{d}X_{s}^{(\mathrm{ind})}\mathrm{d}X^{(\mathrm{cls})}\\
 & = & \frac{1}{h^{d-2r+2d_{\mathrm{cls}}}}\int_{\mathbb{R}^{d_{\mathrm{ind}}}}\int_{\mathbb{R}^{d_{\mathrm{ind}}}}\int_{\mathbb{R}^{d_{\mathrm{ind}}}}\int_{\mathbb{R}^{d_{\mathrm{cls}}}}K^{2}\left(T_{j}\right)K\left(T_{\ell}\right)\sigma\left(hT_{j}^{(\mathrm{ind})},hT_{\ell}^{(\mathrm{ind})};hT^{(\mathrm{cls})}\right)\left(T_{j}^{(q)}\right)^{2r}\\
 &  & \qquad\qquad\times K\left(T_{s}\right)\sigma\left(hT_{j}^{(\mathrm{ind})},hT_{s}^{(\mathrm{ind})};hT^{(\mathrm{cls})}\right)\\
 &  & \qquad\qquad\times f_{3}\left(hT_{j}^{(\mathrm{ind})},hT_{\ell}^{(\mathrm{ind})},hT_{s}^{(\mathrm{ind})};hT^{(\mathrm{cls})}\right)\mathrm{d}T_{j}^{(\mathrm{ind})}\mathrm{d}T_{\ell}^{(\mathrm{ind})}\mathrm{d}T_{s}^{(\mathrm{ind})}\mathrm{d}T^{(\mathrm{cls})}\\
 & = & O\left(h^{2r-d-2d_{\mathrm{cls}}}\right),
\end{eqnarray*}
where the second equality follows from the change of variables $T_{j}^{(\mathrm{ind})}=X_{j}^{(\mathrm{ind})}/h$, $T_{\ell}^{(\mathrm{ind})}=X_{\ell}^{(\mathrm{ind})}/h$, $T_{s}^{(\mathrm{ind})}=X_{s}^{(\mathrm{ind})}/h$ and $T^{(\mathrm{cls})}=X^{(\mathrm{cls})}/h$ (we define $T_{j}=\left(T_{j}^{(\mathrm{ind})\top},T^{(\mathrm{cls})\top}\right)^{\top}$, $T_{\ell}=\left(T_{\ell}^{(\mathrm{ind})\top},T^{(\mathrm{cls})\top}\right)^{\top}$, $T_{s}=\left(T_{s}^{(\mathrm{ind})\top},T^{(\mathrm{cls})\top}\right)^{\top}$). Thus,
\begin{eqnarray*}
 &  & \operatorname{Cov}\left[K_{h}\left(X_{j}\right)K_{h}\left(X_{\ell}\right)\sigma\left(X_{j}^{(\mathrm{ind})},X_{\ell}^{(\mathrm{ind})};X^{(\mathrm{cls})}\right)\left(X_{j}^{(q)}\right)^{r},\right.\\
 &  & \qquad\qquad\qquad\qquad\qquad\qquad\left.K_{h}\left(X_{j}\right)K_{h}\left(X_{s}\right)\sigma\left(X_{j}^{(\mathrm{ind})},X_{s}^{(\mathrm{ind})};X^{(\mathrm{cls})}\right)\left(X_{j}^{(q)}\right)^{r}\right]\\
 & = & O\left(h^{2r-d-2d_{\mathrm{cls}}}\right).
\end{eqnarray*}
}{\small\par}

{\small{}(iii) When $j\neq t,\ell=s$,
\begin{eqnarray}
 &  & \operatorname{Cov}\left[K_{h}\left(X_{j}\right)K_{h}\left(X_{\ell}\right)\sigma\left(X_{j}^{(\mathrm{ind})},X_{\ell}^{(\mathrm{ind})};X^{(\mathrm{cls})}\right)\left(X_{j}^{(q)}\right)^{r},\right.\nonumber \\
 &  & \qquad\qquad\qquad\qquad\qquad\qquad\left.K_{h}\left(X_{t}\right)K_{h}\left(X_{\ell}\right)\sigma\left(X_{t}^{(\mathrm{ind})},X_{\ell}^{(\mathrm{ind})};X^{(\mathrm{cls})}\right)\left(X_{t}^{(q)}\right)^{r}\right]\nonumber \\
 &  & \mathbb{E}\left[K_{h}\left(X_{j}\right)K_{h}^{2}\left(X_{\ell}\right)\sigma\left(X_{j}^{(\mathrm{ind})},X_{\ell}^{(\mathrm{ind})};X^{(\mathrm{cls})}\right)\left(X_{j}^{(q)}\right)^{r}\right.\nonumber \\
 &  & \qquad\qquad\times\left.K_{h}\left(X_{t}\right)\sigma\left(X_{t}^{(\mathrm{ind})},X_{\ell}^{(\mathrm{ind})};X^{(\mathrm{cls})}\right)\left(X_{t}^{(q)}\right)^{r}\right]\label{eq:var_Ir_e2}\\
 &  & -\underbrace{\mathbb{E}\left[K_{h}\left(X_{j}\right)K_{h}\left(X_{\ell}\right)\sigma\left(X_{j}^{(\mathrm{ind})},X_{\ell}^{(\mathrm{ind})};X^{(\mathrm{cls})}\right)\left(X_{j}^{(q)}\right)^{r}\right]^{2}}_{=\left\{ \eqref{eq:e_Ir_e1}\right\} ^{2}}\nonumber \\
 & = & O\left(h^{2r-d-2d_{\mathrm{cls}}}\right)\nonumber 
\end{eqnarray}
by a similar derivation to the case (ii).}{\small\par}

{\small{}(iv) When $j\neq t,\ell\neq s$,
\begin{eqnarray}
 &  & \operatorname{Cov}\left[K_{h}\left(X_{j}\right)K_{h}\left(X_{\ell}\right)\sigma\left(X_{j}^{(\mathrm{ind})},X_{\ell}^{(\mathrm{ind})};X^{(\mathrm{cls})}\right)\left(X_{j}^{(q)}\right)^{r},\right.\nonumber \\
 &  & \qquad\qquad\qquad\qquad\qquad\qquad\left.K_{h}\left(X_{t}\right)K_{h}\left(X_{s}\right)\sigma\left(X_{t}^{(\mathrm{ind})},X_{s}^{(\mathrm{ind})};X^{(\mathrm{cls})}\right)\left(X_{t}^{(q)}\right)^{r}\right]\nonumber \\
 & = & \mathbb{E}\left[K_{h}\left(X_{j}\right)K_{h}\left(X_{\ell}\right)\sigma\left(X_{j}^{(\mathrm{ind})},X_{\ell}^{(\mathrm{ind})};X^{(\mathrm{cls})}\right)\left(X_{j}^{(q)}\right)^{r}\right.\nonumber \\
 &  & \qquad\qquad\times\left.K_{h}\left(X_{t}\right)K_{h}\left(X_{s}\right)\sigma\left(X_{t}^{(\mathrm{ind})},X_{s}^{(\mathrm{ind})};X^{(\mathrm{cls})}\right)\left(X_{t}^{(q)}\right)^{r}\right]\label{eq:var_Ir_e3}\\
 &  & -\underbrace{\mathbb{E}\left[K_{h}\left(X_{j}\right)K_{h}\left(X_{\ell}\right)\sigma\left(X_{j}^{(\mathrm{ind})},X_{\ell}^{(\mathrm{ind})};X^{(\mathrm{cls})}\right)\left(X_{j}^{(q)}\right)^{r}\right]^{2}}_{=\left\{ \eqref{eq:e_Ir_e1}\right\} ^{2}}.\nonumber 
\end{eqnarray}
We can evaluate an expectation as}{\small\par}

{\small{}
\begin{eqnarray*}
 & \eqref{eq:var_Ir_e3}: & \mathbb{E}\left[K_{h}\left(X_{j}\right)K_{h}\left(X_{\ell}\right)\sigma\left(X_{j}^{(\mathrm{ind})},X_{\ell}^{(\mathrm{ind})};X^{(\mathrm{cls})}\right)\left(X_{j}^{(q)}\right)^{r}\right.\\
 &  & \qquad\qquad\times\left.K_{h}\left(X_{t}\right)K_{h}\left(X_{s}\right)\sigma\left(X_{t}^{(\mathrm{ind})},X_{s}^{(\mathrm{ind})};X^{(\mathrm{cls})}\right)\left(X_{t}^{(q)}\right)^{r}\right]\\
 & = & \frac{1}{h^{4d}}\int_{\mathbb{R}^{d_{\mathrm{ind}}}}\int_{\mathbb{R}^{d_{\mathrm{ind}}}}\int_{\mathbb{R}^{d_{\mathrm{ind}}}}\int_{\mathbb{R}^{d_{\mathrm{cls}}}}\int_{\mathbb{R}^{d_{\mathrm{ind}}}}\\
 &  & \qquad\qquad\times K\left(\frac{X_{j}}{h}\right)K\left(\frac{X_{\ell}}{h}\right)\sigma\left(X_{j}^{(\mathrm{ind})},X_{\ell}^{(\mathrm{ind})};X^{(\mathrm{cls})}\right)\left(X_{j}^{(q)}\right)^{r}\\
 &  & \qquad\qquad\times K\left(\frac{X_{t}}{h}\right)K\left(\frac{X_{s}}{h}\right)\sigma\left(X_{t}^{(\mathrm{ind})},X_{s}^{(\mathrm{ind})};X^{(\mathrm{cls})}\right)\left(X_{t}^{(q)}\right)^{r}\\
 &  & \qquad\qquad\times f_{4}\left(X_{j}^{(\mathrm{ind})},X_{\ell}^{(\mathrm{ind})},X_{t}^{(\mathrm{ind})},X_{s}^{(\mathrm{ind})};X^{(\mathrm{cls})}\right)\mathrm{d}X_{j}^{(\mathrm{ind})}\mathrm{d}X_{\ell}^{(\mathrm{ind})}\mathrm{d}X_{t}^{(\mathrm{ind})}\mathrm{d}X_{s}^{(\mathrm{ind})}\mathrm{d}X^{(\mathrm{cls})}\\
 & = & h^{2r-3d_{\mathrm{cls}}}\int_{\mathbb{R}^{d_{\mathrm{ind}}}}\int_{\mathbb{R}^{d_{\mathrm{ind}}}}\int_{\mathbb{R}^{d_{\mathrm{ind}}}}\int_{\mathbb{R}^{d_{\mathrm{cls}}}}\int_{\mathbb{R}^{d_{\mathrm{ind}}}}\\
 &  & \qquad\qquad\times K\left(T_{j}\right)K\left(T_{\ell}\right)\sigma\left(hT_{j}^{(\mathrm{ind})},hT_{\ell}^{(\mathrm{ind})};hT^{(\mathrm{cls})}\right)\left(T_{j}^{(q)}\right)^{r}\\
 &  & \qquad\qquad\times K\left(T_{t}\right)K\left(T_{s}\right)\sigma\left(hT_{t}^{(\mathrm{ind})},hT_{s}^{(\mathrm{ind})};hT^{(\mathrm{cls})}\right)\left(T_{t}^{(q)}\right)^{r}\\
 &  & \qquad\qquad\times f_{4}\left(hT_{j}^{(\mathrm{ind})},hT_{\ell}^{(\mathrm{ind})},hT_{t}^{(\mathrm{ind})},hT_{s}^{(\mathrm{ind})};hT^{(\mathrm{cls})}\right)\mathrm{d}T_{j}^{(\mathrm{ind})}\mathrm{d}T_{\ell}^{(\mathrm{ind})}\mathrm{d}T_{t}^{(\mathrm{ind})}\mathrm{d}T_{s}^{(\mathrm{ind})}\mathrm{d}T^{(\mathrm{cls})}\\
 & = & O\left(h^{2r-3d_{\mathrm{cls}}}\right),
\end{eqnarray*}
where the second equality follows from the change of variables $T_{j}^{(\mathrm{ind})}=X_{j}^{(\mathrm{ind})}/h$, $T_{\ell}^{(\mathrm{ind})}=X_{\ell}^{(\mathrm{ind})}/h$, $T_{t}^{(\mathrm{ind})}=X_{t}^{(\mathrm{ind})}/h$, $T_{s}^{(\mathrm{ind})}=X_{s}^{(\mathrm{ind})}/h$ and $T^{(\mathrm{cls})}=X^{(\mathrm{cls})}/h$ (we define $T_{j}=\left(T_{j}^{(\mathrm{ind})\top},T^{(\mathrm{cls})\top}\right)^{\top}$, $T_{\ell}=\left(T_{\ell}^{(\mathrm{ind})\top},T^{(\mathrm{cls})\top}\right)^{\top}$, $T_{t}=\left(T_{t}^{(\mathrm{ind})\top},T^{(\mathrm{cls})\top}\right)^{\top}$, $T_{s}=\left(T_{s}^{(\mathrm{ind})\top},T^{(\mathrm{cls})\top}\right)^{\top}$). Thus,
\begin{eqnarray*}
 &  & \operatorname{Cov}\left[K_{h}\left(X_{j}\right)K_{h}\left(X_{\ell}\right)\sigma\left(X_{j}^{(\mathrm{ind})},X_{\ell}^{(\mathrm{ind})};X^{(\mathrm{cls})}\right)\left(X_{j}^{(q)}\right)^{r},\right.\\
 &  & \qquad\qquad\qquad\qquad\qquad\qquad\left.K_{h}\left(X_{t}\right)K_{h}\left(X_{s}\right)\sigma\left(X_{t}^{(\mathrm{ind})},X_{s}^{(\mathrm{ind})};X^{(\mathrm{cls})}\right)\left(X_{t}^{(q)}\right)^{r}\right]\\
 & = & O\left(h^{2r-3d_{\mathrm{cls}}}\right).
\end{eqnarray*}
}{\small\par}

Thus, by counting cases (i)-(iv),
\begin{eqnarray*}
 &  & \operatorname{Var}\left[I_{r}^{(q)}\right]\\
 & \leq & \frac{1}{n^{2}}\sum_{g=1}^{G}\sum_{1\leq j<\ell\leq n_{g}}\sum_{1\leq t<s\leq n_{g}}\operatorname{Cov}\left[K_{h}\left(X_{gj}\right)K_{h}\left(X_{g\ell}\right)\sigma\left(X_{gj}^{(\mathrm{ind})},X_{g\ell}^{(\mathrm{ind})};X_{g}^{(\mathrm{cls})}\right)\left(X_{gj}^{(q)}\right)^{r},\right.\\
 &  & \qquad\qquad\qquad\qquad\qquad\qquad\qquad\qquad\qquad\left.K_{h}\left(X_{gt}\right)K_{h}\left(X_{gs}\right)\sigma\left(X_{gt}^{(\mathrm{ind})},X_{gs}^{(\mathrm{ind})};X_{g}^{(\mathrm{cls})}\right)\left(X_{gt}^{(q)}\right)^{r}\right]\\
 & \leq & \frac{1}{n^{2}}\sum_{g=1}^{G}\left\{ n_{g}^{2}O\left(h^{2r-2d-d_{\mathrm{cls}}}\right)\ensuremath{+n_{g}^{3}O\left(h^{2r-d-2d_{\mathrm{cls}}}\right)}+n_{g}^{3}O\left(h^{2r-d-2d_{\mathrm{cls}}}\right)+n_{g}^{4}O\left(h^{2r-3d_{\mathrm{cls}}}\right)\right\} \\
 & \leq & \frac{\left(\max_{g}n_{g}\right)h^{-d_{\mathrm{cls}}}}{n}\left\{ O\left(h^{2r-2d}\right)\ensuremath{+\left(\max_{g}n_{g}\right)h^{-d_{\mathrm{cls}}}O\left(h^{2r-d}\right)}+\left(\max_{g}n_{g}\right)^{2}h^{-2d_{\mathrm{cls}}}O\left(h^{2r}\right)\right\} \\
 & = & o\left(1\right)\left\{ O\left(h^{2r-2d}\right)\ensuremath{+\underbrace{\left(\max_{g}n_{g}\right)h^{d_{\mathrm{ind}}}}_{O(1)}O\left(h^{2r-2d}\right)}+\underbrace{\left(\max_{g}n_{g}\right)^{2}h^{2d_{\mathrm{ind}}}}_{=O(1)}O\left(h^{2r-2d}\right)\right\} =o\left(h^{2r-2d}\right).
\end{eqnarray*}

Similarly,
\begin{eqnarray*}
 &  & \operatorname{Var}\left[I^{(p,q)}\right]\\
 & \leq & \frac{1}{n^{2}}\sum_{g=1}^{G}\sum_{1\leq j<\ell\leq n_{g}}\sum_{1\leq t<s\leq n_{g}}\operatorname{Cov}\left[K_{h}\left(X_{gj}\right)K_{h}\left(X_{g\ell}\right)\sigma\left(X_{gj}^{(\mathrm{ind})},X_{g\ell}^{(\mathrm{ind})};X_{g}^{(\mathrm{cls})}\right)\left(X_{gj}^{(p)}\right)\left(X_{gj}^{(q)}\right),\right.\\
 &  & \qquad\qquad\qquad\qquad\qquad\qquad\left.K_{h}\left(X_{gt}\right)K_{h}\left(X_{gs}\right)\sigma\left(X_{gt}^{(\mathrm{ind})},X_{gs}^{(\mathrm{ind})};X_{g}^{(\mathrm{cls})}\right)\left(X_{gt}^{(p)}\right)\left(X_{gs}^{(q)}\right)\right]\\
 & \leq & \frac{1}{n^{2}}\sum_{g=1}^{G}\left\{ n_{g}^{2}O\left(h^{4-2d-d_{\mathrm{cls}}}\right)\ensuremath{+n_{g}^{3}O\left(h^{4-d-2d_{\mathrm{cls}}}\right)}+n_{g}^{3}O\left(h^{4-d-2d_{\mathrm{cls}}}\right)+n_{g}^{4}O\left(h^{4-3d_{\mathrm{cls}}}\right)\right\} \leq o\left(h^{4-2d}\right).
\end{eqnarray*}

Therefore, by Markov's inequality, $I_{0}^{(q)}=h^{-d}\left\{ \frac{\lambda}{2}R_{k}^{d_{\mathrm{cls}}}\sigma\left(0,0;0\right)f_{2}\left(0,0;0\right)+o_{p}\left(1\right)\right\} $, $I_{1}^{(q)}=o_{p}\left(h^{-d+1}\right)$, and $I^{(p,q)}=O_{p}\left(h^{-d+2}\right)$. We conclude by element-wise comparisons.
\end{proof}

\subsection{Proof for Lemma $\text{\ref{lem:Er}}$}
\begin{proof}
Define $\mathcal{E}_{r}^{(q)}=\frac{1}{n}\sum_{g=1}^{G}\sum_{1\leq j<\ell\leq n_{g}}K_{h}\left(X_{gj}\right)\left(X_{gj}^{(q)}\right)^{r}e_{gj}$ for $r=0,1$. We have
\begin{align*}
\mathbb{E}\left[\mathcal{E}_{r}^{(q)}\right] & =\mathbb{E}\left[\frac{1}{n}\sum_{g=1}^{G}\sum_{j=1}^{n_{g}}K_{h}\left(X_{gj}\right)\left(X_{gj}^{(q)}\right)^{r}e_{gj}\right]\\
 & =\mathbb{E}\left[K_{h}\left(X_{gj}\right)\left(X_{gj}^{(q)}\right)^{r}\mathbb{E}\left[e_{gj}\mid X_{gj}\right]\right]\\
 & =0
\end{align*}
by the law of iterated expectations.

For variances,
\begin{eqnarray*}
 &  & \operatorname{Var}\left[\mathcal{E}_{r}^{(q)}\right]\\
 & = & \operatorname{Var}\left[\frac{1}{n}\sum_{g=1}^{G}\sum_{j=1}^{n_{g}}K_{h}\left(X_{gj}\right)\left(X_{gj}^{(q)}\right)^{r}e_{gj}\right]\\
 & = & \mathbb{E}\left[\left\{ \frac{1}{n}\sum_{g=1}^{G}\sum_{j=1}^{n_{g}}K_{h}\left(X_{gj}\right)\left(X_{gj}^{(q)}\right)^{r}e_{gj}\right\} ^{2}\right]-\underbrace{\mathbb{E}\left[\frac{1}{n}\sum_{g=1}^{G}\sum_{j=1}^{n_{g}}K_{h}\left(X_{gj}\right)\left(X_{gj}^{(q)}\right)^{r}e_{gj}\right]^{2}}_{=0}\\
 & = & \frac{1}{n^{2}}\sum_{g=1}^{G}\mathbb{E}\left[\mathbb{E}\left[\left\{ \sum_{j=1}^{n_{g}}K_{h}\left(X_{gj}\right)\left(X_{gj}^{(q)}\right)^{r}e_{gj}\right\} ^{2}\mid X_{g}\right]\right]\\
 & = & \frac{1}{n^{2}}\sum_{g=1}^{G}\sum_{j=1}^{n_{g}}\mathbb{E}\left[K_{h}^{2}\left(X_{gj}\right)\sigma^{2}\left(X_{gj}\right)\left(X_{gj}^{(q)}\right)^{2r}\right]\\
 &  & +2\frac{1}{n^{2}}\sum_{g=1}^{G}\sum_{1\leq j<\ell\leq n_{g}}\mathbb{E}\left[K_{h}\left(X_{gj}\right)K_{h}\left(X_{g\ell}\right)\left(X_{gj}^{(q)}\right)^{r}\left(X_{g\ell}^{(q)}\right)^{r}\sigma\left(X_{gj}^{(\mathrm{ind})},X_{g\ell}^{(\mathrm{ind})};X_{g}^{(\mathrm{cls})}\right)\right]\\
 & = & \begin{cases}
\frac{1}{n}\mathbb{E}\left[H_{0}^{(q)}+2I_{0}^{(q)}\right] & \text{ if }r=0\\
\frac{1}{n}\mathbb{E}\left[H_{2}^{(q)}+2I^{(q,q)}\right] & \text{ if }r=1
\end{cases}\\
 & = & \begin{cases}
O\left(\frac{1}{nh^{d}}\right) & \text{ if }r=0\\
O\left(\frac{1}{nh^{d-2}}\right) & \text{ if }r=1
\end{cases},
\end{eqnarray*}
where the third equality follows from the mutual independence between clusters.

Therefore, by Markov's inequality, $\mathcal{E}_{0}^{(q)}=O_{p}\left(\sqrt{\frac{1}{nh^{d}}}\right)$ and $\mathcal{E}_{1}^{(q)}=O_{p}\left(\sqrt{\frac{1}{nh^{d-2}}}\right)$ We conclude by element-wise comparisons. 
\end{proof}

\subsection{Proof for Lemma $\text{\ref{lem:nabla_m_order}}$}
\begin{proof}
By the proof of Lemma $\text{\ref{lem:Fr}}$,
\begin{eqnarray*}
 &  & \frac{1}{n}\sum_{g=1}^{G}\sum_{j=1}^{n_{g}}K_{h}\left(X_{gj}\right)\left\{ X_{gj}^{\top}\nabla^{2}m(0)X_{gj}\right\} \\
 & = & \frac{1}{n}\sum_{g=1}^{G}\sum_{j=1}^{n_{g}}K_{h}\left(X_{gj}\right)\sum_{p=1}^{d}\sum_{q=1}^{d}\partial_{pq}m(0)X_{gj}^{(p)}X_{gj}^{(q)}\\
 & = & \sum_{q=1}^{d}\partial_{qq}m(0)F_{2}^{(q)}+2\sum_{1\leq p<q\leq d}\partial_{pq}m(0)F^{(p,q)}\\
 & = & \sum_{q=1}^{d}\partial_{qq}m(0)\left\{ h^{2}f(0)\kappa_{2}+o_{p}\left(h^{2}\right)\right\} +2\sum_{1\leq p<q\leq d}\partial_{pq}m(0)o_{p}\left(h^{2}\right)\\
 & = & h^{2}\kappa_{2}\sum_{q=1}^{d}\partial_{qq}m(0)f(0)+o_{p}\left(h^{2}\right).
\end{eqnarray*}
Next, we will evaluate
\begin{eqnarray*}
 &  & \frac{1}{n}\sum_{g=1}^{G}\sum_{j=1}^{n_{g}}K_{h}\left(X_{gj}\right)X_{gj}\left\{ X_{gj}^{\top}\nabla^{2}m(0)X_{gj}\right\} \\
 & = & \frac{1}{n}\sum_{g=1}^{G}\sum_{j=1}^{n_{g}}K_{h}\left(X_{gj}\right)X_{gj}\left\{ \sum_{q=1}^{d}\partial_{qq}m(0)\left(X_{gj}^{(q)}\right)^{2}+2\sum_{1\leq p<q\leq d}\partial_{pq}m(0)X_{gj}^{(p)}X_{gj}^{(q)}\right\} 
\end{eqnarray*}
The compact support of the kernel function implies
\begin{align*}
\partial_{qq}m(0)\frac{1}{n}\sum_{g=1}^{G}\sum_{j=1}^{n_{g}}K_{h}\left(X_{gj}\right)\left(X_{gj}^{(q)}\right)^{3} & =O_{p}\left(h^{3}\right),\\
\partial_{qq}m(0)\frac{1}{n}\sum_{g=1}^{G}\sum_{j=1}^{n_{g}}K_{h}\left(X_{gj}\right)X_{gj}^{(p)}\left(X_{gj}^{(q)}\right)^{2} & =O_{p}\left(h^{3}\right),\\
\partial_{pq}m(0)\frac{1}{n}\sum_{g=1}^{G}\sum_{j=1}^{n_{g}}K_{h}\left(X_{gj}\right)X_{gj}^{(p)}\left(X_{gj}^{(q)}\right)^{2} & =O_{p}\left(h^{3}\right),\\
\partial_{pq}m(0)\frac{1}{n}\sum_{g=1}^{G}\sum_{j=1}^{n_{g}}K_{h}\left(X_{gj}\right)X_{gj}^{(p)}X_{gj}^{(q)}X_{gj}^{(q^{\prime})} & =O_{p}\left(h^{3}\right).
\end{align*}
Thus,
\[
\frac{1}{n}\sum_{g=1}^{G}\sum_{j=1}^{n_{g}}K_{h}\left(X_{gj}\right)X_{gj}\left\{ X_{gj}^{\top}\nabla^{2}m(0)X_{gj}\right\} =O_{p}\left(h^{3}\right)\mathbf{1}_{d}.
\]
\end{proof}

\section{\textbf{Technical discussion}\label{sec:Technical-discussion}}

As we mentioned in Remark $\text{\ref{rem:joint_density}}$, we can relax the identical distribution assumptions for joint densities if we strengthen the continuity assumption for them. For example, we can together replace Assumption $\text{\ref{assu:dgp}}$ (iii) and Assumption $\text{\ref{assu:nw}}$ (ii) by the following assumptions to show the theorems on Section $\text{\ref{sec:Nadaraya-Watson-estimator}}$.
\begin{itemize}
\item \textbf{Assumption} $\text{\ref{assu:dgp}}$ (iii'): \textit{$X_{gj}$ are identically distributed across all $g$ and $j$ with common marginal density $f(x)$. For any cluster $g$ with $n_{g}\geq2$, $\left(X_{gj_{1}}^{(\mathrm{ind})},X_{gj_{2}}^{(\mathrm{ind})};X_{g}^{(\mathrm{cls})}\right)$ are identically distributed across all $g$, $j_{1}$, and $j_{2}$ with common joint density 
\[
f_{2}\left(x_{1}^{(\mathrm{ind})},x_{2}^{(\mathrm{ind})};x^{(\mathrm{cls})}\right).
\]
For any $\underline{n}_{g}\in\{3,4\}$ and for any cluster $g$ with $n_{g}\geq\underline{n}_{g}$, $\left(X_{gj_{1}}^{(\mathrm{ind})},\cdots,X_{gj_{\underline{n}_{g}}}^{(\mathrm{ind})};X_{g}^{(\mathrm{cls})}\right)$ with $j_{1}<j_{2}<\dots<j_{\underline{n}_{g}}$ has the joint density 
\[
f_{\left(j_{1},j_{2},\dots,j_{\underline{n}_{g}};g\right)}\left(x_{1}^{(\mathrm{ind})},x_{2}^{(\mathrm{ind})},\cdots,x_{\underline{n}_{g}}^{(\mathrm{ind})};x^{(\mathrm{cls})}\right).
\]
}
\item \textbf{Assumption} $\text{\ref{assu:nw}}$ (ii'):\textit{ There exists some neighborhood $\mathcal{N}$ of $x=\left(x^{(\mathrm{ind})\top},x^{\mathrm{(cls)}\top}\right)^{\top}$ such that $m(x)$ and $f(x)$ are twice continuously differentiable, $f_{2}\left(x^{\mathrm{(ind)}},x^{\mathrm{(ind)}};x^{\mathrm{(cls)}}\right)$ is continuously differentiable, and $\sigma^{2}(x)$, and $\sigma\left(x^{\mathrm{(ind)}},x^{\mathrm{(ind)}};x^{\mathrm{(cls)}}\right)$ are continuous. Moreover, for any $\underline{n}_{g}\in\{3,4\}$,
\[
\left\{ \left\{ f_{\left(j_{1},j_{2},\dots,j_{\underline{n}_{g}};g\right)}\left(x_{1}^{(\mathrm{ind})},x_{2}^{(\mathrm{ind})},\cdots,x_{\underline{n}_{g}}^{(\mathrm{ind})};x^{(\mathrm{cls})}\right)\right\} _{1\leq j_{1}<j_{2}<\dots<j_{\underline{n}_{g}}\leq n_{g}}\right\} _{g:n_{g}\geq\underline{n}_{g}}
\]
 is equicontinuous in the neighborhood $\mathcal{N}$.}
\end{itemize}

\section{\textbf{Additional simulations\label{sec:Add_sim}}}

\subsection{Simulation results for the Nadaraya-Watson estimator}

In this subsection, we will provide simulation results of the Nadaraya-Watson estimator for bandwidth selection and inference.

\subsubsection{Bandwidth selection}

The data-generating processes and the calculations are the same as in Section $\text{\ref{sec:Monte-Carlo-simulation}}$. As we did for local linear estimators in Section $\text{\ref{sec:Monte-Carlo-simulation}}$, we will compare four methods of bandwidth choice. The performance is evaluated by
\[
\operatorname{ASE}(h)=\frac{1}{n_{\text{grid }}}\sum_{k=1}^{n_{\text{grid }}}\left\{ \widehat{m}_{\mathrm{nw}}\left(u_{k},h\right)-m\left(u_{k}\right)\right\} ^{2},
\]
where $\widehat{m}_{\mathrm{nw}}\left(u_{k},h\right)$ is the Nadaraya-Watson estimator with the bandwidth $h$.

\begin{figure}[H]
\includegraphics[scale=0.6]{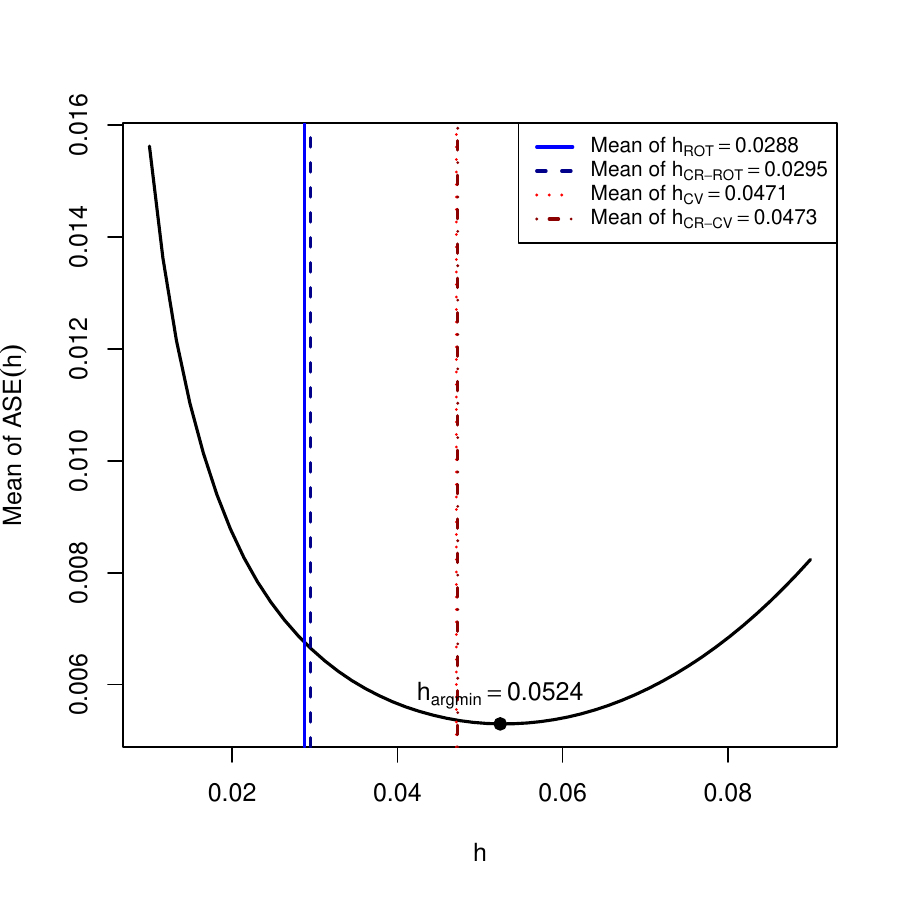}\caption{Mean of $\mathrm{ASE}(h)$ for $m_{\mathrm{nw}}$ in Setup 1 with $\max_{g\protect\leq G}n_{g}=100$ and $\rho_{X}=\rho_{e}=0.5$\label{fig:bw_nw1}}
\end{figure}

\begin{figure}[H]
\includegraphics[scale=0.6]{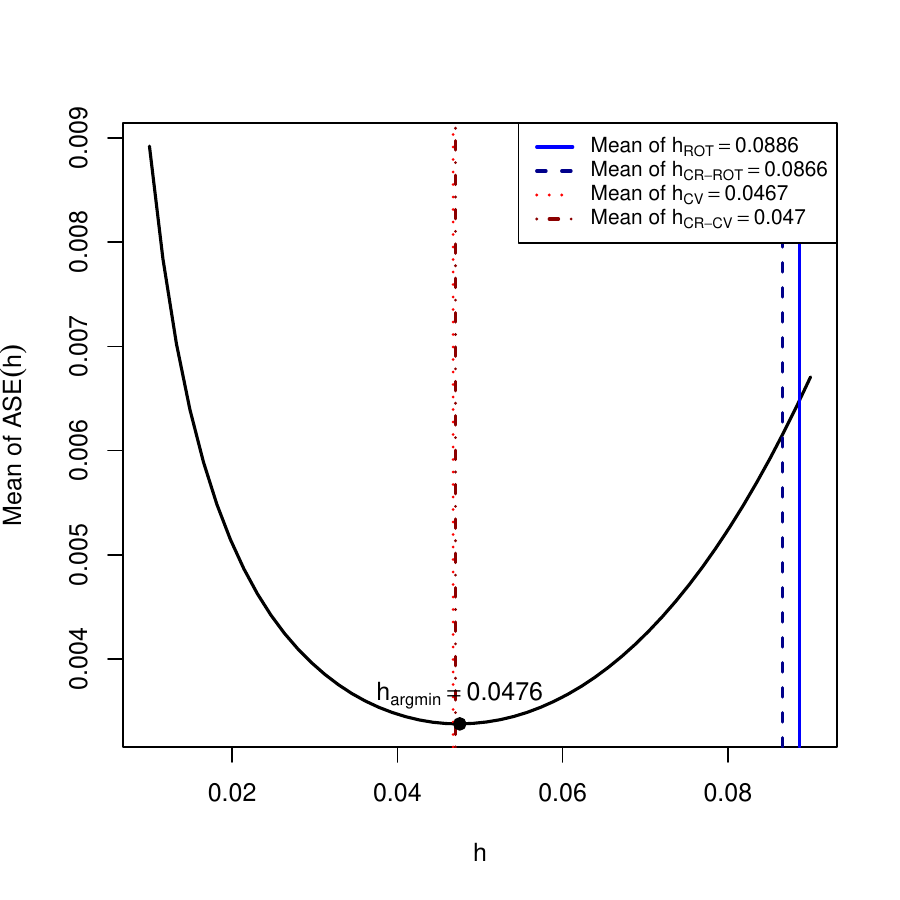}\caption{Mean of $\mathrm{ASE}(h)$ for $m_{\mathrm{nw}}$ in Setup 2 with $\max_{g\protect\leq G}n_{g}=100$ and $\rho_{X}=\rho_{e}=0.5$\label{fig:bw_nw2}}
\end{figure}

Tables $\text{\ref{tab:bw_nw1}}$ and $\text{\ref{tab:bw_nw2}}$ show means of ASEs for the Nadaraya-Watson estimator and means of selected bandwidths (in curly brackets) across each simulation draw for Setup 1 and 2, respectively. Figures $\text{\ref{fig:bw_nw1}}$ and $\text{\ref{fig:bw_nw2}}$ plot values of the bandwidth $h$ in the $x$-axis and means of the function $\mathrm{ASE}(h)$ in the $y$-axis, which are calculated from simulation draws for Setups 1 and 2, respectively. We found almost the same implications as in Section $\text{\ref{subsec:sim_bw}}$, and the detailed explanations are omitted.

\begin{table}[H]
\caption{Mean of ASE and mean of selected bandwidth ($m_{\mathrm{nw}}$, Setup 1)\label{tab:bw_nw1}}

\centering
\begin{center} 
\begin{threeparttable}
{\small

\begin{tabular}{l c c c c c c c c}
\hline
 & \multicolumn{4}{c}{$\max n_g=20$} & \multicolumn{4}{c}{$\max n_g=100$} \\
\cline{2-5} \cline{6-9}
 & $h_{\text{ROT}}$ & $h_{\text{CR-ROT}}$ & $h_{\text{CV}}$ & $h_{\text{CR-CV}}$ & $h_{\text{ROT}}$ & $h_{\text{CR-ROT}}$ & $h_{\text{CV}}$ & $h_{\text{CR-CV}}$ \\
\hline
$(\rho_{X},\rho_{e})$=(0.2,0.2) & $0.0053$   & $0.0053$   & $0.0042$   & $0.0042$   & $0.0053$   & $0.0053$   & $0.0042$   & $0.0042$   \\
                                & $\{0.0297\}$ & $\{0.0302\}$ & $\{0.0471\}$ & $\{0.0471\}$ & $\{0.0292\}$ & $\{0.0297\}$ & $\{0.0467\}$ & $\{0.0468\}$ \\
$(\rho_{X},\rho_{e})$=(0.2,0.5) & $0.0062$   & $0.0061$   & $0.0050$   & $0.0050$   & $0.0063$   & $0.0062$   & $0.0051$   & $0.0051$   \\
                                & $\{0.0297\}$ & $\{0.0302\}$ & $\{0.0471\}$ & $\{0.0472\}$ & $\{0.0292\}$ & $\{0.0297\}$ & $\{0.0467\}$ & $\{0.0468\}$ \\
$(\rho_{X},\rho_{e})$=(0.5,0.2) & $0.0055$   & $0.0054$   & $0.0043$   & $0.0043$   & $0.0055$   & $0.0055$   & $0.0043$   & $0.0043$   \\
                                & $\{0.0292\}$ & $\{0.0300\}$ & $\{0.0473\}$ & $\{0.0476\}$ & $\{0.0288\}$ & $\{0.0295\}$ & $\{0.0472\}$ & $\{0.0474\}$ \\
$(\rho_{X},\rho_{e})$=(0.5,0.5) & $0.0066$   & $0.0065$   & $0.0054$   & $0.0054$   & $0.0068$   & $0.0067$   & $0.0056$   & $0.0056$   \\
                                & $\{0.0292\}$ & $\{0.0300\}$ & $\{0.0475\}$ & $\{0.0477\}$ & $\{0.0288\}$ & $\{0.0295\}$ & $\{0.0471\}$ & $\{0.0473\}$ \\
\hline
\end{tabular}
}
\begin{tablenotes}
\vspace{-0.5cm}
\footnotesize
\item \hspace{-0.5cm} \textit{Note: Means of selected bandwidths are shown in curly brackets.}
\end{tablenotes}\end{threeparttable}\end{center}
\end{table}

\begin{table}[H]
\caption{Mean of ASE and mean of selected bandwidth ($m_{\mathrm{nw}}$, Setup 2)\label{tab:bw_nw2}}

\centering
\begin{center} 
\begin{threeparttable}
{\small

\begin{tabular}{l c c c c c c c c}
\hline
 & \multicolumn{4}{c}{$\max n_g=20$} & \multicolumn{4}{c}{$\max n_g=100$} \\
\cline{2-5} \cline{6-9}
 & $h_{\text{ROT}}$ & $h_{\text{CR-ROT}}$ & $h_{\text{CV}}$ & $h_{\text{CR-CV}}$ & $h_{\text{ROT}}$ & $h_{\text{CR-ROT}}$ & $h_{\text{CV}}$ & $h_{\text{CR-CV}}$ \\
\hline
$(\rho_{X},\rho_{e})$=(0.2,0.2) & $0.0086$   & $0.0072$   & $0.0028$   & $0.0028$   & $0.0081$   & $0.0069$   & $0.0028$   & $0.0028$   \\
                                & $\{0.0890\}$ & $\{0.0865\}$ & $\{0.0467\}$ & $\{0.0468\}$ & $\{0.0876\}$ & $\{0.0853\}$ & $\{0.0463\}$ & $\{0.0464\}$ \\
$(\rho_{X},\rho_{e})$=(0.2,0.5) & $0.0094$   & $0.0079$   & $0.0033$   & $0.0033$   & $0.0089$   & $0.0076$   & $0.0034$   & $0.0034$   \\
                                & $\{0.0893\}$ & $\{0.0868\}$ & $\{0.0467\}$ & $\{0.0468\}$ & $\{0.0878\}$ & $\{0.0855\}$ & $\{0.0465\}$ & $\{0.0467\}$ \\
$(\rho_{X},\rho_{e})$=(0.5,0.2) & $0.0088$   & $0.0077$   & $0.0029$   & $0.0029$   & $0.0087$   & $0.0075$   & $0.0029$   & $0.0029$   \\
                                & $\{0.0896\}$ & $\{0.0877\}$ & $\{0.0474\}$ & $\{0.0475\}$ & $\{0.0889\}$ & $\{0.0869\}$ & $\{0.0469\}$ & $\{0.0471\}$ \\
$(\rho_{X},\rho_{e})$=(0.5,0.5) & $0.0094$   & $0.0083$   & $0.0036$   & $0.0036$   & $0.0094$   & $0.0082$   & $0.0037$   & $0.0037$   \\
                                & $\{0.0892\}$ & $\{0.0874\}$ & $\{0.0471\}$ & $\{0.0474\}$ & $\{0.0886\}$ & $\{0.0866\}$ & $\{0.0467\}$ & $\{0.0470\}$ \\
\hline
\end{tabular}
}
\begin{tablenotes}
\vspace{-0.5cm}
\footnotesize
\item \hspace{-0.5cm} \textit{Note: Means of selected bandwidths are shown in curly brackets.}
\end{tablenotes}\end{threeparttable}\end{center}
\end{table}

\subsubsection{Inference}

The data-generating processes and the calculations are the same as in Section $\text{\ref{sec:Monte-Carlo-simulation}}$. Tables $\text{\ref{tab:CI_nw1}}$-$\text{\ref{tab:CI_nw2_0.4}}$ show the coverage ratio for the Nadaraya-Watson estimator and means of the length of confidence intervals (in curly brackets) across each simulation draw for Setup 1, Setup 2 with $x=0.8$, and Setup 2 with $x=0.4$ , respectively. We found almost the same implications as in Section $\text{\ref{subsec:sim_CI}}$, and the detailed explanations are omitted.

\begin{table}
\caption{Coverage and mean of length of 95\% CI for each standard error ($m_{\mathrm{nw}}$, Setup 1, undersmoothing)\label{tab:CI_nw1}}

\centering
\begin{center} 
\begin{threeparttable}
{\small

\begin{tabular}{l c c c c c c}
\hline
 & \multicolumn{3}{c}{$\max n_g=20$} & \multicolumn{3}{c}{$\max n_g=100$} \\
\cline{2-4} \cline{5-7}
& $CI$ & $CI_{\text{CR}}$ & $CI_{\lambda}$ & $CI$ & $CI_{\text{CR}}$ & $CI_{\lambda}$ \\
\hline
$(\rho_{X},\rho_{e})$=(0.2,0.2) & $0.932$   & $0.939$   & $0.948$   & $0.931$   & $0.937$   & $0.954$   \\
                                & $\{0.265\}$ & $\{0.272\}$ & $\{0.288\}$ & $\{0.260\}$ & $\{0.267\}$ & $\{0.286\}$ \\
$(\rho_{X},\rho_{e})$=(0.2,0.5) & $0.911$   & $0.922$   & $0.957$   & $0.905$   & $0.914$   & $0.952$   \\
                                & $\{0.264\}$ & $\{0.272\}$ & $\{0.310\}$ & $\{0.260\}$ & $\{0.268\}$ & $\{0.312\}$ \\
$(\rho_{X},\rho_{e})$=(0.5,0.2) & $0.924$   & $0.931$   & $0.952$   & $0.921$   & $0.930$   & $0.951$   \\
                                & $\{0.263\}$ & $\{0.270\}$ & $\{0.295\}$ & $\{0.259\}$ & $\{0.266\}$ & $\{0.294\}$ \\
$(\rho_{X},\rho_{e})$=(0.5,0.5) & $0.901$   & $0.910$   & $0.960$   & $0.894$   & $0.901$   & $0.958$   \\
                                & $\{0.263\}$ & $\{0.271\}$ & $\{0.322\}$ & $\{0.259\}$ & $\{0.267\}$ & $\{0.326\}$ \\
\hline
\end{tabular}
}
\begin{tablenotes}
\vspace{-0.5cm}
\footnotesize
\item \hspace{-0.5cm} \textit{Note: Lengths of confidence intervals are shown in curly brackets.}
\end{tablenotes}\end{threeparttable}\end{center}
\end{table}

\begin{table}
\caption{Coverage and mean of length of 95\% CI for each standard error ($m_{\mathrm{nw}}$, Setup 2, $x=0.8$, undersmoothing)\label{tab:CI_nw2_0.8}}

\centering
\begin{center} 
\begin{threeparttable}
{\small

\begin{tabular}{l c c c c c c}
\hline
 & \multicolumn{3}{c}{$\max n_g=20$} & \multicolumn{3}{c}{$\max n_g=100$} \\
\cline{2-4} \cline{5-7}
& $CI$ & $CI_{\text{CR}}$ & $CI_{\lambda}$ & $CI$ & $CI_{\text{CR}}$ & $CI_{\lambda}$ \\
\hline
$(\rho_{X},\rho_{e})$=(0.2,0.2) & $0.903$   & $0.913$   & $0.927$   & $0.900$   & $0.909$   & $0.924$   \\
                                & $\{0.230\}$ & $\{0.237\}$ & $\{0.249\}$ & $\{0.227\}$ & $\{0.234\}$ & $\{0.248\}$ \\
$(\rho_{X},\rho_{e})$=(0.2,0.5) & $0.872$   & $0.882$   & $0.921$   & $0.869$   & $0.880$   & $0.921$   \\
                                & $\{0.229\}$ & $\{0.237\}$ & $\{0.266\}$ & $\{0.226\}$ & $\{0.233\}$ & $\{0.267\}$ \\
$(\rho_{X},\rho_{e})$=(0.5,0.2) & $0.887$   & $0.896$   & $0.922$   & $0.876$   & $0.889$   & $0.922$   \\
                                & $\{0.228\}$ & $\{0.236\}$ & $\{0.252\}$ & $\{0.226\}$ & $\{0.233\}$ & $\{0.251\}$ \\
$(\rho_{X},\rho_{e})$=(0.5,0.5) & $0.854$   & $0.875$   & $0.924$   & $0.838$   & $0.853$   & $0.919$   \\
                                & $\{0.228\}$ & $\{0.237\}$ & $\{0.273\}$ & $\{0.225\}$ & $\{0.233\}$ & $\{0.276\}$ \\
\hline
\end{tabular}
}
\begin{tablenotes}
\vspace{-0.5cm}
\footnotesize
\item \hspace{-0.5cm} \textit{Note: Lengths of confidence intervals are shown in curly brackets.}
\end{tablenotes}\end{threeparttable}\end{center}
\end{table}

\begin{table}
\caption{Coverage and mean of length of 95\% CI for each standard error ($m_{\mathrm{nw}}$, Setup 2, $x=0.4$, undersmoothing)\label{tab:CI_nw2_0.4}}

\centering
\begin{center} 
\begin{threeparttable}
{\small

\begin{tabular}{l c c c c c c}
\hline
 & \multicolumn{3}{c}{$\max n_g=20$} & \multicolumn{3}{c}{$\max n_g=100$} \\
\cline{2-4} \cline{5-7}
& $CI$ & $CI_{\text{CR}}$ & $CI_{\lambda}$ & $CI$ & $CI_{\text{CR}}$ & $CI_{\lambda}$ \\
\hline
$(\rho_{X},\rho_{e})$=(0.2,0.2) & $0.994$   & $0.995$   & $0.998$   & $0.995$   & $0.996$   & $0.999$   \\
                                & $\{0.187\}$ & $\{0.192\}$ & $\{0.206\}$ & $\{0.184\}$ & $\{0.188\}$ & $\{0.205\}$ \\
$(\rho_{X},\rho_{e})$=(0.2,0.5) & $0.989$   & $0.991$   & $0.998$   & $0.988$   & $0.990$   & $0.999$   \\
                                & $\{0.187\}$ & $\{0.192\}$ & $\{0.225\}$ & $\{0.183\}$ & $\{0.188\}$ & $\{0.226\}$ \\
$(\rho_{X},\rho_{e})$=(0.5,0.2) & $0.995$   & $0.996$   & $0.998$   & $0.996$   & $0.996$   & $0.999$   \\
                                & $\{0.186\}$ & $\{0.190\}$ & $\{0.207\}$ & $\{0.183\}$ & $\{0.187\}$ & $\{0.207\}$ \\
$(\rho_{X},\rho_{e})$=(0.5,0.5) & $0.988$   & $0.990$   & $0.999$   & $0.987$   & $0.990$   & $0.999$   \\
                                & $\{0.186\}$ & $\{0.191\}$ & $\{0.230\}$ & $\{0.182\}$ & $\{0.187\}$ & $\{0.232\}$ \\
\hline
\end{tabular}
}
\begin{tablenotes}
\vspace{-0.5cm}
\footnotesize
\item \hspace{-0.5cm} \textit{Note: Lengths of confidence intervals are shown in curly brackets.}
\end{tablenotes}\end{threeparttable}\end{center}
\end{table}

\subsection{Inference without undersmoothing}

The data-generating processes and calculations follow those in Section $\text{\ref{sec:Monte-Carlo-simulation}}$, with the exception of bandwidth choice. Here, we set $h_{m}=h_{\text{CR-CV}}$ rather than using the undersmoothed choice $h_{m}=h_{\text{CR-CV}}\times n^{1/5}\times n^{-2/7}$. With this bandwidth, the bias term is not asymptotically dominated. In terms of the bias, we take two approaches: an infeasible analytical bias correction using knowledge of the DGP (Section $\text{\ref{subsubsec:Inference-bc}}$), and simply ignoring the bias term (Section $\text{\ref{subsubsec:Inference-with-bias}}$). We present results for both Nadaraya-Watson and local linear estimators.

\subsubsection{With infeasible analytical bias correction\label{subsubsec:Inference-bc}}

This subsection presents simulation results for inference methods with analytically derived bias corrections. Tables $\text{\ref{tab:CI_LL1-bc}}$-$\text{\ref{tab:CI_nw2_0.4-bc}}$ show the coverage ratio and means of the length of confidence intervals (in curly brackets) across each simulation draw. Overall, our proposed confidence interval $CI_{\lambda}$ performs well. For example, in Setup 1 (Table $\text{\ref{tab:CI_LL1-bc}}$, homoskedastic errors), $CI_{\mathrm{CR}}$ has slightly better coverages than $CI$ does although both confidence intervals have severe under-coverage values when $\rho_{e}=0.5$. These confidence intervals work more poorly for the case $\max_{g\leq G}n_{g}=100$. On the other hand, $CI_{\lambda}$ performs the best among the three methods. It has accurate coverage (95\%-96\%) for every data-generating process. 

\begin{table}
\caption{Coverage and mean of length of 95\% CI for each standard error ($m_{\mathrm{LL}}$, Setup 1, debiased)\label{tab:CI_LL1-bc}}

\centering
\begin{center} 
\begin{threeparttable}
{\small

\begin{tabular}{l c c c c c c}
\hline
 & \multicolumn{3}{c}{$\max n_g=20$} & \multicolumn{3}{c}{$\max n_g=100$} \\
\cline{2-4} \cline{5-7}
 & $CI$ & $CI_{\text{CR}}$ & $CI_{\lambda}$ & $CI$ & $CI_{\text{CR}}$ & $CI_{\lambda}$ \\
\hline
$(\rho_{X},\rho_{e})$=(0.2,0.2) & $0.923$   & $0.926$   & $0.953$   & $0.914$   & $0.916$   & $0.950$   \\
                                & $\{0.190\}$ & $\{0.193\}$ & $\{0.215\}$ & $\{0.187\}$ & $\{0.189\}$ & $\{0.215\}$ \\
$(\rho_{X},\rho_{e})$=(0.2,0.5) & $0.875$   & $0.880$   & $0.959$   & $0.859$   & $0.864$   & $0.953$   \\
                                & $\{0.189\}$ & $\{0.192\}$ & $\{0.244\}$ & $\{0.186\}$ & $\{0.189\}$ & $\{0.248\}$ \\
$(\rho_{X},\rho_{e})$=(0.5,0.2) & $0.915$   & $0.921$   & $0.956$   & $0.906$   & $0.909$   & $0.953$   \\
                                & $\{0.189\}$ & $\{0.192\}$ & $\{0.225\}$ & $\{0.186\}$ & $\{0.189\}$ & $\{0.226\}$ \\
$(\rho_{X},\rho_{e})$=(0.5,0.5) & $0.858$   & $0.868$   & $0.960$   & $0.836$   & $0.848$   & $0.959$   \\
                                & $\{0.189\}$ & $\{0.192\}$ & $\{0.260\}$ & $\{0.185\}$ & $\{0.189\}$ & $\{0.265\}$ \\
\hline
\end{tabular}
}
\begin{tablenotes}
\vspace{-0.5cm}
\footnotesize
\item \hspace{-0.5cm} \textit{Note: Lengths of confidence intervals are shown in curly brackets.}
\end{tablenotes}\end{threeparttable}\end{center}
\end{table}

\begin{table}[H]
\caption{Coverage and average length of 95\% CI for each standard error ($m_{\mathrm{nw}}$, Setup 1, debiased)\label{tab:CI_nw1-bc}}

\centering
\begin{center} 
\begin{threeparttable}
{\small

\begin{tabular}{l c c c c c c}
\hline
 & \multicolumn{3}{c}{$\max n_g=20$} & \multicolumn{3}{c}{$\max n_g=100$} \\
\cline{2-4} \cline{5-7}
 & $CI$ & $CI_{\text{CR}}$ & $CI_{\lambda}$ & $CI$ & $CI_{\text{CR}}$ & $CI_{\lambda}$ \\
\hline
$(\rho_{X},\rho_{e})$=(0.2,0.2) & $0.925$   & $0.931$   & $0.954$   & $0.915$   & $0.920$   & $0.952$   \\
                                & $\{0.192\}$ & $\{0.195\}$ & $\{0.217\}$ & $\{0.189\}$ & $\{0.192\}$ & $\{0.217\}$ \\
$(\rho_{X},\rho_{e})$=(0.2,0.5) & $0.879$   & $0.886$   & $0.960$   & $0.861$   & $0.869$   & $0.951$   \\
                                & $\{0.192\}$ & $\{0.195\}$ & $\{0.246\}$ & $\{0.188\}$ & $\{0.192\}$ & $\{0.250\}$ \\
$(\rho_{X},\rho_{e})$=(0.5,0.2) & $0.920$   & $0.925$   & $0.956$   & $0.906$   & $0.908$   & $0.954$   \\
                                & $\{0.191\}$ & $\{0.194\}$ & $\{0.227\}$ & $\{0.188\}$ & $\{0.191\}$ & $\{0.228\}$ \\
$(\rho_{X},\rho_{e})$=(0.5,0.5) & $0.857$   & $0.867$   & $0.964$   & $0.833$   & $0.844$   & $0.957$   \\
                                & $\{0.191\}$ & $\{0.195\}$ & $\{0.261\}$ & $\{0.188\}$ & $\{0.191\}$ & $\{0.267\}$ \\
\hline
\end{tabular}
}
\begin{tablenotes}
\vspace{-0.5cm}
\footnotesize
\item \hspace{-0.5cm} \textit{Note: Lengths of confidence intervals are shown in curly brackets.}
\end{tablenotes}\end{threeparttable}\end{center}
\end{table}

\begin{table}
\caption{Coverage and mean of length of 95\% CI for each standard error ($m_{\mathrm{LL}}$, Setup 2, $x=0.8$, debiased)\label{tab:CI_LL2_0.8-bc}}

\centering
\begin{center} 
\begin{threeparttable}
{\small

\begin{tabular}{l c c c c c c}
\hline
 & \multicolumn{3}{c}{$\max n_g=20$} & \multicolumn{3}{c}{$\max n_g=100$} \\
\cline{2-4} \cline{5-7}
 & $CI$ & $CI_{\text{CR}}$ & $CI_{\lambda}$ & $CI$ & $CI_{\text{CR}}$ & $CI_{\lambda}$ \\
\hline
$(\rho_{X},\rho_{e})$=(0.2,0.2) & $0.893$   & $0.899$   & $0.931$   & $0.884$   & $0.892$   & $0.927$   \\
                                & $\{0.168\}$ & $\{0.171\}$ & $\{0.187\}$ & $\{0.166\}$ & $\{0.169\}$ & $\{0.187\}$ \\
$(\rho_{X},\rho_{e})$=(0.2,0.5) & $0.842$   & $0.852$   & $0.919$   & $0.827$   & $0.835$   & $0.926$   \\
                                & $\{0.168\}$ & $\{0.171\}$ & $\{0.209\}$ & $\{0.165\}$ & $\{0.168\}$ & $\{0.212\}$ \\
$(\rho_{X},\rho_{e})$=(0.5,0.2) & $0.898$   & $0.905$   & $0.934$   & $0.873$   & $0.879$   & $0.925$   \\
                                & $\{0.167\}$ & $\{0.171\}$ & $\{0.192\}$ & $\{0.165\}$ & $\{0.168\}$ & $\{0.193\}$ \\
$(\rho_{X},\rho_{e})$=(0.5,0.5) & $0.826$   & $0.835$   & $0.930$   & $0.802$   & $0.809$   & $0.925$   \\
                                & $\{0.167\}$ & $\{0.171\}$ & $\{0.219\}$ & $\{0.164\}$ & $\{0.168\}$ & $\{0.223\}$ \\
\hline
\end{tabular}
}
\begin{tablenotes}
\vspace{-0.5cm}
\footnotesize
\item \hspace{-0.5cm} \textit{Note: Lengths of confidence intervals are shown in curly brackets.}
\end{tablenotes}\end{threeparttable}\end{center}
\end{table}

\begin{table}[H]
\caption{Coverage and average length of 95\% CI for each standard error ($m_{\mathrm{nw}}$, Setup 2, $x=0.8$, debiased)\label{tab:CI_nw2_0.8-bc}}

\centering
\begin{center} 
\begin{threeparttable}
{\small

\begin{tabular}{l c c c c c c}
\hline
 & \multicolumn{3}{c}{$\max n_g=20$} & \multicolumn{3}{c}{$\max n_g=100$} \\
\cline{2-4} \cline{5-7}
 & $CI$ & $CI_{\text{CR}}$ & $CI_{\lambda}$ & $CI$ & $CI_{\text{CR}}$ & $CI_{\lambda}$ \\
\hline
$(\rho_{X},\rho_{e})$=(0.2,0.2) & $0.893$   & $0.897$   & $0.931$   & $0.882$   & $0.886$   & $0.923$   \\
                                & $\{0.167\}$ & $\{0.170\}$ & $\{0.187\}$ & $\{0.165\}$ & $\{0.168\}$ & $\{0.187\}$ \\
$(\rho_{X},\rho_{e})$=(0.2,0.5) & $0.844$   & $0.850$   & $0.918$   & $0.831$   & $0.836$   & $0.924$   \\
                                & $\{0.167\}$ & $\{0.171\}$ & $\{0.209\}$ & $\{0.164\}$ & $\{0.168\}$ & $\{0.211\}$ \\
$(\rho_{X},\rho_{e})$=(0.5,0.2) & $0.903$   & $0.909$   & $0.936$   & $0.878$   & $0.884$   & $0.927$   \\
                                & $\{0.166\}$ & $\{0.170\}$ & $\{0.191\}$ & $\{0.164\}$ & $\{0.167\}$ & $\{0.192\}$ \\
$(\rho_{X},\rho_{e})$=(0.5,0.5) & $0.826$   & $0.837$   & $0.932$   & $0.806$   & $0.816$   & $0.924$   \\
                                & $\{0.166\}$ & $\{0.170\}$ & $\{0.218\}$ & $\{0.164\}$ & $\{0.167\}$ & $\{0.223\}$ \\
\hline
\end{tabular}
}
\begin{tablenotes}
\vspace{-0.5cm}
\footnotesize
\item \hspace{-0.5cm} \textit{Note: Lengths of confidence intervals are shown in curly brackets.}
\end{tablenotes}\end{threeparttable}\end{center}
\end{table}

\begin{table}
\caption{Coverage and mean of length of 95\% CI for each standard error ($m_{\mathrm{LL}}$, Setup 2, $x=0.4$, debiased)\label{tab:CI_LL2_0.4-bc}}

\centering
\begin{center} 
\begin{threeparttable}
{\small

\begin{tabular}{l c c c c c c}
\hline
 & \multicolumn{3}{c}{$\max n_g=20$} & \multicolumn{3}{c}{$\max n_g=100$} \\
\cline{2-4} \cline{5-7}
 & $CI$ & $CI_{\text{CR}}$ & $CI_{\lambda}$ & $CI$ & $CI_{\text{CR}}$ & $CI_{\lambda}$ \\
\hline
$(\rho_{X},\rho_{e})$=(0.2,0.2) & $0.991$   & $0.992$   & $0.997$   & $0.988$   & $0.989$   & $0.998$   \\
                                & $\{0.137\}$ & $\{0.138\}$ & $\{0.157\}$ & $\{0.134\}$ & $\{0.136\}$ & $\{0.158\}$ \\
$(\rho_{X},\rho_{e})$=(0.2,0.5) & $0.975$   & $0.978$   & $0.999$   & $0.969$   & $0.972$   & $1.000$   \\
                                & $\{0.136\}$ & $\{0.139\}$ & $\{0.182\}$ & $\{0.134\}$ & $\{0.136\}$ & $\{0.184\}$ \\
$(\rho_{X},\rho_{e})$=(0.5,0.2) & $0.990$   & $0.991$   & $0.998$   & $0.991$   & $0.992$   & $0.998$   \\
                                & $\{0.136\}$ & $\{0.138\}$ & $\{0.160\}$ & $\{0.133\}$ & $\{0.135\}$ & $\{0.161\}$ \\
$(\rho_{X},\rho_{e})$=(0.5,0.5) & $0.973$   & $0.976$   & $0.998$   & $0.962$   & $0.967$   & $0.999$   \\
                                & $\{0.136\}$ & $\{0.138\}$ & $\{0.188\}$ & $\{0.133\}$ & $\{0.135\}$ & $\{0.192\}$ \\
\hline
\end{tabular}
}
\begin{tablenotes}
\vspace{-0.5cm}
\footnotesize
\item \hspace{-0.5cm} \textit{Note: Lengths of confidence intervals are shown in curly brackets.}
\end{tablenotes}\end{threeparttable}\end{center}
\end{table}

\begin{table}[H]
\caption{Coverage and average length of 95\% CI for each standard error ($m_{\mathrm{nw}}$, Setup 2, $x=0.4$, debiased)\label{tab:CI_nw2_0.4-bc}}

\centering
\begin{center} 
\begin{threeparttable}
{\small

\begin{tabular}{l c c c c c c}
\hline
 & \multicolumn{3}{c}{$\max n_g=20$} & \multicolumn{3}{c}{$\max n_g=100$} \\
\cline{2-4} \cline{5-7}
 & $CI$ & $CI_{\text{CR}}$ & $CI_{\lambda}$ & $CI$ & $CI_{\text{CR}}$ & $CI_{\lambda}$ \\
\hline
$(\rho_{X},\rho_{e})$=(0.2,0.2) & $0.985$   & $0.986$   & $0.995$   & $0.983$   & $0.985$   & $0.995$   \\
                                & $\{0.136\}$ & $\{0.138\}$ & $\{0.157\}$ & $\{0.134\}$ & $\{0.135\}$ & $\{0.157\}$ \\
$(\rho_{X},\rho_{e})$=(0.2,0.5) & $0.969$   & $0.971$   & $0.998$   & $0.965$   & $0.968$   & $0.998$   \\
                                & $\{0.136\}$ & $\{0.138\}$ & $\{0.181\}$ & $\{0.133\}$ & $\{0.135\}$ & $\{0.184\}$ \\
$(\rho_{X},\rho_{e})$=(0.5,0.2) & $0.982$   & $0.983$   & $0.996$   & $0.980$   & $0.983$   & $0.998$   \\
                                & $\{0.135\}$ & $\{0.137\}$ & $\{0.160\}$ & $\{0.133\}$ & $\{0.134\}$ & $\{0.161\}$ \\
$(\rho_{X},\rho_{e})$=(0.5,0.5) & $0.962$   & $0.964$   & $0.996$   & $0.961$   & $0.963$   & $0.997$   \\
                                & $\{0.135\}$ & $\{0.137\}$ & $\{0.188\}$ & $\{0.132\}$ & $\{0.134\}$ & $\{0.192\}$ \\
\hline
\end{tabular}
}
\begin{tablenotes}
\vspace{-0.5cm}
\footnotesize
\item \hspace{-0.5cm} \textit{Note: Lengths of confidence intervals are shown in curly brackets.}
\end{tablenotes}\end{threeparttable}\end{center}
\end{table}

\subsubsection{\textbf{W}ithout bias corrections\label{subsubsec:Inference-with-bias}}

This subsection presents simulation results for inference methods that do not incorporate bias corrections. Tables $\text{\ref{tab:CI_LL1-bias}}$-$\text{\ref{tab:CI_nw2_0.4-bias}}$ show the coverage ratio and means of the length of confidence intervals (in curly brackets) across each simulation draw. These results are the feasible version of our previous inference results. Notably, among the evaluated methods, our $CI_{\lambda}$ confidence intervals exhibit superior performance.

\begin{table}[H]
\caption{Coverage and mean of length of 95\% CI for each standard error ($m_{\mathrm{LL}}$, Setup 1, with bias)\label{tab:CI_LL1-bias}}

\centering
\begin{center} 
\begin{threeparttable}
{\small

\begin{tabular}{l c c c c c c}
\hline
 & \multicolumn{3}{c}{$\max n_g=20$} & \multicolumn{3}{c}{$\max n_g=100$} \\
\cline{2-4} \cline{5-7}
 & $CI$ & $CI_{\text{CR}}$ & $CI_{\lambda}$ & $CI$ & $CI_{\text{CR}}$ & $CI_{\lambda}$ \\
\hline
$(\rho_{X},\rho_{e})$=(0.2,0.2) & $0.917$   & $0.921$   & $0.952$   & $0.908$   & $0.914$   & $0.951$   \\
                                & $\{0.190\}$ & $\{0.193\}$ & $\{0.215\}$ & $\{0.187\}$ & $\{0.189\}$ & $\{0.215\}$ \\
$(\rho_{X},\rho_{e})$=(0.2,0.5) & $0.876$   & $0.886$   & $0.958$   & $0.860$   & $0.869$   & $0.951$   \\
                                & $\{0.189\}$ & $\{0.192\}$ & $\{0.244\}$ & $\{0.186\}$ & $\{0.189\}$ & $\{0.248\}$ \\
$(\rho_{X},\rho_{e})$=(0.5,0.2) & $0.914$   & $0.919$   & $0.955$   & $0.905$   & $0.910$   & $0.949$   \\
                                & $\{0.189\}$ & $\{0.192\}$ & $\{0.225\}$ & $\{0.186\}$ & $\{0.189\}$ & $\{0.226\}$ \\
$(\rho_{X},\rho_{e})$=(0.5,0.5) & $0.858$   & $0.864$   & $0.961$   & $0.835$   & $0.842$   & $0.957$   \\
                                & $\{0.189\}$ & $\{0.192\}$ & $\{0.260\}$ & $\{0.185\}$ & $\{0.189\}$ & $\{0.265\}$ \\
\hline
\end{tabular}
}
\begin{tablenotes}
\vspace{-0.5cm}
\footnotesize
\item \hspace{-0.5cm} \textit{Note: Lengths of confidence intervals are shown in curly brackets.}
\end{tablenotes}\end{threeparttable}\end{center}
\end{table}
\begin{table}[H]
\caption{Coverage and average length of 95\% CI for each standard error ($m_{\mathrm{nw}}$, Setup 1, with bias)\label{tab:CI_nw1-bias}}

\centering
\begin{center} 
\begin{threeparttable}
{\small

\begin{tabular}{l c c c c c c}
\hline
 & \multicolumn{3}{c}{$\max n_g=20$} & \multicolumn{3}{c}{$\max n_g=100$} \\
\cline{2-4} \cline{5-7}
 & $CI$ & $CI_{\text{CR}}$ & $CI_{\lambda}$ & $CI$ & $CI_{\text{CR}}$ & $CI_{\lambda}$ \\
\hline
$(\rho_{X},\rho_{e})$=(0.2,0.2) & $0.918$   & $0.925$   & $0.953$   & $0.907$   & $0.914$   & $0.948$   \\
                                & $\{0.192\}$ & $\{0.195\}$ & $\{0.217\}$ & $\{0.189\}$ & $\{0.192\}$ & $\{0.217\}$ \\
$(\rho_{X},\rho_{e})$=(0.2,0.5) & $0.880$   & $0.888$   & $0.956$   & $0.861$   & $0.868$   & $0.949$   \\
                                & $\{0.192\}$ & $\{0.195\}$ & $\{0.246\}$ & $\{0.188\}$ & $\{0.192\}$ & $\{0.250\}$ \\
$(\rho_{X},\rho_{e})$=(0.5,0.2) & $0.916$   & $0.921$   & $0.956$   & $0.906$   & $0.910$   & $0.951$   \\
                                & $\{0.191\}$ & $\{0.194\}$ & $\{0.227\}$ & $\{0.188\}$ & $\{0.191\}$ & $\{0.228\}$ \\
$(\rho_{X},\rho_{e})$=(0.5,0.5) & $0.859$   & $0.865$   & $0.960$   & $0.837$   & $0.845$   & $0.955$   \\
                                & $\{0.191\}$ & $\{0.195\}$ & $\{0.261\}$ & $\{0.188\}$ & $\{0.191\}$ & $\{0.267\}$ \\
\hline
\end{tabular}
}
\begin{tablenotes}
\vspace{-0.5cm}
\footnotesize
\item \hspace{-0.5cm} \textit{Note: Lengths of confidence intervals are shown in curly brackets.}
\end{tablenotes}\end{threeparttable}\end{center}
\end{table}
\begin{table}[H]
\caption{Coverage and mean of length of 95\% CI for each standard error ($m_{\mathrm{LL}}$, Setup 2, $x=0.8$, with bias)\label{tab:CI_LL2_0.8-bias}}

\centering
\begin{center} 
\begin{threeparttable}
{\small

\begin{tabular}{l c c c c c c}
\hline
 & \multicolumn{3}{c}{$\max n_g=20$} & \multicolumn{3}{c}{$\max n_g=100$} \\
\cline{2-4} \cline{5-7}
 & $CI$ & $CI_{\text{CR}}$ & $CI_{\lambda}$ & $CI$ & $CI_{\text{CR}}$ & $CI_{\lambda}$ \\
\hline
$(\rho_{X},\rho_{e})$=(0.2,0.2) & $0.776$   & $0.788$   & $0.827$   & $0.780$   & $0.787$   & $0.843$   \\
                                & $\{0.168\}$ & $\{0.171\}$ & $\{0.187\}$ & $\{0.166\}$ & $\{0.169\}$ & $\{0.187\}$ \\
$(\rho_{X},\rho_{e})$=(0.2,0.5) & $0.735$   & $0.745$   & $0.846$   & $0.737$   & $0.746$   & $0.856$   \\
                                & $\{0.168\}$ & $\{0.171\}$ & $\{0.209\}$ & $\{0.165\}$ & $\{0.168\}$ & $\{0.212\}$ \\
$(\rho_{X},\rho_{e})$=(0.5,0.2) & $0.753$   & $0.764$   & $0.819$   & $0.755$   & $0.761$   & $0.832$   \\
                                & $\{0.167\}$ & $\{0.171\}$ & $\{0.192\}$ & $\{0.165\}$ & $\{0.168\}$ & $\{0.193\}$ \\
$(\rho_{X},\rho_{e})$=(0.5,0.5) & $0.706$   & $0.721$   & $0.851$   & $0.715$   & $0.725$   & $0.855$   \\
                                & $\{0.167\}$ & $\{0.171\}$ & $\{0.219\}$ & $\{0.164\}$ & $\{0.168\}$ & $\{0.223\}$ \\
\hline
\end{tabular}
}
\begin{tablenotes}
\vspace{-0.5cm}
\footnotesize
\item \hspace{-0.5cm} \textit{Note: Lengths of confidence intervals are shown in curly brackets.}
\end{tablenotes}\end{threeparttable}\end{center}
\end{table}
\begin{table}[H]
\caption{Coverage and average length of 95\% CI for each standard error ($m_{\mathrm{nw}}$, Setup 2, $x=0.8$, with bias)\label{tab:CI_nw2_0.8-bias}}

\centering
\begin{center} 
\begin{threeparttable}
{\small

\begin{tabular}{l c c c c c c}
\hline
 & \multicolumn{3}{c}{$\max n_g=20$} & \multicolumn{3}{c}{$\max n_g=100$} \\
\cline{2-4} \cline{5-7}
 & $CI$ & $CI_{\text{CR}}$ & $CI_{\lambda}$ & $CI$ & $CI_{\text{CR}}$ & $CI_{\lambda}$ \\
\hline
$(\rho_{X},\rho_{e})$=(0.2,0.2) & $0.772$   & $0.783$   & $0.821$   & $0.782$   & $0.791$   & $0.840$   \\
                                & $\{0.167\}$ & $\{0.170\}$ & $\{0.187\}$ & $\{0.165\}$ & $\{0.168\}$ & $\{0.187\}$ \\
$(\rho_{X},\rho_{e})$=(0.2,0.5) & $0.737$   & $0.745$   & $0.842$   & $0.734$   & $0.743$   & $0.852$   \\
                                & $\{0.167\}$ & $\{0.171\}$ & $\{0.209\}$ & $\{0.164\}$ & $\{0.168\}$ & $\{0.211\}$ \\
$(\rho_{X},\rho_{e})$=(0.5,0.2) & $0.748$   & $0.756$   & $0.819$   & $0.752$   & $0.758$   & $0.829$   \\
                                & $\{0.166\}$ & $\{0.170\}$ & $\{0.191\}$ & $\{0.164\}$ & $\{0.167\}$ & $\{0.192\}$ \\
$(\rho_{X},\rho_{e})$=(0.5,0.5) & $0.701$   & $0.714$   & $0.846$   & $0.707$   & $0.718$   & $0.853$   \\
                                & $\{0.166\}$ & $\{0.170\}$ & $\{0.218\}$ & $\{0.164\}$ & $\{0.167\}$ & $\{0.223\}$ \\
\hline
\end{tabular}
}
\begin{tablenotes}
\vspace{-0.5cm}
\footnotesize
\item \hspace{-0.5cm} \textit{Note: Lengths of confidence intervals are shown in curly brackets.}
\end{tablenotes}\end{threeparttable}\end{center}
\end{table}
\begin{table}[H]
\caption{Coverage and mean of length of 95\% CI for each standard error ($m_{\mathrm{LL}}$, Setup 2, $x=0.4$, with bias)\label{tab:CI_LL2_0.4-bias}}

\centering
\begin{center} 
\begin{threeparttable}
{\small

\begin{tabular}{l c c c c c c}
\hline
 & \multicolumn{3}{c}{$\max n_g=20$} & \multicolumn{3}{c}{$\max n_g=100$} \\
\cline{2-4} \cline{5-7}
 & $CI$ & $CI_{\text{CR}}$ & $CI_{\lambda}$ & $CI$ & $CI_{\text{CR}}$ & $CI_{\lambda}$ \\
\hline
$(\rho_{X},\rho_{e})$=(0.2,0.2) & $0.954$   & $0.957$   & $0.980$   & $0.945$   & $0.949$   & $0.982$   \\
                                & $\{0.137\}$ & $\{0.138\}$ & $\{0.157\}$ & $\{0.134\}$ & $\{0.136\}$ & $\{0.158\}$ \\
$(\rho_{X},\rho_{e})$=(0.2,0.5) & $0.935$   & $0.939$   & $0.988$   & $0.922$   & $0.926$   & $0.985$   \\
                                & $\{0.136\}$ & $\{0.139\}$ & $\{0.182\}$ & $\{0.134\}$ & $\{0.136\}$ & $\{0.184\}$ \\
$(\rho_{X},\rho_{e})$=(0.5,0.2) & $0.945$   & $0.948$   & $0.981$   & $0.943$   & $0.945$   & $0.984$   \\
                                & $\{0.136\}$ & $\{0.138\}$ & $\{0.160\}$ & $\{0.133\}$ & $\{0.135\}$ & $\{0.161\}$ \\
$(\rho_{X},\rho_{e})$=(0.5,0.5) & $0.924$   & $0.929$   & $0.989$   & $0.909$   & $0.915$   & $0.987$   \\
                                & $\{0.136\}$ & $\{0.138\}$ & $\{0.188\}$ & $\{0.133\}$ & $\{0.135\}$ & $\{0.192\}$ \\
\hline
\end{tabular}
}
\begin{tablenotes}
\vspace{-0.5cm}
\footnotesize
\item \hspace{-0.5cm} \textit{Note: Lengths of confidence intervals are shown in curly brackets.}
\end{tablenotes}\end{threeparttable}\end{center}
\end{table}
\begin{table}[H]
\caption{Coverage and average length of 95\% CI for each standard error ($m_{\mathrm{nw}}$, Setup 2, $x=0.4$, with bias)\label{tab:CI_nw2_0.4-bias}}

\centering
\begin{center} 
\begin{threeparttable}
{\small

\begin{tabular}{l c c c c c c}
\hline
 & \multicolumn{3}{c}{$\max n_g=20$} & \multicolumn{3}{c}{$\max n_g=100$} \\
\cline{2-4} \cline{5-7}
 & $CI$ & $CI_{\text{CR}}$ & $CI_{\lambda}$ & $CI$ & $CI_{\text{CR}}$ & $CI_{\lambda}$ \\
\hline
$(\rho_{X},\rho_{e})$=(0.2,0.2) & $0.951$   & $0.953$   & $0.978$   & $0.941$   & $0.944$   & $0.978$   \\
                                & $\{0.136\}$ & $\{0.138\}$ & $\{0.157\}$ & $\{0.134\}$ & $\{0.135\}$ & $\{0.157\}$ \\
$(\rho_{X},\rho_{e})$=(0.2,0.5) & $0.928$   & $0.934$   & $0.982$   & $0.915$   & $0.918$   & $0.986$   \\
                                & $\{0.136\}$ & $\{0.138\}$ & $\{0.181\}$ & $\{0.133\}$ & $\{0.135\}$ & $\{0.184\}$ \\
$(\rho_{X},\rho_{e})$=(0.5,0.2) & $0.937$   & $0.942$   & $0.985$   & $0.935$   & $0.938$   & $0.980$   \\
                                & $\{0.135\}$ & $\{0.137\}$ & $\{0.160\}$ & $\{0.133\}$ & $\{0.134\}$ & $\{0.161\}$ \\
$(\rho_{X},\rho_{e})$=(0.5,0.5) & $0.919$   & $0.924$   & $0.988$   & $0.910$   & $0.915$   & $0.987$   \\
                                & $\{0.135\}$ & $\{0.137\}$ & $\{0.188\}$ & $\{0.132\}$ & $\{0.134\}$ & $\{0.192\}$ \\
\hline
\end{tabular}
}
\begin{tablenotes}
\vspace{-0.5cm}
\footnotesize
\item \hspace{-0.5cm} \textit{Note: Lengths of confidence intervals are shown in curly brackets.}
\end{tablenotes}\end{threeparttable}\end{center}
\end{table}

\bibliographystyle{apecon}
\phantomsection\addcontentsline{toc}{section}{\refname}\bibliography{listb}

\end{document}